\documentclass{jfm}
\usepackage{subcaption}
\usepackage{graphicx}
\usepackage{amsmath}
\usepackage{epstopdf, epsfig}
\usepackage{soul}
\usepackage{mathtools}
\DeclarePairedDelimiter\absfrac{\lvert}{\rvert}

\usepackage{color}
\newcommand{\sid}[1]{\textcolor{black}{{#1}}}

\newcommand{\lvec}{\mathbf{L}}
\newcommand{\lsvec}{\mathbf{L^s}}
\newcommand{\gs}{\mathbf{G^s}}
\newcommand{\g}{\mathbf{G}}

\newcommand{\wrms}{\left\langle\omega^2\right\rangle^{1/2}}
\newcommand{\rij}{\mathcal{R}_{ij}(\mathbf{x},\mathbf{r}_{\alpha\beta})}
\newcommand{\hp}{\mathbf{H^p}}
\newcommand{\hpp}{{H^p}^\prime}

\newcommand{\h}{\mathbf{H}}
\newcommand{\ang}[1]{\ensuremath{\left\langle {#1} \right\rangle}}

\newcommand{\kmax}{k_\mathrm{max}}

\newcommand{\utildebf}{\mathbf{\widetilde{u}}}
\newcommand{\utilde}{\widetilde{u}}
\newcommand*\diff{\mathop{}\!\mathrm{d}}

\shorttitle{Identifying and disentangling flow structures in turbulence}
\shortauthor{Mukherjee S., Mascini, M. \& Portela, L.M.}

\title{Correlation and decomposition framework for identifying and disentangling flow structures: canonical examples and application to isotropic turbulence}

\author{Siddhartha Mukherjee\aff{1},
  Merlijn Mascini\aff{1}
 \and Luis M. Portela\aff{1}\corresp{\email{L.Portela@tudelft.nl}}}

\affiliation{\aff{1}Section of Transport Phenomena, Department of Chemical Engineering, Delft University of Technology, 2629 HZ, Delft, Netherlands}

\begin{document}

\maketitle

\begin{abstract}
Turbulence has long been held synonymous to structure. The description of its phenomenology often invokes the concept of spatial ``coherent structures'' arising in its vector fields. Despite advances in structure eduction techniques, the organization of turbulence fields has resisted clear description\textemdash primarily, due to a lack of tools to identify instantaneous spatial organization, aggravated by an obfuscating scale superposition. We present a generalized correlation framework, and introduce correlation measures that identify structures as instantaneous patterns in vector fields; coupled with a paradigm, using Helmholtz decomposition concepts, to disentangle these structures. After testing the correlations using simple canonical flows, we apply them to realizations from direct numerical simulations of homogeneous isotropic turbulence. We find that regions of high kinetic energy manifest as localized velocity jets, which intersperse the velocity field, contrary to the prevalent view of high kinetic energy regions as large swirling structures (eddies). We confirm that regions of high enstrophy form small vorticity jets, invariably associated with a surrounding region of swirling velocity. Correlation field statistics viz-a-vis turbulence fields shows that the high kinetic energy jets and high enstrophy swirls are mostly spatially exclusive. Decomposing the velocity field, using the Biot-Savart law, into contributions from different levels and regions of the vorticity field, reveals the organization of these structures. High kinetic energy jets are neither self-inducing (due to low vorticity contents), nor significantly induced by strong vorticity; they are almost entirely induced by, non-local, intermediate range vorticity ($\sim \omega^\prime$, i.e. the rms vorticity), which permeates the volume. High enstrophy swirls, on the other hand, are a superposition of self-induced swirling motion along with a background-induced flow. Intermediate vorticity, moreover, has the highest contribution to the induction of the velocity field everywhere. This suggests that turbulence organization could emerge from non-local and non-linear field interactions, dominated by permeating intermediate vorticity, leading to an alternative description of turbulence, contrary to the notion of a strict hierarchy of coherent structures. The tools presented in this paper can be readily applied to study generic vector and scalar fields associated with diverse phenomena.
\end{abstract}

\section{Introduction}
``Structure'' in a field can be defined as a certain distribution of the properties of the field in a region, characterized by a small number of parameters, which can be described (deterministically) in a ``simple way''. For instance, in a velocity field, swirling motion can be considered as a kind of structure, which brings to mind examples such as a tornado, cyclone or a simple bathtub vortex. The concept of structure in flow fields immediately also invokes the notion of ``coherent motion'', one interpretation of which is: regions of the flow that have a certain spatial pattern (for instance a swirling motion). This idea of structure can also be understood by considering its opposite, i.e. a structure-less field, which mathematically may be defined as random.

The structure in a general field, and in particular in a velocity field, can be the result of (arbitrary) choices in constructing the field and of the (intrinsic) dynamics of the field. For example, the addition of a translation or a rotation generate a ``coherent motion'' that is not related with the intrinsic dynamics of the velocity field. The pattern of the field at infinity can be seen as the result of these arbitrary choices, and it can be ``removed'', in order to obtain patterns associated with the (intrinsic) dynamics of the field. In classical Newtonian mechanics, this is equivalent to observing motion with respect to the ``distant stars''. \sid{The use of correlation and Helmholtz decomposition concepts allows the generalization of these ideas, which are essential for deciphering instrinsic field structures. For instance, the spatial correlation of a field over a sphere with an infinite radius can be made zero by performing an opposite transformation in the field (eg. a translation or a rotation). From a Helmholtz decomposition perspective, this is equivalent to making a transformation in the field such that the generalized contribution of any region of the infinity-field (far-field contribution from ``large distances'') becomes equal to zero. We apply these techniques to study turbulence, to investigate both the structures that arise in turbulent flows, and the composition of these structures from a Helmholtz decomposition perspective.}

Turbulent flows have been found to be very rich in structure across different representational spaces, so much so that turbulence has been held synonymous to structure \citep{tsinober2014essence}. Moreover, turbulent flow fields are intriguing due to the superposition of structure and randomness across scales; uncovering and characterizing which has garnered profound interest over the past decades. In describing velocity field structures in turbulence, a key idea often used, albeit ill-defined, is that of the ``eddy'', which also refers to coherent regions of swirling motion. The superposition of eddies (or coherent motion across all scales) has served as the conceptual background upon which most of turbulence theory has been built \citep{frisch1995turbulence,dubrulle2019beyond}. How these coherent structures arise across all scales, and what they look like, however, is not fully known. In this paper, we are interested in finding out whether the finite-sized spatial structures comprising turbulent flow fields can be identified and isolated in the vector fields where they \sid{are believed to} arise. Further, we are interested in considering instantaneous structures which are continuously produced and destroyed, and are not the result of an \sid{averaging procedure}. \sid{According to the conventional ``cascade'' perspective, these structures may range from the largest scales that contain most of the kinetic energy and ``drive'' the dynamics, to the small scales associated with the dissipation of kinetic energy.} In this framework, it should be noted that the smallest scales are merely a \textit{consequence} of the turbulence dynamics, and are hence not dynamically significant in determining the overall flow \citep{tsinober2014essence}.

There have been various approaches aimed at identifying coherent structures in different contexts that are prevalent in the turbulence literature. Most widely used are techniques based upon the velocity gradient tensor $A_{ij}=\partial u_i/\partial x_j$ and its symmetric ($S_{ij}$) and skew-symemtric ($\Omega_{ij}$) parts. For instance, \cite{jeong1995identification} define a criterion (called $\lambda_2$) based on the eigenvalues of the local pressure Hessian, which is related to $S_{ij}$ and $\Omega_{ij}$. \cite{dubief2000coherent} used the second and third invariants ($Q$ and $R$) of $A_{ij}$, originally used to characterize the topology of point flow patterns \citep{chong1990general}, and \cite{haller2005objective} used the strain acceleration tensor along fluid trajectories. \citet{farge2001coherent} used a wavelet decomposition to identify coherent and incoherent vorticity structures, \cite{hussain1986coherent} and \cite{sirovich1987turbulence} studied statistically emerging lower dimensional attractors, while others have extensively studied Lagrangian structures crucial for material transport \citep{peacock2010introduction,peacock2013lagrangian,haller2015lagrangian}. Non-linear equilibrium solutions have also been classified as exact coherent structures \citep{waleffe1997self,waleffe2001exact,deguchi2014canonical}. \cite{lozano2014time} studied spatio-temporally coherent vortical structures, while \cite{she1990intermittent} and \cite{jimenez1993structure} investigated the structure of strong vorticity (worms) in homogeneous isotropic turbulence, and \cite{moisy2004geometry} \sid{quantified the large-scale spatial distribution of small, localized, intense vorticity worms}.

\sid{These (and many other) studies and techniques have greatly informed our understanding of coherent structures in turbulence. Several of these studies use a ``functional decomposition'' approach (eg. wavelet and spectral decomposition) to study the coherent structures, their relations and ``hierarchy''; eg. \cite{argoul1989wavelet} and \cite{alexakis2018cascades} address the ``cascade'' concept \citep{richardson1922weather} using wavelet and spectral decompositions, respectively. However, when using a ``functional decomposition'' approach the turbulence fields are separated into ``classes'' and this ``class perspective'' does not represent individual coherent structures, and their relations, occuring in the actual physical space.}

\sid{Many basic concepts associated with coherent structures, like the existence of a hierarchy of coherent structures (as invoked, for instance, in the \cite{richardson1922weather} ``cascade''), or the energetic interaction of eddies \citep{waleffe1992nature} and eddy breakups, have remained intractable in the physical space, where these ideas were first envisioned. Part of this disconnect is due to the lack of tools designed to identify instantaneous spatial structures, which may be driving these processes. The other issue is extracting these structures from their obfuscating scale superposition, in order to study their form and dynamics. To address these issues we use correlation concepts and Helmholtz decomposition concepts, which enable us to identify and extract individual flow structures from turbulence fields. In this study, we deal with incompressible, homogeneous isotropic turbulence, with a zero mean velocity, hence the removal of a velocity pattern associated with an ``artificial frame of reference'' is not an issue. We approach the concept of coherent structures with a focus on the following key aspects:}

\begin{enumerate}
\item \textbf{Finite structure size} - \sid{We consider a ``coherent structure'' to be a finite, spatial structure, which represents a unit of coherent motion (eg. an ``eddy'' is a ``coherent structure'' in which the coherent motion is a swirling veloity field). It must, hence, have a spatial form, that is to say, it cannot be completely irregular.} ``Coherence'', in this context, becomes almost synonymous to ``correlation'', as an ordered spatial structure must comprise of a neighbourhood of vectors that are strongly correlated (either positively or negatively). Here, it becomes important to highlight the distinction from \textit{point-criteria} used for educing structures, which are based on the velocity gradient tensor (or derivatives thereof, like $Q$, $R$, etc.). These techniques describe point-structures, reasoning from the Taylor expansion perspective of the velocity field in the infinitesimal neighbourhood of each point in the flow field. Structures in the flow field, however, are \textit{finite regions} of spatio-temporal order, and may not necessarily be related to local velocity gradients, \sid{as the velocity at a point results from spatial integrals of the velocity derivatives. We study coherent structures in velocity and vorticity spaces by developing correlation measures designed to seek out particular spatial order in these structures.}

\item \textbf{Instantaneity} - \sid{The spatial structures described above} exist instantaneously, and are not consequences of averaging procedures. In fact, a structure will have an entire lifecycle, from generation until destruction (driven by the dynamics of the Navier-Stokes equations). While ignoring the temporal evolution of the structures, in this work we limit ourselves to identifying structures in instantaneous realizations (i.e. snapshots) of turbulence fields. Hence, we consider only the geometry of structures and not their kinematics or dynamics.

\item \textbf{Disentangling structures} - \sid{Part of the complexity of turbulence fields comes from the superposition of structures, which makes it difficult to even define, let alone extract and study an individual structure. The reason for this superposition of (velocity field) structures can be understood from the Biot-Savart law, which gives that the velocity at each point in space is generated by the \textit{spatial} integrals of quantities associated with the gradients of the velocity. The integrals account for both \textit{near-field} and \textit{far-field} contributions. The summative nature of the Biot-Savart law, hence, provides a paradigm for \textit{disentangling} the contributions that generate any given velocity field pattern or structure, as different contributions may be isolated by employing suitable conditional sampling criteria on the velocity reconstruction.} We use this method to identify regions of the velocity gradient field which `generate'\textemdash in a Biot-Savart sense\textemdash a particular velocity structure.
\end{enumerate}

The tools developed in this study, namely a set of generalized correlation measures, along with velocity reconstruction using the Biot-Savart law, allow us to look at turbulence fields from a different perspective; for instance they enable us to identify the structure of high kinetic energy and high enstrophy regions. Reconstructing the velocity field using the Biot-Savart law, further, reveals the distribution of the vorticity contributions in the generation of these structures, along with the generation of the total velocity field. \sid{This paves the way for studying turbulence as a dynamical system of interacting structures that arise in its physical fields, the interplay between which manifests as the dynamics. Moreover, our results give novel insights into turbulence, in particular, regarding the emergence of flow organization.}

The layout of the paper is as follows. We begin by proposing different instantaneous correlation measures in section \ref{sec:CorrelationDefinitions}, which are designed to identify simple vector-field structures, based upon a generalization of the correlation tensor, along with correlations associated with the Biot-Savart law in section \ref{sec:BiotSavart}. These correlations are first applied to canonical flows in section \ref{sec:CanonicalFlows}, where some of their features are highlighted. In section \ref{sec:Turbulence}, the correlations are applied to incompressible, homogeneous isotropic turbulence flow fields, where the particular flow structures associated with high kinetic energy and high enstrophy regions are identified. \sid{In section \ref{sec:FlowStructure} we first show instances of individual flow structures. These results, obtained using an in-house code are shown to be essentially similar to those obtained upon using a reference dataset in Appendix \ref{app:JHTD-Validation}. Further in section \ref{sec:FlowStructure}, we perform the velocity field reconstruction using the Biot-Savart law, and both qualitatively show and quantify, the vorticity composition of velocity field structures, following which we end with the conclusions of this study where we describe the picture of emergence of structures in turbulence.}

\section{Generalized correlation}\label{sec:CorrelationDefinitions}
\begin{figure}
  \centerline{\includegraphics[width=0.65\linewidth]{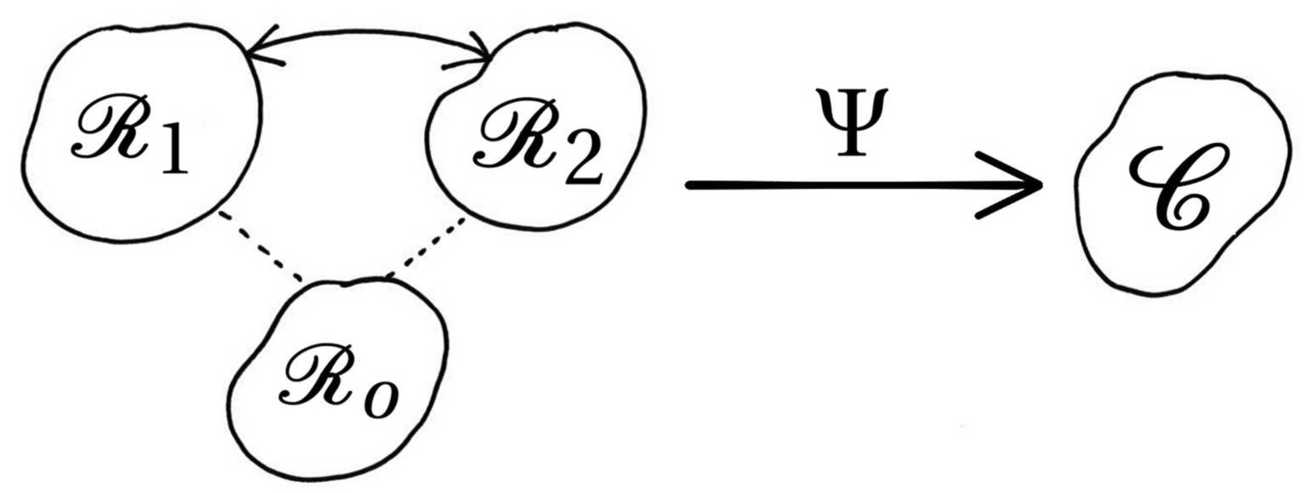}}

  \caption{\sid{Correlation between phase-space regions $\mathcal{R}_1$ and $\mathcal{R}_2$, as observed from phase-space region $\mathcal{R}_0$; each region is defined over a space-set $\mathcal{S}$ and a time-set $\mathcal{T}$. Using a function $\psi$, these regions are mapped onto a correlation space, producing a region shown here as $\mathcal{C}$.}}
\label{fig:correlationGeneralSchematic}
\end{figure}

\sid{Correlation, in its most general form, can be interpreted as the relation between one region of a phase-space (or a field) with another; the two regions and their relation being defined based upon certain rules, when viewed from another phase-space region (the region of observation). This can be expressed as the relation between $\mathcal{R}_1(\mathcal{S}_1,\mathcal{T}_1)$ and $\mathcal{R}_2(\mathcal{S}_2,\mathcal{T}_2)$ as viewed from $\mathcal{R}_0(\mathcal{S}_0,\mathcal{T}_0)$, as illustrated in figure \ref{fig:correlationGeneralSchematic}. Here $\mathcal{S}$ denotes a space-set (eg. a bounded continuous region or a set of points), $\mathcal{T}$ denotes a time-set (eg. a continuous time interval or a set of time instances), and $\mathcal{R}$ denotes a phase-space-set defined over $(\mathcal{S},\mathcal{T})$ (which could be defined, for example, using the velocity field or the pressure field). Based upon a set of rules given by any function $\psi$, defined over $\mathcal{S}_1, \mathcal{S}_2, \mathcal{S}_o, \mathcal{T}_1, \mathcal{T}_2, \mathcal{T}_o$, the original phase-space-sets can be mapped to a correlation-set $\mathcal{C}$, with appropriate dimensions, based upon the definition of $\psi$.}

\sid{The usual definition of the two-point correlation tensor for turbulent flows can be seen as a particular case of this generalized definition, where the definition additionally also involves \textit{statistical} (averaging) concepts, with the phase-space-sets being composed of an ensemble of different realizations; making it a statistical measure. We first frame the usual definition of the two-point correlation tensor in the context of this generalized definition. Then, via analogy, we will define a \textit{deterministic} two-point correlation, in order to characterize the structure of individual fields. Note that since the usual definition of the two-point correlation tensor is a statistical-concept, applied to an ensemble of different field realizations, it is not necessarily a good representation of the structure of each \textit{individual} field. On the other hand, the deterministic two-point correlation that we will define contains a (simplified) characterization of the structure of each individual field.}

\sid{The usual two-point correlation tensor for turbulent flows is defined as}
\begin{equation}
\mathcal{R}_{ij}(\mathbf{x},\mathbf{r},t,\Delta t) = \left\langle u_i(\mathbf{x},t)u_j(\mathbf{x}+\mathbf{r},t+\Delta t) \right\rangle
\label{eq:twoPointCorrelation}
\end{equation}

\sid{where $\mathbf{x}$ denotes a position, $\mathbf{r}$ the separation between two positions, $t$ a time, $\Delta t$ the difference between two times, $\mathbf{u} = (u_x, u_y, u_z)$ the velocity at a given position and time and $\ang{\cdot}$ ensemble averaging. In the context of the generalized correlation, this definition can be framed as:}
\begin{itemize}
\item \sid{$\mathcal{S}_1$: spatial region of interest, containing $\mathbf{x}$}
\item \sid{$\mathcal{S}_2$: spatial region of interest, containing $\mathbf{x}+\mathbf{r}$}
\item \sid{$\mathcal{T}_1$: time interval of interest, containing $t$}
\item \sid{$\mathcal{T}_2$: time interval of interest, containing $t+\Delta t$}
\item \sid{$\mathcal{R}_1$: phase-space-set, composed of the ensemble of velocity fields defined over $\mathcal{S}_1$ and $\mathcal{T}_1$, i.e. $\mathcal{R}_1(\mathcal{S}_1,\mathcal{T}_1) = \cup_n \mathbf{u}_n(\mathcal{S}_1,\mathcal{T}_1)$}
\item \sid{$\mathcal{R}_2$: phase-space-set, composed of the ensemble of velocity fields defined over $\mathcal{S}_2$ and $\mathcal{T}_2$, i.e. $\mathcal{R}_2(\mathcal{S}_2,\mathcal{T}_2) = \cup_n \mathbf{u}_n(\mathcal{S}_2,\mathcal{T}_2)$}
\item \sid{$\mathcal{S}_0 = \mathcal{S}_1$, $\mathcal{T}_0 = \mathcal{T}_1$, $\mathcal{R}_0 = \mathcal{R}_1$}
\item \sid{$\psi$, a function composed of a deterministic part, defined over the individual velocity fields of the ensemble, $\mathbf{u}_n = (u_{x,n}, u_{y,n}, u_{z,n})$, and a statistical part:}
\begin{itemize}
\item \sid{Deterministic part: $\psi_n (\mathbf{x},\mathbf{r},t,t+\Delta t) = u_{i,n}(\mathbf{x},t)u_{j,n}(\mathbf{x}+\mathbf{r},t+\Delta t)$}
\item \sid{Statistical part: $\overline{\psi}(\mathbf{x},\mathbf{r},t,t+\Delta t) = \ang{\psi_n(\mathbf{x},\mathbf{r},t,t+\Delta t)}$\\
where $\ang{\cdot}$ denotes the ensemble averaging operator, which is usually a linear operator (eg. arithmetic averaging); however, different (non-linear) operations could also be used, leading to definitions different from the usual two-point two-time correlation tensor.}
\end{itemize}
\end{itemize}

\sid{Therefore, in the context of the generalized definition, it results that the usual two-point two-time correlation can be defined as $\mathcal{C}(\mathcal{R}_1,\mathcal{R}_2)$, whose elements are given by the correlation tensor}
\begin{equation}
\mathcal{R}_{ij}(\mathbf{x},\mathbf{r},t,\Delta t) = \overline{\psi}(\mathbf{x},\mathbf{r},t,\Delta t)
\end{equation}
\sid{By considering $\Delta t = 0$, i.e. $\mathcal{T}_1 = \mathcal{T}_2$, the two-point two-time correlation tensor is reduced to the usual two-point correlation tensor:}
\begin{equation}
\mathcal{R}_{ij}(\mathbf{x},\mathbf{r},t) = \ang{u_i(\mathbf{x},t)u_j(\mathbf{x}+\mathbf{r},t)}
\end{equation}

\sid{This is a tensor with 9 components, which for a generic turbulent flow is a function of $\mathbf{x}$, $\mathbf{r}$ and $t$. It contains a lot of information (from a statistical perspective), however, for some turbulent flows, the information needed for its characterization can be significantly reduced. In particular, for homogeneous isotropic turbulence, the dependence on $\mathbf{x}$ disappears, and the dependence on $\mathbf{r}$ reduces to a function of the radial distance $r=|\mathbf{r}|$ alone, regardless of the orientation of the vector $\mathbf{r}$. As the choice of the orientation of $\mathbf{r}$ is arbitrary and spans all directions of space, the velocity components $u_i$ can be considered along three orthogonal directions, which can be given as $\mathbf{e}_1 \parallel \mathbf{r}$, $\mathbf{e}_2 \bot \mathbf{r}$ and $\mathbf{e}_3 \bot \left(\mathbf{r}, \mathbf{e}_2 \right)$. This yield three correlation functions $f(r,t)$ (longitudinal), $g_1(r,t)$ and $g_2(r,t)$ (transverse), respectively. For homogeneous isotropic turbulence, these three correlation functions completely characterize the usual two-point correlation tensor; if, additionally, the turbulence is also statistically-steady, the three correlations depend only on $r$. A simplified characterization of these correlation functions can be given by using an integral measure of them, which can be obtained by integrating over $r$, to get the integral lengths.}

\sid{The usual two-point correlation is a statistical concept, which mixes measures of the structure of individual fields with a measure of the structure of their ensemble. In general, the structure of the ensemble does not represent the structure of individual fields; actually, it can be completely different. For example, homogeneous isotropic turbulence refers to the ensemble; the individual fields are often far from being homogeneous and isotropic. The characterization of the structure using the usual two-point correlation, even of individual field realizations, only holds \textit{statistically} (i.e. upon suitable spatial averaging). We propose a deterministic characterization of the individual (instantaneous) fields using a correlation definition  similar to the one employed for the usual two-point correlation, but considering only the \textit{deterministic} part of $\psi$. This will be supplemented by a simplified characterization of the correlation, using integral measures, which, since the individual fields are not homogeneous and isotropic, are different and more elaborate that the usual integral measures for homogeneous isotropic turbulence. This simplified characterization, even though incomplete, is more manageable.}

\sid{We propose $\mathcal{C}(\mathcal{R}_1,\mathcal{R}_2)$ as a definition for the correlation of individual (instantaneous) vector fields, which, for the sake of concreteness, is illustrated here for the velocity field. Here again $\mathcal{S}_1$ is the spatial region of interest containing $\mathbf{x}$, $\mathcal{S}_2$ is the spatial region of interest containing $\mathbf{x}+\mathbf{r}$, $\mathcal{T}_1 = \mathcal{T}_2$ is the time interval of interest containing $t$. The elements of the phase-space-sets are the velocity fields and the phase-space-set of observation, $\mathcal{R}_0$, is equivalent to $\mathcal{R}_1$. The elements of $\mathcal{C}(\mathcal{R}_1,\mathcal{R}_2)$ are the correlation tensor fields}

\begin{align}
\mathcal{R}_{ij}\left( \mathbf{x},\mathbf{r},t \right) &= \psi\left( \mathbf{x},\mathbf{r},t \right) \\
\psi\left( \mathbf{x},\mathbf{r},t \right) &= u_i\left( \mathbf{x},t \right)u_j\left( \mathbf{x}+\mathbf{r},t \right)
\end{align}

\sid{The separation vector $\mathbf{r}$ can be represented by a scalar, $r=|\mathbf{r}|$ denoting the separation distance, and a direction. In 3D space, this direction can be specified by two angles (and with only one angle in 2D) i.e. the azimuthal angle $\alpha$ and the elevation angle $\beta$. We can define the separation vector $\mathbf{r}_{\alpha\beta}$ as a vector of length $r$ which points along the direction specified by $\alpha$ and $\beta$, while being placed at point $\mathbf{x}$. Hence, the correlation tensor can be written as $\mathcal{R}_{ij}(\mathbf{x},r_{\alpha\beta},t)$, which is a function of seven variables, namely $x,y,z,r,\alpha,\beta,t$, where $x,y,z$ are the Cartesian components of $\mathbf{x}$. Since in this work we limit ourselves to identifying structures in instantaneous field realizations, for the remainder of this paper we will omit the time dependence $t$. The correlation tensor can then be expressed in matrix form as}

\begin{equation}
\setlength{\arraycolsep}{5pt}
\renewcommand{\arraystretch}{1.3}
\mathcal{R}_{ij}(\mathbf{x},\mathbf{r}_{\alpha\beta}) = \left[
\begin{array}{ccc}
  u_x(\mathbf{x})u_x(\mathbf{x}+\mathbf{r}_{\alpha\beta}) & u_x(\mathbf{x})u_y(\mathbf{x}+\mathbf{r}_{\alpha\beta}) & u_x(\mathbf{x})u_z(\mathbf{x}+\mathbf{r}_{\alpha\beta})  \\
  \displaystyle
u_y(\mathbf{x})u_x(\mathbf{x}+\mathbf{r}_{\alpha\beta}) & u_y(\mathbf{x})u_y(\mathbf{x}+\mathbf{r}_{\alpha\beta}) & u_y(\mathbf{x})u_z(\mathbf{x}+\mathbf{r}_{\alpha\beta})  \\
  \displaystyle
u_z(\mathbf{x})u_x(\mathbf{x}+\mathbf{r}_{\alpha\beta}) & u_z(\mathbf{x})u_y(\mathbf{x}+\mathbf{r}_{\alpha\beta}) & u_z(\mathbf{x})u_z(\mathbf{x}+\mathbf{r}_{\alpha\beta})  \\
\end{array}  \right]
\label{eq:corrMatrix}
\end{equation}

This contains a lot of information, which needs to be reduced for practical reasons. So, instead of considering the entire $\mathcal{R}$ matrix, we work with one of its invariants, the trace, which is written as
\begin{equation}
\mathrm{tr}\left(\mathcal{R}_{ij}(\mathbf{x},\mathbf{r}_{\alpha\beta})\right) = u_x(\mathbf{x})u_x(\mathbf{x}+\mathbf{r}_{\alpha\beta}) + u_y(\mathbf{x})u_y(\mathbf{x}+\mathbf{r}_{\alpha\beta}) + u_z(\mathbf{x})u_z(\mathbf{x}+\mathbf{r}_{\alpha\beta}) 
\end{equation}

This is \sid{easier to conceptualize, as the trace is also} the dot product between the velocities at points $\mathbf{x}$ and $\mathbf{x} + \mathbf{r}_{\alpha\beta}$
\begin{equation}
\mathrm{tr}\left(\mathcal{R}_{ij}(\mathbf{x},\mathbf{r}_{\alpha\beta})\right) = \mathbf{u}(\mathbf{x})\cdot \mathbf{u}(\mathbf{x}+\mathbf{r}_{\alpha\beta})
\end{equation}

A further reduction can be performed by integrating this quantity along directions specified by $\mathbf{r}_{\alpha\beta}$, to associate an integral measure $L_{\alpha\beta}(\mathbf{x},\Lambda)$ along each direction as
\begin{equation}
L_{\alpha\beta}(\mathbf{x},\Lambda) = \int_0^\Lambda\mathbf{u}({\mathbf{x}})\cdot \mathbf{u}({\mathbf{x}+\mathbf{r}_{\alpha\beta})} \mathrm{d}r_{\alpha\beta}
\label{eq:Lmanifold}
\end{equation}

\begin{figure}
  \centerline{\includegraphics[width=0.5\linewidth]{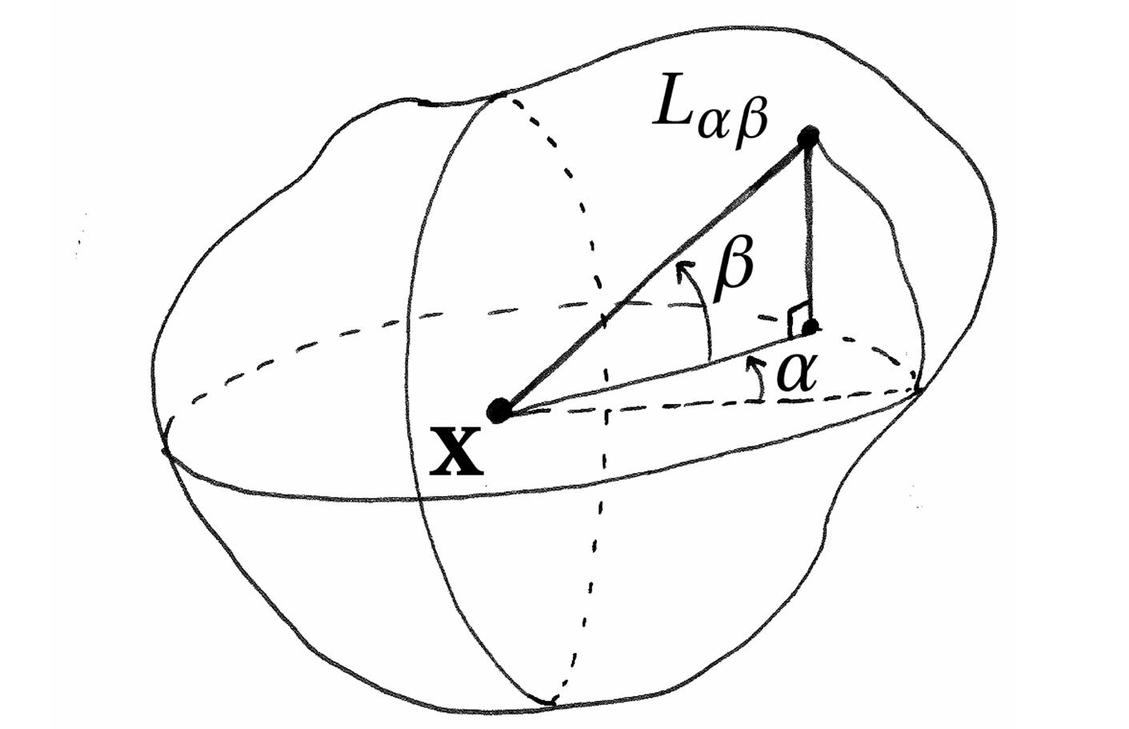}}

  \caption{A correlation surface around point $\mathbf{x}$, defined by the measure $L_{\alpha\beta}$ found by integrating $\mathrm{tr}(\rij)$ upto distance $\Lambda$, along a direction specified by the angles $\alpha$ (azimuth) and $\beta$ (elevation).}
\label{fig:correlationSurfaceSchematic}
\end{figure}

\sid{The correlation tensor field is hence reduced to a two-dimensional manifold around each point, as illustrated in figure \ref{fig:correlationSurfaceSchematic}. This 2D manifold gives a simplified characterization of the structure of the field around $\mathbf{x}$. For a given $\Lambda$, its shape gives an integral measure of how this structure depends on the direction. For instance, the direction of maximum $L_{\alpha\beta}$ could be determined as a function of position; however, even though interesting in itself, this will not be explored here. Further, this manifold is invariant under translation and rotation of the original coordinate axes, along with being invariant under reflection (similarly to an axial vector).}

\sid{In principle, such a manifold can be calculated for any vector field, leading, for a given $\Lambda$, to a correlation surface for each point in space. This still contains a lot of information, which, in general, can pose difficulties to represent in a compact form and interpret. Also, since we shall utilize numerical datasets, the calculation of the manifold requires binning the angles $\alpha$ and $\beta$ into discrete increments. The resolution of these angles will depend significantly upon the resolution of the data, where high resolution simulations will be required to acurately describe even a small subset of angles, along with demanding computational requirement to calculate the correlation manifold at each point in space.}

\sid{Therefore, we perform a final simplification, where instead of the entire manifold $L_{\alpha\beta}(\mathbf{x},\Lambda)$, we represent it by a three-tuple $\mathbf{L}(\mathbf{x},\Lambda)$, along three arbitrary orthogonal directions $x,y,z$ (forming an orthonormal base), which can be summarized as}

\begin{equation}
\setlength{\arraycolsep}{5pt}
\renewcommand{\arraystretch}{1.3}
\mathbf{L}(\mathbf{x},\Lambda) = \left[
\begin{array}{c}
\displaystyle
L_x(\mathbf{x},\Lambda) \\
\displaystyle
L_y(\mathbf{x},\Lambda) \\
\displaystyle
L_z(\mathbf{x},\Lambda) \\
\end{array} \right], \quad \mathrm{where}\quad L_i(\mathbf{x},\Lambda) = \int_{-\Lambda}^{\Lambda} \mathbf{u}(\mathbf{x})\cdot \mathbf{u}(\mathbf{x}+\mathbf{r}_i) \mathrm{d}r_i
\label{eq:LVector}
\end{equation}
\sid{where $i \in \left\lbrace x,y,z\right\rbrace$ represents the three spatial directions. Note that this three-tuple, being a simplifcation of the manifold, is also invariant to rotation, translation and reflection. Hence, $\mathbf{L}$ at each point $\mathbf{x}$ depends only on the choice of the arbitrary $x,y,z$ directions used to ``sample'' the manifold.}

\sid{This correlation measure, $\mathbf{L}$, can be expected to yield large absolute values at points $\mathbf{x}$ that are surrounded by large regions in which (i) the local flow streamlines are well-aligned (such that the directions of $\mathbf{u}(\mathbf{x})$ and $\mathbf{u}(\mathbf{x}+\mathbf{r})$ are similar) and (ii) the magnitude of these vectors is high. The definition of $\mathbf{L}$ provides a combined measure of the organization and size of the structure, and of the magnitude of the field in that region. The definition does not include an implicit normalization, which, for instance, could be achieved by dividing $L_i(\mathbf{x},\Lambda)$ by the integral of the kinetic energy along the $i$ direction, within the limits $ -\Lambda < \mathbf{r}_i < \Lambda$. The current definition is expected to identify regions of the flow which contain both structural organization (in the manner of well-aligned streamlines), and a large field magnitude. Normalizing the correlation can allow identifying regions with structural organization alone, while disregarding the field magnitude. Note that, since the current definition also includes the size of the structure in the correlation measure, a (very) small region with a high field magnitude and a high structural organization will not have a large absolute value of $\lvec$. A separate consideration of the size of the structure could be achieved by analyzing the influence of $\Lambda$ on $\mathbf{L}$. The value of $\Lambda$ limits the size of the structures, and the limit $\Lambda \to 0$ leads to a point-criterion. For ``large'' values of $\Lambda$, structures of all sizes can contribute to $\lvec$, with the larger ones being associated with larger absolute values of $\lvec$, hence, the analysis of the variation of $\lvec$ with $\Lambda$ allows to separate the effect of the size from the structural organization and field magnitude effects. Different forms of the correlation measures can be defined, to educe different aspects of structural organization, including, or not, in different ways, the size of the structure and field magnitude effects. For the present study, we do not normalize the correlation measures. Also, we limit ourselves to a value of $\Lambda$ that is ``large enough'' to include all the ``relevant'' structure sizes. In section \ref{sec:ChoiceOfLambda} we perform a limited study on the influence of the choice of $\Lambda$ and show that for the situation considered in this study (homogeneous isotropic turbulence) the Taylor microscale is a good choice for $\Lambda$.}

\begin{figure}
	\begin{subfigure}{\linewidth}
	\centerline{\includegraphics[width=\linewidth]{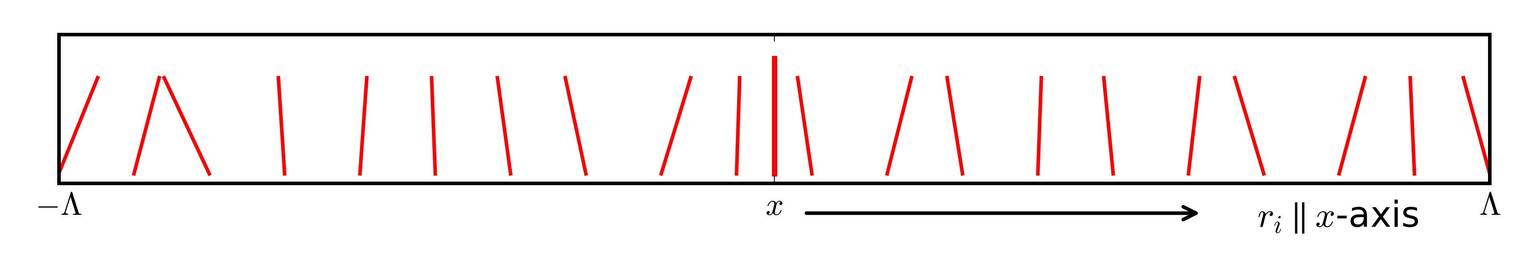}}

	\caption{In the $\mathbf{L}$ correlation measure, the velocity at point $\mathbf{x}$ is correlated to the velocities between points $-\Lambda < \mathbf{r}_i< \Lambda$.}
	\label{fig:1}
	\end{subfigure} 
	\begin{subfigure}{\linewidth}
  \centerline{\includegraphics[width=\linewidth]{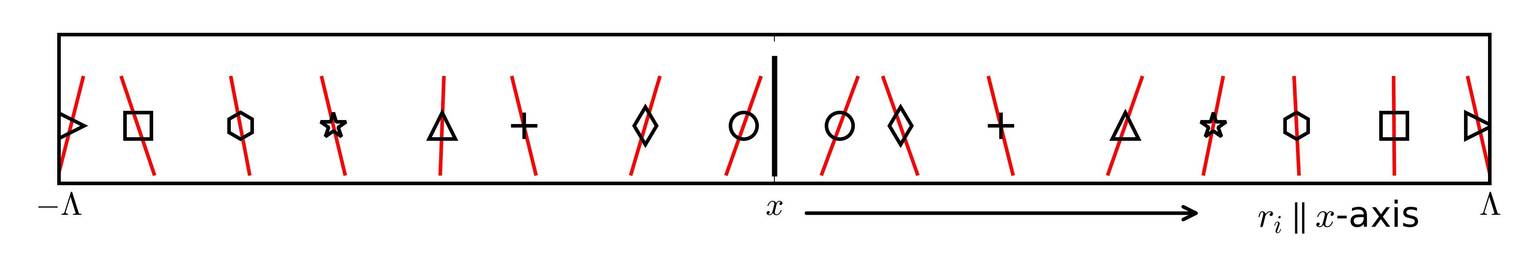}}

  	\caption{In the $\mathbf{L^s}$ correlation measure, velocity pairs equidistant from the point $\mathbf{x}$ are correlated, between $0 < |\mathbf{r}_i| < \Lambda$. These pairs are marked by the same symbol.}
	\label{fig:2}
  \end{subfigure}
  \caption{Schematic of the $\mathbf{L}$ and $\mathbf{L^s}$ correlation measures, shown along the $x$ direction.}
\label{fig:schematicLCorr}
\end{figure}

A different way of constructing the $\mathbf{L}$ correlation measure can be
\begin{equation}
\setlength{\arraycolsep}{5pt}
\renewcommand{\arraystretch}{1.3}
\mathbf{L^s}(\mathbf{x},\Lambda) = \left[
\begin{array}{c}
\displaystyle
L^s_x(\mathbf{x},\Lambda) \\
\displaystyle
L^s_y(\mathbf{x},\Lambda) \\
\displaystyle
L^s_z(\mathbf{x},\Lambda) \\
\end{array} \right], \quad \mathrm{where}\quad L^s_i(\mathbf{x},\Lambda) = \int_{0}^{\Lambda} \mathbf{u}(\mathbf{x}-\mathbf{r}_i)\cdot \mathbf{u}(\mathbf{x}+\mathbf{r}_i) \mathrm{d}r_i
\label{eq:LSymmVector}
\end{equation}
\sid{Here, the correlation measure along an axis is constructed using the dot product between velocity pairs equidistant from $\mathbf{x}$, symmetrically, along a given direction (hence the notation $\mathbf{L^s}$ for $\mathbf{L}$-symmetric). This measure is also expected to yield high values when the flow streamlines are parallel (or anti-parallel) in the $\Lambda-$neighbourhood of $\mathbf{x}$, and when the magnitude of the vectors is high. Moreover, this measure will be more sensitive to the larger symmetries and anti-symmetries in the field. These two correlation measures are illustrated in figure \ref{fig:schematicLCorr}.}

These measures can be applied to any vector field. We define \sid{correlation measures} $\mathbf{G}(\mathbf{x},\Lambda)$ and $\mathbf{G^s}(\mathbf{x},\Lambda)$ for the vorticity field, which by analogy are given as 
\begin{equation}
\setlength{\arraycolsep}{5pt}
\renewcommand{\arraystretch}{1.3}
\mathbf{G}(\mathbf{x},\Lambda), \quad \mathrm{where}\quad G_i(\mathbf{x},\Lambda) = \int_{-\Lambda}^{\Lambda} \boldsymbol{\omega}(\mathbf{x})\cdot \boldsymbol{\omega}(\mathbf{x}+\mathbf{r}_i) \mathrm{d}r_i
\label{eq:GVector}
\end{equation}
and
\begin{equation}
\setlength{\arraycolsep}{5pt}
\renewcommand{\arraystretch}{1.3}
\mathbf{G^s}(\mathbf{x},\Lambda), \quad \mathrm{where}\quad G^s_i(\mathbf{x},\Lambda) = \int_{0}^{\Lambda} \boldsymbol{\omega}(\mathbf{x}-\mathbf{r}_i)\cdot \boldsymbol{\omega}(\mathbf{x}+\mathbf{r}_i) \mathrm{d}r_i
\label{eq:GSVector}
\end{equation}

\sid{The correlation measure $\g$ is expected to yield large absolute values at points $\mathbf{x}$ that are surrounded by large regions in which (i) the \textit{vorticity streamlines} are well-aligned and (ii) the magnitude of the vorticity is high. The correlation measure $\gs$ is the vorticity field equivalent of $\lsvec$, and is expected to be more sensitive to the symmetries and anti-symmetries in the vorticity field, along with being sensitive to the vorticity magnitude.}

\sid{Note that, in general, structures in the velocity and vorticity fields can be very different, with very different sizes and magnitudes, hence, the concept of ``large regions'' and ``large values'' are relative and need to be interpreted in the individual context of the different correlation measures. In homogeneous isotropic turbulence, regions of high vorticity magnitude are related to the smaller scales of turbulence; they correspond to the long tails of the statistical distribution of the vorticity (i.e. $\omega \gg \omega^\prime$ where $\omega^\prime = \wrms$) and occur intermittently. The correlation measures $\lvec$ (and $\lsvec$) and $\g$ (and $\gs$) give a sepearate (simplified) characterization of the structure of the individual (instantaneous) velocity and vorticity fields. For simplicity of language, here onward in the paper we will refer to them, and other correlation measures, simply, as correlations.}

The correlations defined so far consider the velocity and vorticity fields separately, however, other correlations can be defined, which use both these fields, exploiting the relation between the velocity and vorticity. The vorticity $\boldsymbol{\omega}$ is defined as $\boldsymbol{\omega} = \nabla \times \mathbf{u}$. The velocity field, in turn, can be reconstructed from the vorticity field using the Biot-Savart law. This serves as an important tool to identify, as well as disentangle, structures, and is briefly described below.

\section{Biot-Savart reconstruction and associated correlations}\label{sec:BiotSavart}
\subsection{Biot-Savart reconstruction}
\sid{We start with the Helmholtz decomposition, which states that a sufficiently smooth (twice continuously differentiable) vector field, defined on a bounded or an unbounded domain, can be uniquely decomposed into three components : (i) an irrotational vector field $\mathbf{D}$, (ii) a solenoidal vector field $\mathbf{C}$ and (iii) a harmonic vector field $\mathbf{B}$. Applied to the velocity field $\mathbf{u}$, this can be written as}
\begin{equation}
\mathbf{u} = \mathbf{D} + \mathbf{C} + \mathbf{B}
\label{eq:Helmholtz}
\end{equation}
where
\begin{subequations}
\begin{align}
\mathbf{D} &= \nabla \phi \\
\mathbf{C} &= \nabla \times \boldsymbol{\varphi} \\
\mathbf{B} &= \nabla \psi
\end{align}
\end{subequations}
and
\begin{subequations}
\begin{align}
\nabla^2 \phi &= \nabla \cdot \mathbf{u} \\
\nabla^2 \boldsymbol{\varphi} &= - \nabla \times \mathbf{u} = -\boldsymbol{\omega} \\
\nabla^2 \psi &= 0
\end{align}
\end{subequations}

\sid{In a bounded domain with a volume $V$ and a bounding surface $S$, the three components can be written as a generalized Biot-Savart law (see for instance \cite{wu2007vorticity})}

\begin{subequations}
\begin{align}
\mathbf{D} &= \frac{1}{4\pi}\int_V \frac{\nabla \cdot \mathbf{u}}{|\mathbf{r}|^3} \diff V^\prime \label{eq:genBS-D}\\
\mathbf{C} &= \frac{1}{4\pi}\int_V \frac{\boldsymbol{\omega} \times \mathbf{r}}{|\mathbf{r}|^3} \diff V^\prime \label{eq:genBS-C}\\
\mathbf{B} = -\frac{1}{4\pi}\int_S \frac{ \left(\mathbf{n}\cdot\mathbf{u} \right)\mathbf{r} + \left(\mathbf{n}\times\mathbf{u} \right)\times\mathbf{r} }{|\mathbf{r}|^3} \diff S^\prime &= -\frac{1}{4\pi}\int_S \frac{ \left(\mathbf{n}\cdot\mathbf{r} \right)\mathbf{u} + \left(\mathbf{n}\times\mathbf{r} \right)\times\mathbf{u} }{|\mathbf{r}|^3} \diff S^\prime \label{eq:genBS-B}
\end{align}
\end{subequations}
\sid{where $\mathbf{r}$ is the position vector, from a point in the volume $V$ (or the surface $S$) to the point $\mathbf{x}$ where the integrals are being evaluated. The integrals over the volume $V$ can be considered the ``near-field'' contribution whereas the surface integral over the bounding surface $S$ can be considered the ``far-field'' contribution. Any region within the domain has a bounding surface that separates it from the rest of the domain. Therefore, the integral over this bounding surface can be seen as a representing the sum of the contributions of (i) the integral over the volume surrounding the region and (ii) the integral over the bounding surface of the whole domain.}

\sid{In an unbounded (infinite) domain, the Helmholtz decomposition reduces to}
\begin{equation}
\mathbf{u} = \mathbf{D} + \mathbf{C}
\end{equation}
\sid{provided that $\mathbf{B}$ goes to zero when $|\mathbf{r}|$ goes to infinity. This happens if the distance over which the velocity field is correlated along the surface grows slower than $|\mathbf{r}|$ when $|\mathbf{r}|$ goes to infinity.}

\sid{This can be illustrated by considering an integral correlation measure $L_\alpha$ defined on the surface of a sphere, similarly to what was done for 3D, which is given as}
\begin{equation}
L_\alpha\left(\mathbf{x},\Lambda \right) = \int_0^\Lambda \mathbf{u}({\mathbf{x}})\cdot \mathbf{u}({\mathbf{x}+\mathbf{r}_{\alpha})} \mathrm{d}r_\alpha
\end{equation}
\sid{where $\mathbf{x}$ is a point on the surface of the sphere. Since this calculation is confined to the surface, the correlation around each point $\mathbf{x}$ is simply a function of an angle $\alpha$ and an integration length $\Lambda$, measured along the direction (on the surface) specified by $\alpha$. In the Helmholtz decomposition, the contribution of the boundary to the velocity field (i.e. the contribution of $\mathbf{B}$ to $\mathbf{u}$) goes to zero if}

\begin{equation}
\lim_{|\mathbf{r}| \to \infty} \frac{L_\alpha \left(\mathbf{x},\Lambda \right)}{|\mathbf{r}|} = 0
\label{eq:surfaceLimit}
\end{equation}

\sid{for any $\mathbf{x}$, $\alpha$ and $\Lambda < \pi|\mathbf{r}|$ (see, for example, \citet{phillips1933vector}). For an infinite domain, the contribution of $\mathbf{B}$ to the velocity field could be the result of an ``organized motion over an infinite distance'', which can be seen as the result of (arbitrary) choices in constructing the field (e.g. the choice of a particular frame of reference). This arbitrary artificial ``coherent motion'', which is the ``extrinsic motion'' can be removed by making a transformation in the velocity field such that the condition given by eq. \ref{eq:surfaceLimit} becomes true for any $\mathbf{x}$, $\alpha$ and $\Lambda < \pi|\mathbf{r}|$; this condition will result in the integrals given by eq. \ref{eq:genBS-B} becoming equal to zero for any surface $S$ in the sphere, when the radius of the sphere goes to infinity.}

\sid{When doing turbulent flow simulations with periodic boundary conditions, any arbitrary artificial ``coherent motion'' can be removed by not considering $\mathbf{B}$, provided that the domain is large enough to take into account any relevant ``intrinsic motion''; i.e. provided that the domain is ``significantly'' larger than the distance over which the velocity field is correlated; this requires that the domain needs to be ``significantly'' larger than the integral length scale of the turbulence, which is our case. If this happens, for $|\mathbf{r}|$ larger than half the domain size the contribution of $\mathbf{B}$ to the velocity approaches a constant, which can be made zero by neglecting $\mathbf{B}$ (which is equivalent to ``choosing the appropriate frame of reference''). Actually, in our case, as usually true for turbulence simulations in triperiodic domains, this constant is already zero, since the mean velocity is equal to zero; as it will be shown, in this case $\mathbf{C}$ (i.e. eq. \ref{eq:genBS-C}) gives a good approximation of $\mathbf{u}$ (since we deal with an incompressible flow, $\mathbf{D} = 0$) for $|\mathbf{r}|$ approaching the domain size.}

\sid{With these considerations, the generalized Helmholtz decomposition reduces to the simplified Biot-Savart law, applicable for incompressible flows over periodic (or infinite) domains, which is given as}
\begin{equation}
\mathbf{u}(\mathbf{x}) = \frac{1}{4\pi}\int_V \frac{\boldsymbol{\omega}\times \mathbf{r}}{|\mathbf{r}|^3}\mathrm{d} V^\prime
\label{eq:BiotSavart}
\end{equation}

\sid{The Biot-Savart law provides a way to disentangle flow structures by isolating the contributions from different vorticity regions to a local velocity structure. For instance, \textit{local} and \textit{non-local} vorticity contributions can be separated using this paradigm, or the vorticity field can be conditionally sampled to identify the contribution of different vorticity levels in generating velocity field structures.}

\begin{figure}
  \centerline{\includegraphics[width=0.8\linewidth]{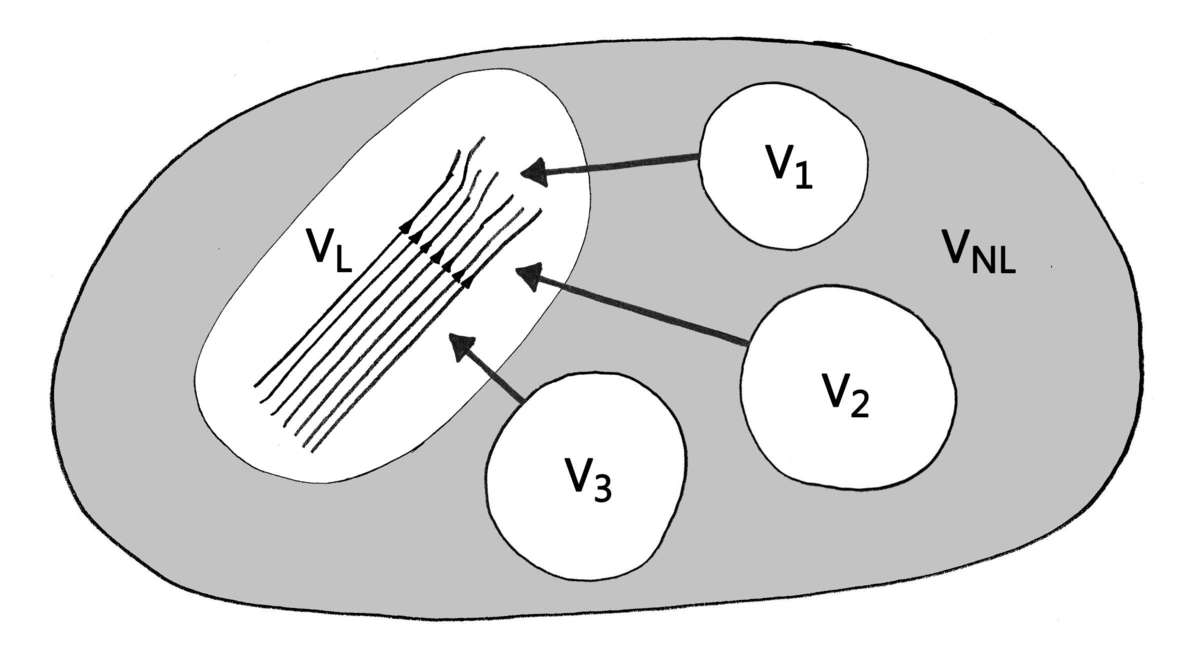}}

  \caption{Schematic of a Biot-Savart reconstruction of the velocity field in a region $V_\mathrm{L}$, from three isolated non-local regions $V_1, V_2$ and $V_3$. When the region $V_\mathrm{L}$ has negligible vorticity of its own, the flow generated within the region by non-local vorticity contributions is a potential flow which can be written as the gradient of a harmonic function $\psi$.}
\label{fig:schematicBiotSavart}
\end{figure}

\sid{Note that, even though far-field vorticity contribution, i.e. $\mathbf{B}$, may be absent, the volumetric region can also be split into an isolated \textit{local} region ($V_\mathrm{L}$), surrounded by the \textit{non-local} region ($V_\mathrm{NL}$), which essentially behaves as a \textit{far-field} for the region $V_\mathrm{L}$. This leads to the consequence that, if the local region has negligible vorticity within $V_\mathrm{L}$, while the non-local region $V_\mathrm{NL}$ \textit{induces} a velocity field within $V_\mathrm{L}$, then this velocity field must be a \textit{potential flow}. This means that the local flow in $V_\mathrm{L}$ can be described by the gradient of a harmonic function $\psi$, i.e. $\mathbf{u}_\mathrm{L} = \nabla \psi$, while $\nabla^2 \psi = 0$. The non-local contributions from $V_\mathrm{NL}$ \textit{cannot} generate vorticity within the local region $V_\mathrm{L}$ (as illustrated in figure \ref{fig:schematicBiotSavart}). A last feature to note regarding the Biot-Savart law is the rapid decay (of $1/r^2$ over a distance $r$) of the vorticity contribution, which means that a small, isolated, vorticity region cannot extend its influence over a large distance beyond its immediate neighbourhood.}

\subsection{Correlations related to the Biot-Savart law}
Ideas associated with the Biot-Savart law can be used to define correlation measures in order to identify, extract and disentangle structures associated with the relation between the velocity and vorticity fields. In regions of strong vorticity associated with swirling-flow \sid{in the orthogonal plane}, the Lamb vector, i.e. $\boldsymbol{\omega}\times \mathbf{u}$, yields high values. Although, this is again a \textit{local} quantity. We propose a correlation which utilizes this idea and extends it to a \textit{non-local} form, where the vorticity at point $\mathbf{x}$ is correlated with the velocity at point $\mathbf{x} + \mathbf{r}_i$, leading to a three-tuple $\mathbf{H}(\mathbf{x},\Lambda)$, which, similarly to $\mathbf{L}(\mathbf{x},\Lambda)$, can be written as follows
\begin{equation}
\setlength{\arraycolsep}{5pt}
\renewcommand{\arraystretch}{1.3}
\mathbf{H}(\mathbf{x},\Lambda) \quad \mathrm{where}\quad H_i(\mathbf{x},\Lambda) = \int_{-\Lambda}^{\Lambda}\left(\frac{\boldsymbol{\omega}(\mathbf{x})\times \mathbf{r}_i}{|\mathbf{r}_i|}\right)\cdot \mathbf{u}(\mathbf{x}+\mathbf{r}_i) \mathrm{d}r_i
\label{eq:HVector}
\end{equation}

\sid{The above correlation has a flavour of the Biot-Savart law and it allows correlating the contribution of the \textit{local} vorticity $\boldsymbol{\omega}(\mathbf{x})$ with the \textit{global} vorticity contribution to its neighbouring velocity field, since the velocity $\mathbf{u}(\mathbf{x}+\mathbf{r}_i)$ can be seen as an integral result of the global vorticity field.} Note that, with the above definition, $\mathbf{H}(\mathbf{x},\Lambda)$ will tend to be \textit{orthogonal} to $\boldsymbol{\omega}(\mathbf{x})$, as $H_i$ (i.e. $H$ along the $i-$direction) will have a high magnitude when the vorticity is large and orthogonal to the $i-$direction.

\begin{figure}
  \centerline{\includegraphics[width=\linewidth]{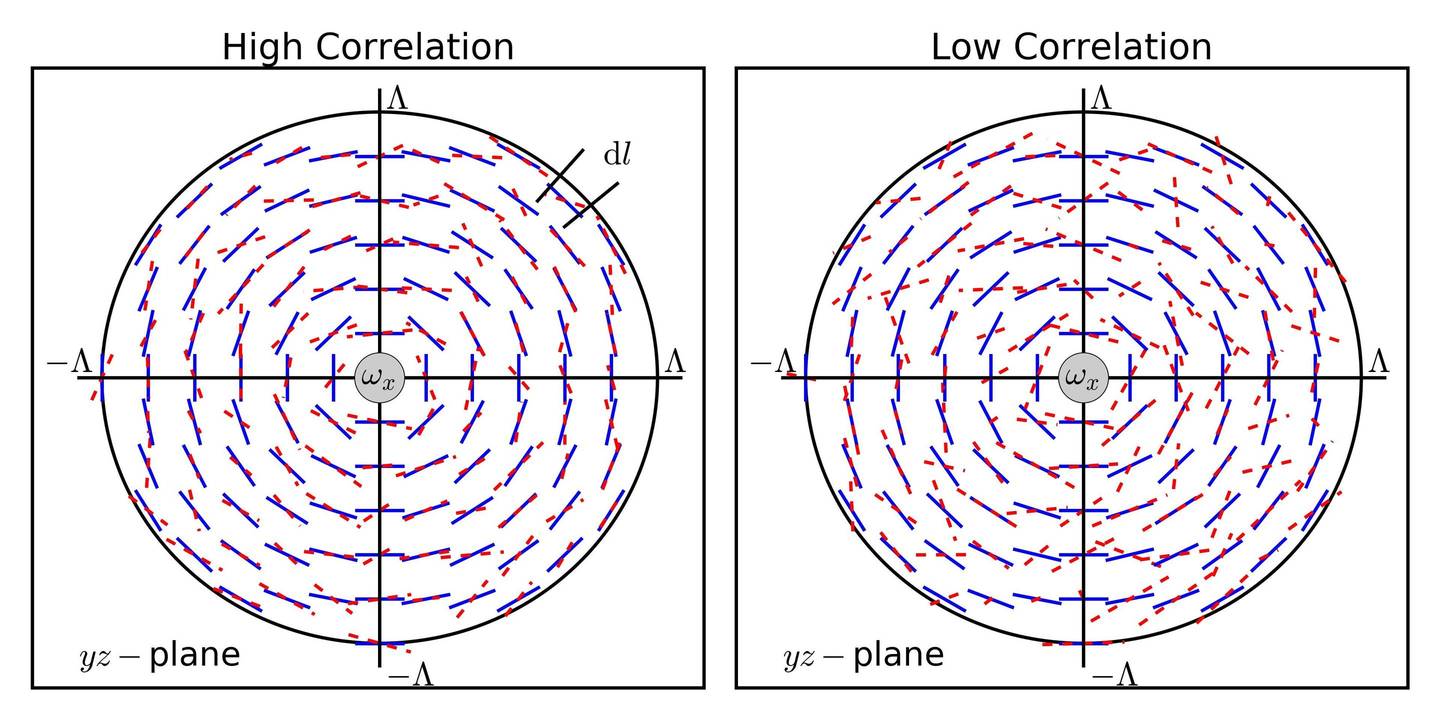}}

  \caption{A schematic of the $\mathbf{H^p}$ correlation. The $\omega_x$ vorticity is used to generate a velocity field in the $yz-$plane (shown in solid-blue lines), which is correlated with the real velocity field (dashed-red lines) within a circular region of radius $\Lambda$. The correlation is integrated around rings ($\mathrm{d}l$ elements) and then along the radial direction ($\mathrm{d}r$). The correlation will be strong when the local vorticity dominates in producing, in a Biot-Savart sense, the velocity field in its neighbourhood (left), and low when the local velocity field is not associated with the central vorticity (right).}
\label{fig:schematicHArea}
\end{figure}

Since the vorticity at a point generates flow, in a Biot-Savart sense, in the plane \textit{orthogonal} to the vorticity vector, a more natural correlation definition is proposed, which takes into account this fact. This is done by correlating the vorticity along a particular direction, say $x$, to the flow in the orthogonal plane, i.e. $yz$. At each point, a velocity field is generated using the $x-$vorticity (i.e. $\omega_x$) in the orthogonal $yz-$plane \sid{within a circular region (along perimeters of circles of radius $0 < r_{yz} \leq \Lambda$). This local velocity field is calculated with a simplification of the Biot-Savart law}, by taking the cross product of the vorticity with unit vectors in the orthogonal plane ($\mathbf{r}_{yz}/|\mathbf{r}_{yz}|$), as done for the $\mathbf{H}$ correlation. This is illustrated in figure \ref{fig:schematicHArea}, where the $\omega_x$ vorticity component generates the velocity field shown in blue (solid lines), while the \textit{real} velocity field generated from the global vorticity contributions is shown in red (dashed lines). The $\mathbf{H^p}$ correlation (for $\mathbf{H}-$\textit{planar}) is calculated as the integral of the dot product between the vorticity-generated velocity vectors (blue) and the real velocity (red), over rings of radius $0<r_{yz}<\Lambda$. Since the length of the rings increases proportionally with the radius $r$, the integral over each ring is further divided by $r$ (i.e. $|\mathbf{r}_{yz}|$), to give an average correlation at a distance $r$, though other definitions can be used. This correlation is given by
\begin{equation}
\setlength{\arraycolsep}{5pt}
\renewcommand{\arraystretch}{1.3}
\mathbf{H^p}(\mathbf{x},\Lambda) \quad \mathrm{where}\ H^p_x(\mathbf{x},\Lambda) = \int_0^\Lambda \frac{1}{|\mathbf{r}_{yz}|}\oint_{\mathcal{L}} \left(\frac{{\omega_x}(\mathbf{x})\hat{i}\times \mathbf{r}_{yz}}{|\mathbf{r}_{yz}|}\right)\cdot \mathbf{u}(\mathbf{x}+\mathbf{r}_{yz})\ \mathrm{d}l\ \mathrm{d}r_{yz}
\label{eq:HVectorArea}
\end{equation}
\sid{where $\hat{i}$ is the unit vector in the $x$-direction}. $H^p_y$ and $H^p_z$ are defined in a similar way. Evidently, this correlation is more computationally expensive to calculate than $\mathbf{H}$, as three planar regions need to be considered for each point in space. This requirement can be relaxed by sampling the rings $0 < r \leq \Lambda$ with a chosen frequency, i.e. using every $n$-th ring such that $r \in \left\lbrace n,2n,3n,\dotsc \right\rbrace$.

\sid{$\hp(\mathbf{x},\Lambda)$ can be seen as an area-integral correlation-measure of the planar (2D) version of $\mathbf{C}$ in eq. \ref{eq:genBS-C} (see, for e.g., \cite{wu2007vorticity}) with $\mathbf{u}$. In other words, it correlates the contribution of $\omega_x$ (the vorticity at $\mathbf{x}$) to $\mathbf{C}$ at $\mathbf{x}+\mathbf{r}$ with $\mathbf{u}$ at $\mathbf{x}+\mathbf{r}$; i.e. it correlates the value of $\mathbf{u}(\mathbf{x}+\mathbf{r})$ \textit{due to} $\omega_x(\mathbf{x})$ with the \textit{actual value} of $\mathbf{u}(\mathbf{x}+\mathbf{r})$. However, contrary to, e.g., $L_{\alpha\beta}(\mathbf{x},\Lambda)$, which is a line-integral correlation-measure over a distance $\Lambda$, $\hp(\mathbf{x},\Lambda)$ is an area-integral correlation-measure over a disc of radius $\Lambda$. Similarly, different combinations of the 2D and 3D versions of $\mathbf{C}$, and line, area and volume integral correlation-measures can be constructed, providing different ``Biot-Savart perspectives'' on the correlation between the vorticity field and the velocity field; however, we do not further explore these possibilities here.}

\section{Correlations applied to simplified canonical flows}\label{sec:CanonicalFlows}
\sid{In this section we test the correlations developed in the two previous sections using simplified flows, which are constructed using simplified canonical flows. In order to illustrate some key features of the different correlations, we consider simplified 1D, 2D and 3D velocity fields.}

\subsection{One-dimensional fields}\label{subsec:Oseen}
\sid{We construct a 1D velocity field along a line}, starting with Oseen vortices, which can be defined by a tangential velocity field and a vorticity field, given as 
\begin{align}
u_\theta(r) &= \frac{\Gamma}{2\pi r} \left[ 1 - \exp\left(-\frac{r^2}{4\nu t}\right) \right] \label{eq:Oseen} \\
\omega_z(r) &= \frac{\Gamma}{4\pi \nu t} \exp\left( -\frac{r^2}{4\nu t}\right) \label{eq:OseenVorticity}
\end{align}
where $r$ is the radial distance from the vortex center, $\Gamma$ is the circulation, $\nu$ is the fluid kinematic viscosity and $t$ is the time. The Oseen vortex comprises a small core region in (near) solid body rotation, within which the velocity increases (approximately) radially as $u_\theta \propto r$ to its maximum value. Beyond this, there exists a (near) potential flow region (where $\omega_z$ is nearly zero) and $u_\theta \propto 1/r$. We add a noise $\zeta$ to the velocity field given by eq. \ref{eq:Oseen}, to generate a ``structure immersed in noise''. The vortex has a certain `reach', which depends on the amplitude of $\zeta$, and is defined as the distance beyond which $\zeta > u_\theta$. The vorticity of the Oseen vortex is calculated in Cartesian coordinates as $\omega_z = \nabla\times\mathbf{u} = \partial u_y/\partial x - \partial u_x/\partial y$, \sid{and not using eq. \ref{eq:OseenVorticity}}, due to the addition of the noise to the velocity field.

\sid{We generate the velocity field by placing the centers of two counter-rotating Oseen vortices along the $x-$axis, separated by a distance greater than the typical reach of either vortex. Along the $x-$axis the velocity only has a $y-$component, $u_y$, and it is this one-dimensional velocity field that we consider. Furthermore, we impose a periodicity in $u_y$, over a length of $N_x$. This field, hence, consists of a large periodic structure, with smaller sub-structures (which are associated with each of the Oseen vortices and their interaction). This creates a pattern of symmetries and anti-symmetries in the velocity and vorticity fields.}

\begin{figure}
  \centerline{\includegraphics[width=0.9\linewidth]{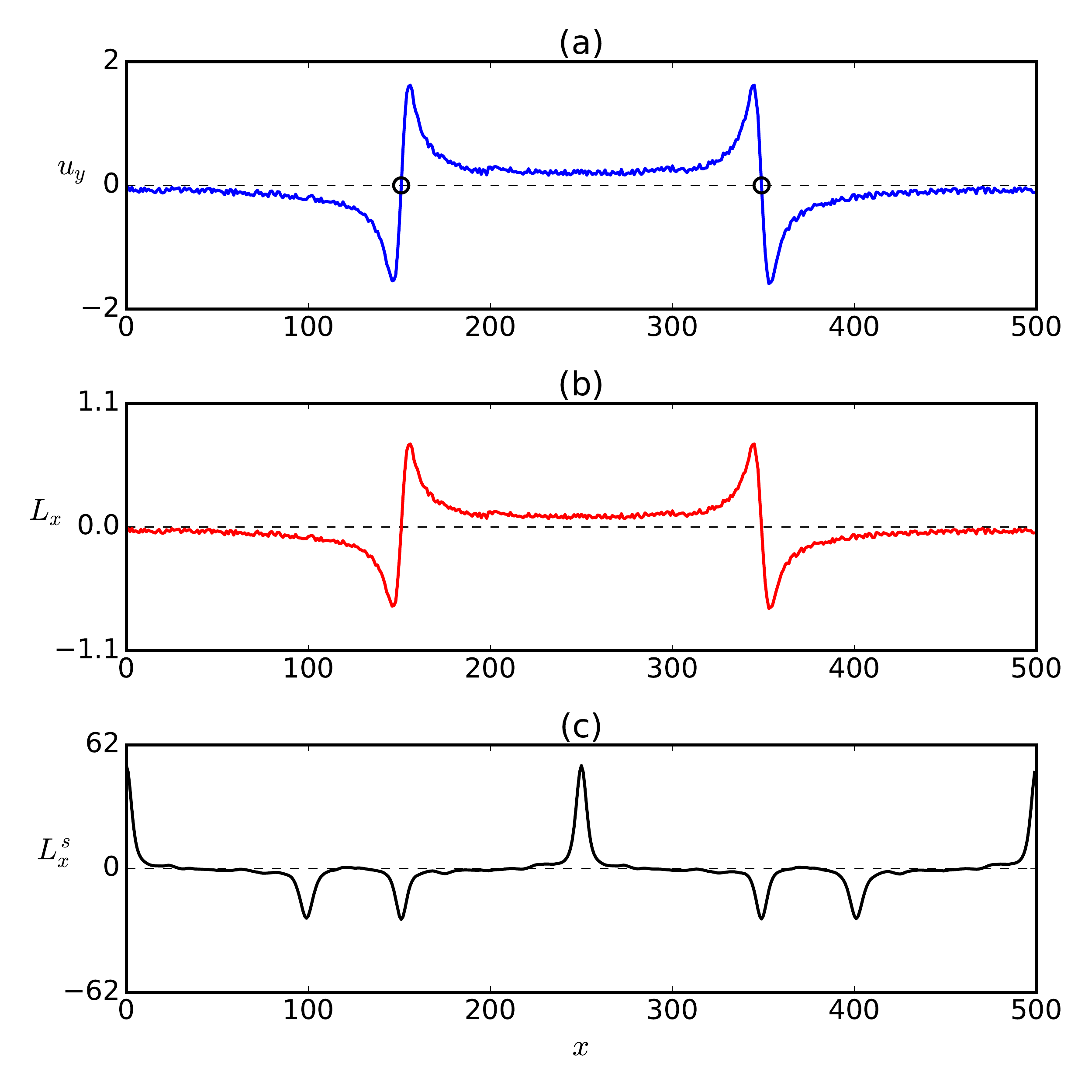}}

  \caption{(a) A one-dimensional velocity field comprising a larger structure composed of two smaller sub-structures (i.e. two counter-rotating Oseen vortices, with their centers marked with circles). Correlations calculated for this field are shown in panels (b) $L_x$ and (c) $L^s_x$, obtained by integrating over $\Lambda=N_x/2$, which spans the entire length of the field, which is periodic over $N_x=500$.}
\label{fig:dotCorrelationExamplesRAW}
\end{figure}

\sid{The two vortices are generated using the same parameters, which are given (in arbitrary units) as: $\Gamma=10$, $\nu = 2.0$ and $t=2.0$. The length of the field is $N_x = 500$ and the centers of the vortices are placed at $150$ and $350$, with a core solid body rotation region extending over roughly $5$ units, and an Oseen velocity pattern extending roughly $50$ units, on either side of the centers. Uniformly distributed random noise ($-1< \zeta <1$), scaled to an amplitude of $1\%$ of the maximum velocity magnitude, was added to the velocity. The velocity reduces to within $2.5\%$ of the maximum value in the range of $x<50$ and $x>450$. As can be seen in the top panel of figure \ref{fig:dotCorrelationExamplesRAW}, the velocity field consists of a ``larger structure'' (approximately in the range $50<x<450$), which is symmetric, except for the noise, with respect to its middle ($x \approx 250$). This larger structure contains the following sub-structures: (i) two Oseen velocity patterns around the center of each vortex (within approximately $100<x<200$ and $300<x<400$), containing a solid body rotation region and a potential flow region, and (ii) an almost uniform velocity region (between roughly $200<x<300$) due to the interaction between the two vortices.}

The middle and bottom panels of figure \ref{fig:dotCorrelationExamplesRAW} show $\mathbf{L}$ and $\mathbf{L^s}$, respectively (which, here, only have a $x-$component, i.e. $L_x$ and $L_x^s$), with the integration length spanning the entire length of the velocity field, i.e. $\Lambda = N_x/2$. A few features of the correlation profiles point at the nature of the correlation definitions, as well as the importance of the choice of $\Lambda$.

First, $L_x$ is found to have a shape similar to the function itself. If we consider the definition of $L_x$, it is the product between the value of the function at a point and its integral over a length, hence, if either of the two is zero, the correlation becomes zero. Therefore, the correlation will have significant, or large, values where the function itself has significant, or large, values, and the structure of the function (i.e. its symmetries and/or anti-symmetries) does not make its integral small. 

\sid{When the function is the sum of several ``basis functions'' (in this case two ``pure'' Oseen vortices and a ``noise''), the $L_x$ correlation at a point involves the product between the local values of the basis functions and the integrals of the basis functions. So, even if the structure of each of the ``basis functions'' makes their integral small (i.e. if $L_x$ of a basis function itself is small), $L_x$ of the total function is not necessarily small. In the particular case presented here, the function $u_y$ is the sum of a ``pure function'', $u_{yp}$, and a ``noise function'', $\zeta$. The pure function $u_{yp}$ is the sum of two ``basis functions'', the Oseen vortices; the interaction between the velocity fields of these two Oseen vortices results in a larger structure with a finite integral. The integral of the noise function, $\zeta$, is itself a ``noise''. Hence, the $L_x$ correlation of $u_y$ has the same shape as $u_y$ itself. Mathematically, this can be seen as}
\begin{align}
L_x(x) &= \int_{-\Lambda}^\Lambda [u_{yp}(x) + \zeta(x)]\cdot [u_{yp}(x+r_x) + \zeta(x+r_x)] \mathrm{d}r_x \nonumber \\
&= [u_{yp}(x) + \zeta(x)]\cdot \int_{-\Lambda}^\Lambda [u_{yp}(x+r_x) + \zeta(x+r_x)] \mathrm{d}r_x \nonumber \\
&= [u_{yp}(x) + \zeta(x)]\cdot \left( \int_{-\Lambda}^\Lambda u_{yp}(x+r_x) \mathrm{d}r_x + \int_{-\Lambda}^\Lambda \zeta(x+r_x) \mathrm{d}r_x\right)
\end{align}
\sid{If either the integral of $u_{yp}$ or the integral of $\zeta$ are non-zero, $L_x$ will be similar to $u_{yp}$. Interestingly, even for a velocity profile which leads to the integral of $u_{yp}$ becoming zero (i.e. $L_x$ of the pure function is zero), the additional noise breaks the overall symmetries and anti-symmetries, such that $L_x$ becomes non-zero, and the shape of the pure velocity field $u_{yp}$ can be extracted from $L_x$. If the noise is removed from the velocity field, and the vortices are placed sufficiently far from each other such that the integral of $u_{yp}$ becomes zero, $L_x$, indeed, goes to zero for $\Lambda = N_x/2$ (not shown here).}

$L^s_x$, in figure \ref{fig:dotCorrelationExamplesRAW}(c), shows different features, starting with a central peak around $x=250$, which is exactly between the two counter-rotating vortices. Although the velocity around this position is small, $L^s_x$ attains a large value since the velocity field is essentially mirrored around this point, hence being perfectly correlated (with only the noise values being different between the mirrored halves), \sid{i.e. the large peak around $x=250$ is associated with the symmetry of the larger structure around this point}. At the core of the two vortices ($x=150$ and $x=350$), where $L_x$ becomes zero, $L^s_x$ shows a large negative peak, since the velocity field on the left and right of these points is anti-correlated up to the reach of each vortex, \sid{i.e. the large magnitude of $L_x^s$ at $x=150$ and $x=350$ is associated with the anti-symmetry of the ``local sub-structures'' around these points}. The part of $L^s_x$ in the region outside of the larger structure (i.e. approximately $x<100$ and $x>400$), is a repetition of the $L^s_x$ profile in the center (i.e. $150<x<350$) due to the periodicity of the velocity field and the integration length spanning the entire length of the field ($\Lambda=N_x/2$). 

\begin{figure}
  \centerline{\includegraphics[width=0.9\linewidth]{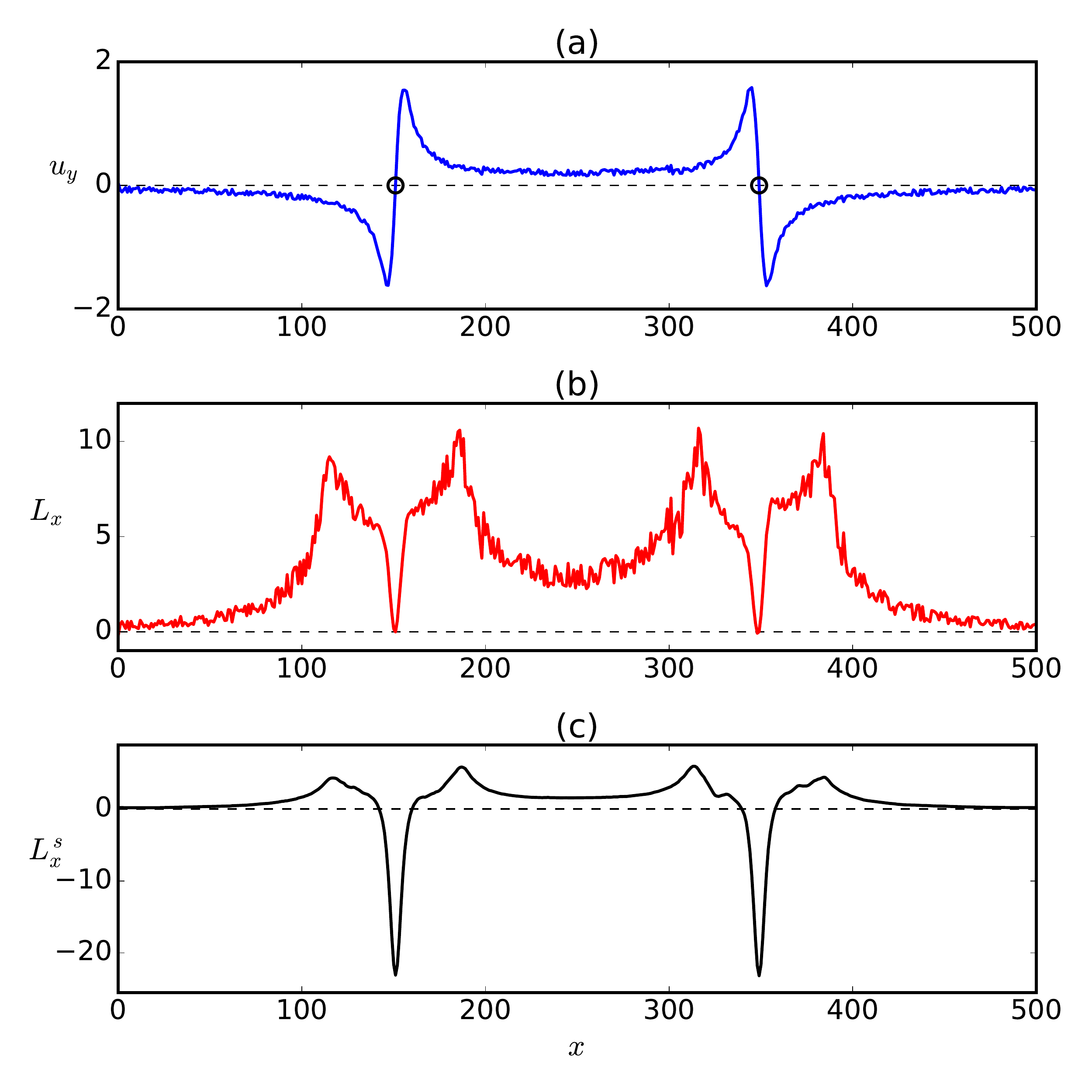}}

  \caption{(a) One-dimensional velocity field and correlations (b) $L_x$ and (c) $L^s_x$ calculated for an integration length of $\Lambda = 35$, which spans the approximate size of the individual `sub-structures', i.e the two counter-rotating Oseen vortices. \sid{Here, at the vortex cores $L_x$ goes to zero while $L^s_x$ yields strong negative peaks. In the potential flow regions of the two vortices and in the region between the vortices, $L_x$ and $L^s_x$ have a similar shape overall, with both correlations yielding positive values.}}
\label{fig:dotCorrelationSmallLambda}
\end{figure}

Figure \ref{fig:dotCorrelationSmallLambda} shows the correlations for a similar velocity field, now integrated over a length of $\Lambda=35$, which corresponds approximately to the size of the sub-structures (i.e. the individual vortices). This is because within a distance of $35$ units from the center of each vortex, the velocity reduces to roughly $10\%$ of its maximum value (note that a slightly lower or higher $\Lambda$ does not change the results significantly). The $L_x$ and $L^s_x$ profiles show a few similarities and differences. \sid{First, at the vortex cores, where the velocity is close to zero, $L_x$ goes to zero, while $L_x^s$ yields strong negative peaks because the velocity is strongly anti-correlated across the core region. In the potential flow regions of the two vortices and in the region between the vortices, the general shape of the two correlations is similar, with both $L_x$ and $L^s_x$ yielding positive correlation values, reflecting that the velocity is mostly uniform in these regions. This shows that in the core regions of vortices $\lvec$ and $\lsvec$ correlations behave very differently while in regions of roughly uniform flow they have a similar behaviour.} $L^s_x$ does not identify the `larger structure', which in this case has a lengthscale larger than the integration length of the correlation. This shows the importance of the choice of $\Lambda$ in identifying larger or smaller symmetries and anti-symmetries of the vector fields. We shall revisit the importance of the choice of $\Lambda$, in the context of turbulence, in section \ref{sec:ChoiceOfLambda}.

Note that it is not only the length of integration $\Lambda$, but also the lower and upper limits of integration which determine the symmetries and structures being identified. For instance, eq. \ref{eq:Lmanifold}, can be changed to, instead, find non-local structures between lengths $\Lambda_1 <r_{\alpha\beta} < \Lambda_2$, while looking around from point $\mathbf{x}$. We show an example using the $\lsvec$ correlation, as follows

\begin{equation}
L^s_i(\mathbf{x},\Lambda_1,\Lambda_2) = \int_{\Lambda_1}^{\Lambda_2} \mathbf{u}(\mathbf{x}-\mathbf{r}_i)\cdot \mathbf{u}(\mathbf{x}+\mathbf{r}_i) \mathrm{d}r_i
\end{equation}

\begin{figure}
  \centerline{\includegraphics[width=0.9\linewidth]{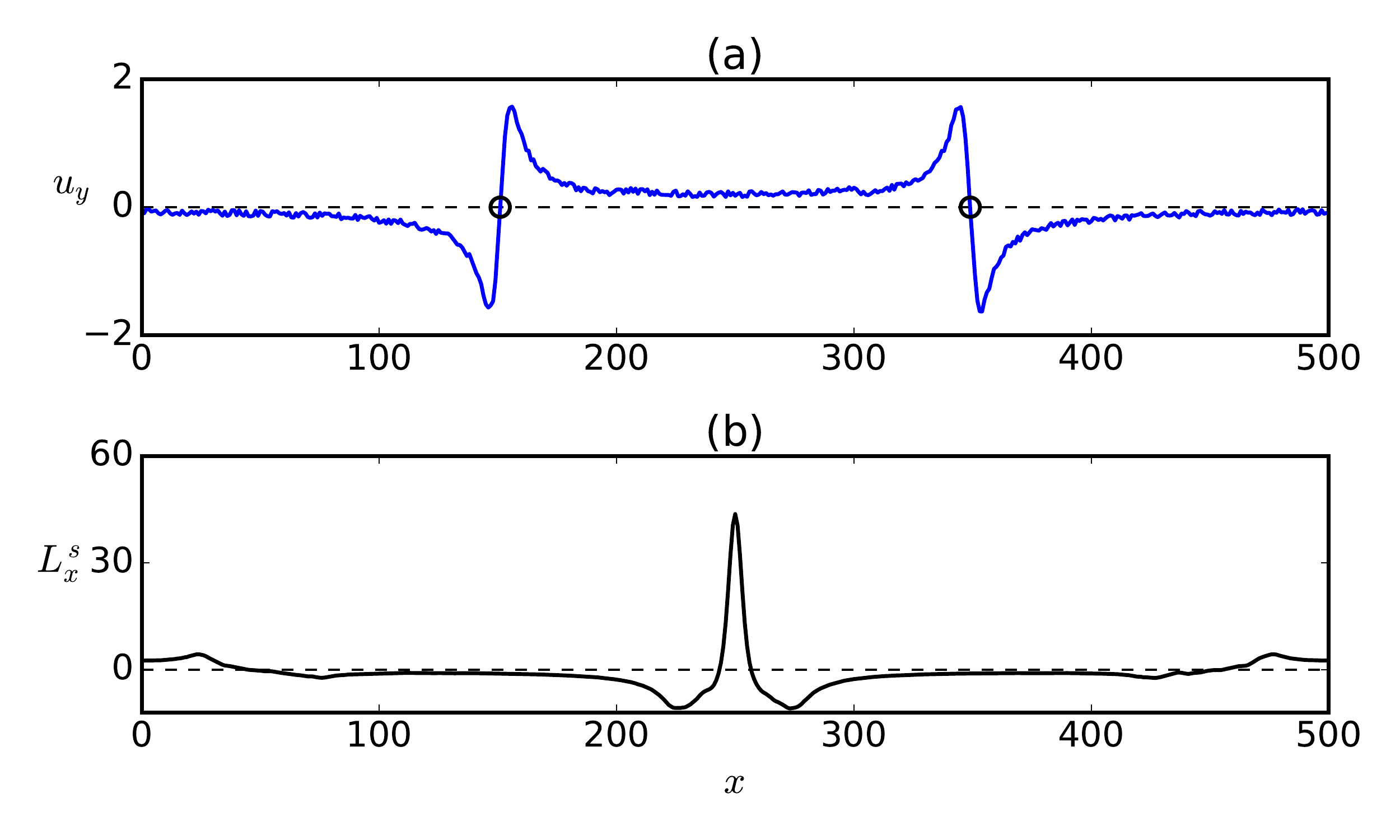}}

  \caption{$L^s_x$ correlation integrated from $\Lambda_1 = 75$ to $\Lambda_2 = 125$ shows how non-local symmetries can be identified by varying the limits of integration.}
\label{fig:dotCorrelationLS-TwoLambdas}
\end{figure}

Recalling the generalized correlation definition, this change in the limits of $\Lambda$ is essentially defining $\mathcal{S}_1$ and $\mathcal{S}_2$ as finite regions going from $x\in [x_0+\Lambda_1,x_0+\Lambda_2]$ and $x\in [x_0-\Lambda_2,x_0-\Lambda_1]$ respectively, while $x_0$ is a point of the region of observation $\mathcal{S}_o$. One example of this is shown in figure \ref{fig:dotCorrelationLS-TwoLambdas}, with $\Lambda_1 = 75$ and $\Lambda_2 = 125$. These integration limits are such that, when $x_0$ corresponds to the middle of the larger structure in the velocity field (i.e. at $x=250$), $\Lambda_1$ and $\Lambda_2$ span across most of the vortex regions. The $L^s_x$ profile, consequently, shows a strong peak in between the two vortices, resulting from the larger structure comprising the two counter-rotating vortices. These particular limits of integration do not identify any significant large symmetries, as observed from other $x$ locations.

\begin{figure}
  \centerline{\includegraphics[width=0.9\linewidth]{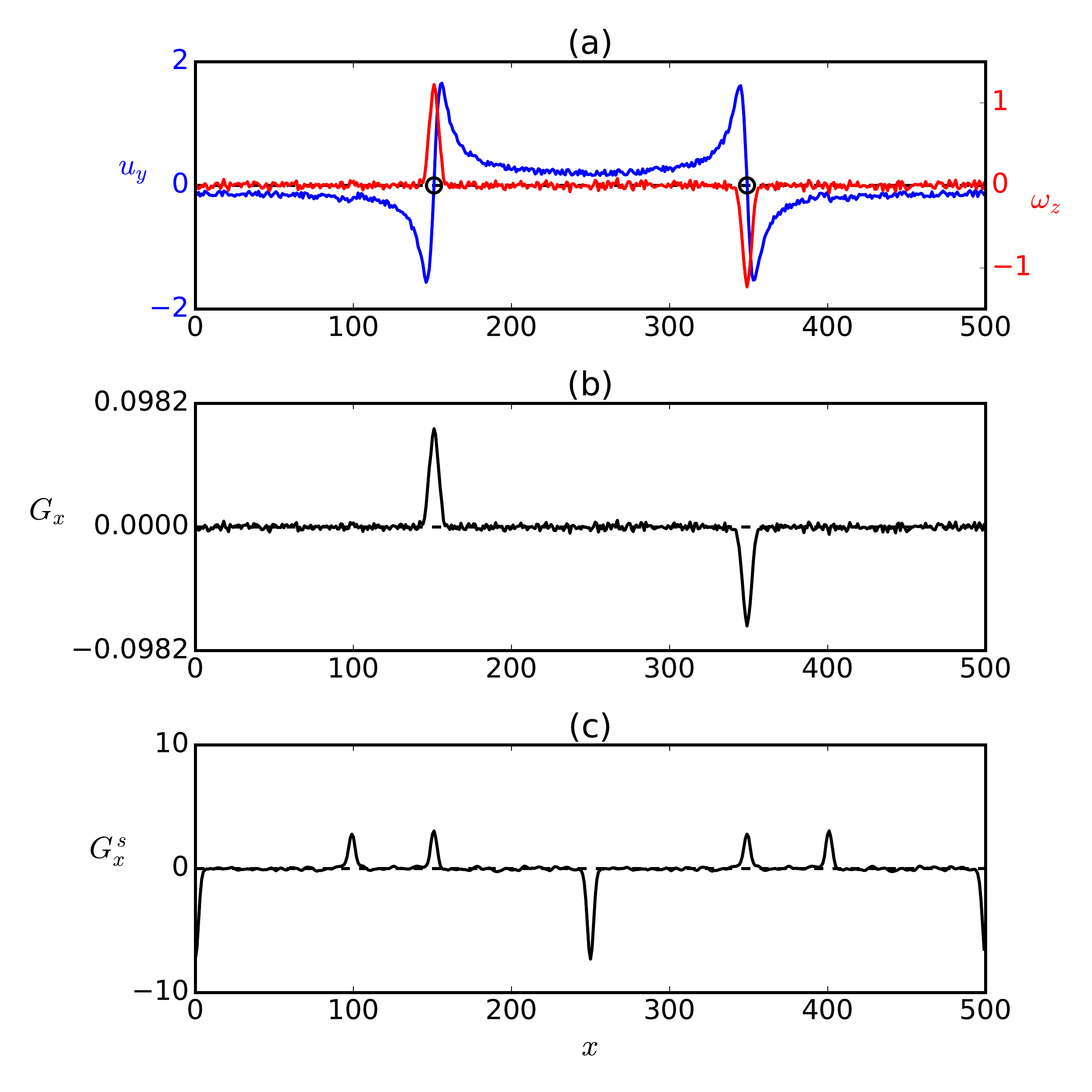}}

  \caption{$G_x$ and $G^s_x$ correlations (integrated up to $\Lambda=N_x/2$) are shown for the Oseen vortex-pair. \sid{Analogously for $L_x$ with $u_y$, the shape of $G_x$ is similar to the shape of $\omega_z$. $G^s_x$ has a profile similar to $L^s_x$ in figure \ref{fig:dotCorrelationExamplesRAW}, but with the opposite sign.}}
  \label{fig:oneDimensionalOseen-GCorr}
\end{figure}

Figure \ref{fig:oneDimensionalOseen-GCorr} shows correlations $G_x$ and $G^s_x$, which are the vorticity field equivalents of $L_x$ and $L^s_x$, integrated over $\Lambda=N_x/2$. The $G_x$ correlation remains mostly zero throughout, with small, noisy fluctuations. This is because, unlike the velocity field, which yields a finite value for the intergal of $u_y(x)$ (over the length $\Lambda=N_x/2$) due to the interaction between the two vortices and the non-zero contribution from the integral of the noise term $\zeta$, the integral of $\omega_z$ remains nearly zero, since the vorticity is localized at the core of the two vortices and has the same magnitude, but opposite sign at the two cores; the only contribution here is from the non-zero integral of the noise term, which breaks the anti-symmetry of the vorticity field. \sid{Similarly for $L_x$ with $u_y$, the break of the anti-symmetry due to the noise, leads to the shape of $G_x$ being similar to the shape of $\omega_z$; however, since here the only contribution is from the non-zero integral of the noise term, the magnitude of $G_x$ is much smaller than the magnitude of $\omega_z$.}

\sid{The $G^s_x$ correlation has a very similar behaviour to $L^s_x$, just with the opposite sign. It yields a large negative peak at the middle of the larger structure ($x=250$), associated with the anti-symmetry of the vorticity of the larger structure around this point (while $L^s_x$ has a large positive peak, associated with the symmetry of the velocity field around this point). It also shows smaller positive peaks at the core of the vortices ($x=150$ and $x=350$), around which the vorticity is high and symmetric (while $L^s_x$ has a negative peak, associated the anti-symmetry of the velocity field around the vortex core). The $G^s_x$ profile is also repeated due to the periodicity of the velocity field and the large integration length, similarly to the $L^s_x$ profile in figure \ref{fig:dotCorrelationExamplesRAW}.}

\begin{figure}
  \centerline{\includegraphics[width=0.9\linewidth]{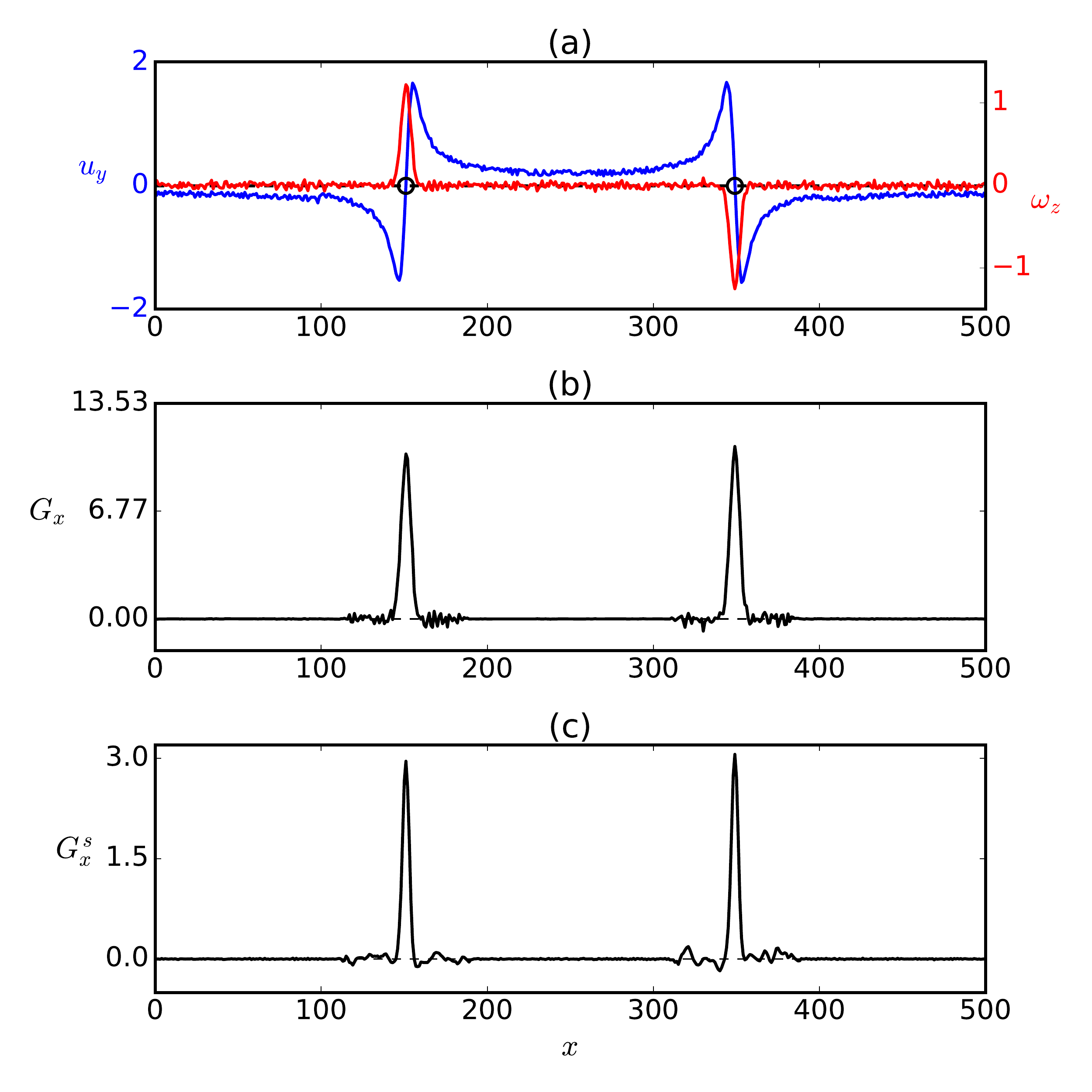}}

  \caption{$G_x$ and $G^s_x$ correlations, integrated up to $\Lambda=35$, are shown for the Oseen vortex-pair. Both correlations are similar and show a strong peak at the vortex cores, while the peak in $G^s_x$ decays more rapidly, since it is more dependent on the symmetry of the vorticity field.}
  \label{fig:oneDimensionalOseen-GCorrLambda}
\end{figure}

Figure \ref{fig:oneDimensionalOseen-GCorrLambda} shows the $G_x$ and $G^s_x$ correlations for an integration length of $\Lambda=35$. Both $G_x$ and $G^s_x$ show a very similar overall shape, yielding sharp positive peaks at the vortex cores, where the vorticity is high and symmetric. The $G_x$ correlation decays with some ``noise'', in the potential flow regions corresponding to the two vortices. The $G^s_x$ correlation, here, gives a sharper and less noisy profile. This is because, at the vortex core, the vorticity values to the left and right are perfectly symmetric (apart from the noise values), which gives a large correlation $G^s_x$ value at the vortex core. Slightly moving away from the core in either direction strongly disturbs this symmetry, leading to a sharper profile for $G^s_x$ than for $G_x$.

\begin{figure}
  \centerline{\includegraphics[width=0.9\linewidth]{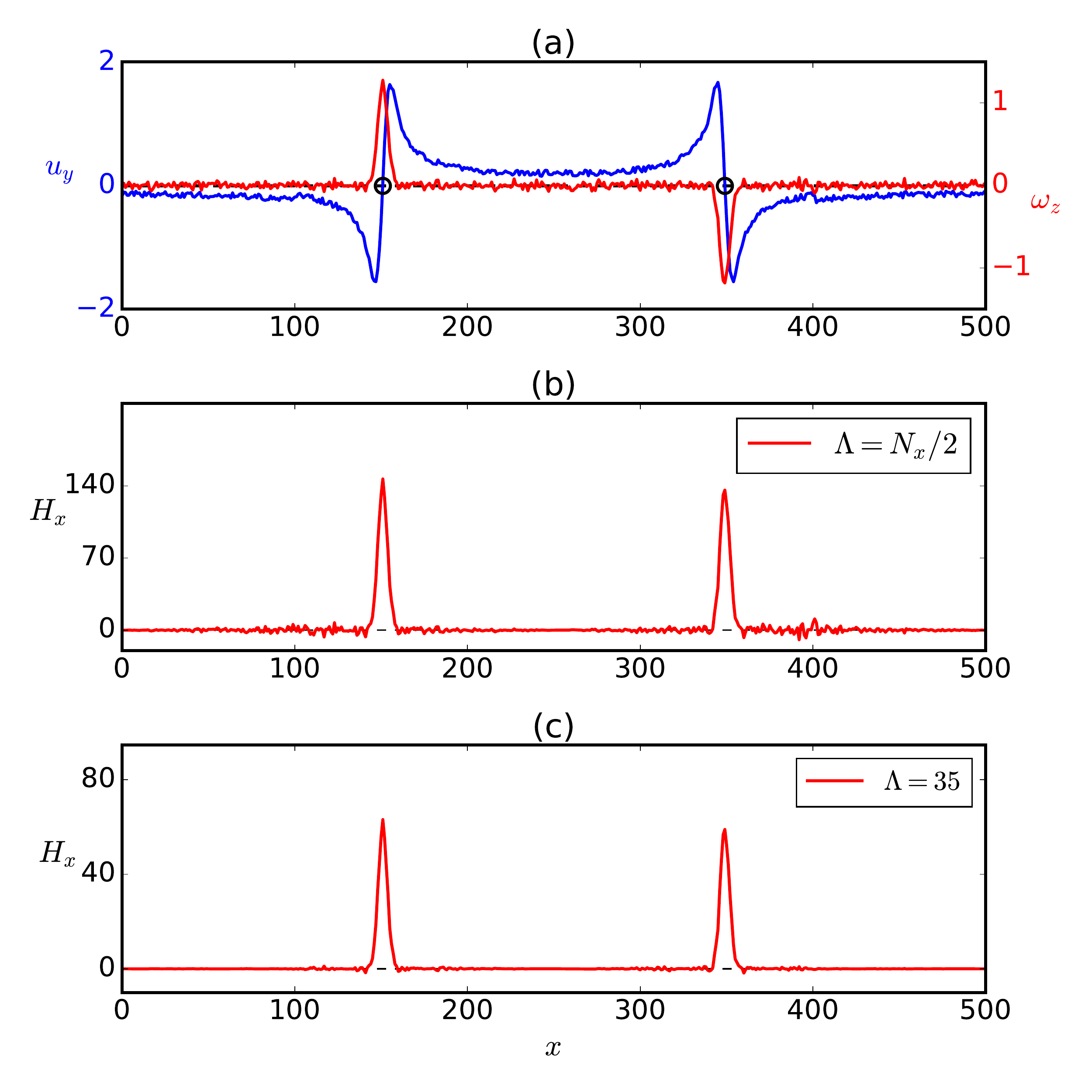}}

  \caption{$H_x$ correlation, integrated up to (b) $\Lambda=N_x/2$ and (c) $\Lambda=35$, is shown for the Oseen vortex-pair. The correlation yields strong positive peaks at the vortex cores, and is found to be independent of the choice of $\Lambda$ (except for a change in the magnitude of the correlation).}
  \label{fig:oneDimensionalOseen-HCorr}
\end{figure}

Finally, figure \ref{fig:oneDimensionalOseen-HCorr} shows the $H_x$ correlation for the Oseen vortex-pair, integrated over lengths of $\Lambda = N_x/2$ (panel b) and $\Lambda = 35$ (panel c). The shape of the $H_x$ correlation is found to be almost insensitive to the choice of $\Lambda$, while a higher $\Lambda$ increases the amplitude of the correlation. This can be understood from the construction of this correlation, which, in a Biot-Savart sense, is designed to identify regions where the angular velocity aligns with the vorticity. In this example, $H_x$ yields large positive values in the core region of the two vortices, since the flow around the vortex cores is generated by the vorticity at the cores. The independence of the choice of $\Lambda$ is because the influence of the local vorticity at a point $x_0$ rapidly decays over distance, such that at larger $\Lambda$ values, the velocity field $u_y(x_0\pm\Lambda)$ is not influenced by $\omega_z(x_0)$. The larger structure in this example is induced by the sum of the two individual vortices, and is, hence, ``externally generated'', which is why $H_x$ remains zero in the middle of the two structures. The results, for a `linear' version of the $H_x^p$ correlation (instead of its planar construction), are identical, and are not additionally shown here.

\sid{The $L_x$ and $L^s_x$ correlations help in identifying the structure of high velocity regions, while their analogues for the vorticity field, the $G_x$ and $G^s_x$ correlations, help in identifying the structure of high vorticity regions. The $H_x$ (and $H^p_x$) correlation gives information on the vorticity-velocity structure. The similarities and differences between these correlation measures gives information on the structure of the velocity and vorticity fields. In this example, where the velocity field is ``induced'' by a pair of vortices, at the core of the vortices, while $L_x$ is close to zero, there exists a large peak in the values of $L^s_x$, $G_x$, $G^s_x$ and $H_x$. On the other hand, in the region outside of the core of the vortices where the velocity is approximately uniform, if the integration length does not allow the capture of the larger structure periodicity, $L_x$ and $L^s_x$ have a similar shape, while $G_x$, $G^s_x$ and $H_x$ remain close to zero.}

\subsection{Two-dimensional Taylor-Green flow}
We next test the correlations on a two-dimensional Taylor-Green flow field, which comprises a set of counter-rotating vortices. Interaction between vortex-pairs generates regions of unidirection flow, seprated by regions of swirling flow. A simplified Taylor-Green flow pattern can be generated by creating a velocity field as follows

\begin{align}
u_x(x,y) &= \sin(x)\cos(y) \nonumber \\
u_y(x,y) &= -\cos(x)\sin(y)
\end{align}

\begin{figure}
  \centerline{\includegraphics[width=0.9\linewidth]{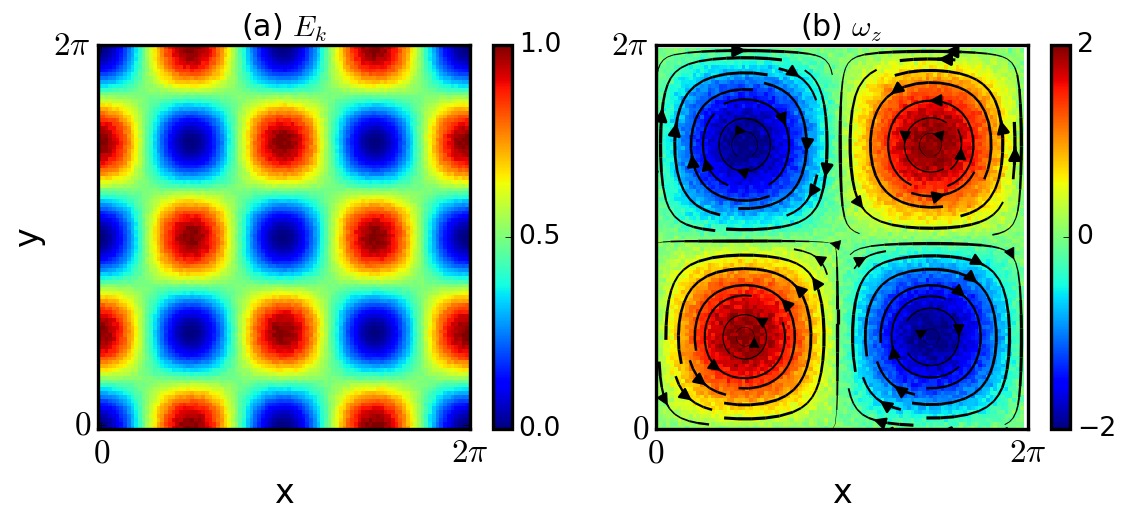}}

  \caption{A Taylor-Green velocity field. Panel (a) shows the kinetic energy $E_k$ field, , while panel (b) shows the vorticity field $\omega_z$, along with the flow streamlines (in black, with thickness varying with the velocity amplitude).}
  \label{fig:taylorGreen-Flow}
\end{figure}

We generate such a velocity field with $0 \leq x < 2\pi$ and $0 \leq y < 2\pi$, in arbitrary units. A uniform noise $\zeta$, of amplitude $1\%$ of the maximum velocity magnitude (which is unity), is added to the velocity field. The vorticity field, which has just one component, i.e. $\omega_z$, is calculated using central differences. In  figure \ref{fig:taylorGreen-Flow}(a), \sid{the kinetic energy $E_k$ is shown, while panel (b) shows the vorticity field $\omega_z$, along with the flow streamlines, where the counter-rotating vortex array can be seen. We would like to highlight that the velocity field structure in Taylor-Green flow is determined by the four large vortices, which are generated by regions of diffused, relatively high vorticity (as opposed to small, localized regions of high vorticity found in turbulence). These vortices have a diameter of approximately $\pi$. Essentially, the velocity field structure consists of: (i) four large vortices with a core region of high vorticity and low kinetic energy, (ii) regions of well-aligned jet-like velocity, with high kinetic energy and low vorticity, in between the large vortices, and (iii) a stagnation flow in the center with low kinetic energy and low vorticity.}

\begin{figure}
  \centerline{\includegraphics[width=0.9\linewidth]{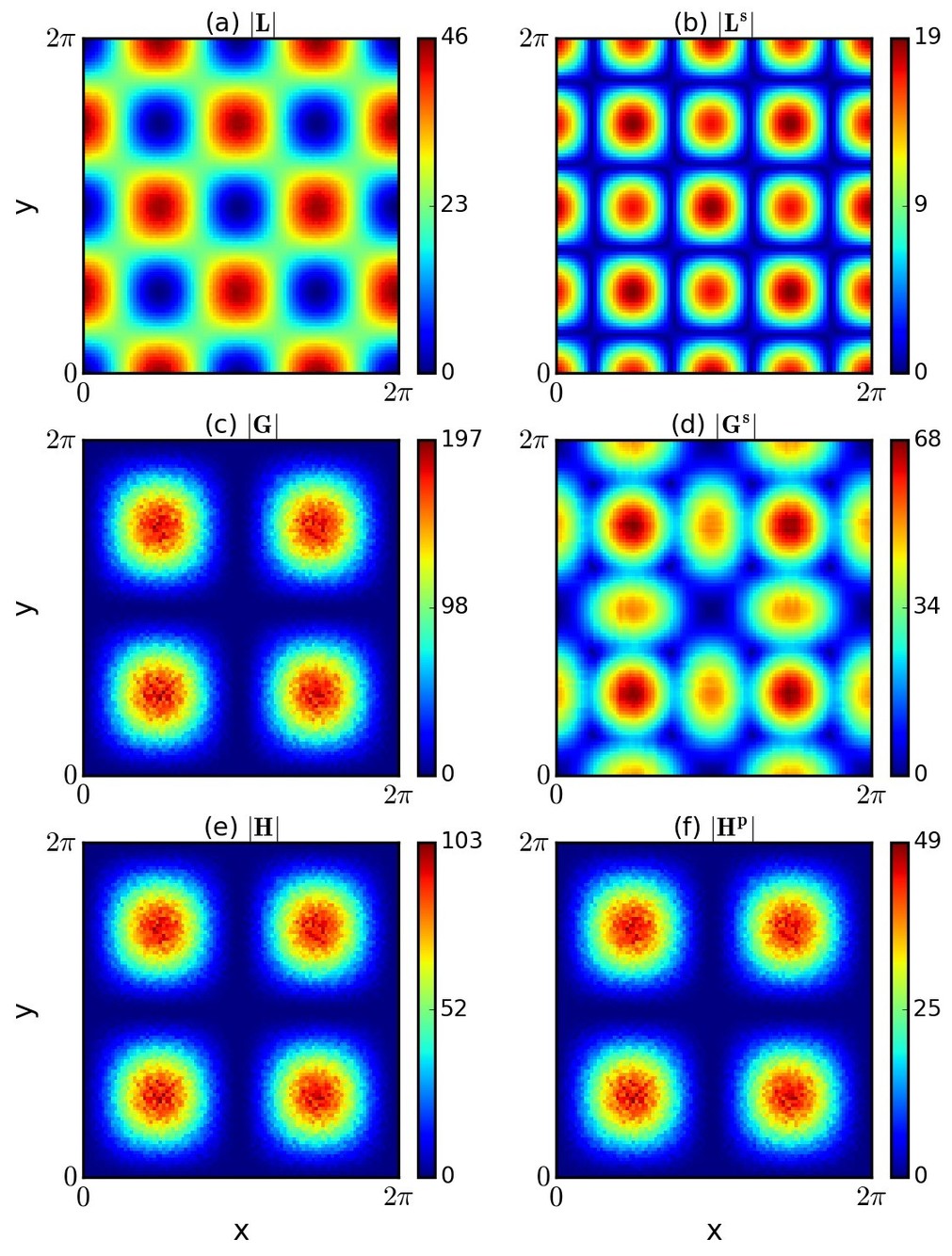}}

  \caption{Correlation field amplitudes for the Taylor-Green flow field, obtained upon integrating the correlations over $\Lambda = \pi/2$.}
  \label{fig:taylorGreen-CorrelationsPiBy2}
\end{figure}

\sid{In figure \ref{fig:taylorGreen-CorrelationsPiBy2}, the correlation fields for the Taylor-Green flow are shown. The correlation two-tuples are for the $x$ and $y$ directions, with an integration length of $\Lambda = \pi/2$. Panel (a) shows $\mathbf{L}$, which is found to have a high magnitude in regions of high kinetic energy $E_k$ (refer to figure \ref{fig:taylorGreen-Flow}a). This is because the $\mathbf{L}$ correlation identifies velocity regions that have simultaneously a high velocity magnitude and a high degree of local vector alignment. At the vortex core regions, $|\lvec|$ yields low values. It can be noted that for this velocity field, the regions of high $|\lvec|$ are separated from regions of high vorticity, showing that jet-like coherent regions can be generated by \textit{non-local} vorticity (since the vorticity within high $|\lvec|$ regions is negligible, see figure \ref{fig:taylorGreen-Flow}b).}

\sid{Figure \ref{fig:taylorGreen-CorrelationsPiBy2}(b) shows $|\lsvec|$, which has a very different structure than $|\lvec|$, since it identifies symmetries and anti-symmetries in the velocity field, which are not just associated with jet-like parallel streamlines. There are regions of high $|\lsvec|$ associated with different types of symmetries/anti-symmetries, other than the jet-like parallel streamlines. First, the regions of highest $|\lsvec|$ are found at the vortex cores, as the vortices have a large swirling region with a velocity that is perfectly anti-correlated across the vortex cores. High $|\lsvec|$ regions are also found coinciding with high $|\lvec|$ regions, where the flow has a symmetry associated with jet-like parallel streamlines. These two regions of high $|\lsvec|$ are separated by thin regions where $|\lsvec|$ goes to zero, due to the larger symmetries in the velocity field which cancel out in the calculation of $\lsvec$. There is also a region of high $|\lsvec|$ at the center, which is a stagnation flow region, which reflects the anti-symmetries in the velocity field of the four large vortices that produce the stagnation region.}

\sid{Panel (c) shows $|\g|$, which yields high values in the vortex core regions (which also are the regions of high vorticity, see figure \ref{fig:taylorGreen-Flow}b). In the regions of jet-like flow between vortex-pairs, $|\g|$ has values very close to zero, which is due to the negligible vorticity in these regions. Panel (d) shows $|\gs|$, which has a very different strucure than $|\g|$, since it identifies the symmetries and anti-symmetries in the vorticity field across every point. First, smaller diffused region of high $|\gs|$ are found corresponding to the vortex cores, similarly to $|\g|$, since the vorticity across the vortex cores is highly correlated. In between counter-rotating vortex-pairs, both in the $x$ and $y$ directions, there appear slightly elongated regions of high $|\gs|$. This is because along lines connecting the centres of counter-rotating vortex-pairs, the vorticity field varies uniformly from positive to negative, and this sign transition coincides with the jet-like velocity region, which itself has negligible vorticity; this strong anti-symmetry in the vorticity field gives the high $|\gs|$. Note that at the center there exists also a strong symmetry in the vorticity field along the diagonals of the square, however, this is not reflected in $\gs$, which is a two-tuple for the $x$ and $y$ directions, along which the vorticity is zero. This indicates that the particular alignment of the structures can lead to different values of the correlation, depending on the directions chosen for calculating the correlations. However, in homogeneous isotropic turbulence, with an arbitrary alignment of the structures, this effect will not appear, in a statistical sense.}

\sid{Lastly, panels (e) and (f) show the $|\h|$ and $|\hp|$ fields. Both correlations have a pattern that closely resembles $|\g|$, with large values at the vortex core regions, since the vortices here are generated by strong vorticity values at the vortex cores. Regions of jet-like flow, which are \textit{externally} induced by the interaction between the velocity fields of adjacent vortices, yield negligible values of $|\h|$ and $|\hp|$.}

\subsection{Three-dimensional Burgers vortices}
As a final example of the correlations applied to canonical flows, we now consider a three-dimensional velocity field generated by superposing two Burgers vortices, which again generates a `large-scale' structure comprising two smaller sub-structures. The Burgers vortex is an exact solution of the Navier-Stokes equation, consisting of a radial velocity component along with a tangential velocity, and can be constructed as
\begin{equation}
u_z = a z,\quad u_r = -\frac{a r}{2},\quad u_\theta = \frac{\Gamma}{2\pi r}\left[ 1 - \exp\left(-\frac{a r^2}{4\nu}\right)\right]
\label{eq:Burgers}
\end{equation}
where $a$ represents the rate of strain, $\Gamma$ the circulation and $\nu$ the kinematic viscosity. Here $u_z$, $u_r$ and $u_\theta$ give velocity components in the axial, radial and tangential directions, which are converted from cylindrical to Cartesian coordinates. We isolate the vortex in space by multiplying the velocity field with a three-dimensional Gaussian function $\mathcal{G}$ to contain the Burgers vortex within a spherical region,
\begin{equation}
\mathcal{G}(x, y, z) = \exp\left[-\left(\frac{x^2 + y^2 + z^2}{\sigma^2} \right) \right]
\end{equation}  
where $x$, $y$ and $z$ are measured with the origin placed at the center of the vortex. The value of $\sigma$ is chosen such that it creates a spherical region circumscribing the axial length of the vortex. This suppresses the strain regions generated by each vortex, far away from its center, such that the velocity field comprises primarily of two swirling-flow regions. The swirling-flow of the Burgers vortex resembles the one-dimensional Oseen vortex (as was described in Section \ref{subsec:Oseen}), with a core in solid-body rotation where $u_\theta \propto r$, followed by a potential flow region with $u_\theta \propto 1/r$. A low amplitude uniform noise, $\zeta$, is added to the final velocity field, over which the correlations are subsequently calculated. \sid{Lastly, the vortices are also rotated at arbitrary azimuthal and elevation angles $(\alpha, \beta)$. This is done to change the orientation of the velocity field symmetries with respect to the orthogonal bases along which the correlations are calculated, to test the applicability of the correlation definitions for arbitrarily aligned structures, as will be encountered in turbulence fields.}

\begin{figure}
	\begin{subfigure}{0.5\linewidth}
	\centerline{\includegraphics[width=\linewidth]{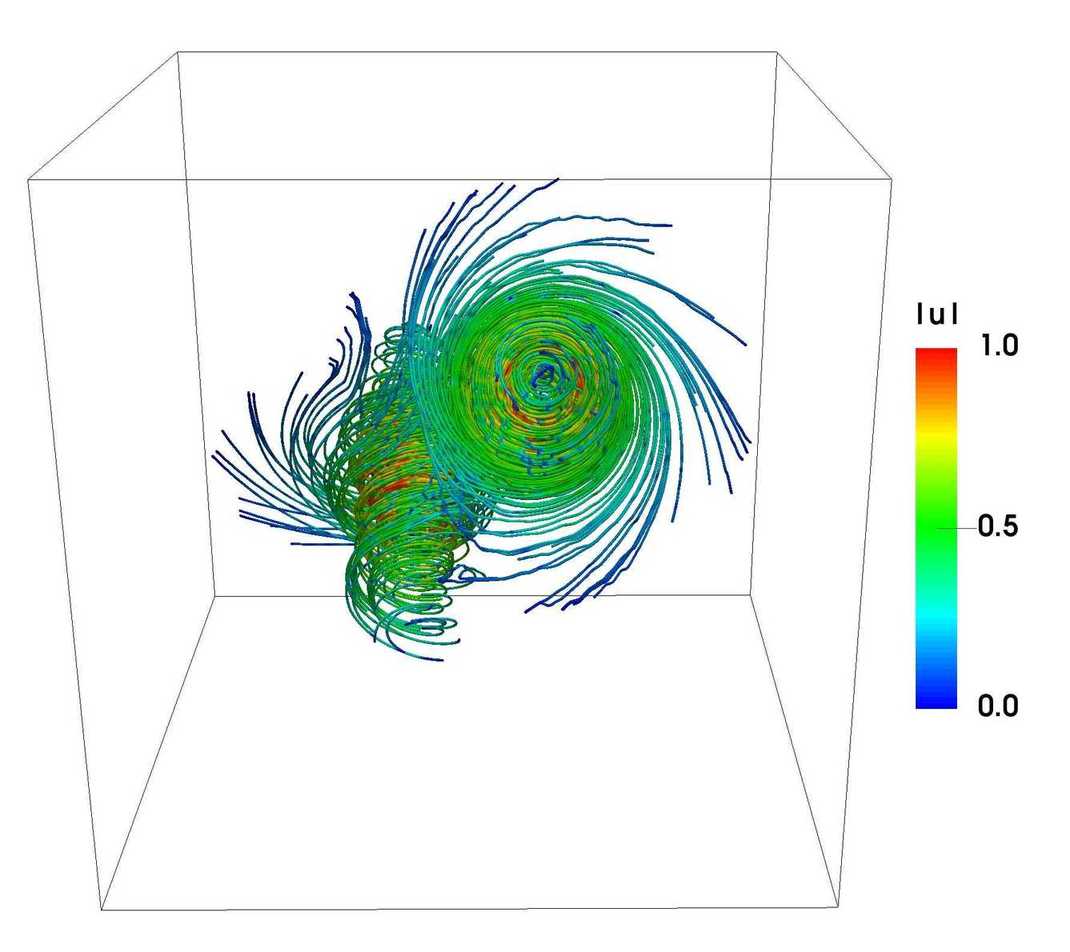}}

	\caption{View 1}
	\end{subfigure} 
	\begin{subfigure}{0.5\linewidth}
	\centerline{\includegraphics[width=\linewidth]{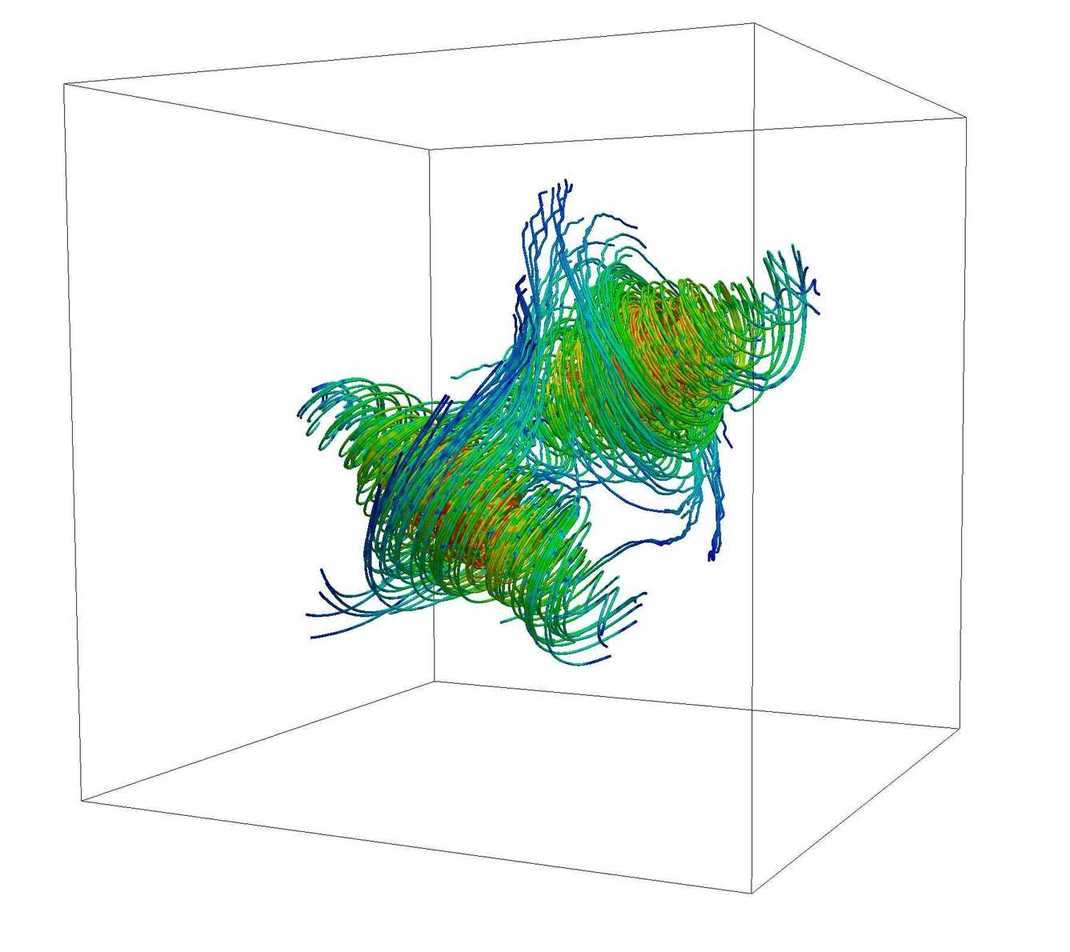}}

	\caption{View 2}
	\end{subfigure} 
	\caption{Two arbitrarily aligned Burgers vortices are shown, with isolated swirling-flow regions. The velocity streamlines are coloured according to the velocity magnitude, normalized beween $0$ and $1$, which show that the two vortices begin to ``interact''. The edges of the cubes shown here run from $\left\lbrace x,y,z \right\rbrace \in [10,90]$.}
	\label{fig:BurgersVortex}
\end{figure}

We generate two vortices, on a grid of $100^3$, with $a=0.1$, $\nu = 0.025$, $\Gamma = 15$ and $\zeta=0.004$ (i.e. $\sim 1\%$ of the maximum velocity magnitude), all quantities being presented in arbitrary units. The actual values being used here are not important, as we simply intend to generate a velocity field with a Burgers vortex \textit{structure}. The resulting vortex has a core region with solid body rotation up to $5$ units (grid cells) from its axis, and the velocity magnitude in the potential flow region (with swirling motion) decays to approximately $40\%$ and $10\%$ of the maximum velocity magnitude within $15$ and $30$ grid cells from the axis, respectively. Both vortices are multiplied with the Gaussian function $\mathcal{G}$, generated with $\sigma = 5$. The vortices are rotated at arbitrary azimuthal and elevation angles of $(0.6, 0.25)$ and $(-0.45, -0.3)$, measured in radians. These vortices are then superposed by adding their velocity fields, with their centres placed at $(40,40,40)$ and $(70,70,70)$. The resulting velocity field is shown in figure \ref{fig:BurgersVortex}. Since the vortices are placed close to each other, their velocity fields begin to entangle and interact. The larger structure of the Burgers vortex-pair, however, is distinct from the Oseen vortex-pair in Section \ref{subsec:Oseen}, and it does not have the same kind of symmetries and anti-symmetries.

\begin{figure}
	\begin{subfigure}{0.5\linewidth}
  \centerline{\includegraphics[width=\linewidth]{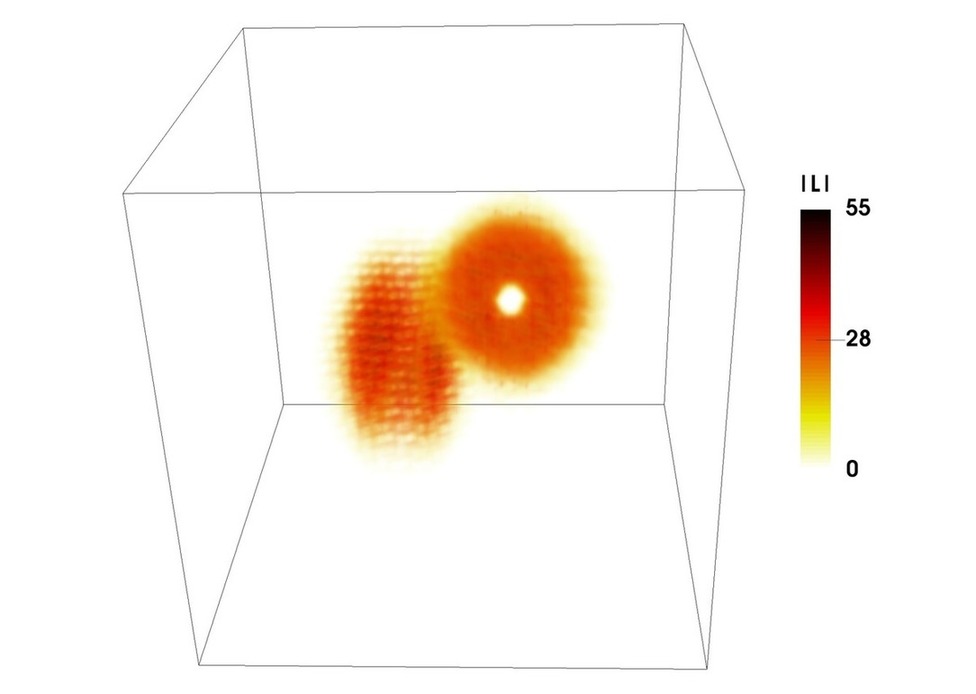}}

  	\caption{$\mathbf{L}$ correlation}
  \end{subfigure}
  	\begin{subfigure}{0.5\linewidth}
  \centerline{\includegraphics[width=\linewidth]{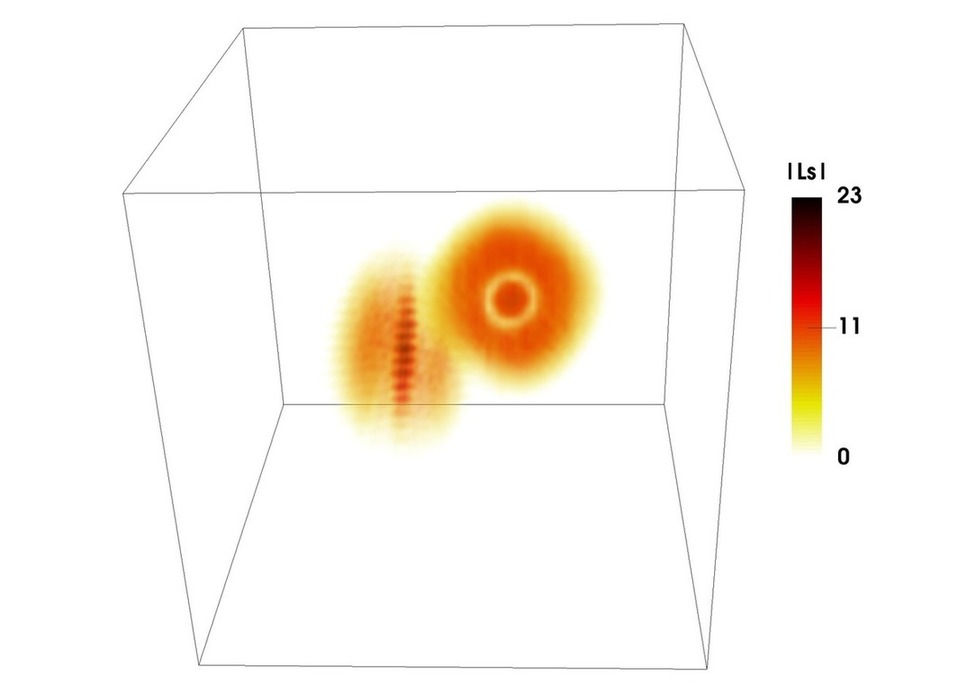}}

  	\caption{$\lsvec$ correlation}
  \end{subfigure}

   	\begin{subfigure}{0.5\linewidth}
  \centerline{\includegraphics[width=\linewidth]{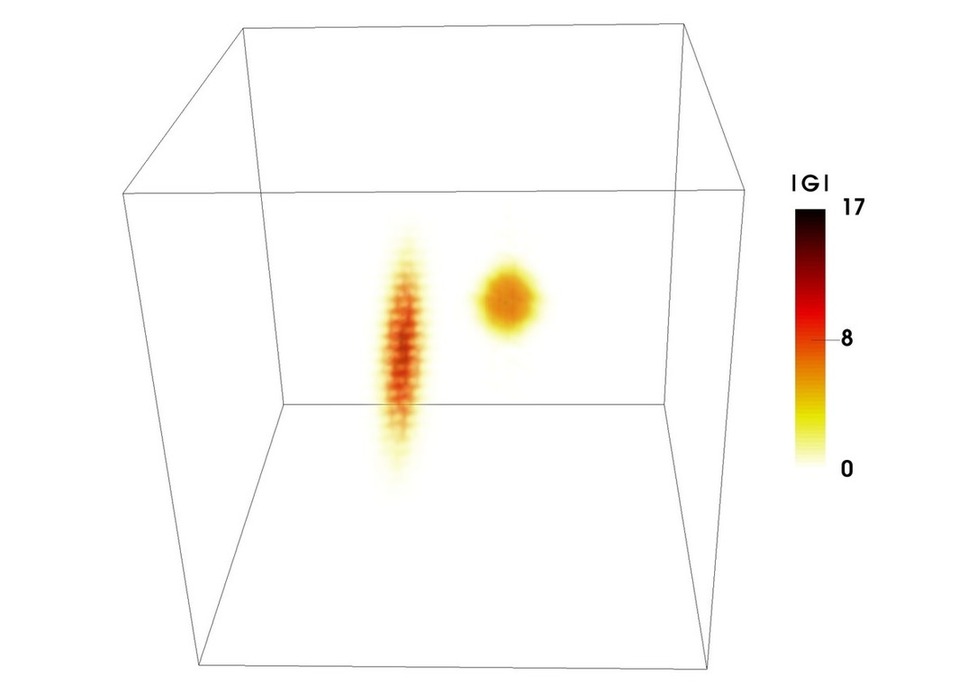}}

  	\caption{$\mathbf{G}$ correlation}
  \end{subfigure}
  \begin{subfigure}{0.5\linewidth}
  \centerline{\includegraphics[width=\linewidth]{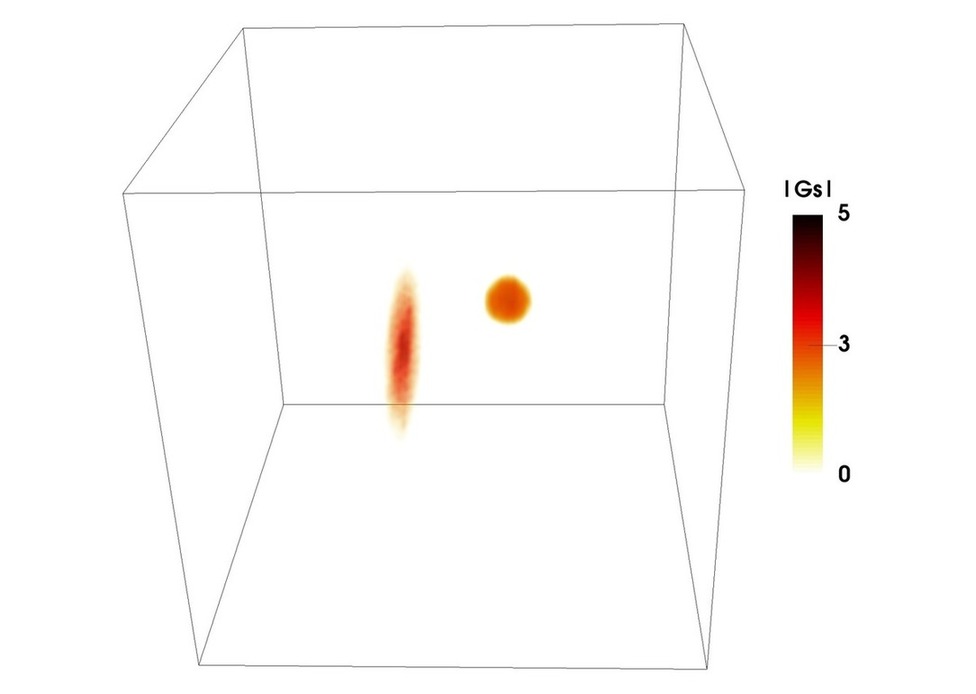}}

  	\caption{$\gs$ correlation}
  \end{subfigure}
  
   	\begin{subfigure}{0.5\linewidth}
  \centerline{\includegraphics[width=\linewidth]{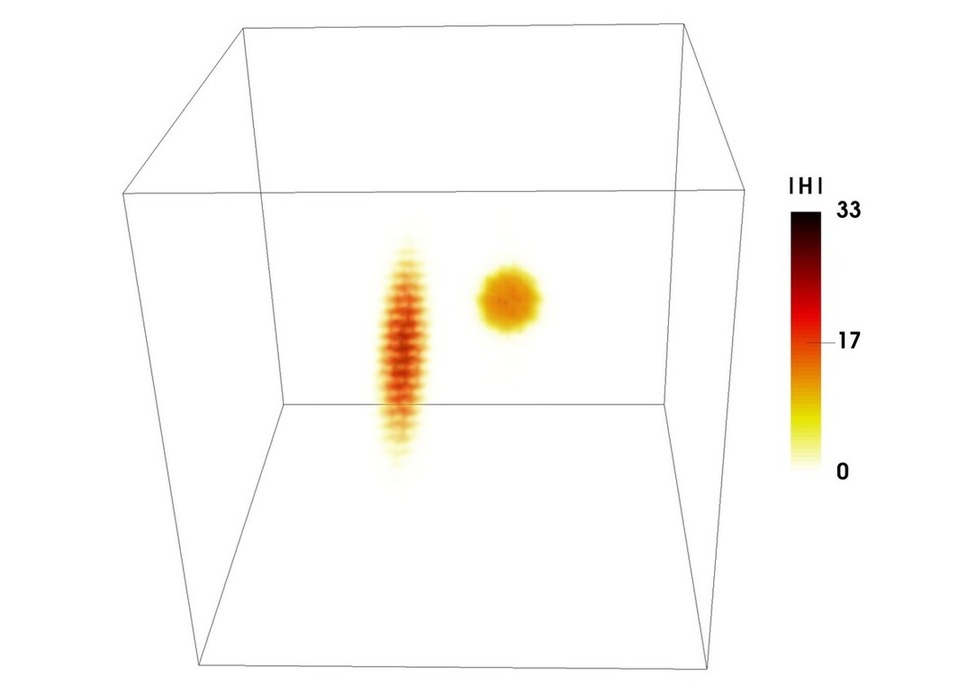}}

  	\caption{$\mathbf{H}$ correlation}
  \end{subfigure}
  	\begin{subfigure}{0.5\linewidth}
  \centerline{\includegraphics[width=\linewidth]{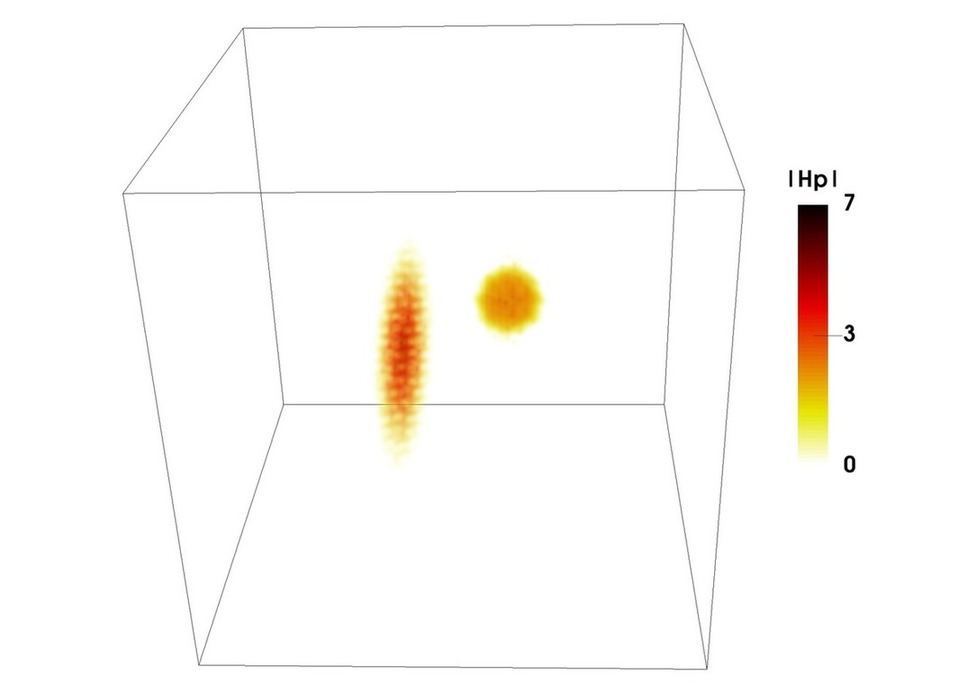}}

  	\caption{$\mathbf{H^p}$ correlation}
  \end{subfigure}
  \caption{Amplitude of all correlations calculated for the two Burgers vortices are shown for view 1 in figure \ref{fig:BurgersVortex}, which gives a simultaneous look at the axis of one vortex and the core region of the other. The edges of the cubes shown here run from $\left\lbrace x,y,z \right\rbrace \in [10,90]$.}
\label{fig:BurgersVortexCorrelationsTop}
\end{figure}

Figure \ref{fig:BurgersVortexCorrelationsTop} shows the amplitude of all the correlations, integrated over a length of $\Lambda=12$ for the Burgers vortex-pair. \sid{The value of $\Lambda$ is large enough to encompass most of the structure of each vortex, and the results were found to remain qualitatively unchanged for $\Lambda=10$ and $\Lambda=15$, which only causes a change in the magnitude of the correlations.} The features of the correlation fields strongly reflect the behaviour of their one-dimensional analogues as was presented for the Oseen vortex-pair \sid{(see figures \ref{fig:dotCorrelationSmallLambda} and \ref{fig:oneDimensionalOseen-GCorrLambda}, where $\Lambda$ also encompasses, roughly, the structure of each vortex)}. First, panels (a) and (b) show the $\lvec$ and $\lsvec$ correlations. The correlation regions are well aligned with the axes of the vortices, and have a size slightly smaller than the extent of the swirling-flow region. $\lvec$ goes to zero at the vortex core where the velocity is also zero. $\lsvec$ yields a large value at the vortex core, across which the flow is highly anti-correlated, surrounding which is a thin region of zero correlation, and an outer region of finite correlation corresponding to the potential flow region of the vortex. \sid{Both these correlations identify the outer swirling-flow region where the streamlines are well-aligned, and locally parallel; however, in the core region of the vortex they are very different. Due to their definitions, near the center of a strong swirling flow $\lsvec$ has a large magnitude whereas $\lvec$ has a magnitude close to zero.} Panels (c) and (d) show the $\g$ and $\gs$ correlations, both of which yield thin, elongated correlation profiles aligned with the axes of the vortices, while the $\gs$ correlation is sharper. This again reflects that, at the core of the vortices, the vorticity vectors are well aligned. Panels (e) and (f) show $\h$ and $\hp$, which also yield strong correlation profiles at the cores of the vortices, aligned with the axes of the vortices. This is because the swirling velocity field is associated with the vorticity at the vortex core regions. 

\sid{We also explored larger integration lengths, assuming a periodic field, similarly to what was done for the one-dimensional flow. We found that for larger integration lengths, of $\Lambda \approx N_x/2$, the correlations also begin to recognize non-local symmetries (due to periodicity) in the twin Burgers vortex velocity field, similarly to the one-dimensional Oseen vortex pair (see figures \ref{fig:dotCorrelationExamplesRAW} and \ref{fig:oneDimensionalOseen-GCorr}), which is not additionally shown here. This shows that the correlations, depending on the integration length, can identify large complex structures, containing smaller sub-structures; however, this will not be further explored here.}

\sid{We note that, similar to the Oseen vortex examples, in the vortex core there exists a strong relation between $\lsvec$, $\g$, $\gs$, $\h$ and $\hp$, with all of them having large magnitudes, whereas $\lvec$ has a magnitude close to zero. On the contrary, in the outer regions of the vortex, there exists a strong relation between $\lsvec$ and $\lvec$, with both of them having high magnitudes, whereas $\g$, $\gs$, $\h$ and $\hp$ have a magnitude close to zero. Essentially, over the entire vortex, where the swirling flow is strong, $\lsvec$ has a large magnitude, and this region of high $|\lsvec|$ encloses the vortex core where $|\lvec|$ is close to zero and $|\g|$, $|\gs|$, $|\h|$ and $|\hp|$ are also high. This happens because the swirling flow region is self-induced by the high vorticity at the core of the vortices leading to a highly anti-correlated velocity field around the center of the vortex, where the velocity is close to zero; moving away from the center of the vortex, the vorticity decays rapidly but the swirling flow is still strong, and the velocity field, locally, is highly correlated due to the parallel streamlines. Note that this is unlike the example of the Taylor-Green flow, where the jet-like high $|\lvec|$ regions are externally-induced by the \textit{interaction} of vortices, leading to symmetries and anti-symmetries in the vorticity field in this region, which can result in high magnitudes of $\gs$ even with low magnitudes of $\g$. The similarities and differences between the different correlation definitions play a key role in the subsequent analysis of turbulence field structures.}

\sid{Overall, we find that the correlation definitions are adept at identifying typical velocity and vorticity field patterns, also in fields arbitrarily aligned with respect to directions along which the correlations are evaluated.}

\section{Correlations applied to homogeneous isotropic turbulence}\label{sec:Turbulence}
After applying the correlations to canonical flows, we now study how these ideas fare for real turbulence vector fields. A turbulence velocity field is considered, typically, to have structures across multiple scales, while the vorticity field mainly comprises structures at the smaller scales. These fields, further, are highly complex and irregular. Applying the correlations to instantaneous snapshots of turbulence vector fields, as obtained from direct numerical simulations of homogeneous isotropic turbulence, reveals their potential in educing coherent structures. In this section, we begin with a brief description of the simulation method used to generate the turbulence data. Since the correlations yield three-tuple fields containing information about the structure of the flow, we first describe how the correlation fields look qualitatively in comparison to the vector fields they are based on, i.e. the velocity and the vorticity fields. We then describe the statistics of the correlation fields, like the PDFs, CDFs, spectral characteristics and their spatial organization. \sid{In the next section, we focus on individual structures and unravel their Biot-Savart composition.}

\subsection{Simulation details and dataset}
For this study, we use a dataset from DNS simulations of homogeneous isotropic turbulence, for which the Navier-Stokes equations with a body force $\mathbf{F}$ (as given below) are solved numerically
\begin{equation}
\frac{\partial \mathbf{u}}{\partial t} + \left(\mathbf{u}\cdot\nabla\right) \mathbf{u} = -\frac{\nabla p}{\rho} + \nu \nabla^2 \mathbf{u} + \frac{\mathbf{F}}{\rho}
\label{eq:NS-Mom}
\end{equation}
\begin{equation}
\nabla \cdot \mathbf{u} = 0
\label{eq:NS-Con}
\end{equation}

Turbulence is generated in a periodic box by means of low wavenumber forcing, which is divergence-free by construction and is concentrated over a range of Fourier modes. It is of the form given by \cite{biferale2011lattice}, and has properties similar to that devised by  \cite{alvelius1999random} and \cite{ten2006application}, which can be written as
\begin{eqnarray}
F_x &= \sum_{k={k_a}}^{{k_b}} \rho A(k) \left[ \sin(2\pi k y + \phi_y(k)) + \sin(2\pi k z + \phi_z(k)) \right] \nonumber \\
F_y &= \sum_{k={k_a}}^{{k_b}} \rho A(k) \left[ \sin(2\pi k x + \phi_x(k)) + \sin(2\pi k z + \phi_z(k)) \right] \nonumber \\
F_z &= \sum_{k={k_a}}^{{k_b}} \rho A(k) \left[ \sin(2\pi k x + \phi_x(k)) + \sin(2\pi k y + \phi_y(k)) \right]
\label{eq:forcing}
\end{eqnarray}
The forcing is stochastic (white noise) in time, which is achieved by varying each $\phi(k)$ randomly, and the force is distributed over a small range of wavenumers, given by $k_a \leq k \leq k_b$ (for this study we fix $k_a = 1, k_b = 8$), and the amplitude $A(k)$ of each of these wavenumbers is a Gaussian distribution in Fourier space, centered around a central forcing wavenumber $k_f$, given as
\begin{equation}
A(k) = A\exp\left( -\frac{\left(k -k_f\right)^2}{c}\right)
\end{equation}
where $c$ sets the width of the distribution ($c=1.25$ here), $k_f = 2$, and $A$ is the forcing amplitude. We solve equations \ref{eq:NS-Mom} and \ref{eq:NS-Con} with a standard lattice-Boltzmann (LB) solver, incorporating the turbulence forcing as per equation \ref{eq:forcing}. This method has been used before for simulating homogeneous isotropic turbulent flows of various kinds \citep{ten2004fully,ten2006application,biferale2011lattice,perlekar2012droplet,mukherjee2019droplet}.

The simulation is performed in a periodic box of size $(2\pi)^3$ resolved over $N^3$ grid points along each direction, all units being dimensionless, hence resolving a range of wavenumbers from $k = 2\pi/N$ (i.e. the largest scale has a length $N\ [lu]$) to $k=2\pi/2=\pi$ (i.e. the smallest scale has a length $2\ [lu]$). Since we simulate homogeneous isotropic turbulence, by definition, all physical quantities are fluctuating and do not have a mean value, i.e. $\mathbf{u}=\mathbf{u}^\prime$ and $\boldsymbol{\omega}=\boldsymbol{\omega}^\prime$. The Kolmogorov scale is defined as $\eta \sim \left( \nu^3/\epsilon\right)^{1/4}$, where $\nu$ and $\epsilon$ are the kinematic viscosity and energy dissipation rate, respectively. We adhere to the criterion for a DNS, as given by \cite{moin1998direct}, i.e. $\kmax\eta > 1$. The Taylor microscale is calculated as 
\begin{equation}
\lambda = \left( \frac{15\nu {u^\prime}^2}{\ang{\epsilon}}\right)^{1/2}
\label{eq:TaylorLambda}
\end{equation}
where $u^\prime$ is the root-mean-square velocity. The average rate of energy dissipation $\ang{\epsilon}$ is calculated as $\ang{\epsilon} = \nu\ang{\omega^2}$, where $\ang{\omega^2}$ is the average enstrophy. Note that the enstrophy $\omega^2 = \boldsymbol{\omega}\cdot\boldsymbol{\omega}$ is analogous to the turbulence kinetic energy $E_k=\mathbf{u}\cdot\mathbf{u}/2$. For homogeneous isotropic turbulence, since $u^\prime=v^\prime=w^\prime$, we have $E_k = 3{u^\prime}^2/2$ or $u^\prime = \sqrt{2E_k/3}$. The root-mean-square vorticity, $\omega^\prime$, is obtained as $\ang{\boldsymbol{\omega}\cdot\boldsymbol{\omega}}^{1/2}$. In general, $E_k$ and $\epsilon$ (apart from $\nu$) are average measures of $u^\prime$ and $\omega^\prime$, respectively. The large eddy turnover timescale is given as $T^\star = \mathcal{L}/u^\prime$, where $\mathcal{L}$ is the forcing lengthscale, given as $\mathcal{L} = N/k_f$. Using $\lambda$, the Taylor Reynolds number is calculated as
\begin{equation}
Re_\lambda = \frac{u^\prime \lambda}{\nu}
\label{eq:TaylorRe}
\end{equation}
and the Kolmogorov timescale is given as
\begin{equation}
\tau_k = \left(\frac{\epsilon}{\nu}\right)^{-1/2}
\end{equation}

The turbulence simulation (parameters given in table \ref{tab:Simulation}) is performed for a fluid  initially at rest, to which the turbulence force is applied. After a brief transient duration, turbulence becomes well developed and attains a statistical steady-state, i.e. with a balance of power input and energy dissipation. The simulation is then run for several additional large eddy timescales ($\sim 20-30 T^\star$), during which around $\sim 20$ field snapshots are retained for analysis, all separated by $50\tau_k$, to give converged statistical results.

\begin{table}
  \begin{center}
\def~{\hphantom{0}}
  \begin{tabular}{lccccccccccccc}
  	 $N^3$ & $k_f$ & $\nu$ & $u^\prime$ & $\omega^\prime$ & $\ang{E_k}$ & $\ang{\epsilon}$ & $\lambda$ & $Re_\lambda$ & $\eta$ & $\tau_k$\\ [3pt]
   	 $256^3$ & $2$ & 0.0047 & 0.034 & 0.0103 & $1.8\times 10^{-3}$ & $5.0\times 10^{-7}$ & 13 & 95 & 0.67 & 97\\
  	 \end{tabular}
  \caption{Simulation details, with all quantities presented in dimensionless lattice units $[lu]$, average kinetic energy $\ang{E_k} = \left(\sum_k E(k)\right)/N$, and the average rate of energy dissipation $\ang{\epsilon}=(\sum_k 2\nu k^2E(k))/N$.}
  \label{tab:Simulation}
  \end{center}
\end{table}

\begin{figure}
  \centerline{\includegraphics[width=0.7\linewidth]{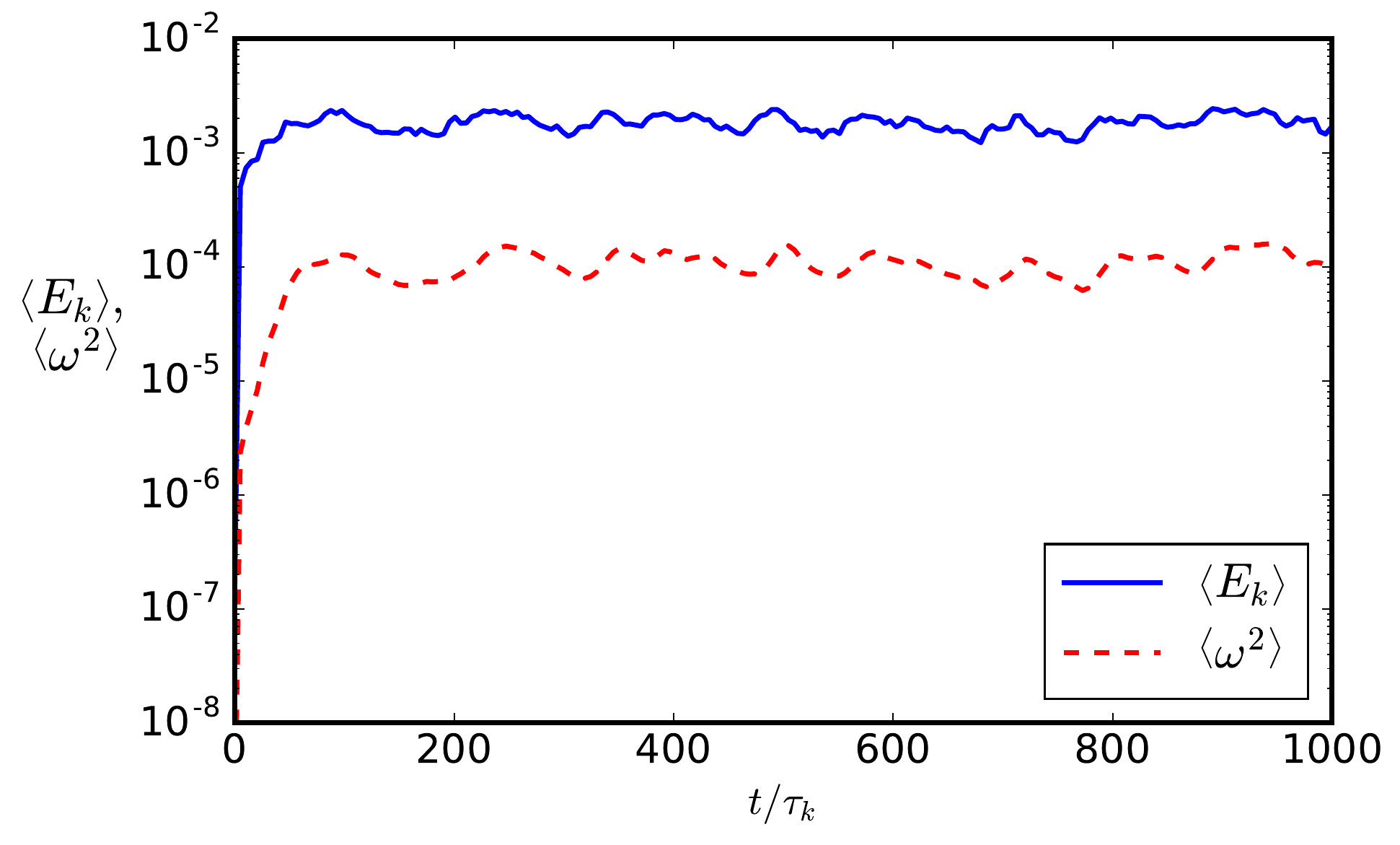}}

  \caption{Time evolution of the volume averaged turbulence kinetic energy $\ang{E_k}$ and enstrophy $\ang{\omega^2}$. Both quantities attain a steady-state value, reflecting a developed turbulence state.}
\label{fig:TurbulenceDataEvolution}
\end{figure}

Figure \ref{fig:TurbulenceDataEvolution} shows the evolution of $\ang{E_k}$ and $\ang{\omega^2}$. Both quantities attain their steady-state values within a short transient phase, $\sim 100\tau_k$, after which they continue to oscillate around their temporal mean values. Beyond $100\tau_k$, turbulence is well developed, with a sufficient separation of scales. The small temporal oscillation of $\ang{E_k}$, due to the finite volume of the simulation, further manifests in the temporal oscillation of $\ang{\omega^2}$, due to the turbulence dynamics \citep{pearson2004delayed,biferale2011lattice}.

\begin{figure}
	\begin{subfigure}{0.5\linewidth}
  \centerline{\includegraphics[width=\linewidth]{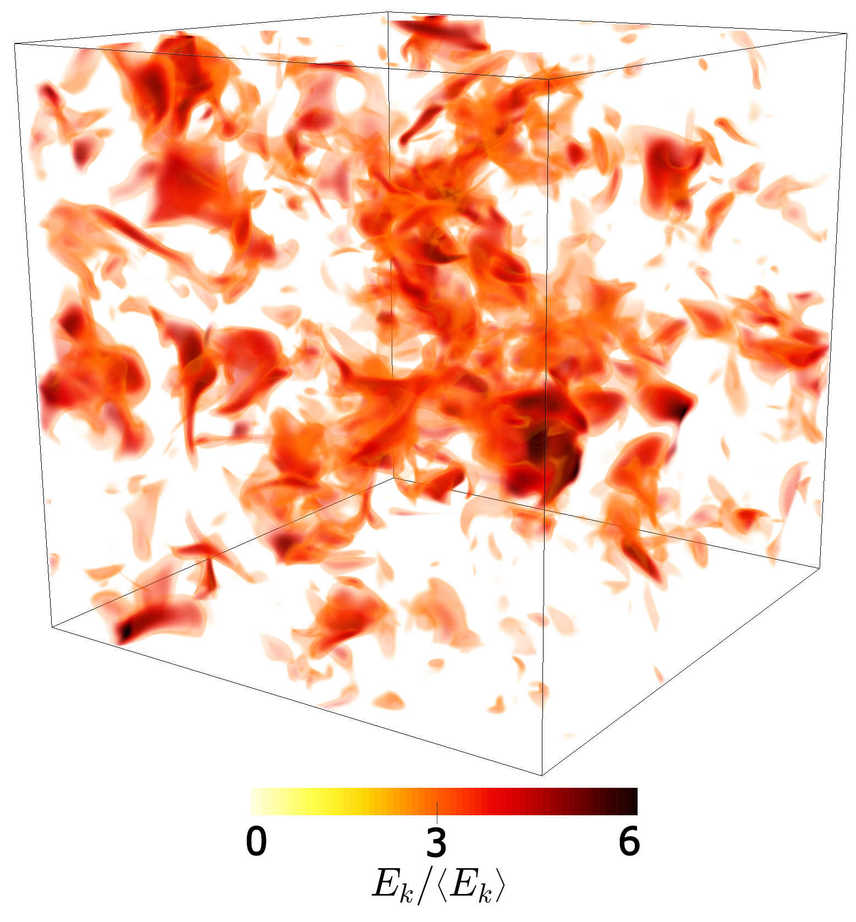}}

  \caption{}
   \end{subfigure}
   	\begin{subfigure}{0.5\linewidth}
  \centerline{\includegraphics[width=\linewidth]{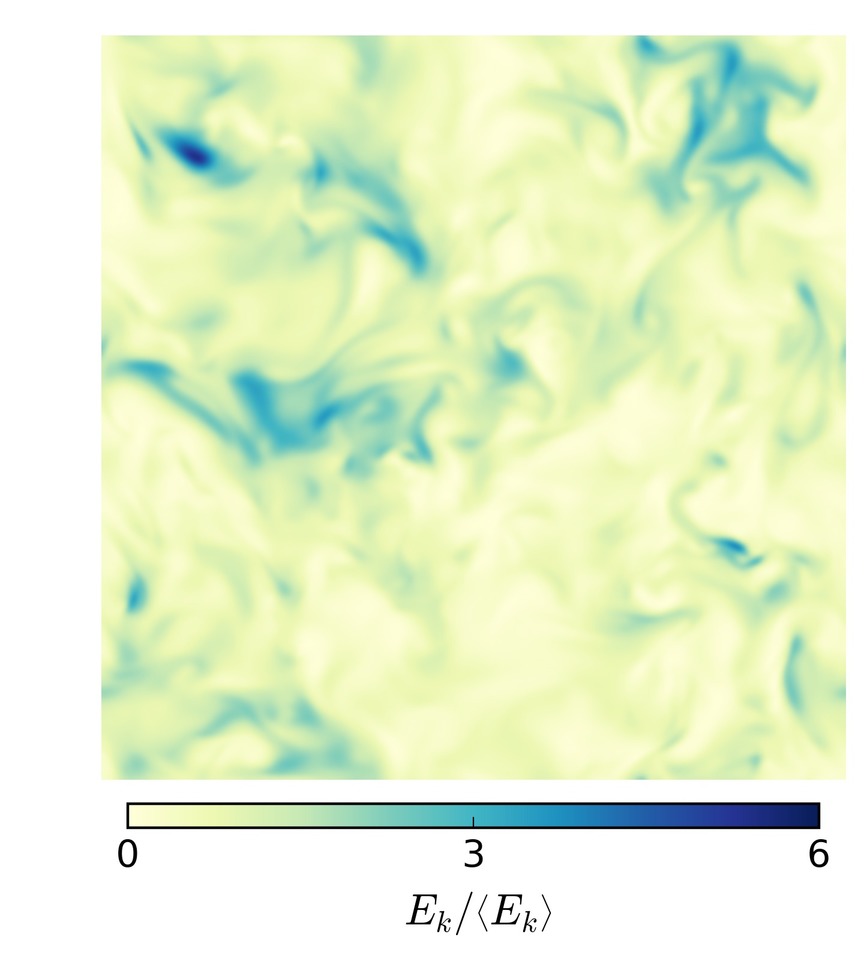}}

  \caption{}
   \end{subfigure}
   
   	\begin{subfigure}{0.5\linewidth}
  \centerline{\includegraphics[width=\linewidth]{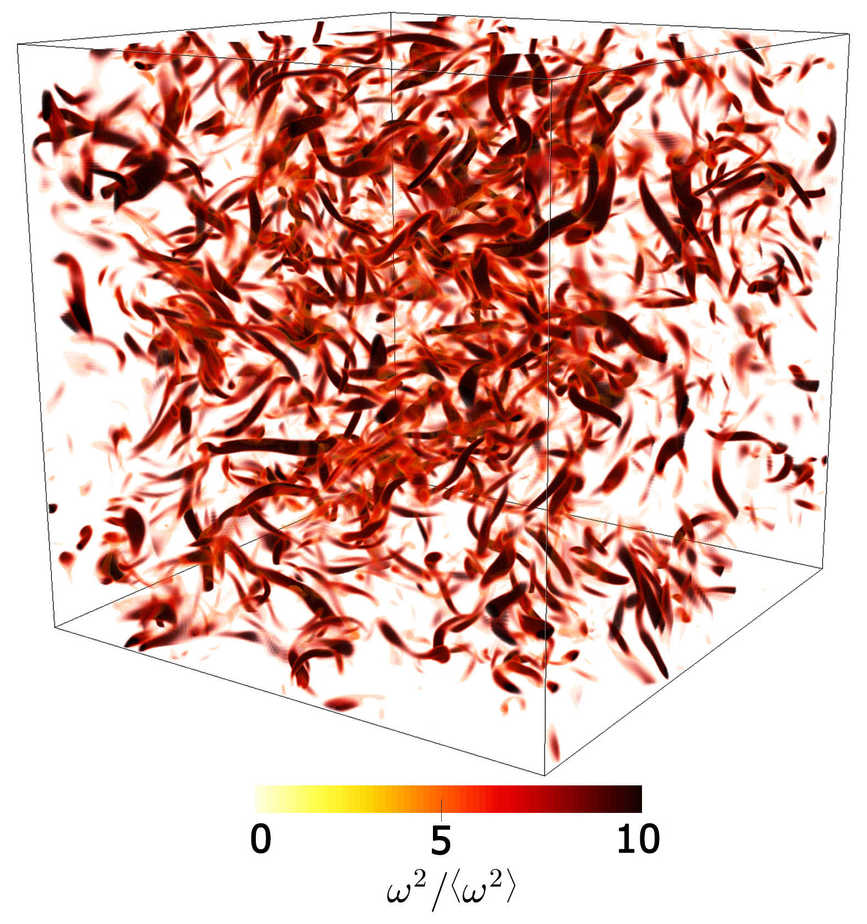}}

  \caption{}
   \end{subfigure}
      	\begin{subfigure}{0.5\linewidth}
  \centerline{\includegraphics[width=\linewidth]{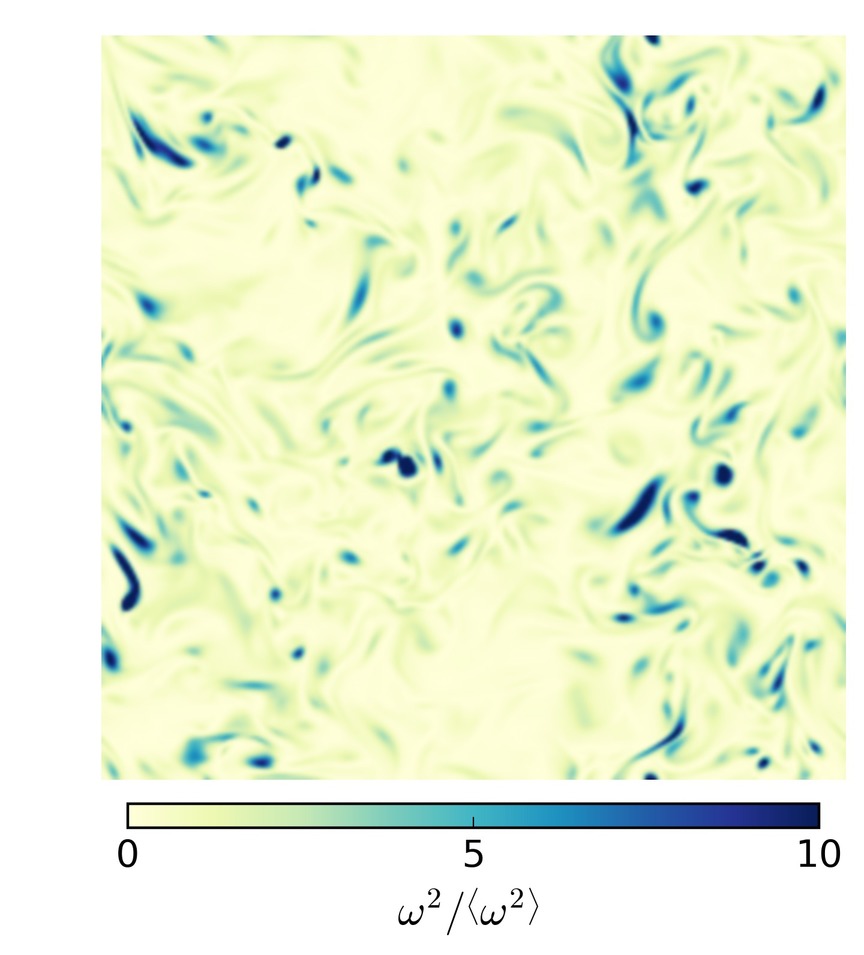}}

  \caption{}
   \end{subfigure}
  \caption{Snapshots of turbulence fields, when the flow is fully developed (at $t=500\tau_k$), show qualitative features of the kinetic energy $E_k$, and the enstrophy, $\omega^2$. Panel (a) shows a 3D volume rendering (at a resolution of $256^3$) of the $E_k$ field, which contains, small to intermediate, irregular structures that are distributed throughout the volume. Panel (b) shows a 2D cross-section of the $E_k$ field (at a resolution of $256^2$) at an arbitrary plane of the 3D simulation domain. Panel (c) shows a volume rendering of the $\omega^2$ field, which is markedly different from the $E_k$ field, as enstrophy is concentrated at the smaller scales, forming tube-like structures (also called ``worms''). Panel (d) shows a planar cross-section of $\omega^2$, which shows that most of the field has low values, interspersed with small regions of concentrated enstrophy. Both the fields have been normalized with their respective volume-averaged quantity, i.e. $\ang{E_k}$ and $\ang{\omega^2}$.}
\label{fig:TurbulenceData}
\end{figure}

Figure \ref{fig:TurbulenceData} shows a snapshot of the turbulence kinetic energy $E_k$ and enstrophy $\omega^2$ fields, as 3D volume renderings and planar cross-sections, at a simulation time of $500\tau_k$. Typical features of the kinetic energy and enstrophy can be seen, where the kinetic energy is distributed over a range of length scales, and forms diffused, small and intermediate sized, irregular structures. Enstrophy (and vorticity in general) is concentrated at the smaller scales, in spatially intermittent tube-like structures, also called ``worms''.

\begin{figure}
  \begin{subfigure}{0.5\linewidth}
  \centerline{\includegraphics[width=\linewidth]{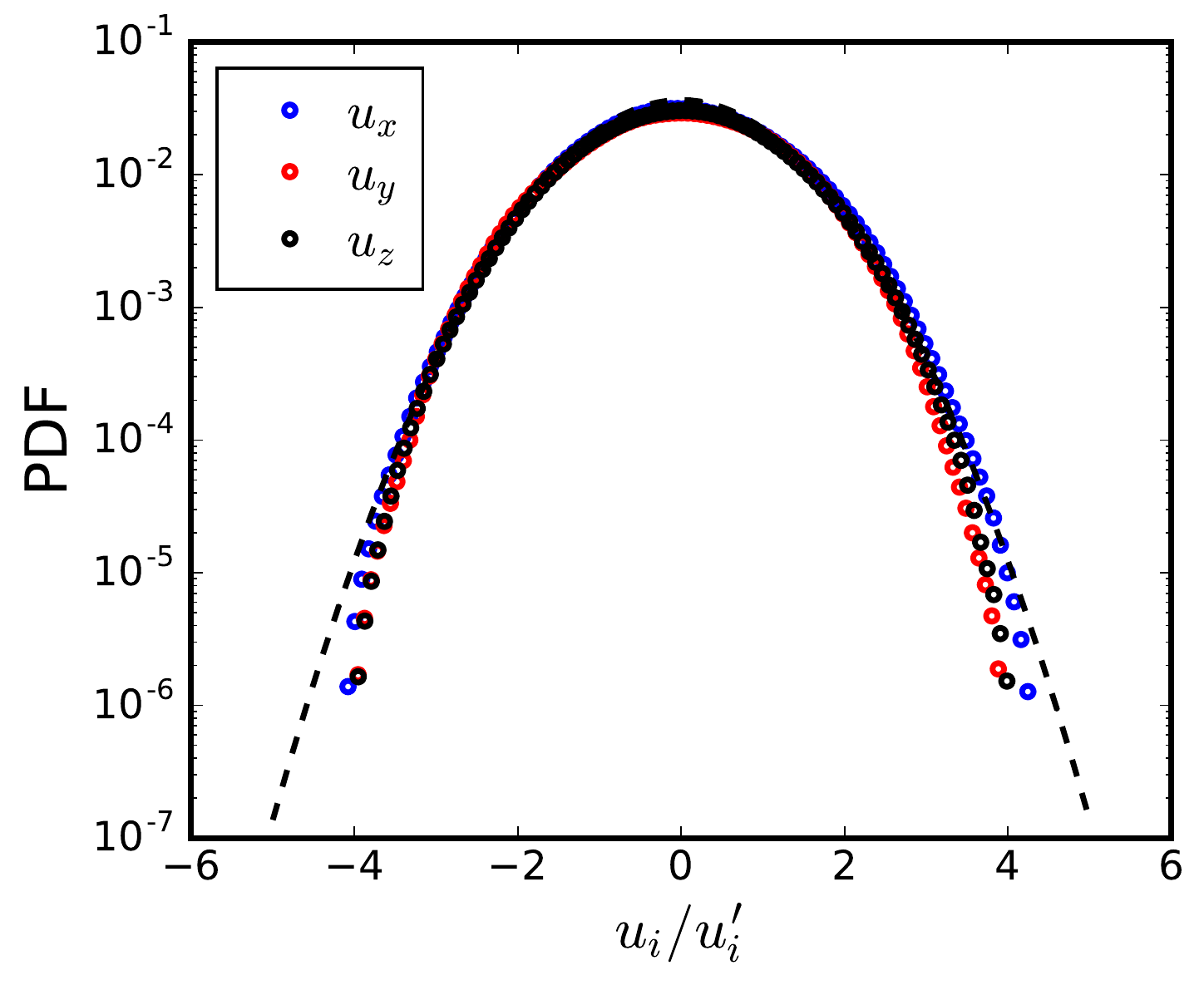}}

  	\caption{}
  \end{subfigure}\quad
  \begin{subfigure}{0.5\linewidth}
  \centerline{\includegraphics[width=\linewidth]{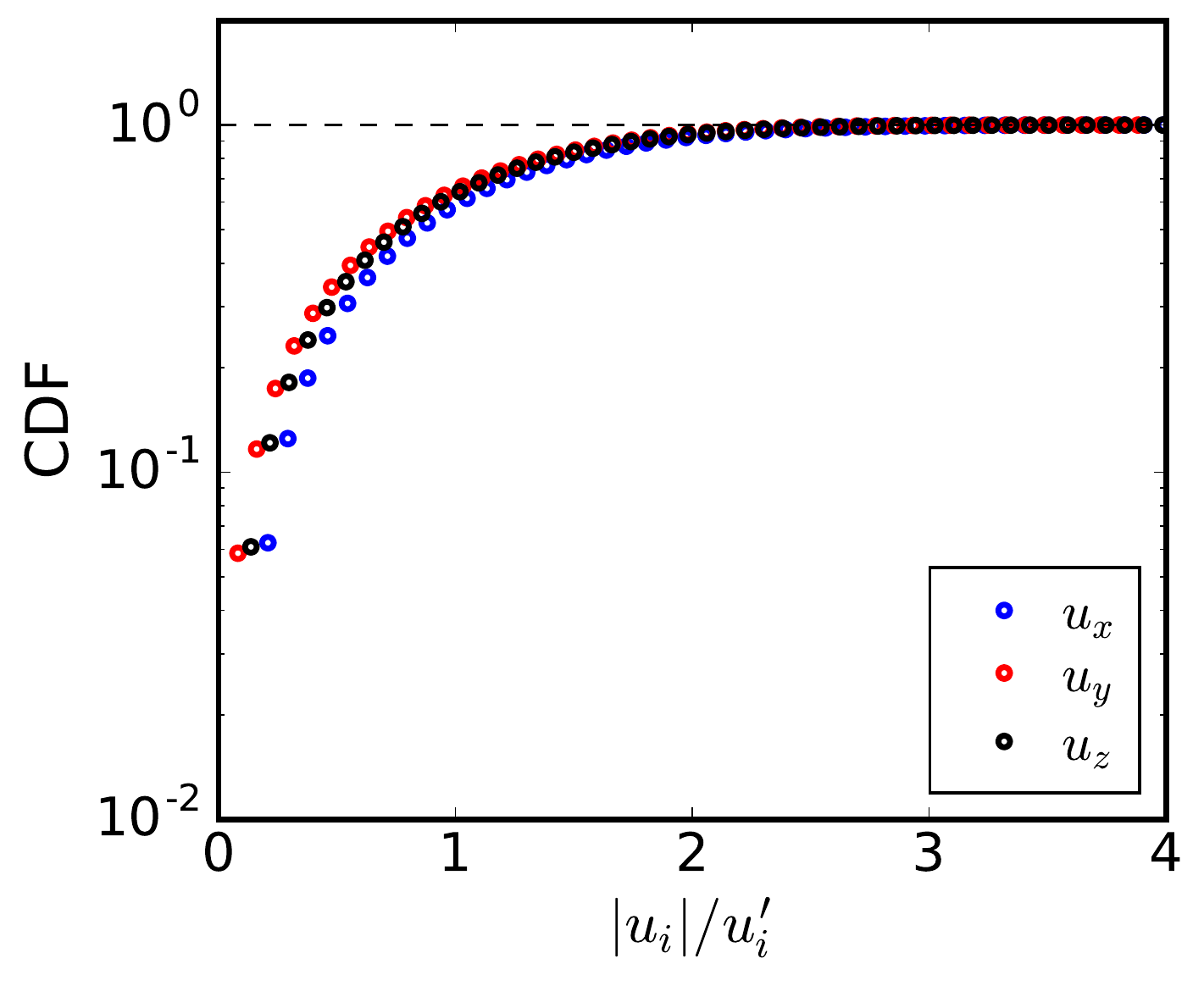}}

  	\caption{}
  \end{subfigure}
  
  \begin{subfigure}{0.5\linewidth}
  \centerline{\includegraphics[width=\linewidth]{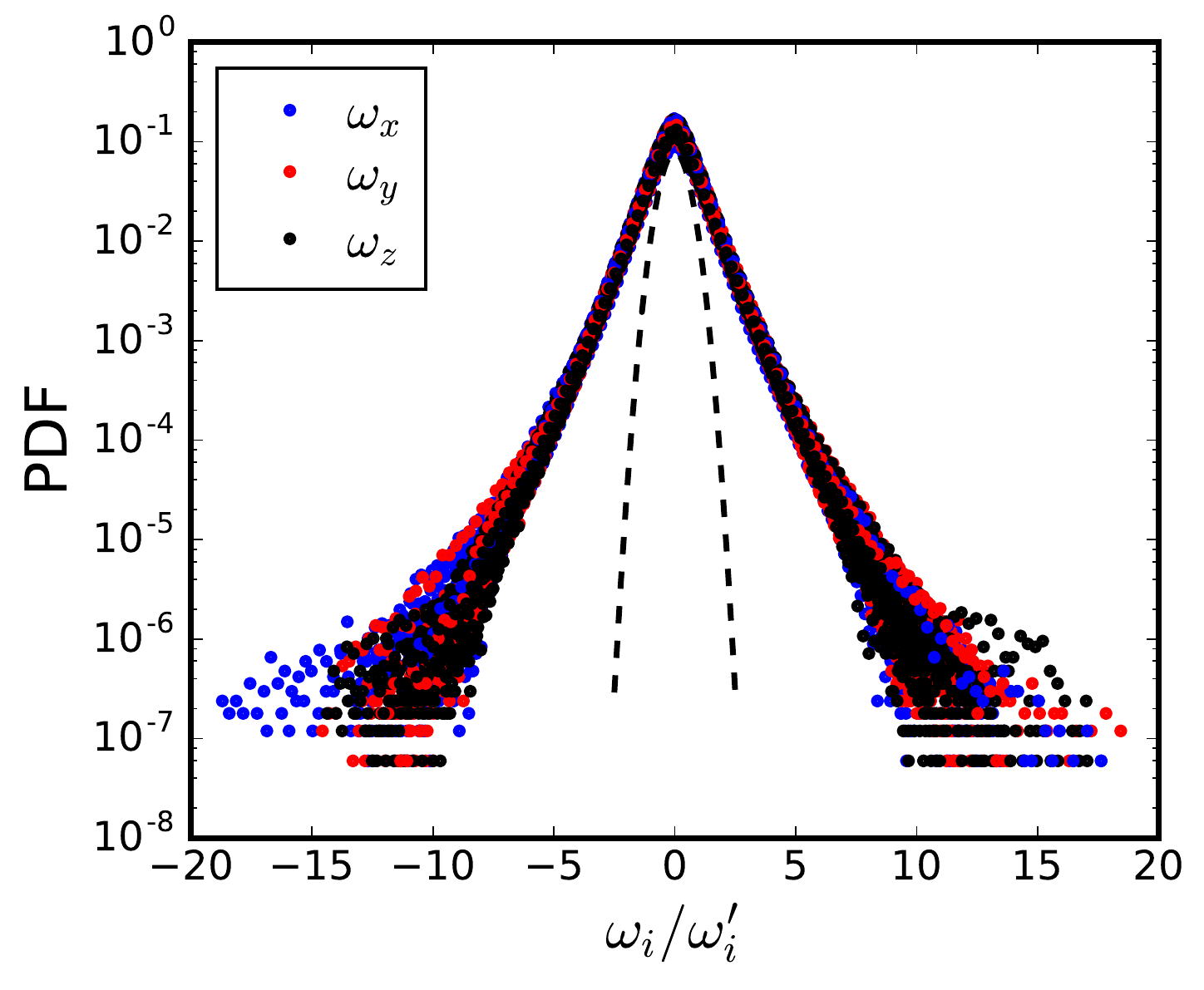}}

  	\caption{}
  \end{subfigure}
  \begin{subfigure}{0.5\linewidth}
  \centerline{\includegraphics[width=\linewidth]{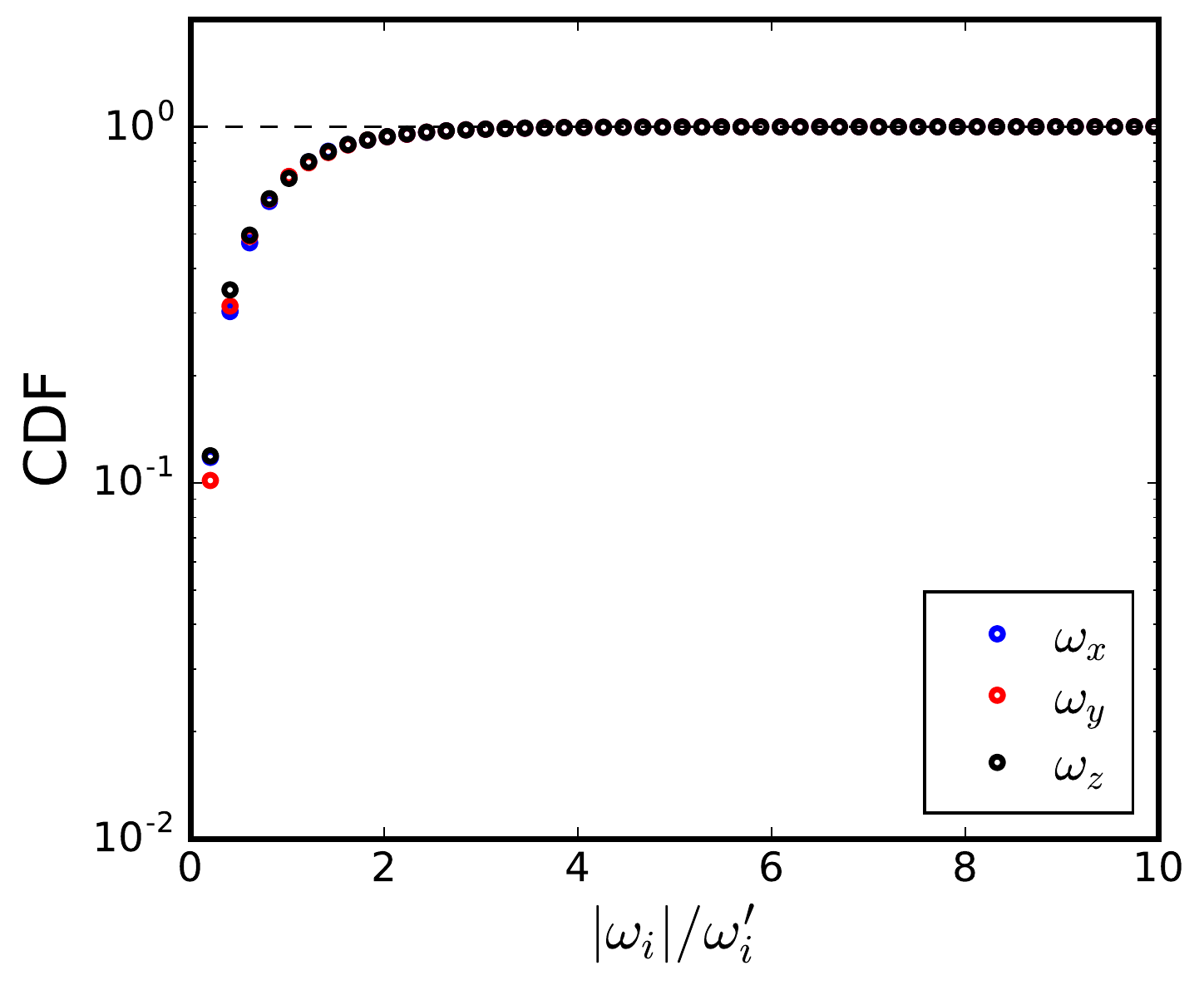}}

  	\caption{}
  \end{subfigure}
  \caption{Probability and Cumulative Distribution Functions (PDF and CDF), shown for the three velocity and vorticity components. The dashed line in panels (a) and (c) show a typical Gaussian distribution.}
\label{fig:TurbulenceData-PDF}
\end{figure}

Figure \ref{fig:TurbulenceData-PDF} shows the probability and cumulative distribution functions (PDFs and CDFs, respectively), of the three velocity and vorticity components. These profiles have been obtained using $20$ field snapshots, all separated by $50\tau_k$. Figure \ref{fig:TurbulenceData-PDF}(a) shows that the velocity components follow a Gaussian distribution (shown as the dashed line), and that the velocity fluctuations are not extreme (here they range from $-4<u_i/u_i^\prime<4$). Figure \ref{fig:TurbulenceData-PDF}(b) shows the CDFs of the velocity components, where $65\%$ and $97\%$ of the velocity has a magnitude below $u_i^\prime$ and $2u_i^\prime$, respectively. Extreme values of the velocity, around $|u_i|>3u_i^\prime$ occupy a very small fraction of the total velocity field. Similarly, figure \ref{fig:TurbulenceData-PDF}(c) and figure \ref{fig:TurbulenceData-PDF}(d) show the PDFs and CDFs of the vorticity components. The PDFs show the typical long-tail distribution of vorticity, which is highly non-Gaussian. The extent of these tails gives a measure of the intermittency in the vorticity field, where increasingly extreme values can occur with a low probability. The CDFs of the vorticity show that most of the vorticity field has a low value, with $70\%$ and $95\%$ of the field below $\omega_i^\prime$ and $2\omega_i^\prime$, respectively. In this regard, the vorticity field has a similar composition as the velocity field; the difference being that the vorticity can also assume much more extreme values (even $\sim 18\omega_i^\prime$ in this case).

\sid{We note that the vorticity field is often classified into a few ``ranges''. \citet{she1990intermittent,she1991structure} proposed a classification where, based upon the amplitude and structure of the vorticity streamlines, the vorticity field is divided into ``low-vorticity'', ``moderate-vorticity'' and ``high-vorticity'' ranges. According to their classification, ``high-vorticity'' ($\omega \gg \omega^\prime$), which occupies a very small fraction of the volume and forms the long tails of the vorticity PDF, forms \textit{vorticity streamlines} that are well-aligned, while the velocity field in the vicinity of these structures has a spiral, swirling motion. ``Moderate-vorticity'' ($\omega > \omega^\prime$), on the other hand, was found to be less organized, the structure of which was described as ``sheet-like'' and ``ribbon-like''. ``Low-vorticity'', at the level of the root-mean-square value ($\omega \sim \omega^\prime$ and $\omega < \omega^\prime$), which occupies most of the volume, was found to form random vorticity streamlines with no apparent structure. In this classification, the well-structured regions of the vorticity field, associated with high $\omega$ values, are highlighted, though they occupy a small fraction of the volume. For our subsequent Biot-Savart analysis in section \ref{sec:BiotSavartStatistics}, where we disentangle the various vorticity contributions to the generation of velocity field structures, we use a similar classification. Since vorticity at the level of the $\omega^\prime$ occupies most of the volume, it is expected to have a significant influence on the Biot-Savart velocity generation. Hence we term this range of vorticity, based upon its amplitude, as ``weak'' and ``intermediate'' background vorticity, since it permeates the volume. We use this classification only as a guideline to interpret our results, as we do not investigate the range of structures of the vorticity field.}

\begin{figure}
  \centerline{\includegraphics[width=0.6\linewidth]{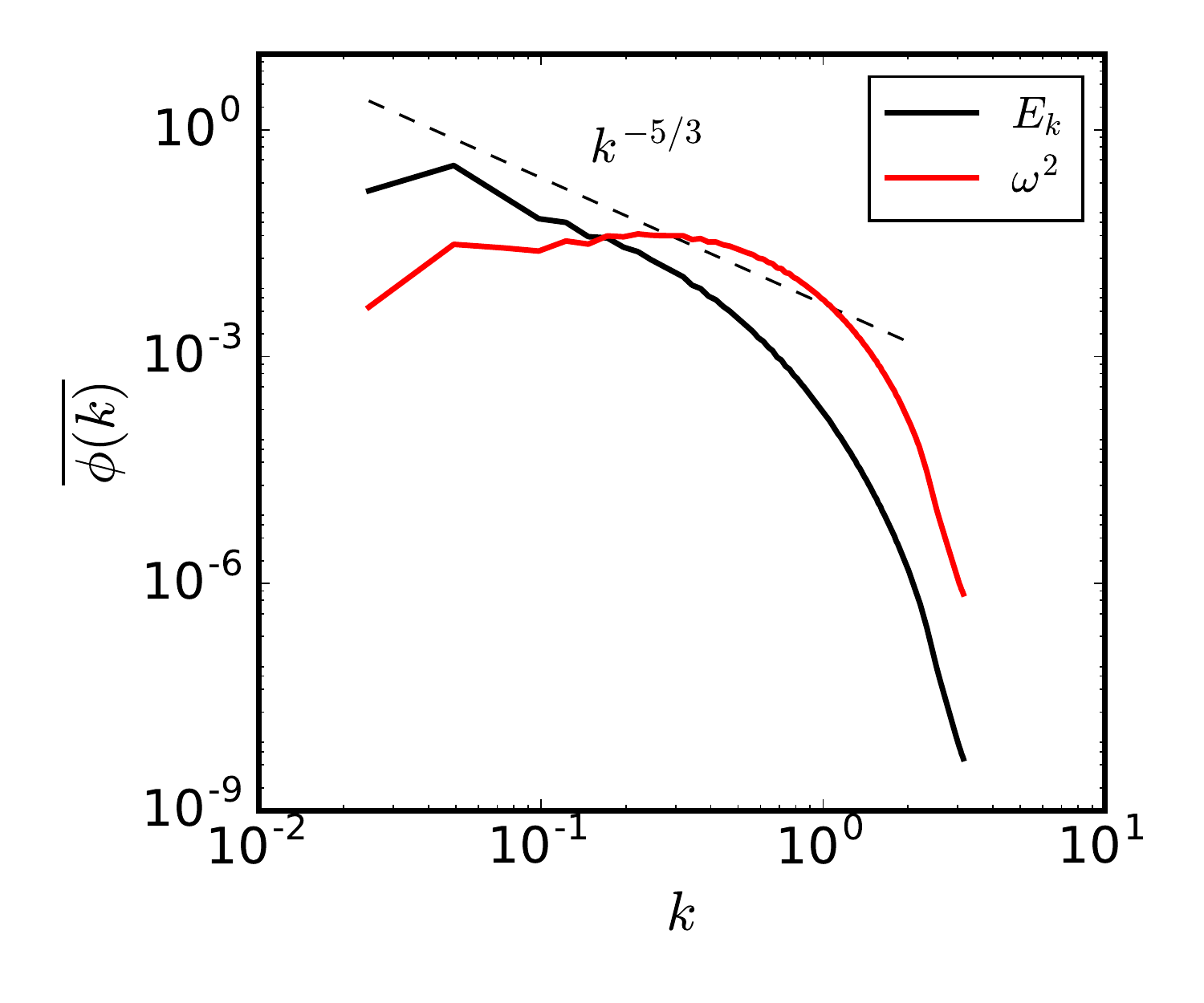}}

  	\caption{Time-averaged, one-dimensional power spectra are shown for the kinetic energy and enstrophy fields.}
  	\label{fig:Turbulence-Spectra}
\end{figure}

The spectra of kinetic energy and enstrophy are calculated using the three-dimensional Fourier transform $\hat{\phi}_{\mathbf{k}}$ of the velocity and vorticity fields, respectively. The three-dimensional spectra are spherically averaged over wavenumber shells $k \in \left[k-1/2, k+1/2\right]$, where $k=\sqrt{\mathbf{k}\cdot\mathbf{k}}$, to give one-dimensional spectra over the scalar wavenumber $k$ as follows
\begin{equation}
\phi(k) = 4\pi k^2\frac{\sum_k |\hat{\phi}_{\mathbf{k}}|^2}{\sum_k 1}
\end{equation}
These one-dimensional spectra are further time averaged over 20 samples separated by $50\tau_k$, to give time-averaged spectral characteristics. Lastly, the spectrum $\phi(k)$ is normalized as $\overline{\phi(k)} = \phi_k/\sum_k \phi(k)$ to facilitate comparison of different quantities, as we are mainly interested in the relative distribution of energy over wavenumber. Figure \ref{fig:Turbulence-Spectra} shows the $E_k$ and $\omega^2$ spectra, where $E_k$ exhibits a well developed inertial range, which follows the $k^{-5/3}$ spectral scaling, while the enstrophy spectra has a small, positive slope, with a broad peak at higher wavenumbers.

\subsection{Qualitative and statistical features of the correlation fields}
The various correlations are calculated for the snapshot of the data presented in figure \ref{fig:TurbulenceData}, for an integration length of $\Lambda = \lambda$. Each correlation field, say $\mathbf{L}$, is normalized by its respective root-mean-square (rms) value, $L^\prime$, which is calculated as $L^\prime = \ang{\mathbf{L}\cdot \mathbf{L}}^{1/2}$. \sid{In general, $\ang{\cdot}$ denotes ensemble averaging, however, in the snapshots of the correlation fields a volume-average was used, in order to normalize the field snapshot with its root-mean-square value at the same instance of time.} The amplitude of a correlation field, say $|\mathbf{L}|$, is simply referred to as $L$. Each of the correlation fields have been shown separately, to highlight their qualitative features, at an arbitrary cross-sectional slice and as a three-dimensional volume rendering. The PDF and CDF of the three components of each correlation have been shown as well, which have been averaged over $20$ field realizations, each separated by $50\tau_k$.

\begin{figure}
	\begin{subfigure}{0.48\linewidth}
  \centerline{\includegraphics[width=\linewidth]{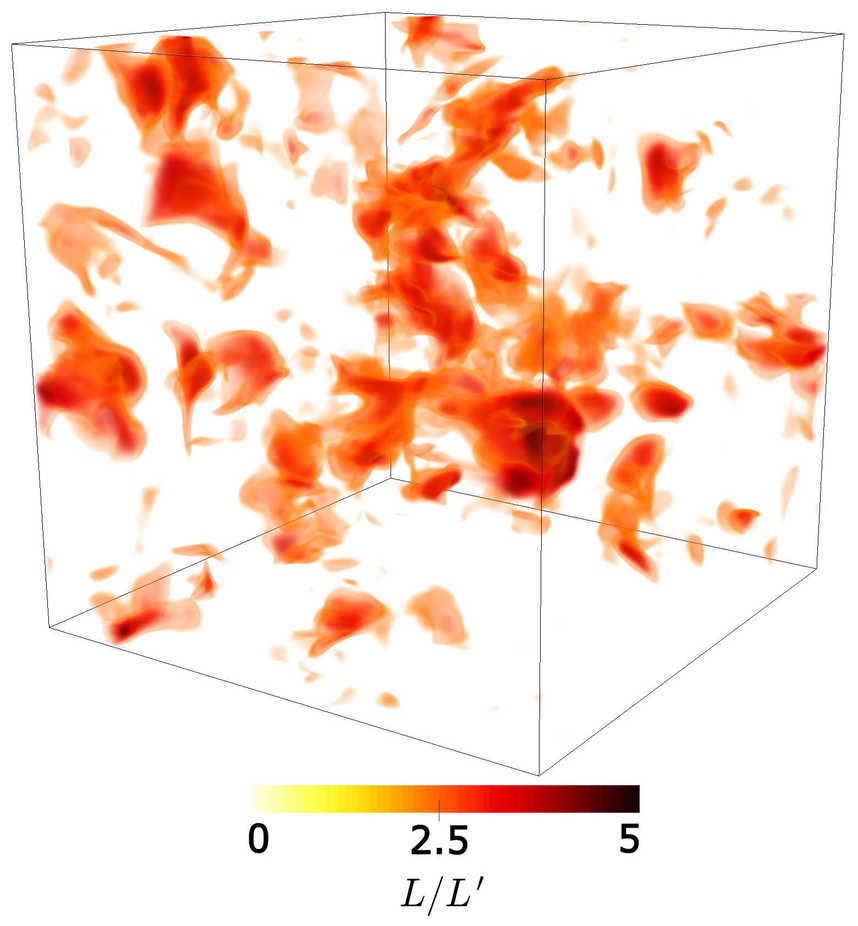}}

  \caption{}
   \end{subfigure}
	\begin{subfigure}{0.48\linewidth}
  \centerline{\includegraphics[width=\linewidth]{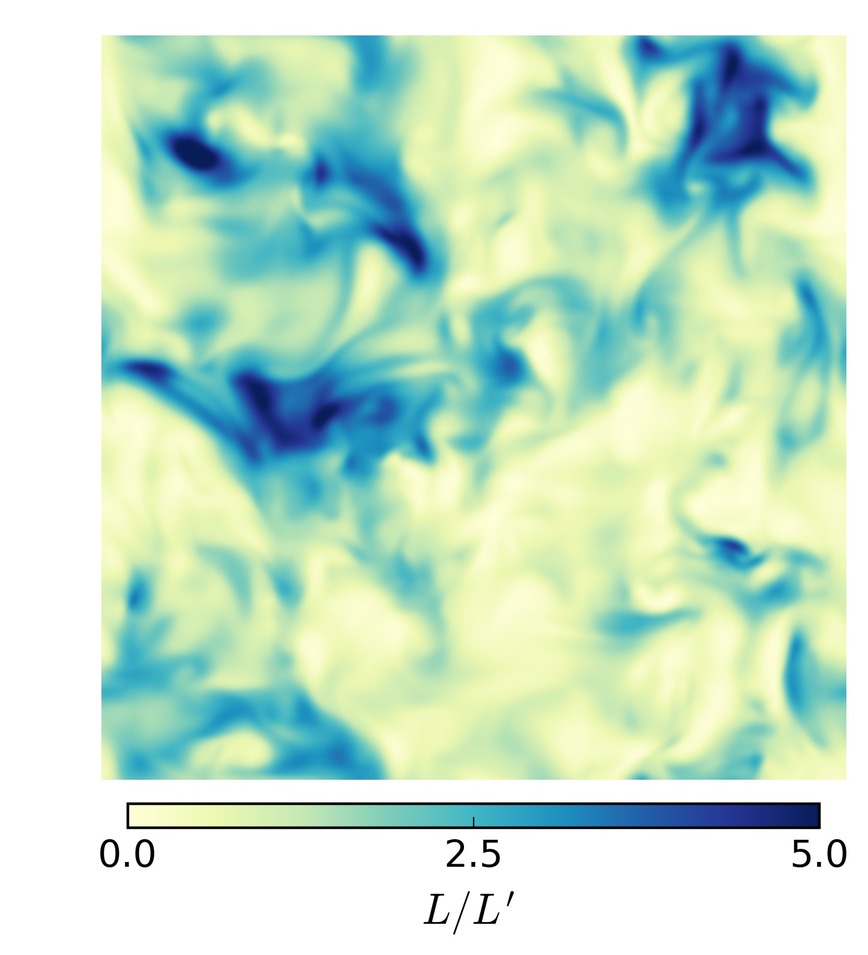}}

  \caption{}
   \end{subfigure}
   
   	\begin{subfigure}{0.48\linewidth}
  \centerline{\includegraphics[width=\linewidth]{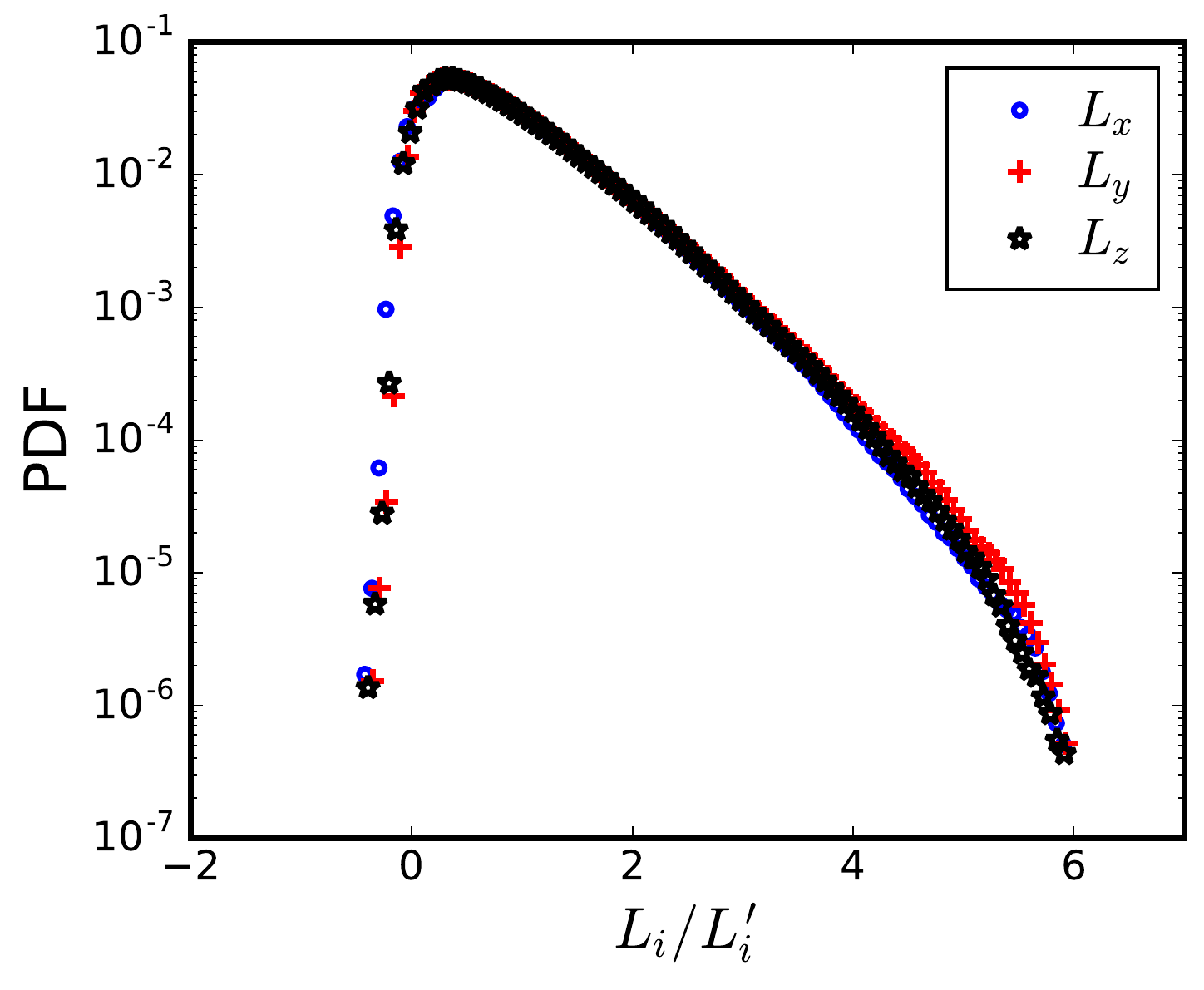}}

  \caption{}
   \end{subfigure}
	\begin{subfigure}{0.48\linewidth}
  \centerline{\includegraphics[width=\linewidth]{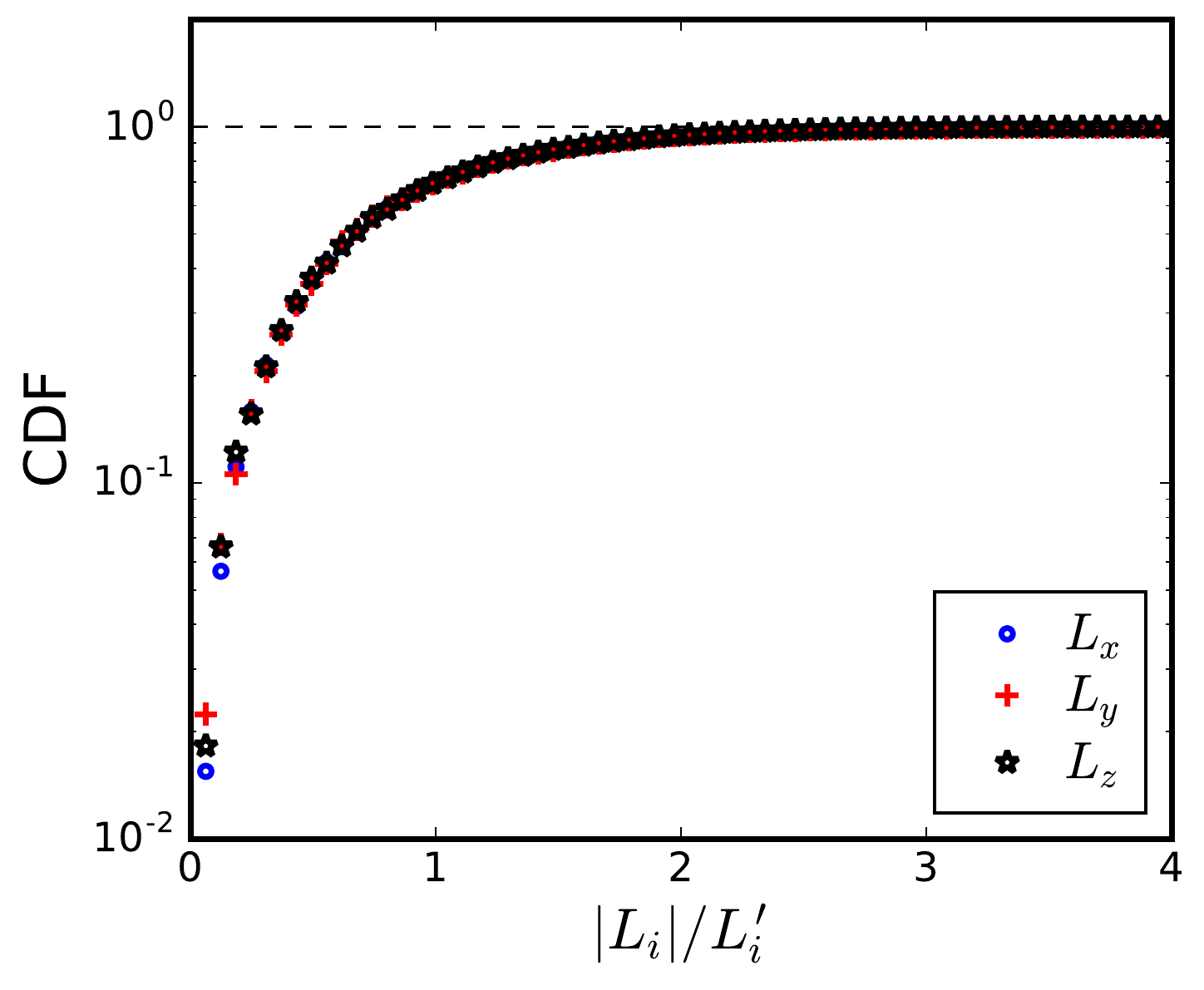}}

  \caption{}
   \end{subfigure}
  \caption{The $\lvec$ correlation is shown for a single field snapshot, at $500\tau_k$. Panel (a) shows the three-dimensional field as a volume rendering, while panel (b) shows the cross-sectional view of $L=\sqrt{\lvec \cdot \lvec}$ at the same plane shown in figure \ref{fig:TurbulenceData}. Panels (c) and (d) show the time-averaged PDFs and CDFs of the three components of $\lvec$.}
\label{fig:TurbulenceCorrelation-LCorr}
\end{figure}

Figure \ref{fig:TurbulenceCorrelation-LCorr} shows the $\lvec$ correlation. \sid{The volume rendering in panel (a) shows the spatial distribution of the correlation field, which shows features across various lengthscales, with diffused regions of high magnitude ranging from intermediate to small sizes. These features, further, are very similar to the features in the $E_k$ field (as seen in figure \ref{fig:TurbulenceData})}. \sid{Note that small, isolated, regions of the correlation field with typically a high magnitude, which we refer to as correlation \textit{kernels}, are measures of the correlation in larger regions of the velocity and vorticity field surrounding them.} Panel (b) shows a cross-sectional view of the correlation field, at the sample plane as that shown  in figure \ref{fig:TurbulenceData}. The PDF of the components of $\lvec$, in panel (c), shows that the correlation is highly positively skewed. \sid{This follows from the definition of $\lvec$, which identifies regions of well-aligned streamlines, i.e. the local velocity $\mathbf{u}({\mathbf{x}})$ is expected to be aligned with the velocity in the neighbourhood $-\Lambda < x_i < \Lambda$, and the product of the two is positive. The strong coincidence of high $L$ with regions of high $E_k$ (which is a \textit{point} quantitiy), reflects that high $E_k$ regions comprise parallel streamlines of \textit{jet-like} flow, hence, not only corresponding to regions of high velocity magnitude, but also exhibiting a high degree of alignment. This correspondence shall be quantified with the subsequent analysis. Further, simlarly to $u_i$, the PDFs of $L_i$ do not extend over a very large range of values, however, contrary to $u_i$, they are strongly non-Gaussian.} Panel (d) shows the CDFs of $L_i$, where approximately $70\%$ and $93\%$ of the $L_i$ fields are below $L_i^\prime$ and $2L_i^\prime$, respectively.

\begin{figure}
	\begin{subfigure}{0.48\linewidth}
  \centerline{\includegraphics[width=\linewidth]{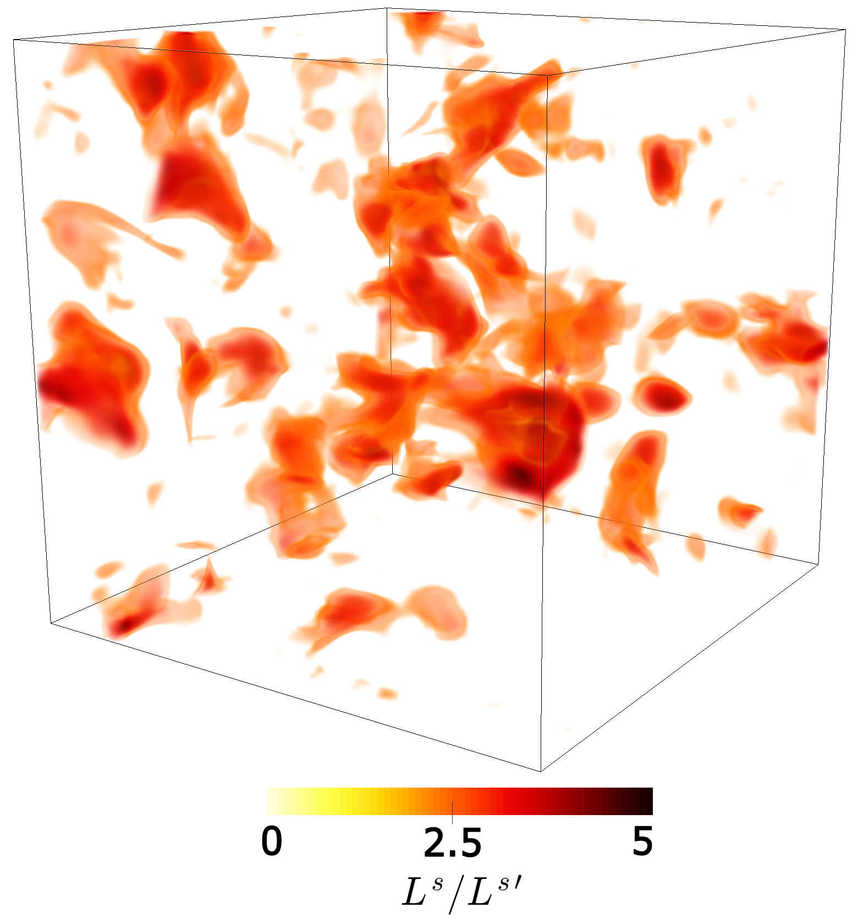}}

  \caption{}
   \end{subfigure}
	\begin{subfigure}{0.48\linewidth}
  \centerline{\includegraphics[width=\linewidth]{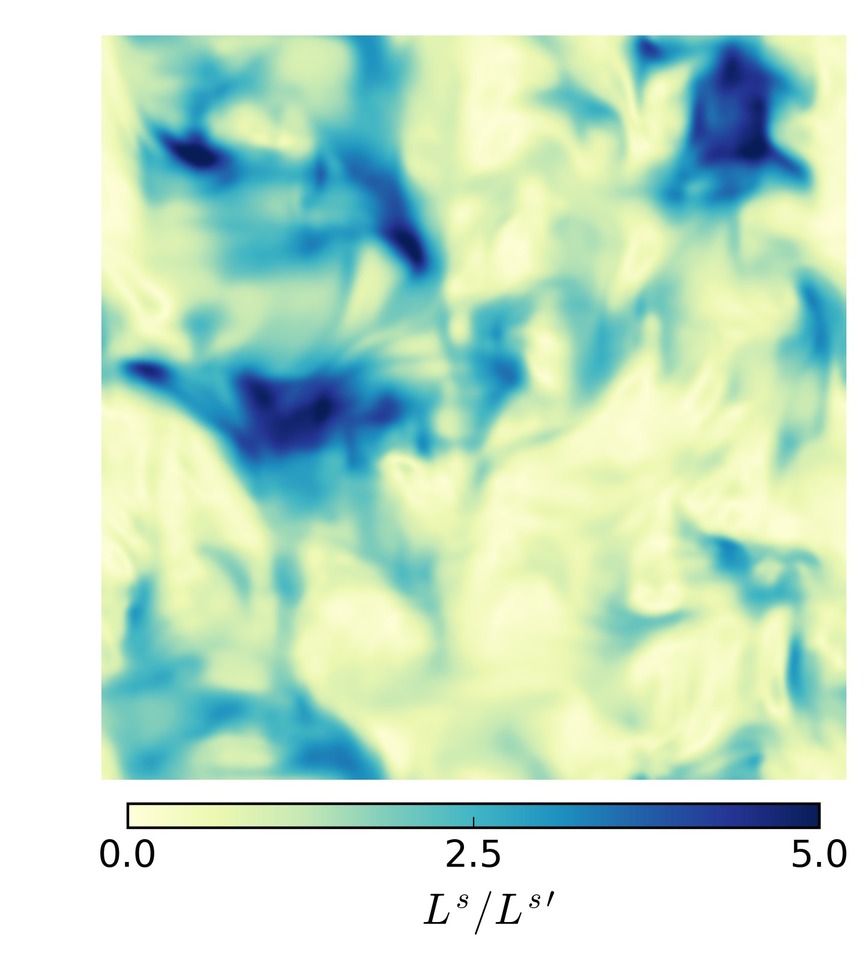}}

  \caption{}
   \end{subfigure}
   
   	\begin{subfigure}{0.48\linewidth}
  \centerline{\includegraphics[width=\linewidth]{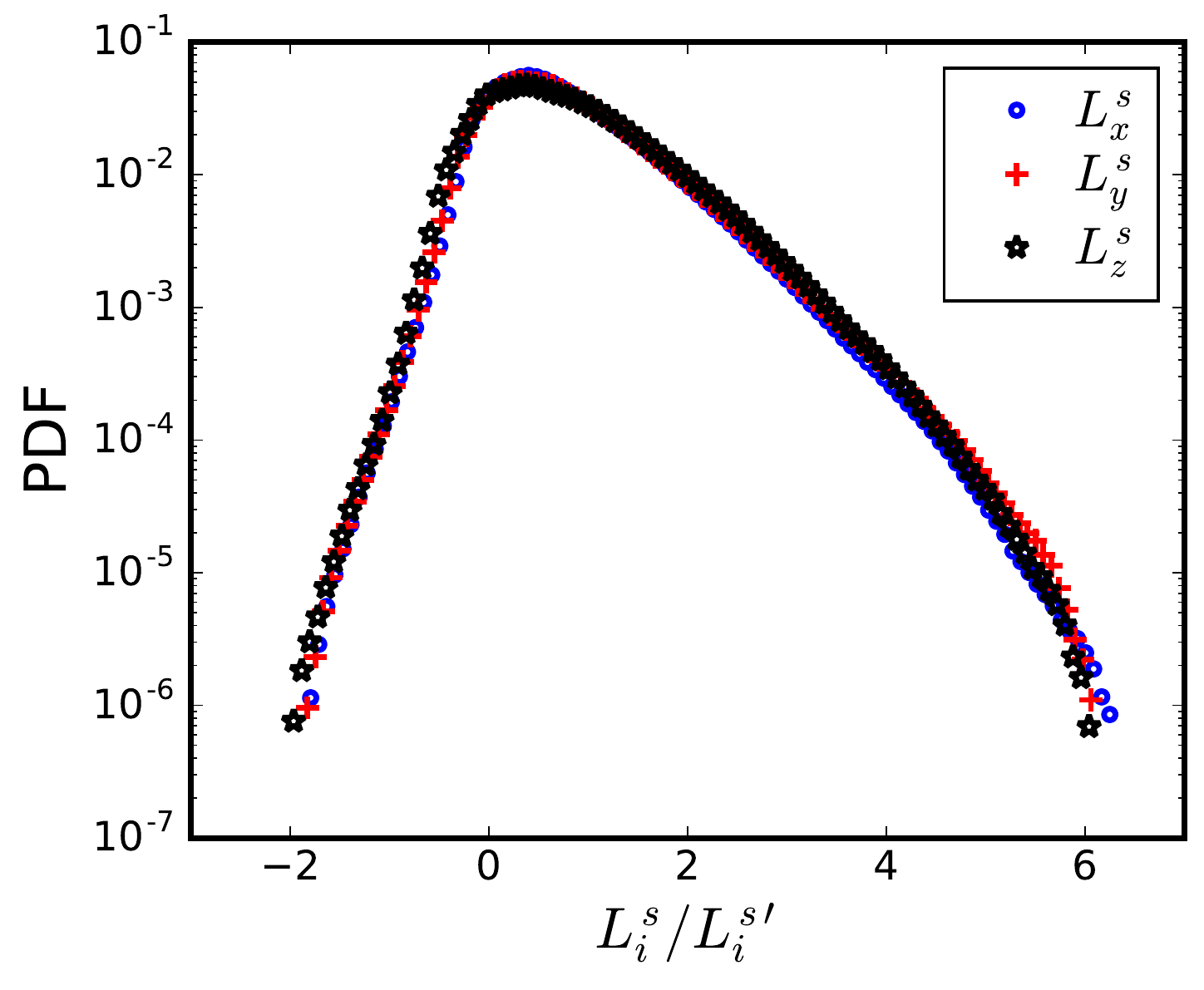}}

  \caption{}
   \end{subfigure}
	\begin{subfigure}{0.48\linewidth}
  \centerline{\includegraphics[width=\linewidth]{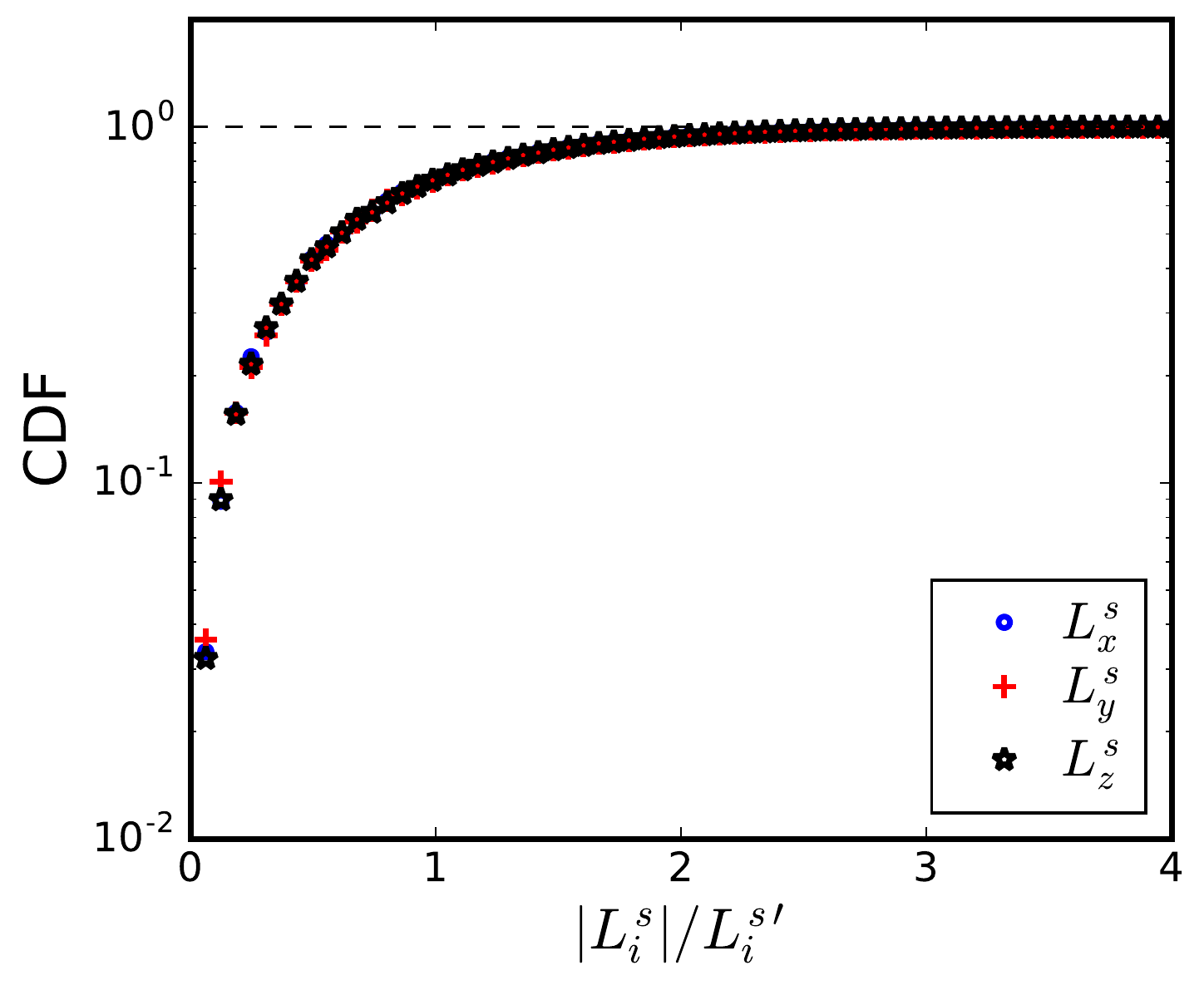}}

  \caption{}
   \end{subfigure}
  \caption{The $\lsvec$ correlation is shown for a single field snapshot, at $500\tau_k$. Panel (a) shows the three-dimensional field as a volume rendering, while panel (b) shows the cross-sectional view of $L^s=\sqrt{\lsvec \cdot \lsvec}$ at the same plane shown in figure \ref{fig:TurbulenceData}. Panels (c) and (d) show the time-averaged PDFs and CDFs of the three components of $\lsvec$.}
\label{fig:TurbulenceCorrelation-LSCorr}
\end{figure}

\sid{Figure \ref{fig:TurbulenceCorrelation-LSCorr} shows the $\lsvec$ correlation. The volumetric distribution in panel (a) and the planar crossection in panel (b) show a striking similarity to the structures and distribution of $L$ (see figure \ref{fig:TurbulenceCorrelation-LCorr}). The $L^s$ field, like $L$, also shows intermediate and small sized diffused regions, found throughout the volume of the flow. While the $\lsvec$ kernels are slightly smaller than the $\lvec$ kernels, the strong correspondence between the two correlation fields is in stark contrast to the canonical flows examples, particularly the Taylor-Green flow pattern (see figure \ref{fig:taylorGreen-CorrelationsPiBy2}). This shows that in (homogeneous isotropic) turbulence velocity fields, there are no large symmetries (within an integration length of $\Lambda = \lambda$), which would, for instance, be associated with ``large eddies'' with a swirling motion. High values of $L^s$, here, arise due to parallel streamlines in jet-like flow regions, which in turn were found to coincide strongly with regions of high $E_k$. This further corroborates that high kinetic energy structures are jet-like. The PDFs of $L^s_i$, in panel (c), point to the same fact, as the distribution is found to be highly positively skewed (note that the $\lsvec$ correlation, by definition, yields large positive values for aligned streamlines, and large negative values for anti-parallel streamlines). Negative values of $L^s$ are also found, albeit with very small probability, which shows that a few regions have anti-parallel streamlines, yielding mild $L^s$ values. Lastly, the CDFs in panel (d) show that, similarly to the $L_i$ fields, approximately $70\%$ and $93\%$ of the $L^s_i$ fields are below ${L^s_i}^\prime$ and $2{L^s_i}^\prime$, respectively.}

\begin{figure}
	\begin{subfigure}{0.48\linewidth}
  \centerline{\includegraphics[width=\linewidth]{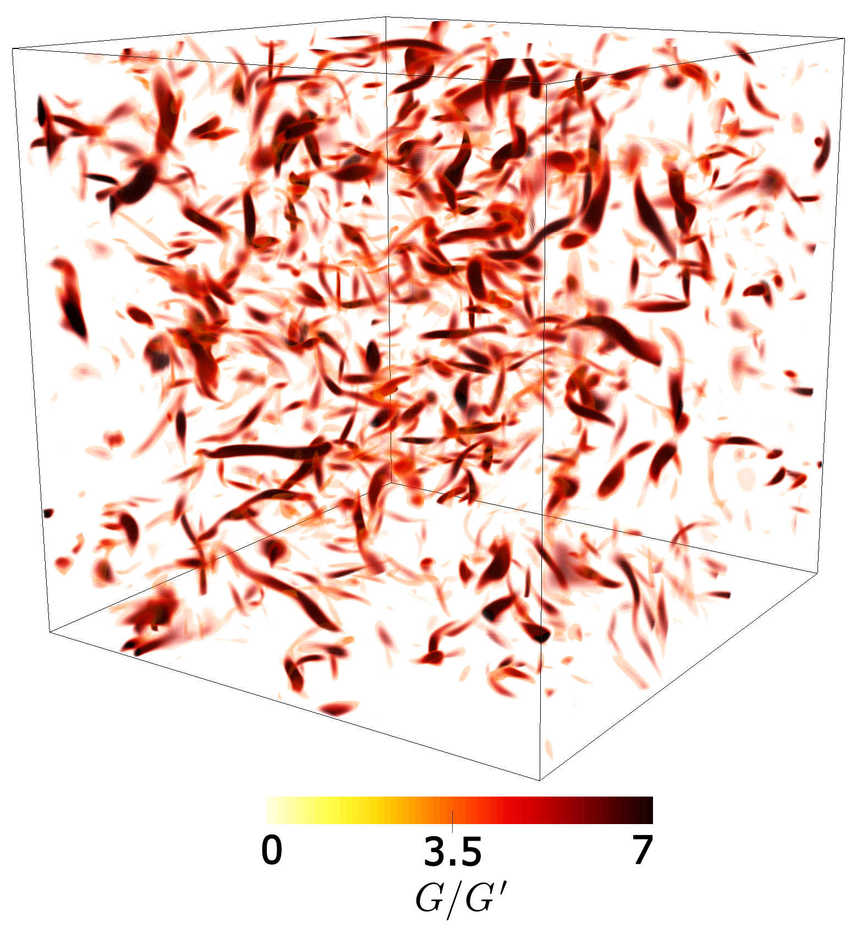}}

  \caption{}
   \end{subfigure}
	\begin{subfigure}{0.48\linewidth}
  \centerline{\includegraphics[width=\linewidth]{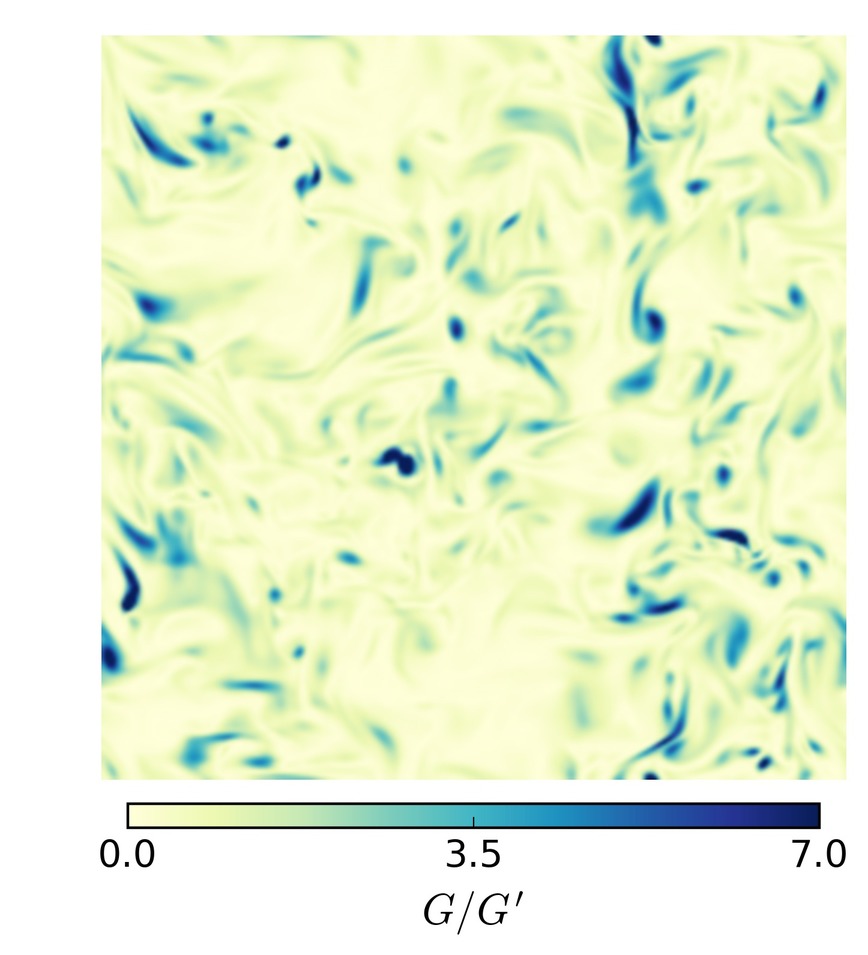}}

  \caption{}
   \end{subfigure}
   
   	\begin{subfigure}{0.48\linewidth}
  \centerline{\includegraphics[width=\linewidth]{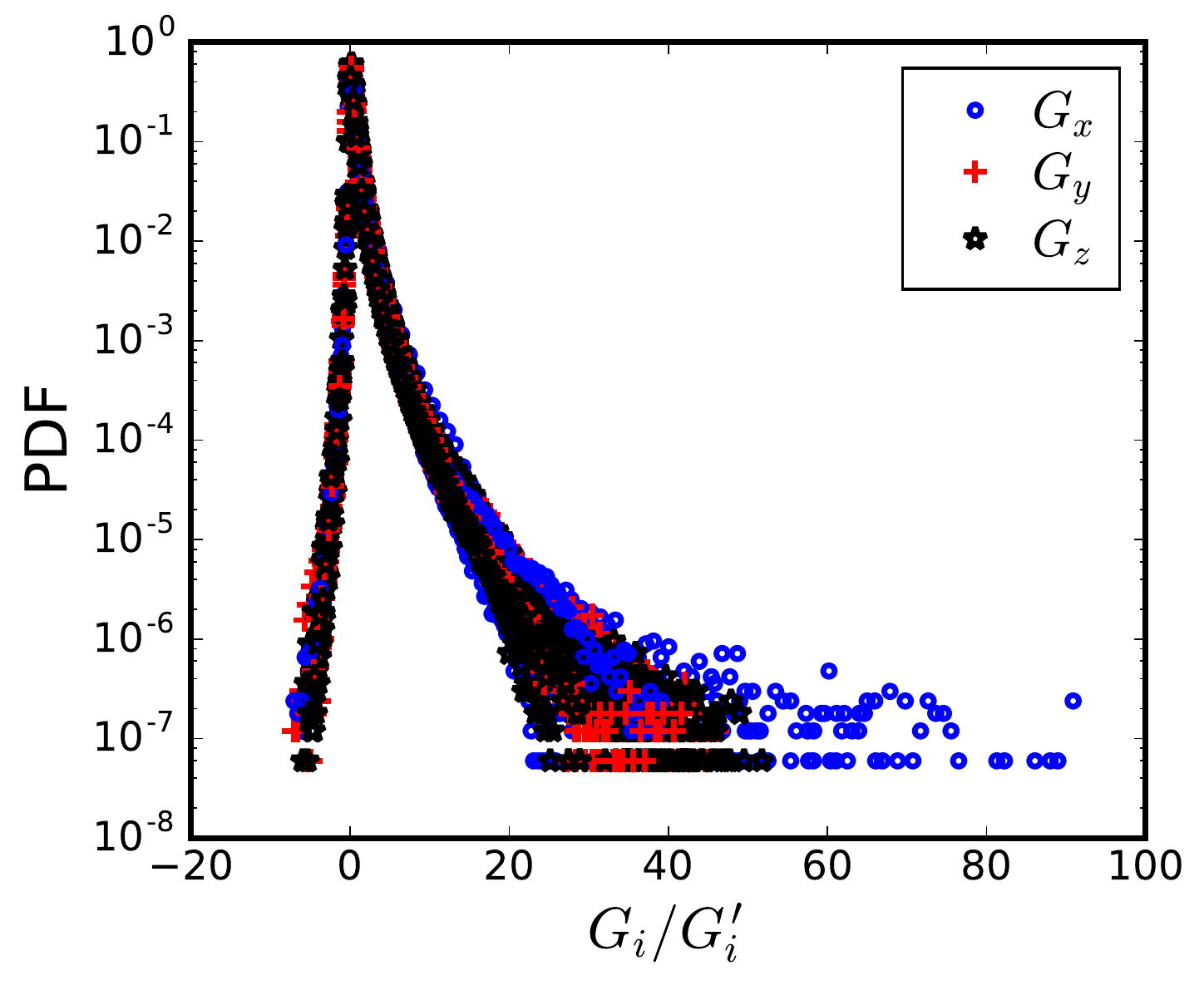}}

  \caption{}
   \end{subfigure}
	\begin{subfigure}{0.48\linewidth}
  \centerline{\includegraphics[width=\linewidth]{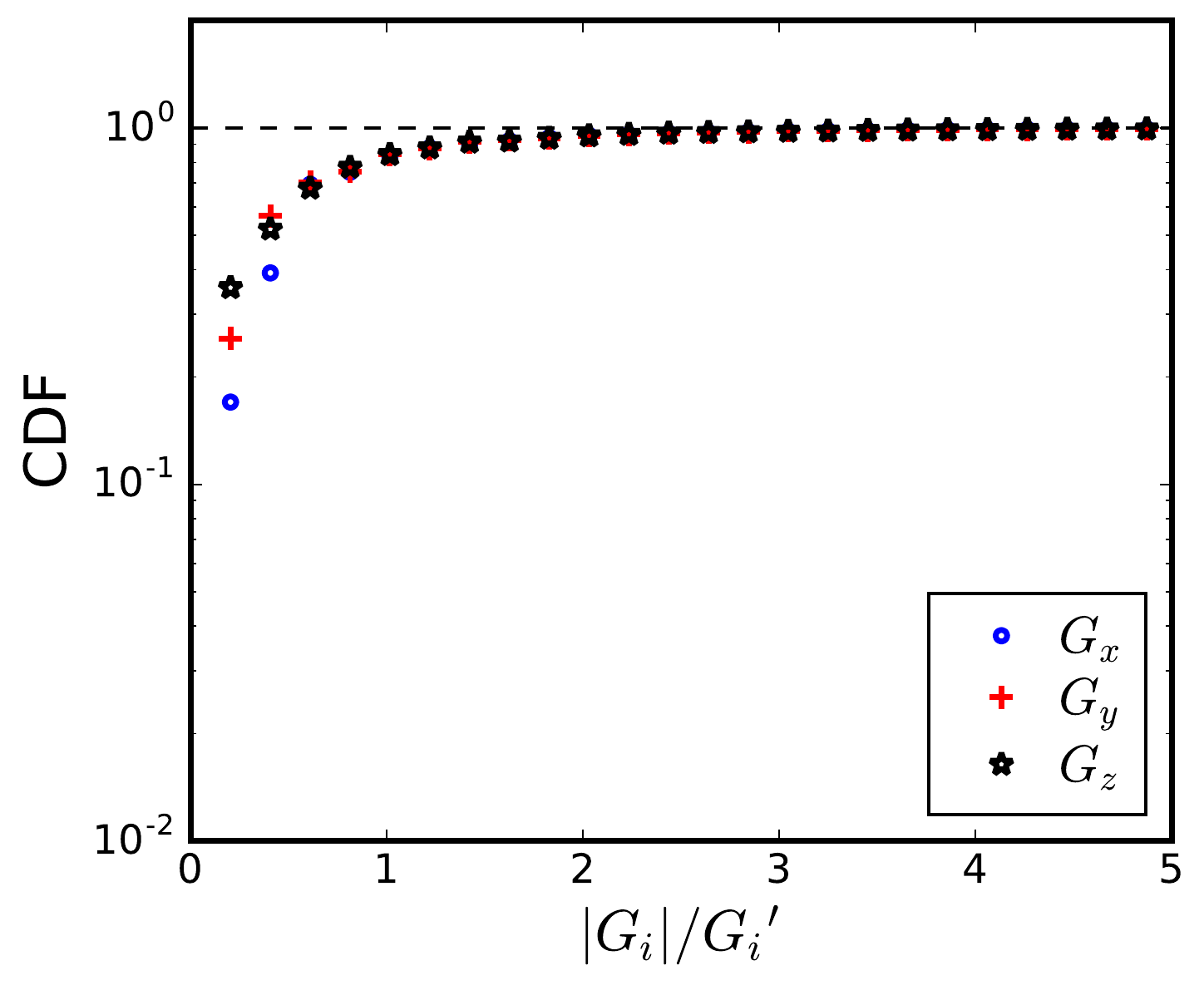}}

  \caption{}
   \end{subfigure}
  \caption{The $\g$ correlation is shown for a single field snapshot, at $500\tau_k$. Panel (a) shows the three-dimensional field as a volume rendering, while panel (b) shows the cross-sectional view of $G=\sqrt{\g \cdot \g}$. Panels (c) and (d) show the time-averaged PDFs and CDFs of the three components of $\g$.}
\label{fig:TurbulenceCorrelation-GCorr}
\end{figure}

Figure \ref{fig:TurbulenceCorrelation-GCorr} shows the $\g$ correlation, which is the equivalent to $\lvec$ for the vorticity field. \sid{The volumetric (panel a) and planar (panel b) distributions of $G$ are found to closely resemble the enstrophy field (as shown in figure \ref{fig:TurbulenceData}), where the $G$ field at high magnitudes also forms worm-like structures. Like $L$, the $G$ field also yields high values in regions which have both high vorticity, and a high degree of alignment of the vorticity streamlines. The strong correspondence with high enstrophy regions shows that these regions form small scale vorticity-jets, the size of which is smaller than the high $E_k$ velocity-jets.} The PDFs of the $G_i$ components show that the correlation yields a long-tailed, positively-skewed distribution, similar to the positive half of the vorticity PDF (figure \ref{fig:TurbulenceData-PDF}c). This again reflects that the vorticity streamlines are well aligned in the core of high enstrophy regions, since the product of $\boldsymbol{\omega}(\mathbf{x})$ and the integral of the vorticity in the neighbourhood $-\Lambda < x_i < \Lambda$, yields positive values, reflecting that the two quantities have the same sign. The CDFs, in panel (d), show that approximately $84\%$ and $94\%$ of the $G_i$ fields are below $G_i^\prime$ and $2G_i^\prime$, respectively. \sid{In comparison, $70\%$ and $95\%$ of the $\omega_i$ fields are belowe $\omega_i^\prime$ and $2\omega_i^\prime$, respectively. This is consistent with the visual impression of figures \ref{fig:TurbulenceData} and \ref{fig:TurbulenceCorrelation-GCorr} and it indicates that the vorticity field in the range $\omega_i^\prime < \omega_i < 2\omega_i^\prime$ is not as closely associated with regions of well-aligned vorticity streamlines as for higher values of $\omega_i$.}

\begin{figure}
	\begin{subfigure}{0.48\linewidth}
  \centerline{\includegraphics[width=\linewidth]{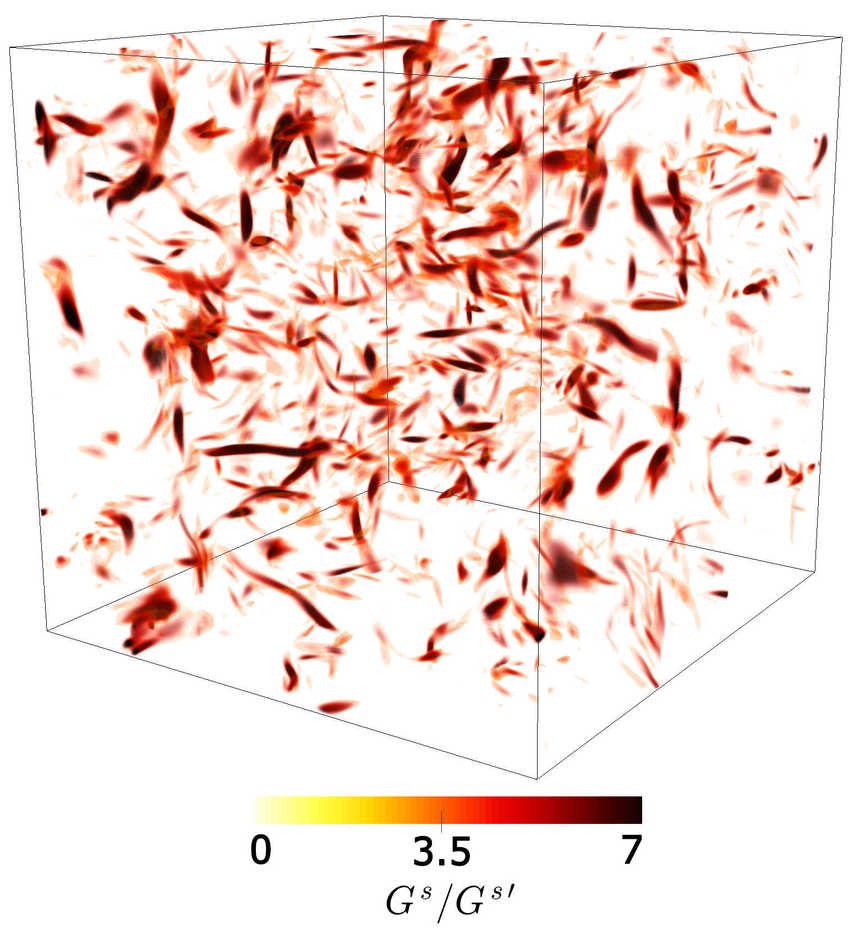}}

  \caption{}
   \end{subfigure}
	\begin{subfigure}{0.48\linewidth}
  \centerline{\includegraphics[width=\linewidth]{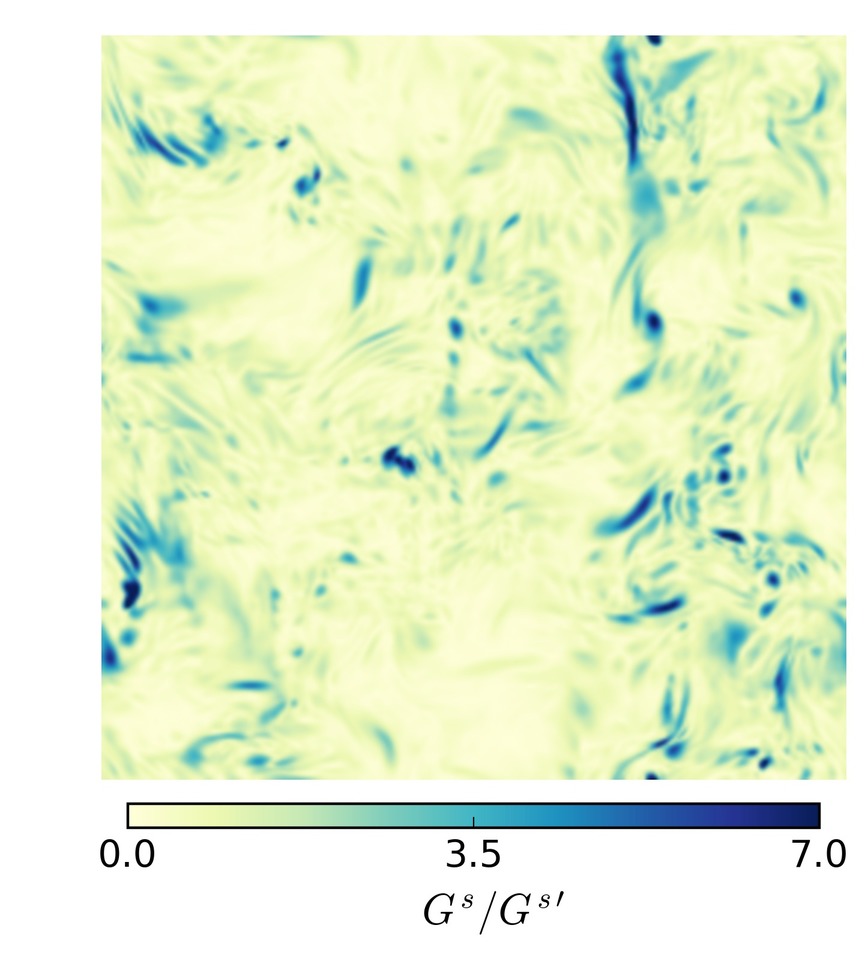}}

  \caption{}
   \end{subfigure}
   
   	\begin{subfigure}{0.48\linewidth}
  \centerline{\includegraphics[width=\linewidth]{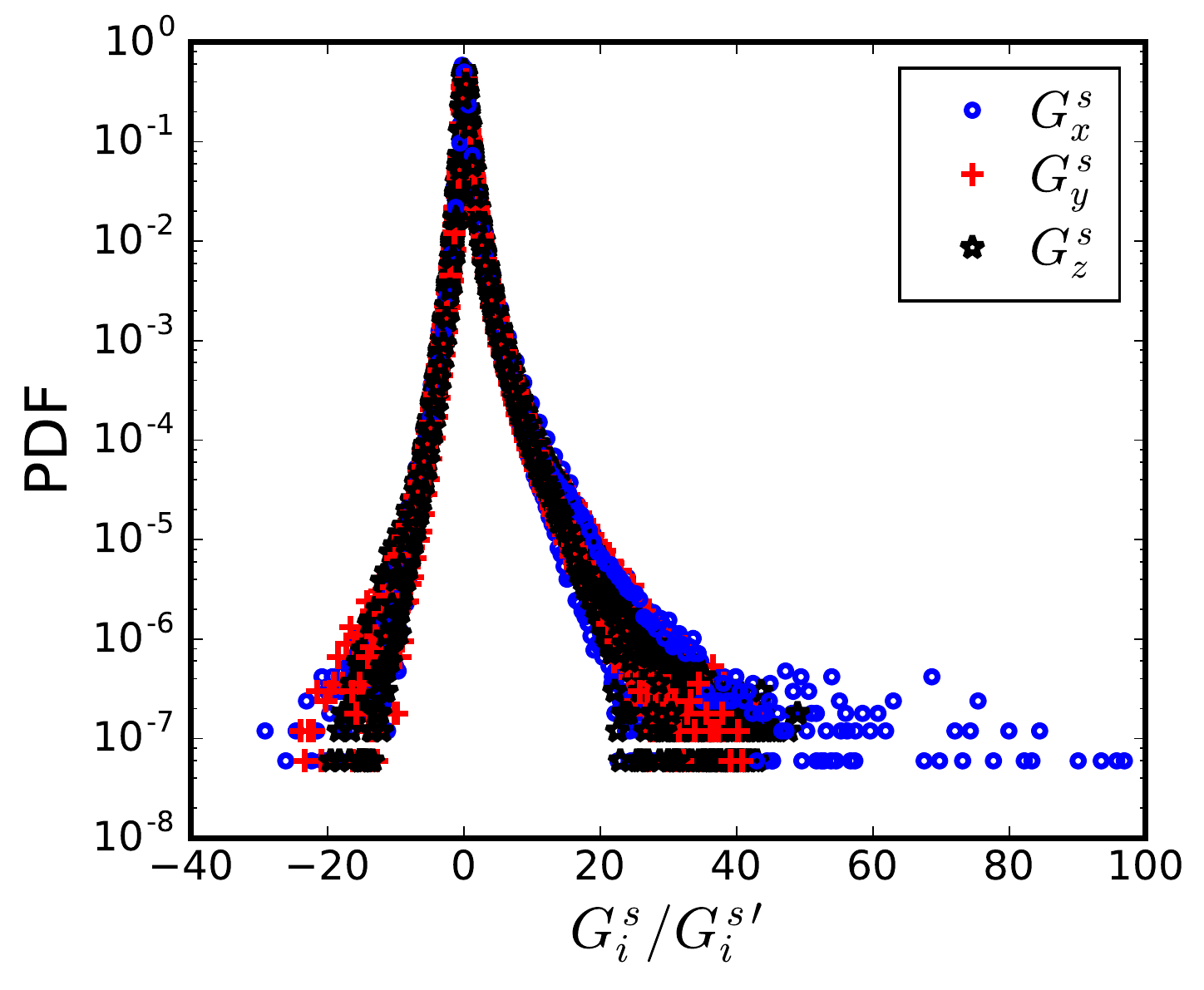}}

  \caption{}
   \end{subfigure}
	\begin{subfigure}{0.48\linewidth}
  \centerline{\includegraphics[width=\linewidth]{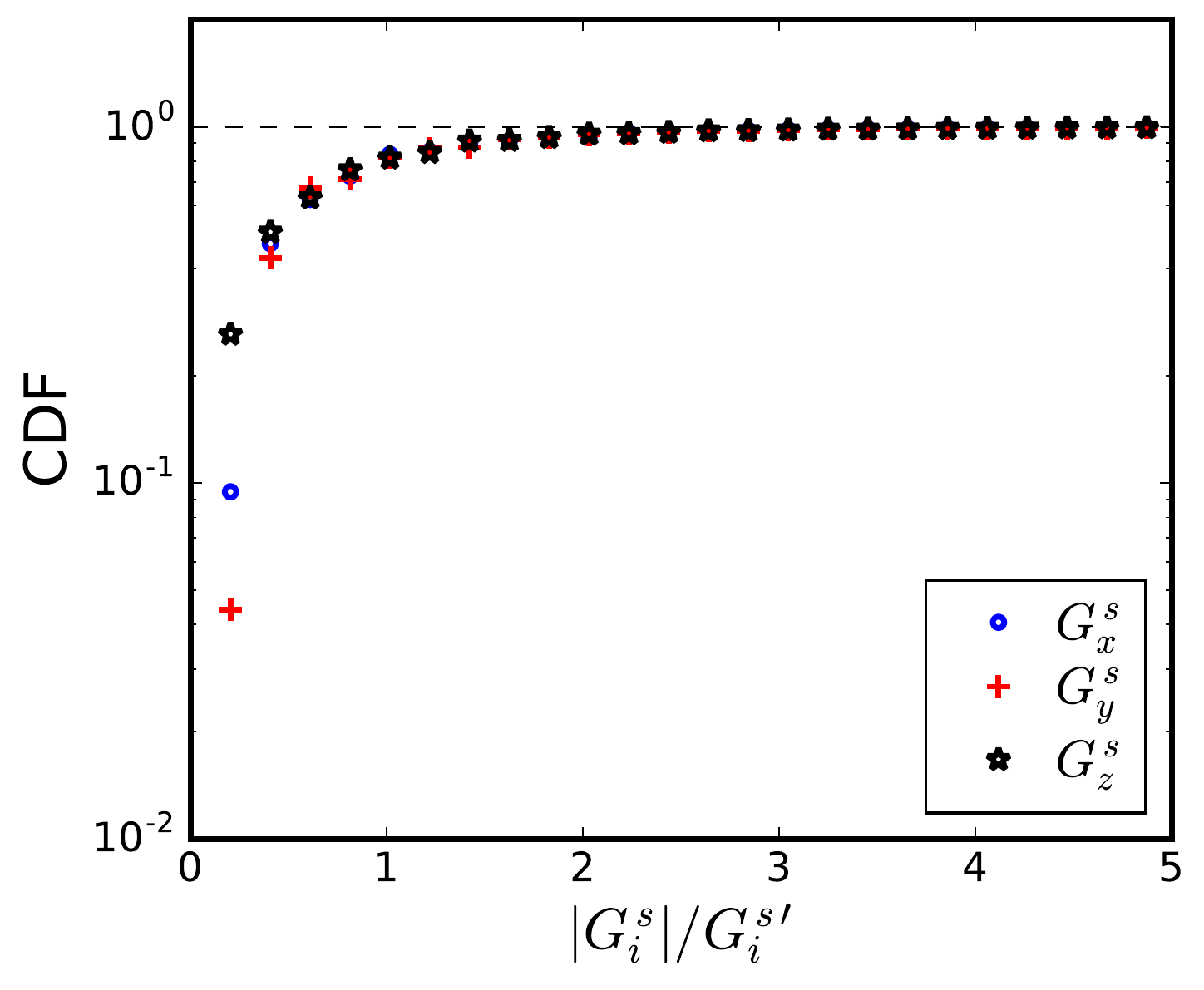}}

  \caption{}
   \end{subfigure}
  \caption{The $\gs$ correlation is shown for a single field snapshot, at $500\tau_k$. Panel (a) shows the three-dimensional field as a volume rendering, while panel (b) shows the cross-sectional view of $G^s=\sqrt{\gs \cdot \gs}$. Panels (c) and (d) show the time-averaged PDFs and CDFs of the three components of $\g$.}
\label{fig:TurbulenceCorrelation-GSCorr}
\end{figure}

Figure \ref{fig:TurbulenceCorrelation-GSCorr} shows the $\gs$ correlation, which is the vorticity field equivalent of $\lsvec$. \sid{The volumetric distribution and planar profile of the correlation field are shown in panels (a) and (b), respectively. The $G^s$ field appears `patchy' in comparison to $\g$, and its kernels are found to be more sparse and fragmented. This is because the $\gs$ correlation is sensitive to the symmetries in the vorticity field (in the $\Lambda-$neighbourhood) along each direction. Visually, we find that there are no high amplitude $G^s$ regions that do not coincide with high amplitude $G$, showing that anti-parallel vorticity streamlines are uncommon.} The PDFs in panel (c) shows that the components $G^s_i$ have a positively-skewed distribution, although negative values of $G^s_i$ are slightly more prevalent than negative values of $G_i$. The positive-skew of the high magnitude $G^s_i$ shows that the stronger vorticity regions have vorticity streamlines that are well aligned. The CDFs in panel (d) show that approximately $80\%$ and $96\%$ of $G^s_i$ is under ${G^s_i}^\prime$ and $2{G^s_i}^\prime$, respectively. \sid{Interestingly, for the same rms thresholds, the CDFs of the $\omega_i$ and $G_i$ fields yield $70\%-95\%$ and $84\%-94\%$, respectively. This suggests that, while high vorticity regions are closely associated with well-aligned vorticity jets, ``intermediate high'' vorticity (roughly, in the range $\omega_i^\prime < \omega_i < 2\omega_i^\prime$) has a less organized, weaker structure but containing more complex local symmetries and anti-symmetries, leading to a more ``patchy'' $G_i^s$ field. This is consistent with the work of \citet{she1990intermittent,she1991structure} and the vorticity ``ranges'' they proposed, however, in this work we will not explore this further.}

\begin{figure}
	\begin{subfigure}{0.48\linewidth}
  \centerline{\includegraphics[width=\linewidth]{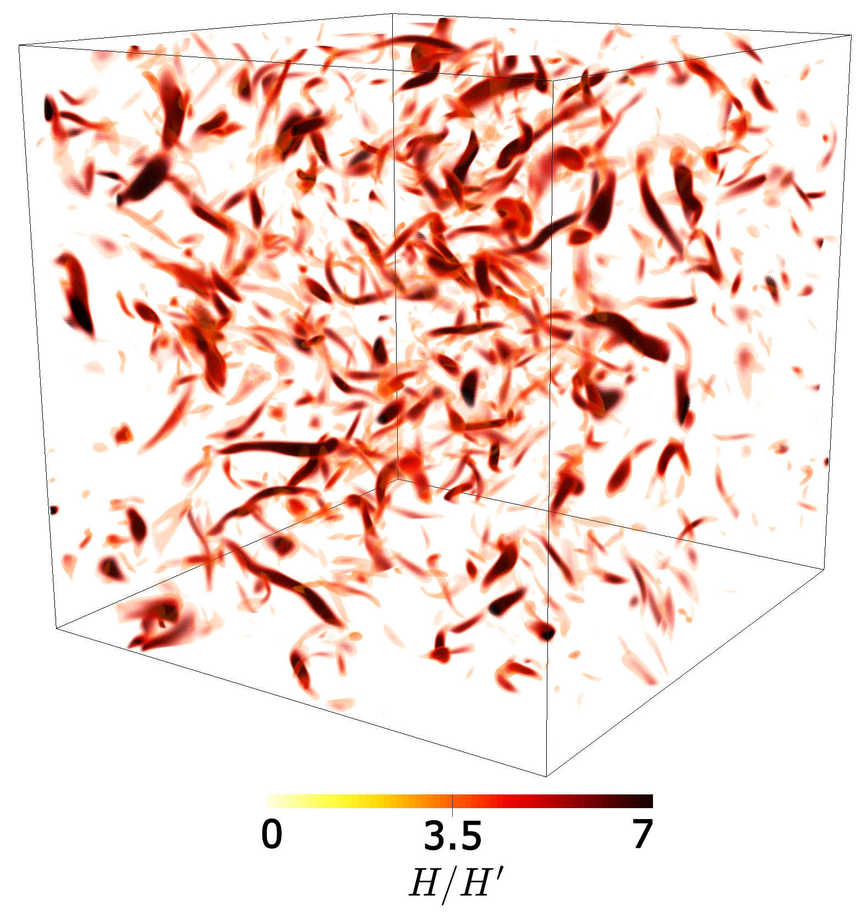}}

  \caption{}
   \end{subfigure}
	\begin{subfigure}{0.48\linewidth}
  \centerline{\includegraphics[width=\linewidth]{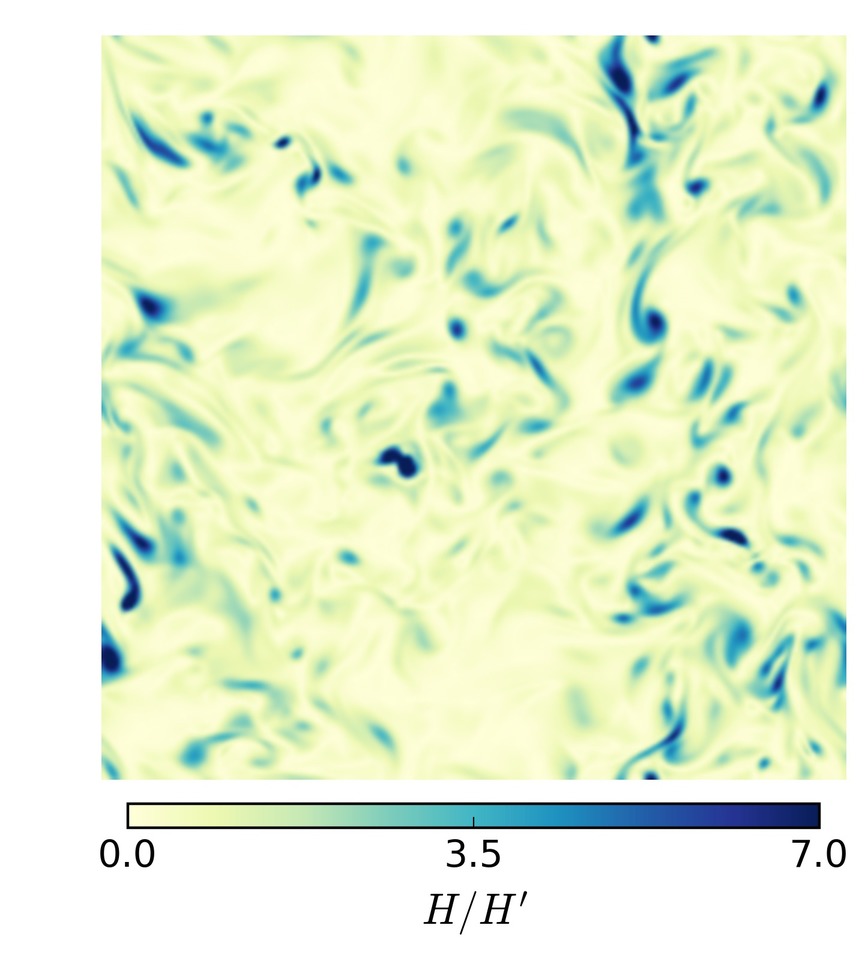}}

  \caption{}
   \end{subfigure}
   
   	\begin{subfigure}{0.48\linewidth}
  \centerline{\includegraphics[width=\linewidth]{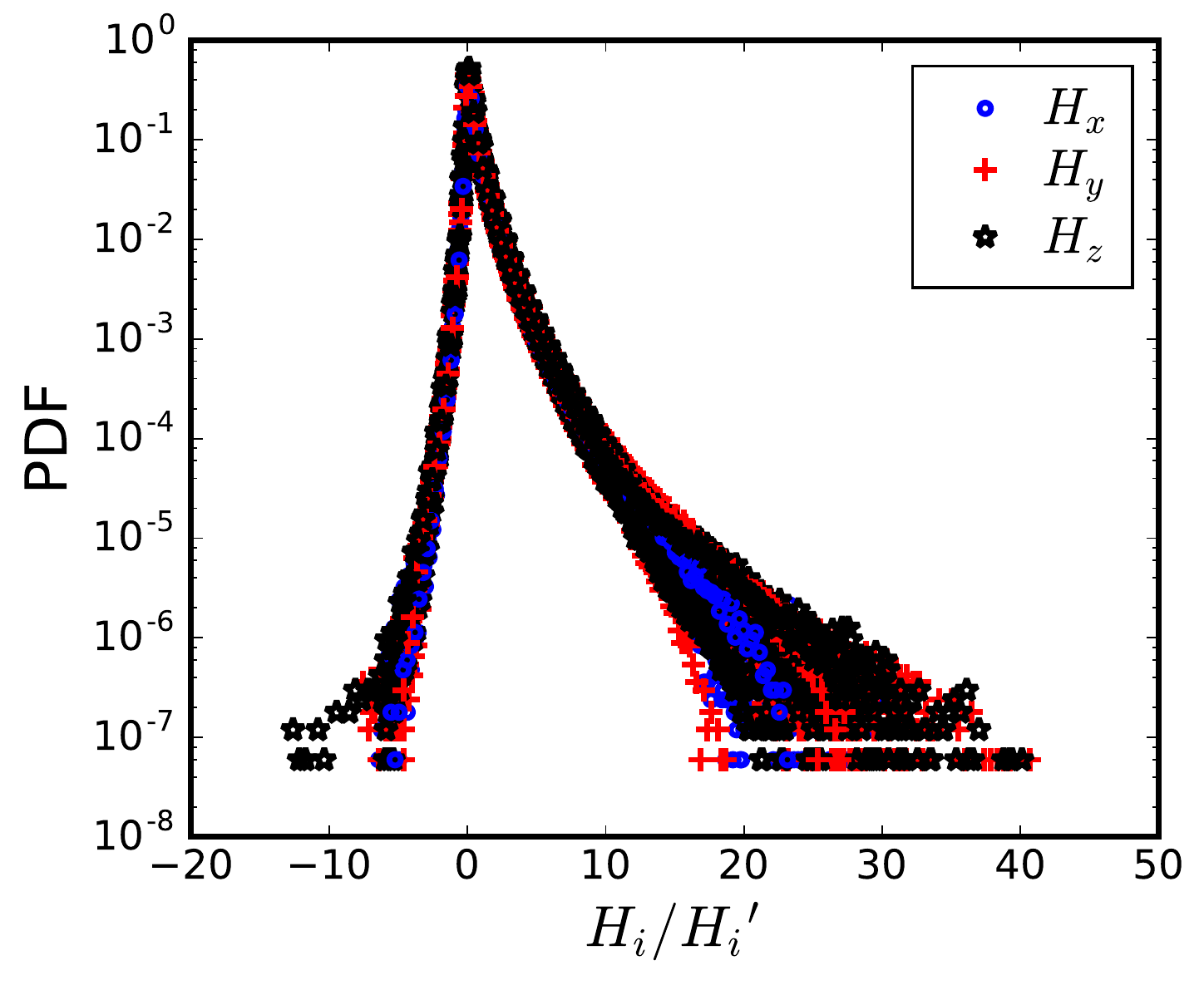}}

  \caption{}
   \end{subfigure}
	\begin{subfigure}{0.48\linewidth}
  \centerline{\includegraphics[width=\linewidth]{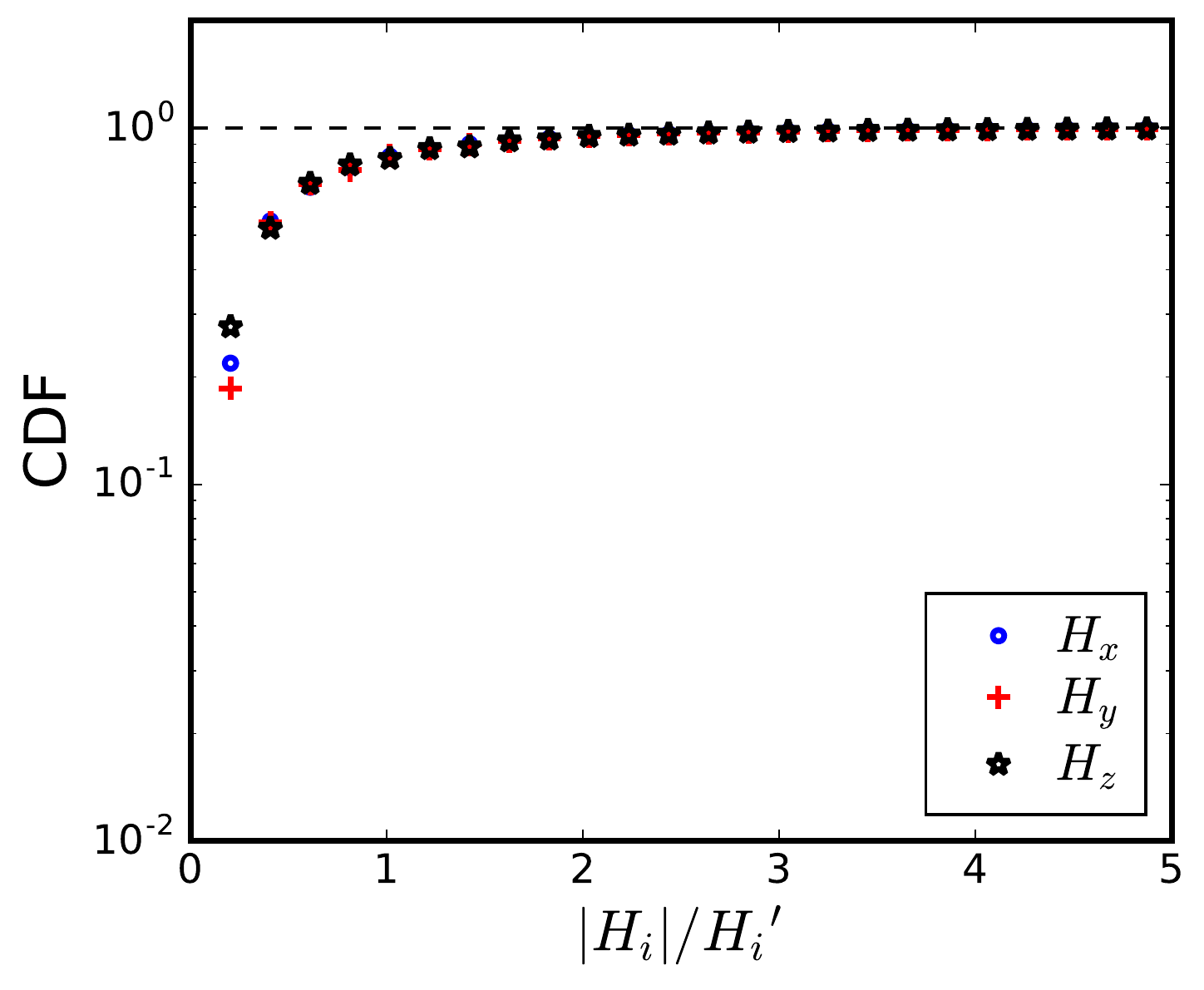}}

  \caption{}
   \end{subfigure}
  \caption{The $\h$ correlation is shown for a single field snapshot, at $500\tau_k$. Panel (a) shows the three-dimensional field as a volume rendering, while panel (b) shows the cross-sectional view of $H=\sqrt{\h \cdot \h}$. Panels (c) and (d) show the time-averaged PDFs and CDFs of the three components of $\h$.}
\label{fig:TurbulenceCorrelation-HCorr}
\end{figure}

\sid{Figure \ref{fig:TurbulenceCorrelation-HCorr} shows the $\h$ correlation, which relates the local vorticity $\boldsymbol{\omega}(\mathbf{x})$ to the velocity field along directions $r_i$. Similarly to the $G$ and $G^s$ fields, the volumetric and planar $H$ fields, in panels (a) and (b), respectively, also closely resemble the enstrophy field in figure \ref{fig:TurbulenceData-PDF}. The PDFs of the $H_i$ components are highly positively-skewed, and have a long-tailed distribution. Together, these results show that the velocity field in the vicinity of strong vorticity regions has an angular velocity closely positively-aligned with the jet-like vorticity streamlines. This is because (i) there exists a strong spatial correspondence between the occurrence of strong enstrophy regions and high magnitude $\mathbf{H}$, and (ii) the distributions of $H_i$ dominantly show positive values, which means that the local flow, in the $\Lambda-$neighbourhood, has an angular velocity well correlated with the vorticity, which tends to form small vorticity jets in the strong vorticity regions. The CDFs of $H_i$ in panel (d) show that approximately $82\%$, $94\%$ and $99\%$ of the $H_i$ fields are below $H_i^\prime$, $2H_i^\prime$ and $3H_i^\prime$, respectively.}

\begin{figure}
	\begin{subfigure}{0.48\linewidth}
  \centerline{\includegraphics[width=\linewidth]{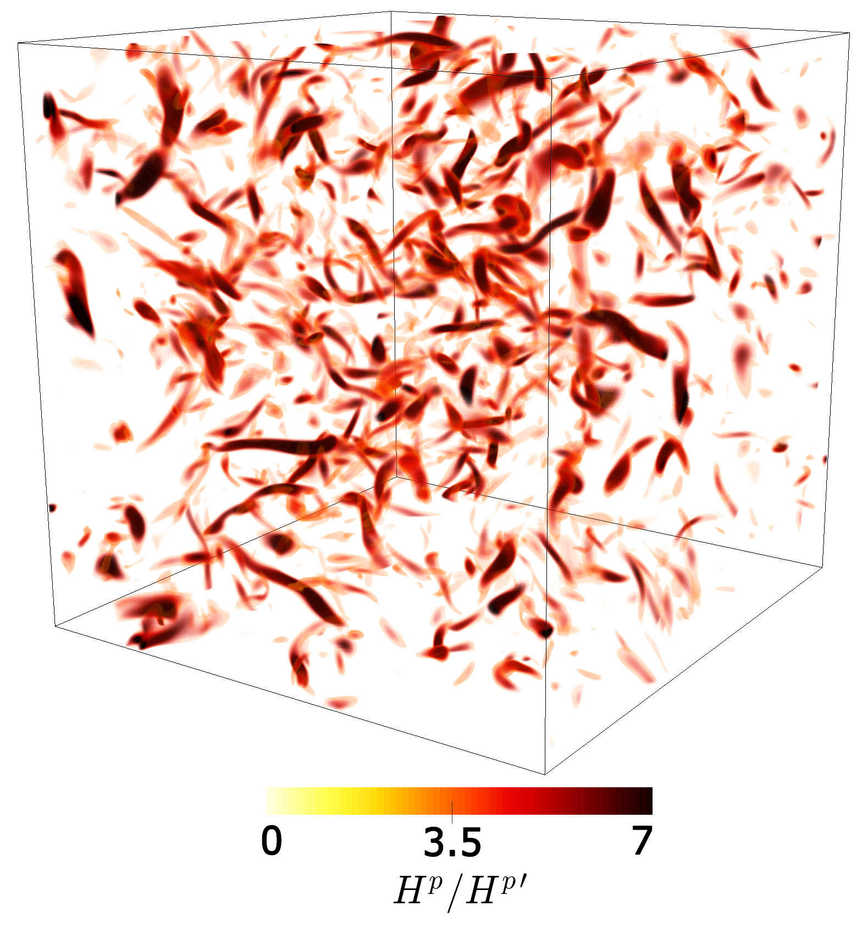}}

  \caption{}
   \end{subfigure}
	\begin{subfigure}{0.48\linewidth}
  \centerline{\includegraphics[width=\linewidth]{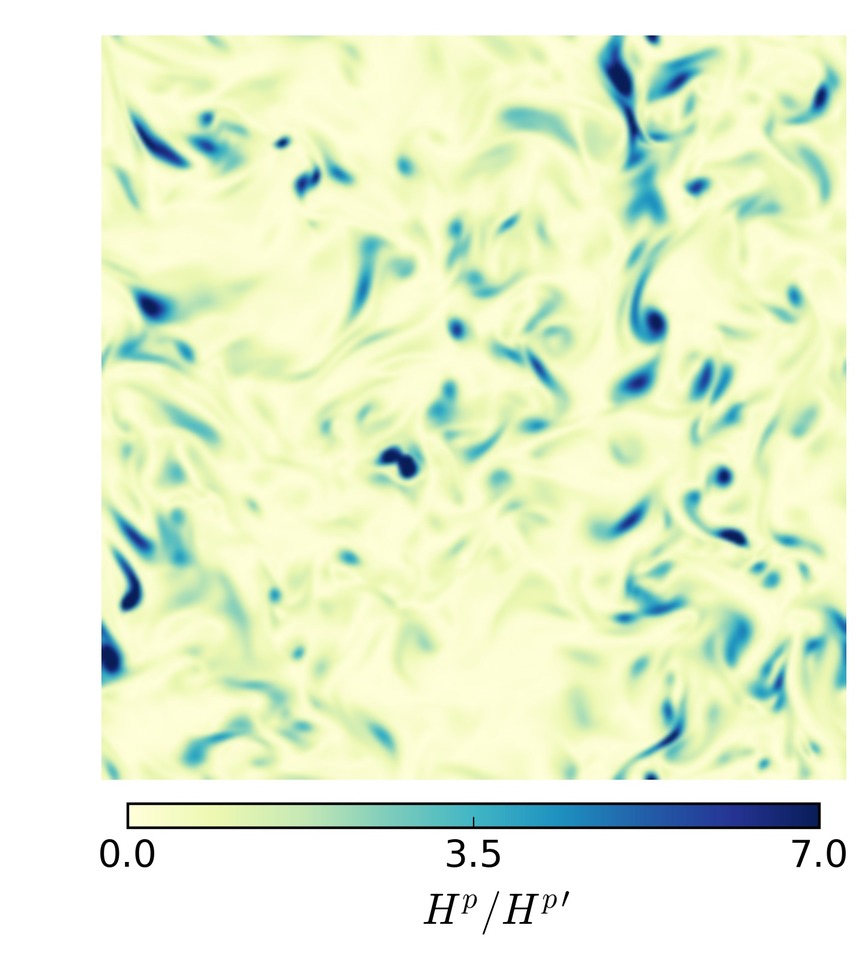}}

  \caption{}
   \end{subfigure}
   
   	\begin{subfigure}{0.48\linewidth}
  \centerline{\includegraphics[width=\linewidth]{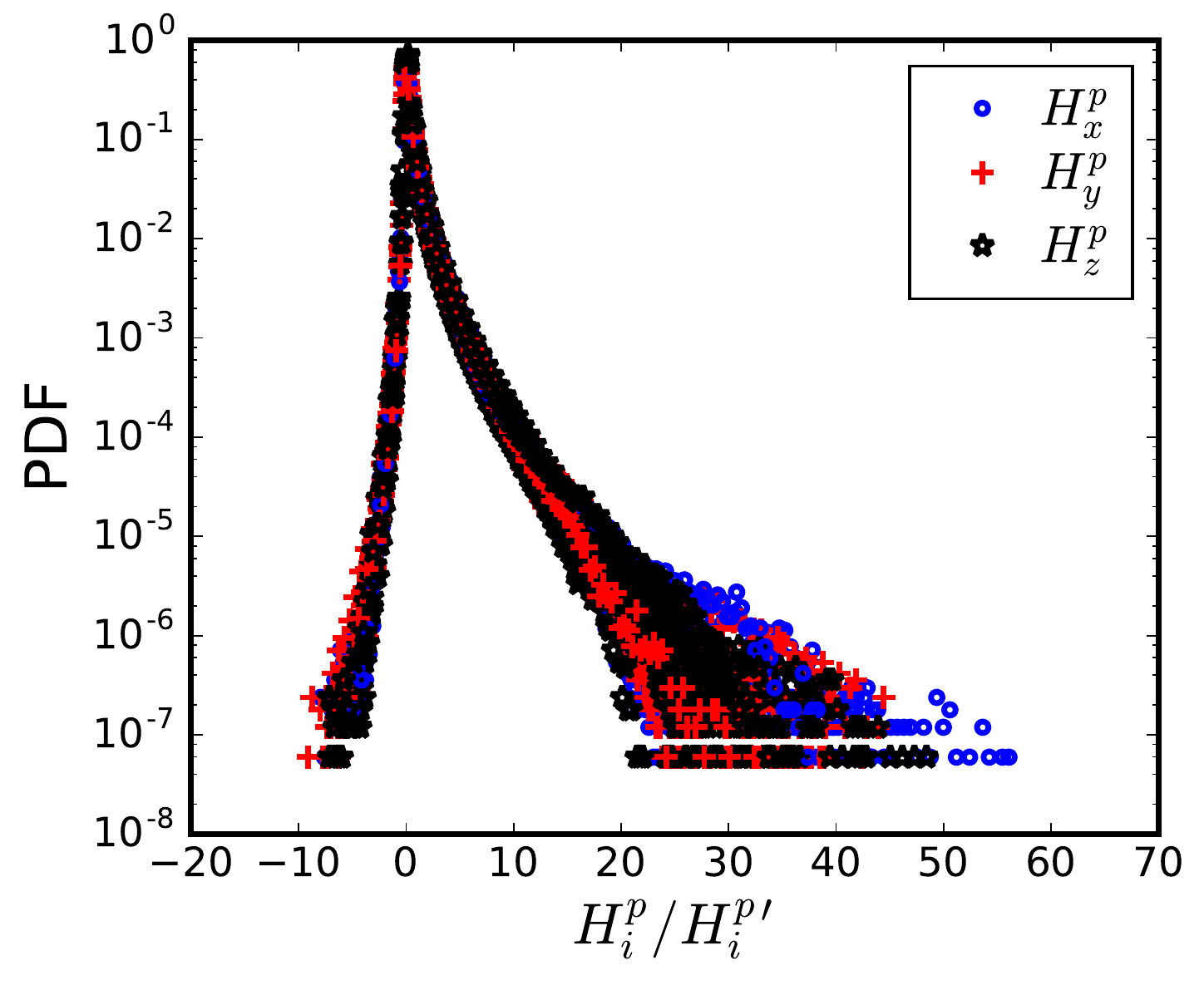}}

  \caption{}
   \end{subfigure}
	\begin{subfigure}{0.48\linewidth}
  \centerline{\includegraphics[width=\linewidth]{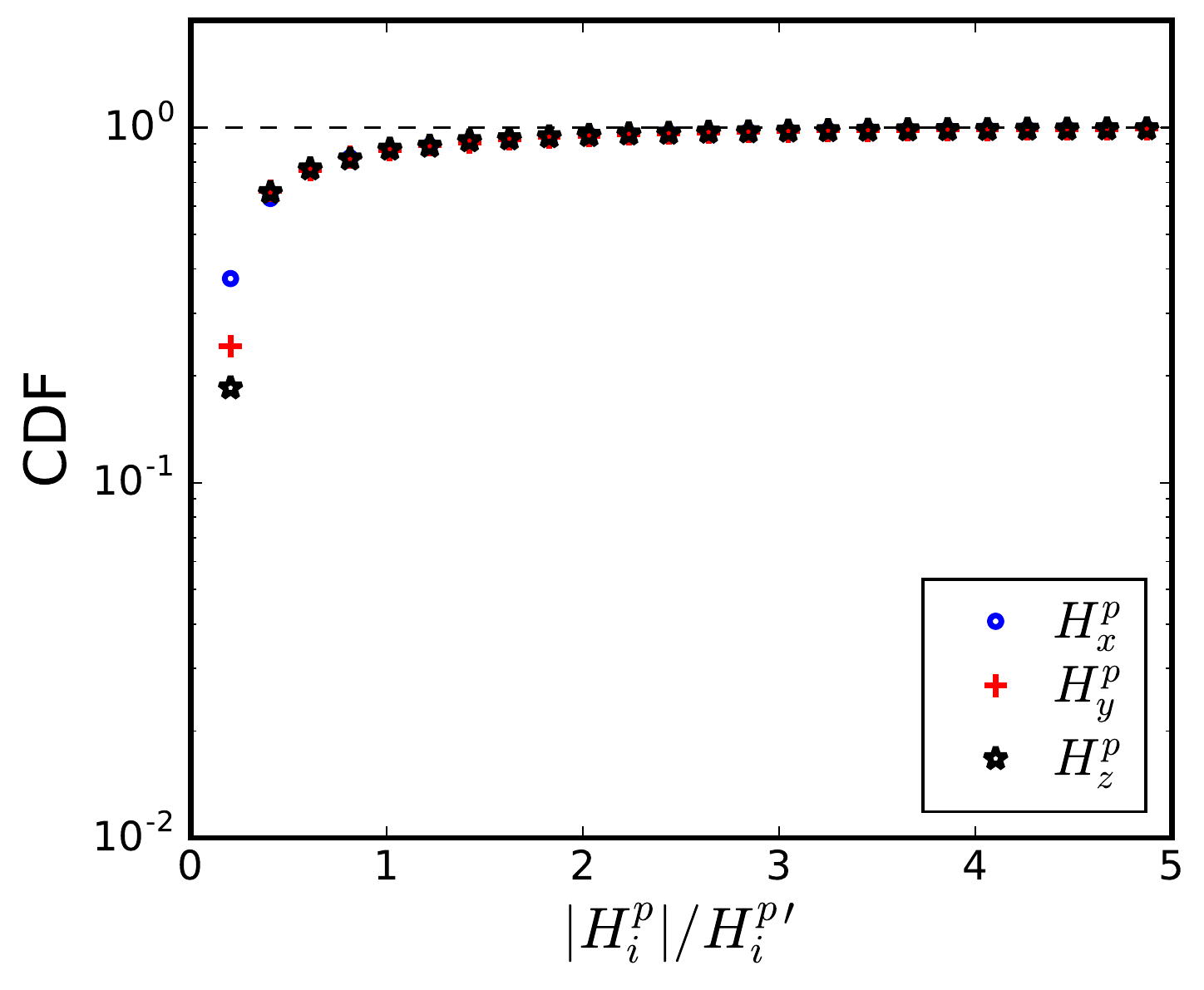}}

  \caption{}
   \end{subfigure}
  \caption{The $\hp$ correlation is shown for a single field snapshot, at $500\tau_k$. Panel (a) shows the three-dimensional field as a volume rendering, while panel (b) shows the cross-sectional view of $H^p=\sqrt{\hp \cdot \hp}$. Panels (c) and (d) show the time-averaged PDFs and CDFs of the three components of $\hp$.}
\label{fig:TurbulenceCorrelation-HCorrArea}
\end{figure}

\sid{Lastly, figure \ref{fig:TurbulenceCorrelation-HCorrArea} shows the $\hp$ correlation, which is very similar to the $\h$ correlation. Similarly to the $G$, $G^s$ and $H$ fields, the volumetric and planar $H^p$ fields, in panels (a) and (b), respectively, also closely resemble the enstrophy field in figure \ref{fig:TurbulenceData}. Similarly to $H_i$, the PDFs of the $H^p_i$ components are highly positively skewed, and have a long-tailed distribution, with slightly more extreme values than $H_i$. The similarity between $H^p$ and $H$, together with the other results, shows that the high enstrophy regions form small scale vorticity jets surrounded by a small region of swirling-flow. Note that the different correlation measures address different aspects of the structure of the velocity and vorticity fields and it is the combination of the different results and the similarities involved that leads to this conclusion (see figure \ref{fig:BurgersVortexCorrelationsTop}, where the correlations are applied to Burgers vortices). For example, a small-scale high vorticity jet is not necessarily surrounded by a swirling-flow region, nor a small region of strong alignment between the angular velocity and the vorticity necessarily implies a swirling-flow. However, comparing the high enstrophy structures ($\omega^2 > 5\ang{\omega^2}$) in figure \ref{fig:TurbulenceData} with regions of high $G$, $G^s$, $H$ and $H^p$, in figures \ref{fig:TurbulenceCorrelation-GCorr}, \ref{fig:TurbulenceCorrelation-GSCorr}, \ref{fig:TurbulenceCorrelation-HCorr} and \ref{fig:TurbulenceCorrelation-HCorrArea}, respectively, we can conclude that strong enstrophy regions are almost invariably associated with a surrounding region of swirling-flow. The CDFs of ${H^p_i}$ show that roughly $86\%$, $94\%$ and $97\%$ of the fields are within $1{H^p_i}^\prime$, $2{H^p_i}^\prime$ and $3{H^p_i}^\prime$, respectively, which are values similar to the relative levels of the $H$ field.}

\sid{We note some overall differences in the correlation fields in comparison to the canonical flows. Unlike both the one-dimensional Oseen vortex examples and the three-dimensional Burgers vortices, in (homogeneous isotropic) turbulence, we do not find high $L$ regions being associated with high $H^p$ at their cores. This shows that there are no large or intermediate scale swirling flow regions, which instead are found only at the smaller scales. Additionally, unlike the Taylor-Green example, the jet-like flow regions in turbulence, which are also associated with high $E_k$, are \textit{not} induced by the interaction of swirling structures. These regions, however, are still externally induced, as our analysis will reveal. This shows that the organization of coherent structures in turbulence is more complex than the simplified canonical flows shown here where there is a clear association between the coherence of high vorticity regions and high kinetic energy regions.}

We also applied these correlations to a reference dataset of homogeneous isotropic turbulence obtained from the Johns Hopkins Turbulence Databases (JHTD) \citep{perlman2007data,li2008public}, which was generated using a pseudo-spectral method on a grid of $1024^3$ at $Re_\lambda = 433$. The results are shown in Appendix \ref{app:JHTD-Validation}; they are found to be essentially similar to those presented here, from our in-house code. We use our own numerical datasets for the remainder of this study, for our ease of access and control over the data.

\subsection{Influence of the choice of $\Lambda$}\label{sec:ChoiceOfLambda}
Before proceeding with further analysis of the correlation fields, it is important to consider the influence of $\Lambda$ on the results. The obvious values of $\Lambda$ that can be disregarded are those extremely small or large. A value of $\Lambda$ that is too small is somewhat meaningless, since we intend to capture non-local structures which have a finite physical size. On the other hand, very large values of $\Lambda$ ($\sim N_x/2$) will introduce periodicity induced artifacts in the correlation fields, which should be avoided. However, there is a wide range of values of $\Lambda$ in $0 < \Lambda < N_x/2$ which are viable. \sid{Here, we do not intend to study in detail the influence of $\Lambda$ on the correlation measures of the field structures, hence, we would like to choose a value of $\Lambda$ large enough to cover most structure sizes, such that the results do not depend strongly on $\Lambda$.}

\sid{The correlations $\g$ and $\gs$ are not expected to vary significantly for different choices of $\Lambda$ (once $\Lambda$ is large enough to cover the size of these structures), since the structures in the vorticity field are small-scaled, and more or less randomly distributed throughout the volume. Similarly, the correlations $\h$ and $\hp$ are expected to be even more insensitive to the choice of $\Lambda$, since the Biot-Savart influence of the vorticity decays with the square of the distance. We consider the $\lvec$ correlation, since it identifies intermediate sized structures in the velocity field, and can potentially be influenced by the choice of $\Lambda$. $\lvec$ has the form}
\begin{equation}
L_i(\mathbf{x},\Lambda) = \int_{-\Lambda}^{\Lambda} \mathbf{u}(\mathbf{x})\cdot \mathbf{u}(\mathbf{x}+\mathbf{r}_i) \mathrm{d}r_i
\end{equation}
where $\mathbf{u}(\mathbf{x})$ can be placed outside the integral as 
\begin{align}
L_i(\mathbf{x},\Lambda) &= \mathbf{u}(\mathbf{x})\cdot \int_{-\Lambda}^{\Lambda} \mathbf{u}(\mathbf{x}+\mathbf{r}_i) \mathrm{d}r_i \nonumber \\
L_i(\mathbf{x},\Lambda) &= 2\Lambda \left( \mathbf{u}(\mathbf{x})\cdot \mathbf{\widetilde{u}}_i \right) 
\end{align}
where 
\begin{equation}
\mathbf{\widetilde{u}}_i = \frac{1}{2\Lambda}\int_{-\Lambda}^{\Lambda} \mathbf{u}(\mathbf{x}+\mathbf{r}_i) \mathrm{d}r_i
\end{equation}

\begin{figure}
  \centerline{\includegraphics[width=\linewidth]{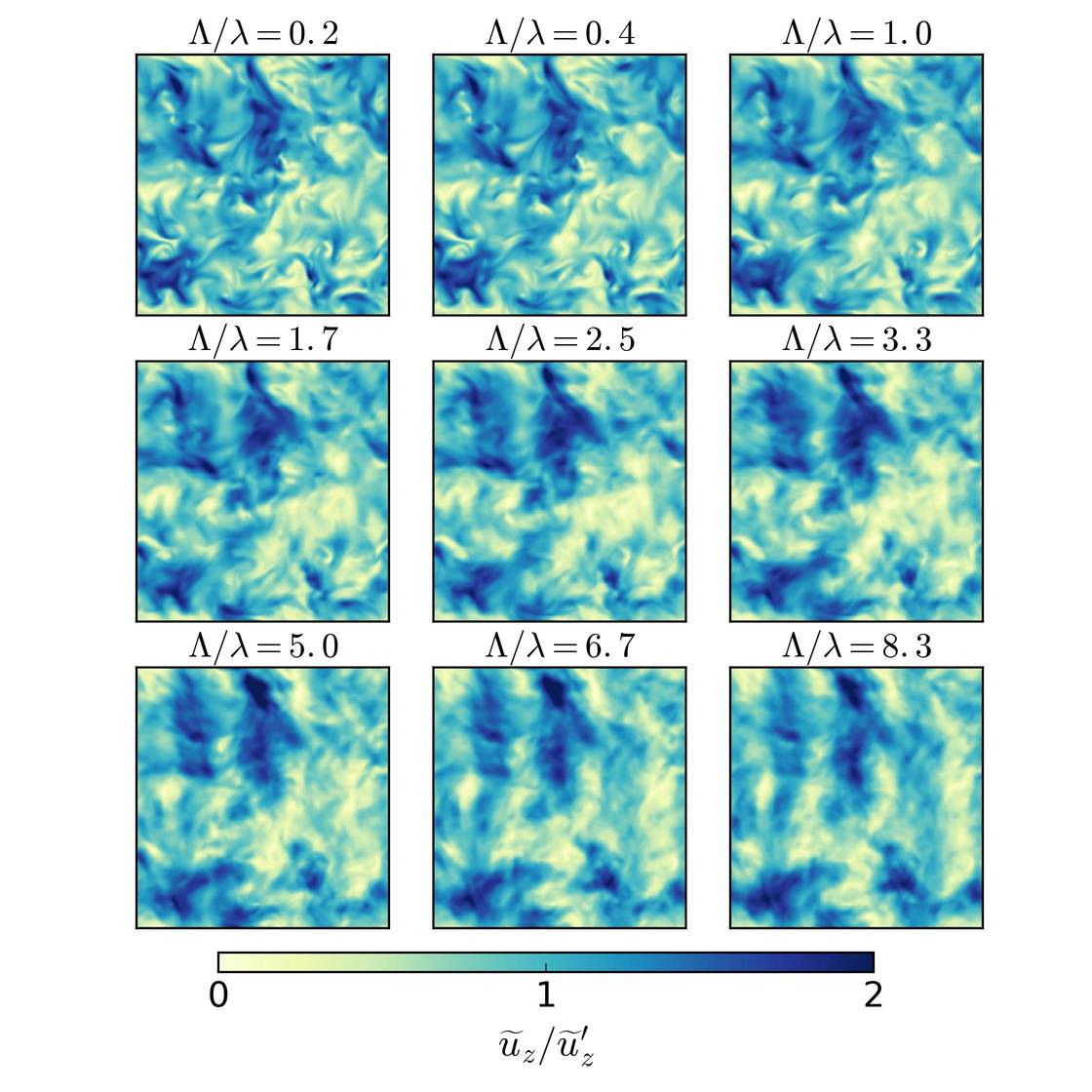}}

  \caption{Crossectional slices of the $\tilde{u}_z$ field for varying $\Lambda/\lambda$.}
\label{fig:UTildeSlices}
\end{figure}

The $\lvec$ correlation is essentially the inner product of the velocity field $\mathbf{u}$ with the $\utildebf$ field (which is a function of $\Lambda$). Hence, if the $\utildebf$ field varies significantly with $\Lambda$, so will $\lvec$. In figure \ref{fig:UTildeSlices}, snapshots of the $\utilde_z = |\utildebf_z|$ field are shown for a wide range of $\Lambda/\lambda$ values, at a planar crossection of the volume (at the same time instance shown in figure \ref{fig:TurbulenceData}). At very small values, $\Lambda/\lambda < 1$, the $\lvec$ field looks very similar to the $E_k$ field, which is expected since the limit $\Lambda \to 0$ reduces $\mathbf{u}\cdot\utildebf$ to $\mathbf{u}\cdot\mathbf{u}$. The $\utilde_z$ field does not appear to change significantly for $1\leq \Lambda/\lambda <7$, which is also true for $\Lambda/\lambda \geq 7$, although those values of $\Lambda$ begin to approach the size of the simulation domain and should be disregarded.

\begin{figure}
  \centerline{\includegraphics[width=0.5\linewidth]{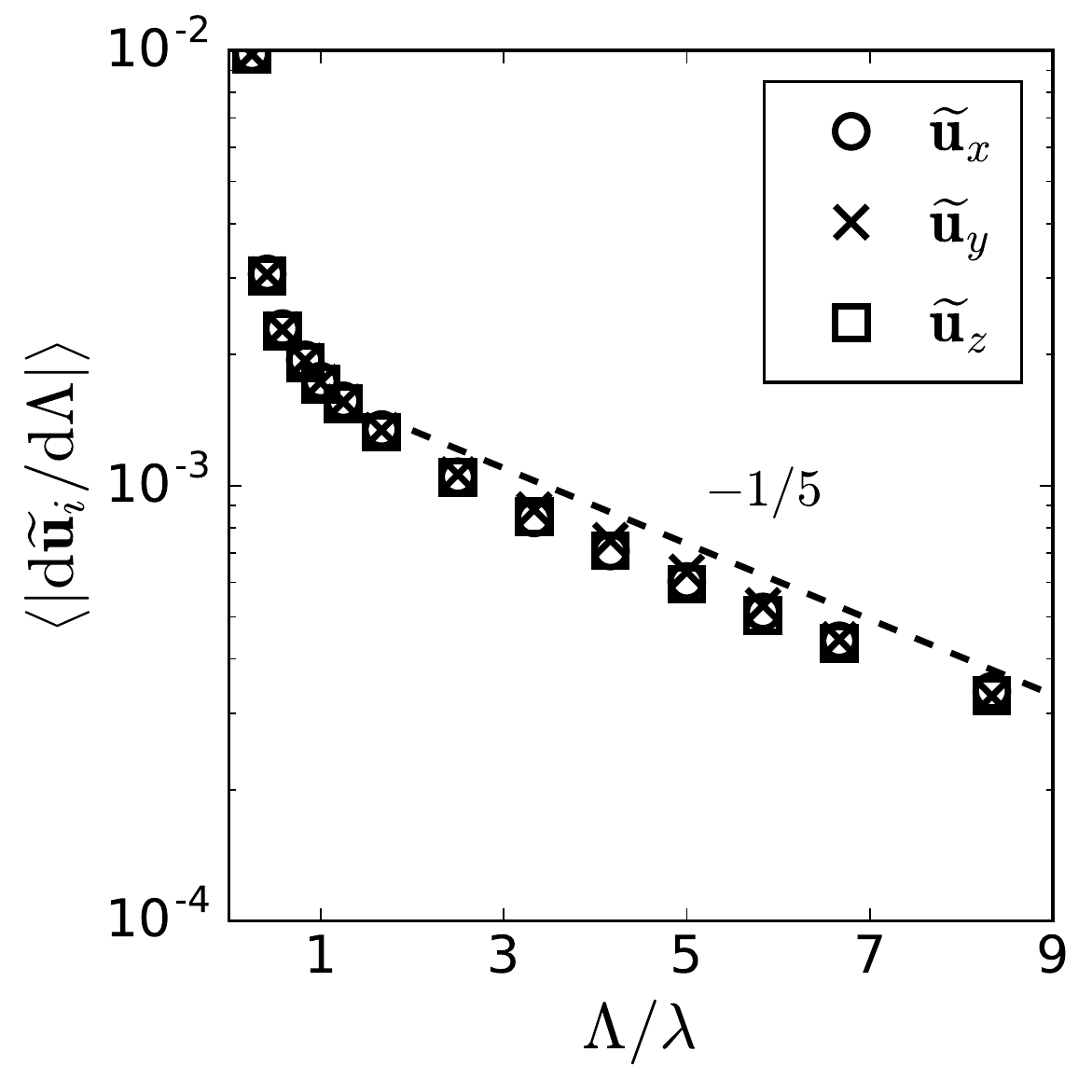}}

  \caption{Rate of change in the $\utildebf_i$ fields with increasing $\Lambda$. Here $\ang{\cdot}$ denotes spatial averaging.}
\label{fig:DUTilde}
\end{figure}

It is interesting that the $\widetilde{u}_z$ field appears to vary slowly for $\Lambda/\lambda > 1$. \sid{The variation of $\mathbf{\widetilde{u}}_i$ with $\Lambda$ can be quantified by calculating}
\begin{equation}
\absfrac*{\frac{\mathrm{d}\utildebf_i}{\mathrm{d}\Lambda}} = \frac{|\utildebf_i^{\Lambda+\Delta\Lambda} - \utildebf_i^{\Lambda}|}{\Delta \Lambda}
\end{equation}
\sid{where $|\cdot|$ is the amplitude of the difference between the two fields. This is shown in figure \ref{fig:DUTilde} for $\utildebf_x, \utildebf_y$ and $\utildebf_z$, where $\ang{\cdot}$ denotes spatial averaging over the entire volumetric domain, and over two independent realizations of $\utildebf_i$ at $t \approx 500\tau_k$ and $1000\tau_k$. The change in $\ang{|\mathrm{d}\utildebf_i/\mathrm{d}\Lambda|}$ is large for $\Lambda/\lambda < 1$. This reflects the fact that most of the velocity structures in the flow are smaller than the Taylor microscale $\lambda$, and they get averaged over in the $\utildebf$ fields for increasing $\Lambda$. Next, $\ang{|\mathrm{d}\utildebf_i/\mathrm{d}\Lambda|}$ seems to decay exponentially for $\Lambda/\lambda>1$, with a slope of approximately $-1/5$. This change of behaviour occurs via a sharp transition around $\Lambda/\lambda \approx 1$.}

\sid{The behaviour of $\ang{|\mathrm{d}\utildebf_i/\mathrm{d}\Lambda|}$ shown in figure \ref{fig:DUTilde} indicates that for large $\Lambda/\lambda$, i.e. in the limit of $\Lambda \to \infty$: (i) $\mathrm{d}\utildebf_i/\mathrm{d}\Lambda \to 0$, (ii) $\utildebf_i \to 0$, (iii) $\mathrm{d}\mathbf{L}/\mathrm{d}\Lambda \to 0$ and (iv) $\lvec$ goes to a constant. This is consistent with equation \ref{eq:surfaceLimit} and the requirement that there should not exist ``an organized motion over an infinite distance''. Also, figure \ref{fig:DUTilde} shows that $\utildebf_i$ varies slowly for $\Lambda/\lambda > 1$ and suggests that $\Lambda|\utildebf_i|$ attains a maximum, roughly, in the range $1 \leq \Lambda/\lambda \leq 4$, hence, the results for $\lvec$ and $\lsvec$ are not expected to vary significantly in this range of $\Lambda$. This, together with the sharp transition around $\Lambda/\lambda \approx 1$, suggests that the Taylor microscale $\lambda$ is a good measure for the integration length of the correlations. For the remainder of this study, we use correlation fields calculated for $\Lambda = \lambda$. This value is also large enough to account for vorticity field structures and vorticity-velocity (i.e. Biot-Savart related) field structures.}

\subsection{Spectral characteristics of the correlation fields}\label{sec:CorrelationSpectra}
\begin{figure}
	\begin{subfigure}{0.5\linewidth}
  \centerline{\includegraphics[width=\linewidth]{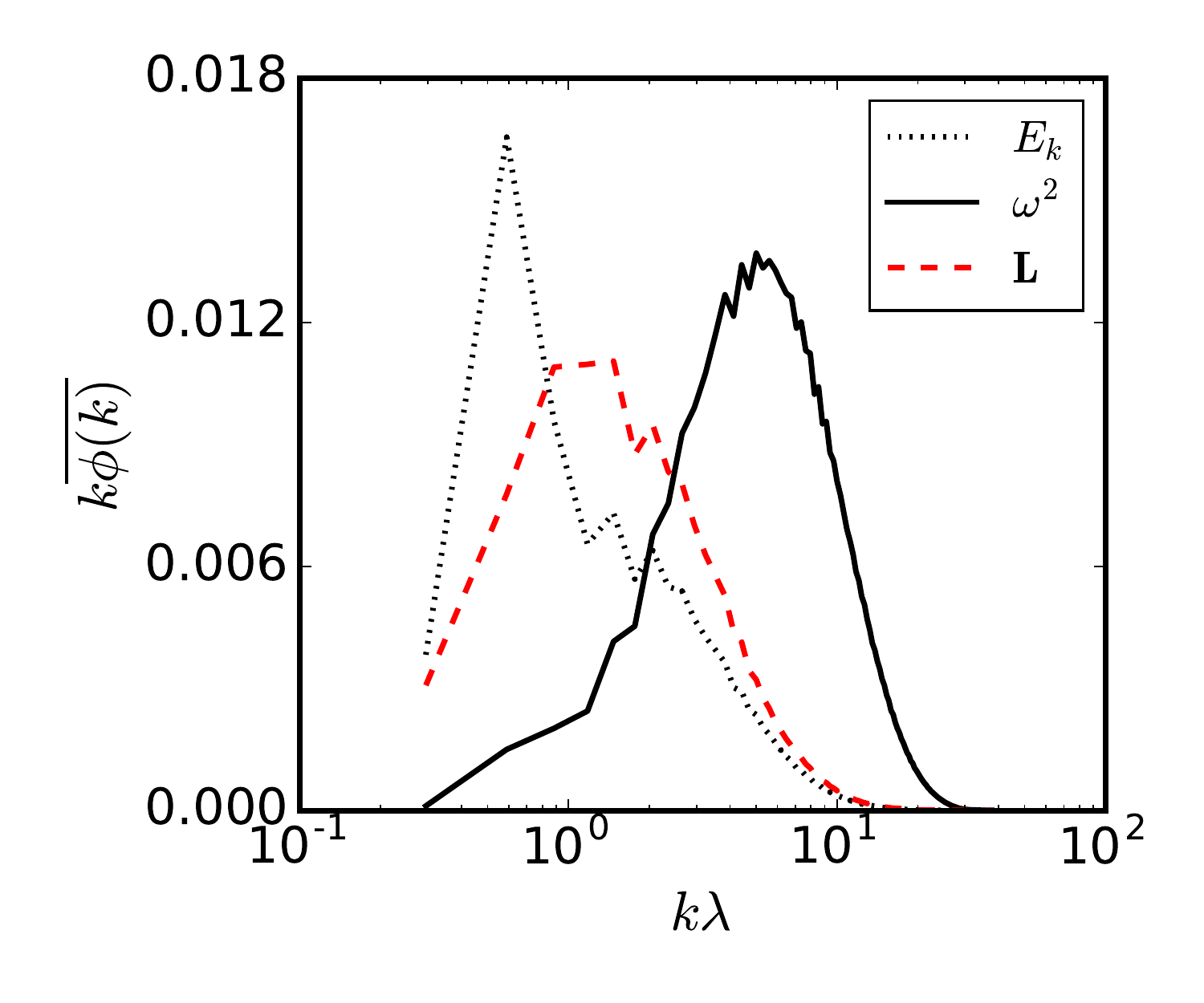}}

  	\caption{}
  \end{subfigure}
  	\begin{subfigure}{0.5\linewidth}
  \centerline{\includegraphics[width=\linewidth]{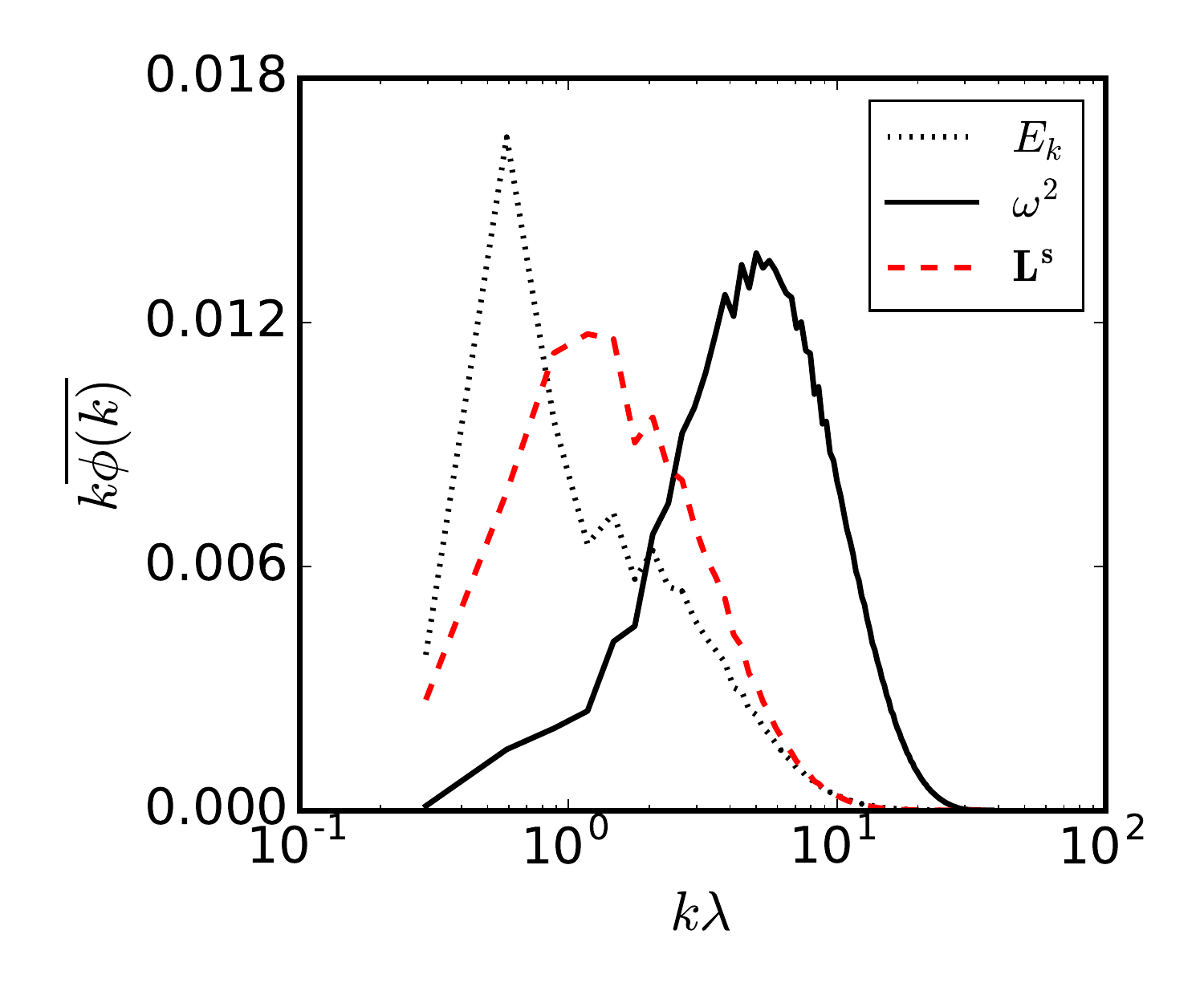}}

  	\caption{}
  \end{subfigure}

  \begin{subfigure}{0.5\linewidth}
  \centerline{\includegraphics[width=\linewidth]{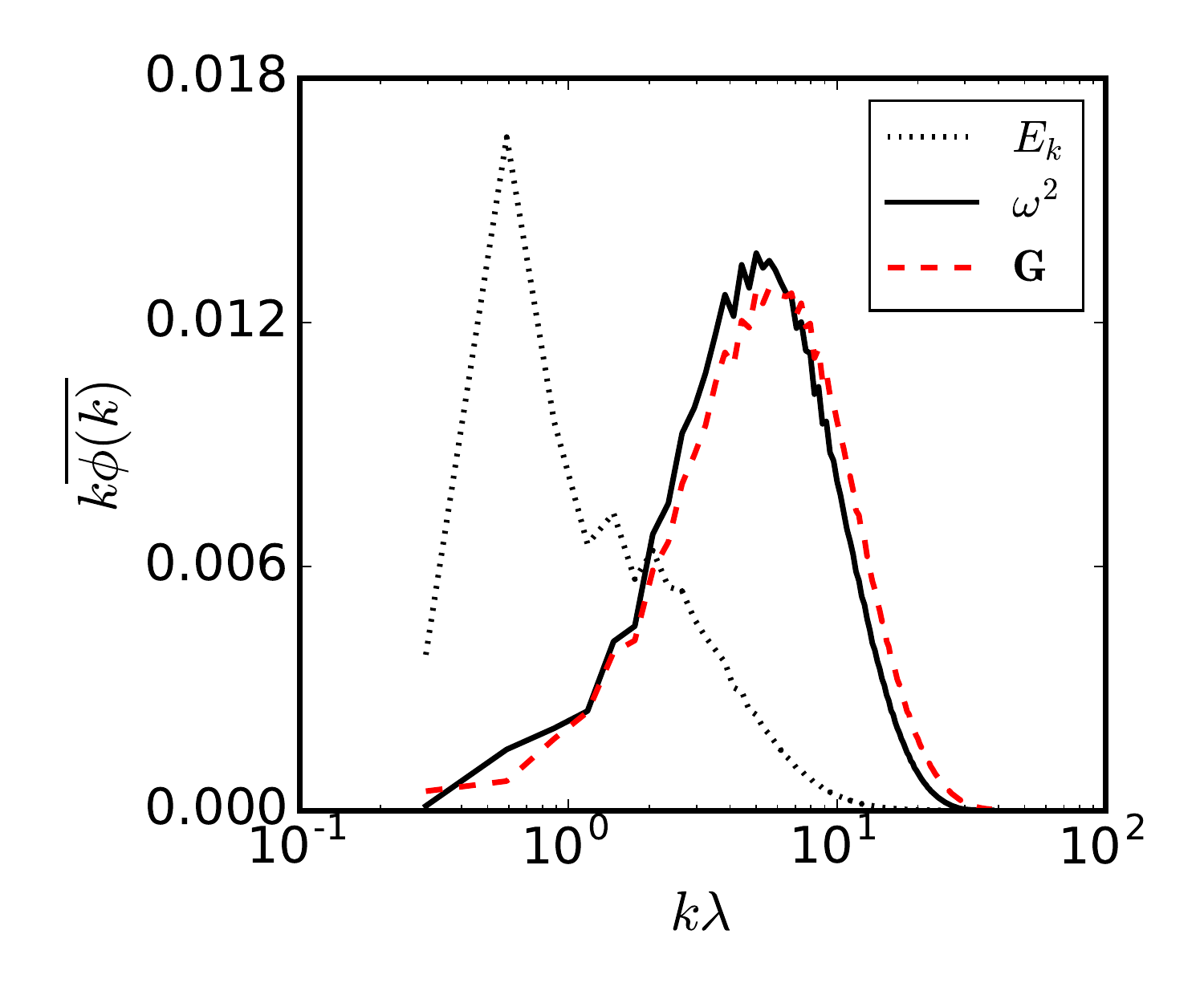}}

  	\caption{}
  \end{subfigure}
  \begin{subfigure}{0.5\linewidth}
  \centerline{\includegraphics[width=\linewidth]{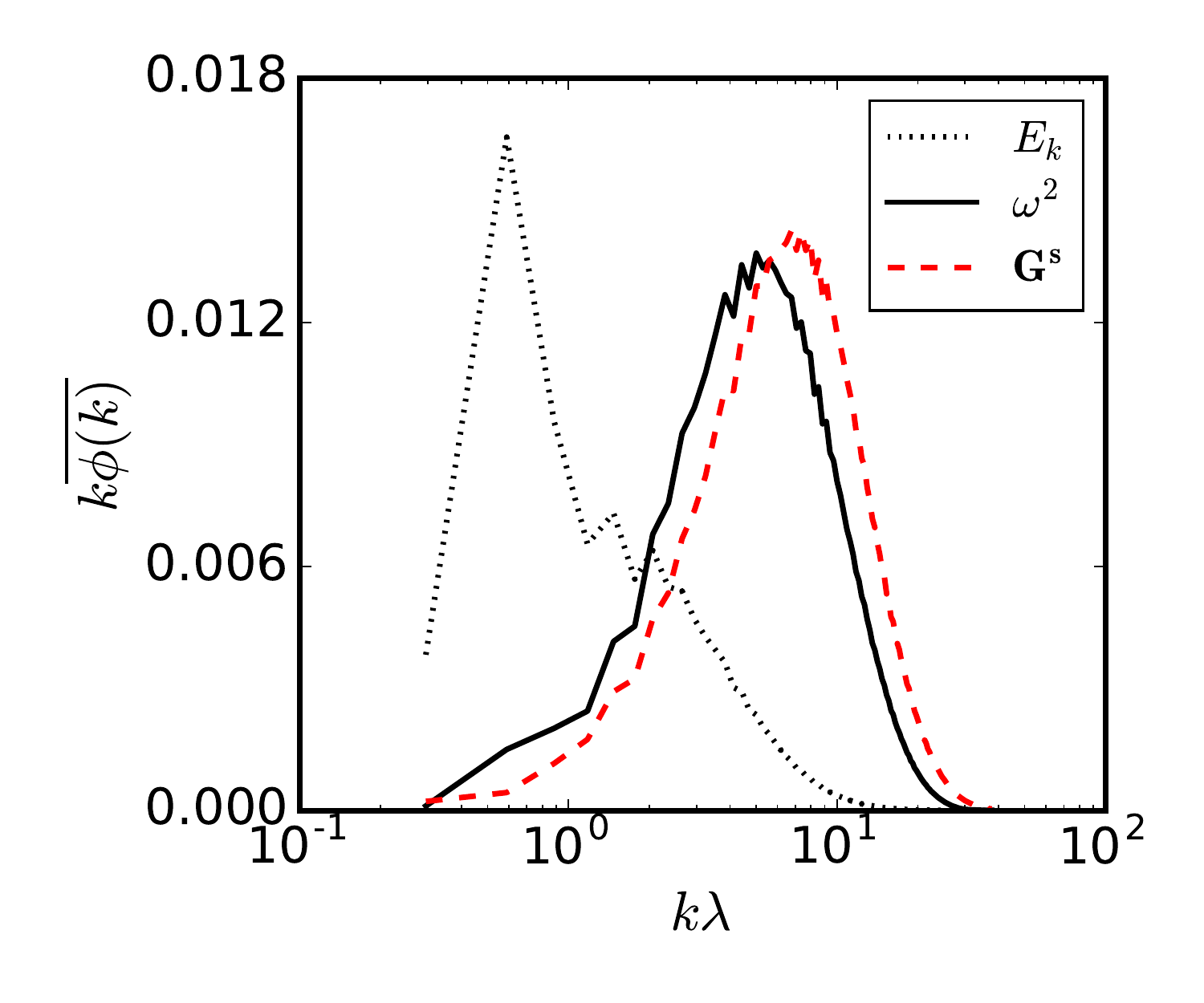}}

  	\caption{}
  \end{subfigure}

  \begin{subfigure}{0.5\linewidth}
  \centerline{\includegraphics[width=\linewidth]{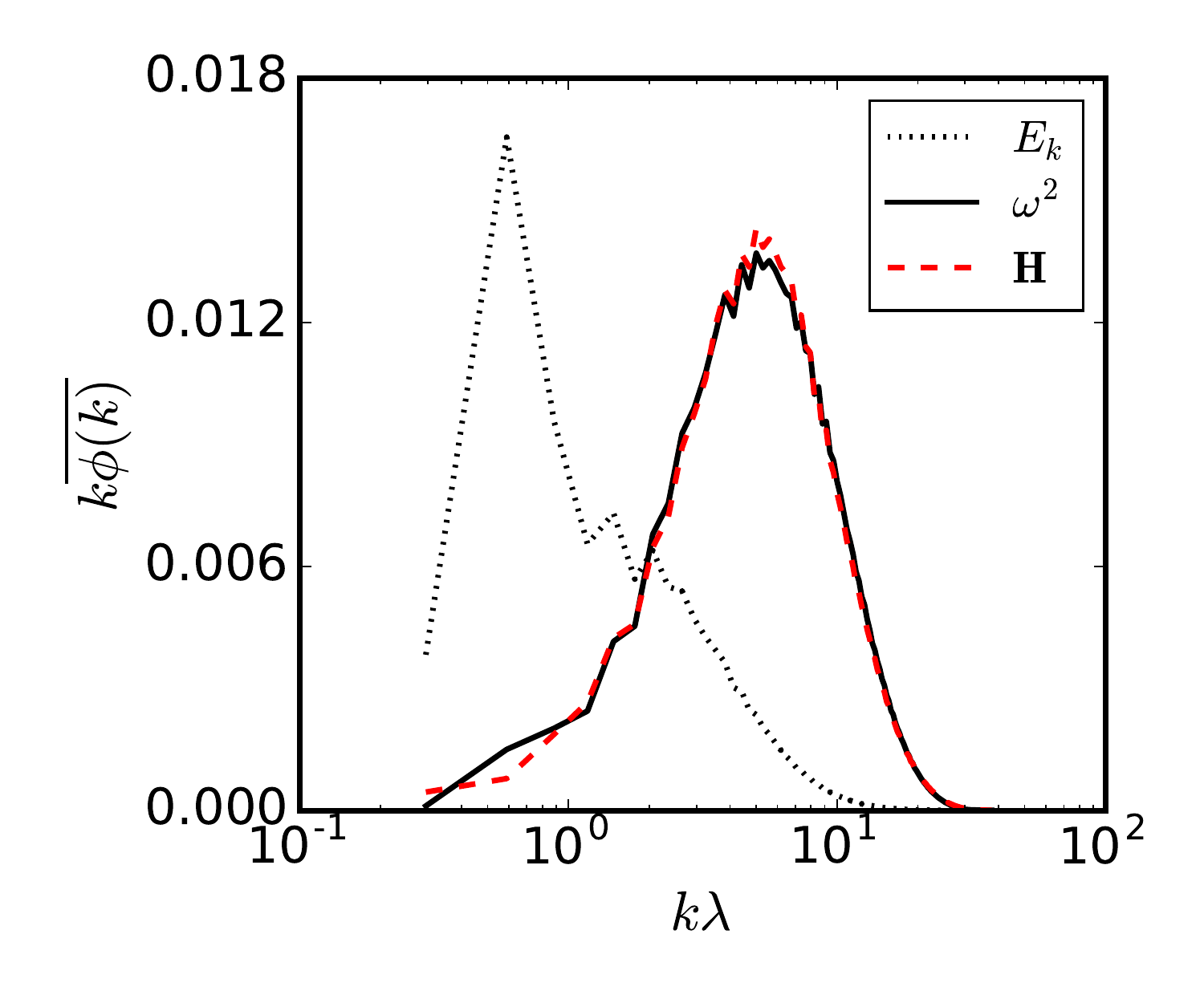}}

  	\caption{}
  \end{subfigure}
  \begin{subfigure}{0.5\linewidth}
  \centerline{\includegraphics[width=\linewidth]{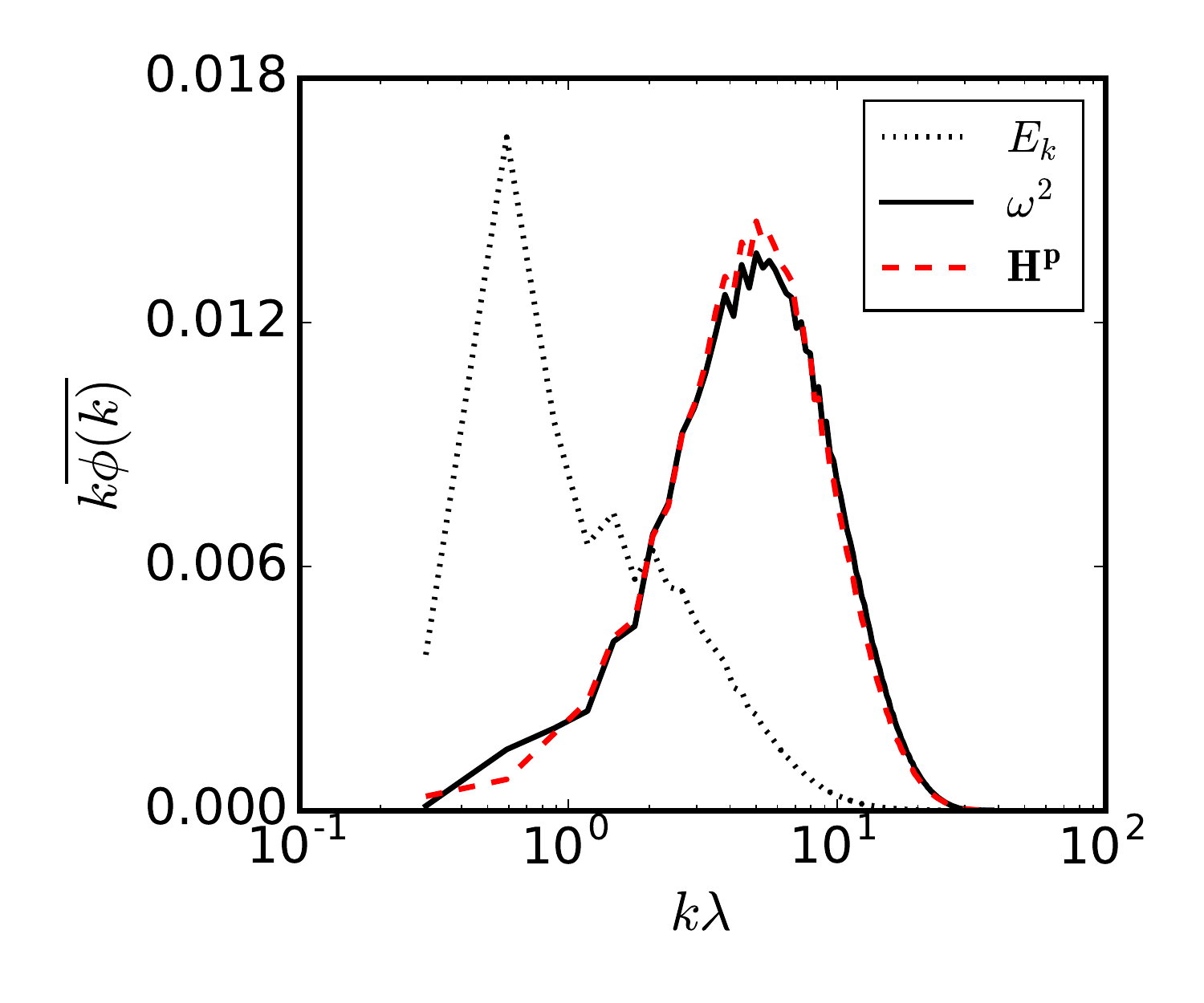}}

  	\caption{}
  \end{subfigure}
  \caption{Time-averaged premultiplied spectra of correlation fields (a) $\lvec$, (b) $\lsvec$, (c) $\g$, (d) $\gs$, (e) $\h$ and (f) $\hp$ shown together with the kinetic energy ($E_k$) and enstrophy ($\omega^2$) premultiplied spectra.}
\label{fig:SpectraCorrelations}
\end{figure}

The spectra of the three-dimensional correlation fields are shown in figure \ref{fig:SpectraCorrelations}, in comparison to the kinetic energy ($E_k$) and enstrophy ($\omega^2$) spectra. The correlation spectra are calculated in the same way as the $E_k$ spectra, where the three-dimensional Fourier transforms of the correlation fields are squared and spherically averaged over wavenumber shells. The spectra have also been time-averaged over $20$ realizations, each separated by $50\tau_k$. Panels (a) and (b) show the $\lvec$ and $\lsvec$ spectrum respectively, that are found to have very similar shapes. This reflects that the spatial distribution of the $\lsvec$ field is very similar to that of the $\lvec$ field, reaffirming that there are no large symmetries or anti-symmetries in the velocity field. The spectra, in comparison to the $E_k$ spectrum, have a shift in the peak to higher wavenumbers. \sid{This is a consequence of the fact that the $\lvec$ and $\lsvec$ spectra reflect the structure of the correlation fields, which comprise of relatively smaller correlation kernels associated with larger flow structures.}

The spectra of $\g$ and $\gs$ in panels (c) and (d), respectively, are found to closely resemble the enstrophy spectrum, with a slight shift in the spectral peak to higher wavenumbers for both cases, which is again due to the correlation kernels being relatively smaller than the vorticity field structures. This shift is more pronounced in the $\gs$ spectrum, which is consistent with the fact that the $\gs$ field has sharper features, as can be seen in both figure \ref{fig:TurbulenceCorrelation-GSCorr}(a) (volume rendered $\gs$ field snapshot) and figure \ref{fig:oneDimensionalOseen-GCorrLambda} (one-dimensional Oseen vortex-pair example). 

The $\h$ and $\hp$ spectra in panels (e) and (f), respectively, are found to have a striking similarity to the enstrophy spectrum. This reflects that the spatial variation of enstrophy consistently yields a similar spatial variation in the $\h$ and $\hp$ fields. \sid{This also indicates that the ``energy containing'' wavelengths of the velocity field do not have any significant swirl, showing that the swirling flow structures are coupled with the vorticity structures, and decoupled from the strong velocity structures.} Changing the correlation integration length $\Lambda$ to different values can possibly influence the spectral characteristics of the correlation fields, \sid{which we do not investigate in this study}.

\sid{Overall, the similarity between $\lvec$ and $\lsvec$ spectra shows that there are no larger symmetries or anti-symmetries in the velocity field. Further, the similarities between $\g$, $\gs$, $\h$ and $\hp$, together with the enstrophy spectrum, shows that high enstrophy regions invariably form concentrated vorticity-jets, surrounded by swirling velocity regions. These results also support the statistics of the correlation fields shown so far. For the remainder of this study, we shall use the $\lvec$ and $\hp$ correlations to identify structures, since the nature of (homogeneous isotropic) turbulence fields leads to a strong correspondence between (i) $\lvec$ and $\lsvec$ correlations and (ii) $\g$, $\gs$, $\h$ and $\hp$ correlations, while in general, this need not be the case.}

\subsection{Spatial distribution of correlation fields}\label{sec:spatialDistribution}
\sid{The different correlation measures identify distinct forms of coherence in the flow. The $\lvec$ correlation identifies intermediate sized regions where the velocity field has a high magnitude and parallel streamlines. The $\hp$ correlation identifies smaller scale high vorticity regions associated with a swirling velocity. The relative spatial distribution of the correlation fields, hence, sheds light on the distribution of the different coherent structures in physical space. In this section, we discuss the spatial distribution statistics of $\lvec$ and $\hp$, in relation to $E_k$ and $\omega^2$.}

\begin{figure}
	\begin{subfigure}{0.5\linewidth}
  \centerline{\includegraphics[width=\linewidth]{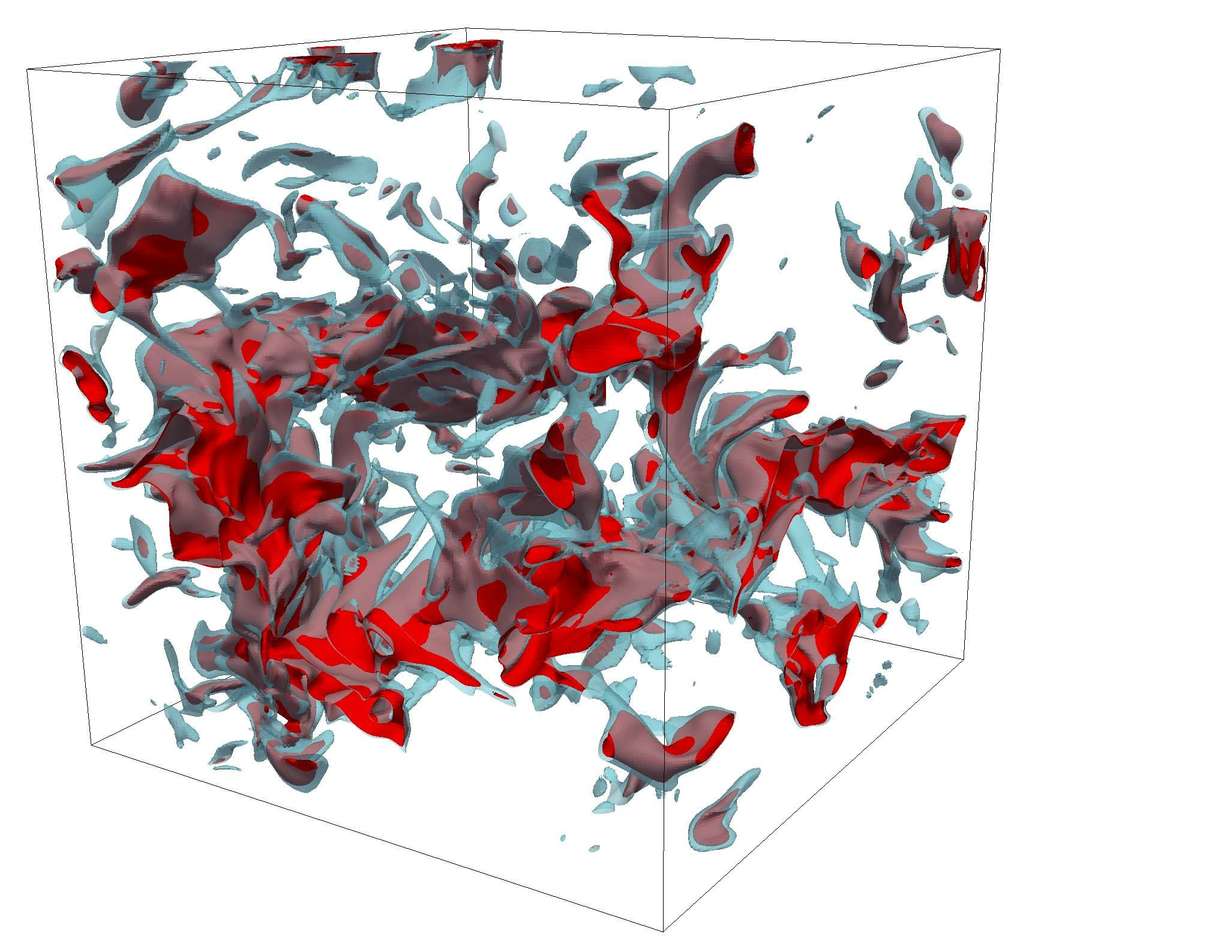}}

  	\caption{Contours of the turbulent kinetic energy $E_k = 2\ang{E_k}$ (blue) shown together with contours of $L = 2 L^\prime$ (red).}
  \end{subfigure}\quad
  	\begin{subfigure}{0.5\linewidth}
  \centerline{\includegraphics[width=\linewidth]{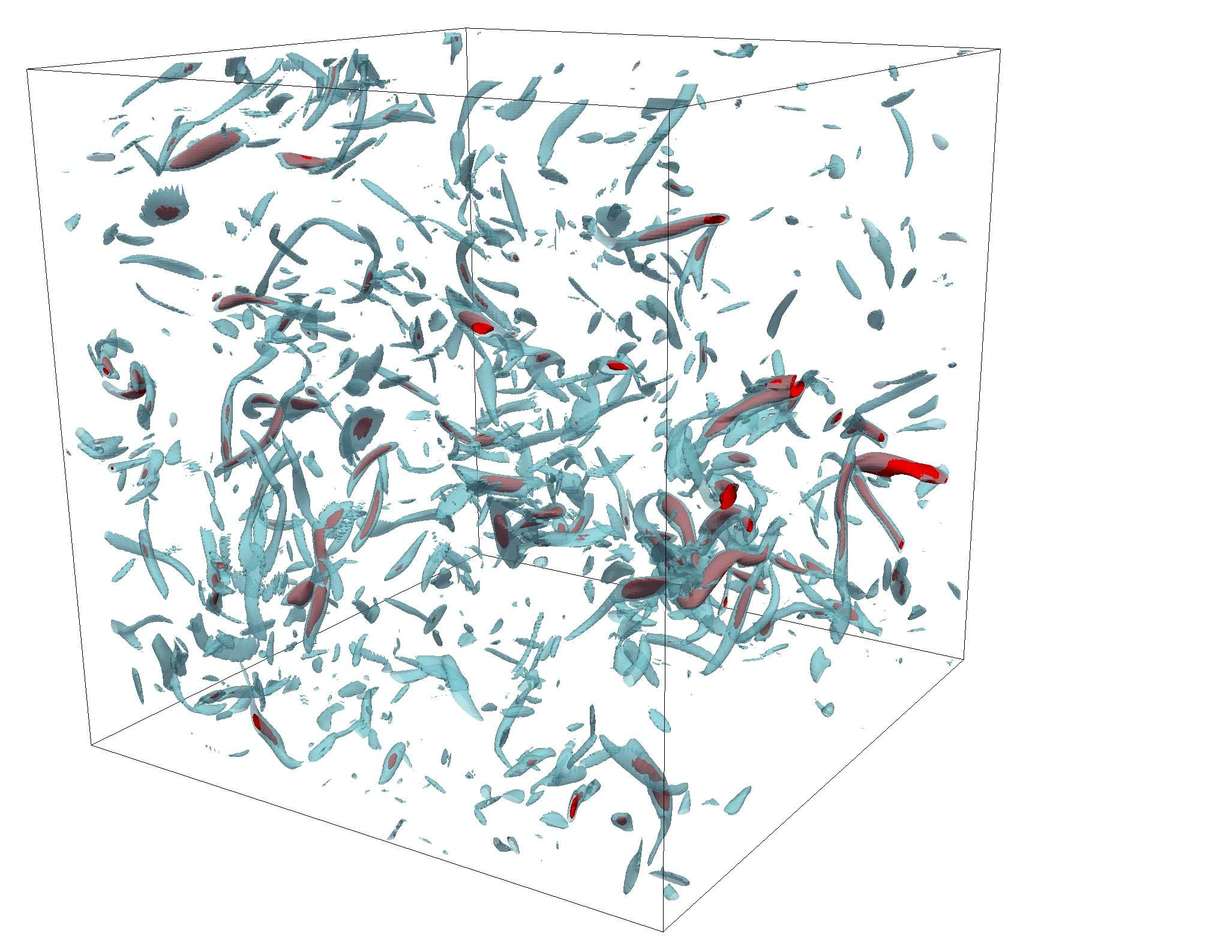}}

  	\caption{Contours of vorticity at $\omega = 3\omega^\prime$ (blue) shown together with contours of $H^p = 8{H^p}^\prime$ (red).}
  	\end{subfigure}
  	
   \begin{subfigure}{0.5\linewidth}
  \centerline{\includegraphics[width=\linewidth]{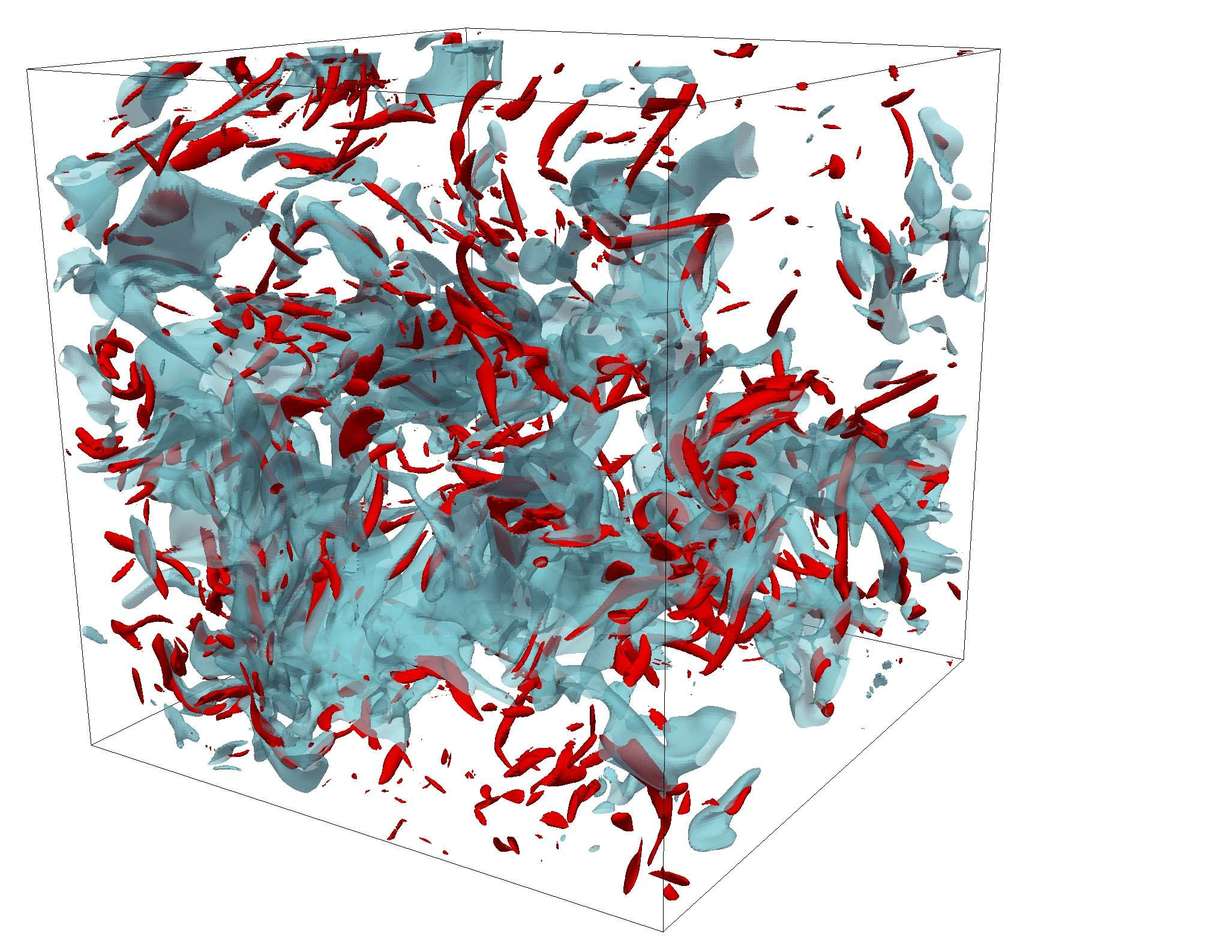}}

  	\caption{Contours of $E_k=2\ang{E_k}$ (blue) along with $\omega=3\omega^\prime$ (red) shows that these fields are spatially exclusive.}
  \end{subfigure}\quad
  	\begin{subfigure}{0.5\linewidth}
  \centerline{\includegraphics[width=\linewidth]{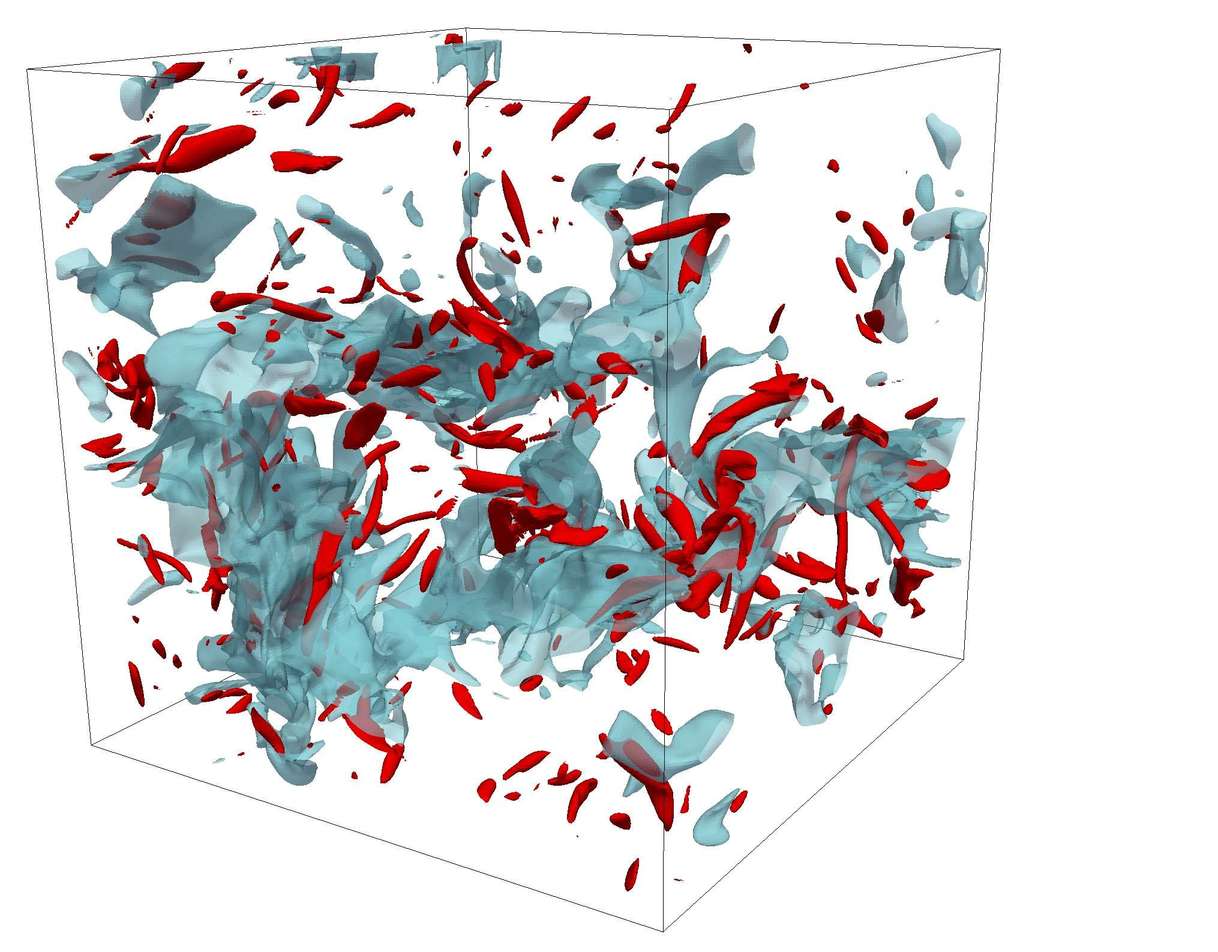}}

  	\caption{Contours of $ L = 2L^\prime$ (blue) along with $H^p = 5{H^p}^\prime$ (red) shows that these correlation fields are spatially exclusive.}
  	\end{subfigure}
 \caption{Spatial distribution of correlations $L$ and $H^p$ in comparison to turbulent kinetic energy $E_k$ and vorticity $\omega$.}
\label{fig:CorrelationsSpatialDistribution}
\end{figure}

First, figure \ref{fig:CorrelationsSpatialDistribution}(a) shows iso-surfaces of $E_k=2\ang{E_k}$ (in blue), together with iso-surfaces of $L=2L^\prime$ (in red). At the chosen threshold levels, the $L$ regions \sid{are mostly contiguous, and} consistently contained within the $E_k$ regions, reflecting the fact that increasingly higher values of the $L$ field occupy successively smaller regions of space. This also shows that high kinetic energy regions, $E_k \geq 2\ang{E_k}$, yield high $L$ values, in the range $L\geq 2L^\prime$. This also suggests that \sid{as $E_k$ gets higher, the level or flow organization increases}.
 
Figure \ref{fig:CorrelationsSpatialDistribution}(b) shows iso-surfaces of vorticity at $\omega = 3\omega^\prime$ (in blue) together with iso-surfaces of $H^p$ at $8{H^p}^\prime$ (in red). The $H^p$ iso-surfaces are shown at a high value to demonstrate that only a fraction of the intermediate vorticity `worms' yield very high $H^p$ values in their core regions. For instance, in panel (d), the $H^p=5{H^p}^\prime$ iso-surfaces shown have a comparable size to the $3\omega^\prime$ iso-contours in panel (b). It is found that $H^p$ occupies equivalent or more volume than $\omega$ at low thresholds, while at increasingly higher thresholds, the $H^p$ field occupies successively smaller fractions of the volume, as shall be quantified with the spatial statistics. 

Figure \ref{fig:CorrelationsSpatialDistribution}(c) shows contours of $E_k$ at $2\ang{E_k}$ (in blue) shown together with contours of vorticity at $\omega = 3\omega^\prime$. It appears that high kinetic energy regions and high vorticity regions are \textit{spatially exclusive} to a large extent. \sid{This spatial distribution of the two quantities might be a consequence of the dynamical separation between them, as high $E_k$ and high $\omega^2$ are influenced by different aspects of the Navier-Stokes dynamics, as also observed by \cite{tsinober2014essence}}. The two fields, as expected, begin to also overlap when the thresholds are lowered, and become more exclusive and distanced at higher thresholds. Lastly, panel (d) shows the distribution of correlations $L$ and $H^p$, where contours of $L = 2L^\prime$ (in blue) are shown together with contours of $H^p = 5{H^p}^\prime$. Since these correlations closely resemble $E_k$ and $\omega$, respectively, they also remain spatially exclusive, at relatively high threshold levels.

\begin{figure}
  \begin{subfigure}{0.49\linewidth}
  \centerline{\includegraphics[width=\linewidth]{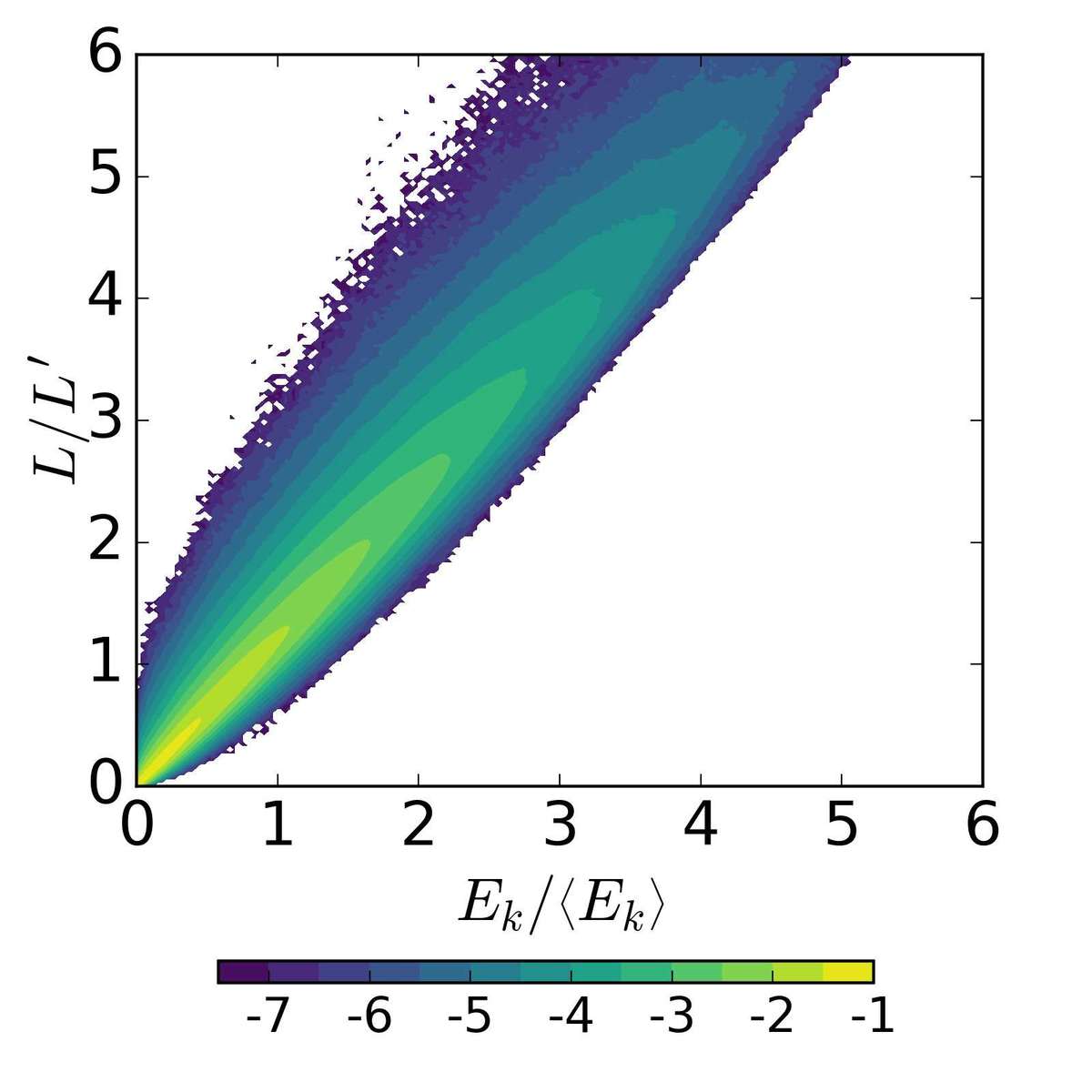}}

  	\caption{}
  \end{subfigure}\quad
\begin{subfigure}{0.49\linewidth}
  \centerline{\includegraphics[width=\linewidth]{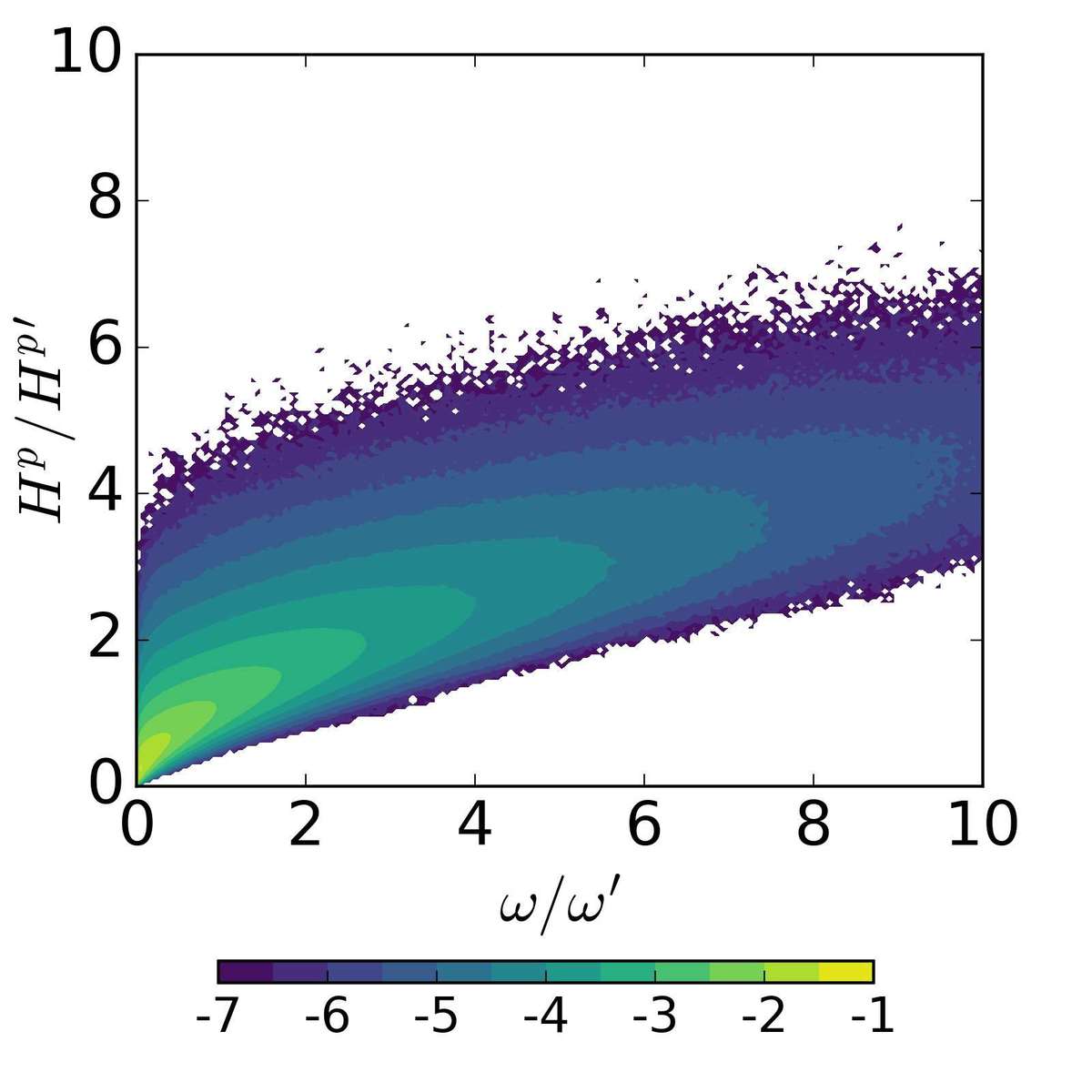}}

  \caption{}
    \end{subfigure}
  
    \begin{subfigure}{0.49\linewidth}
  \centerline{\includegraphics[width=\linewidth]{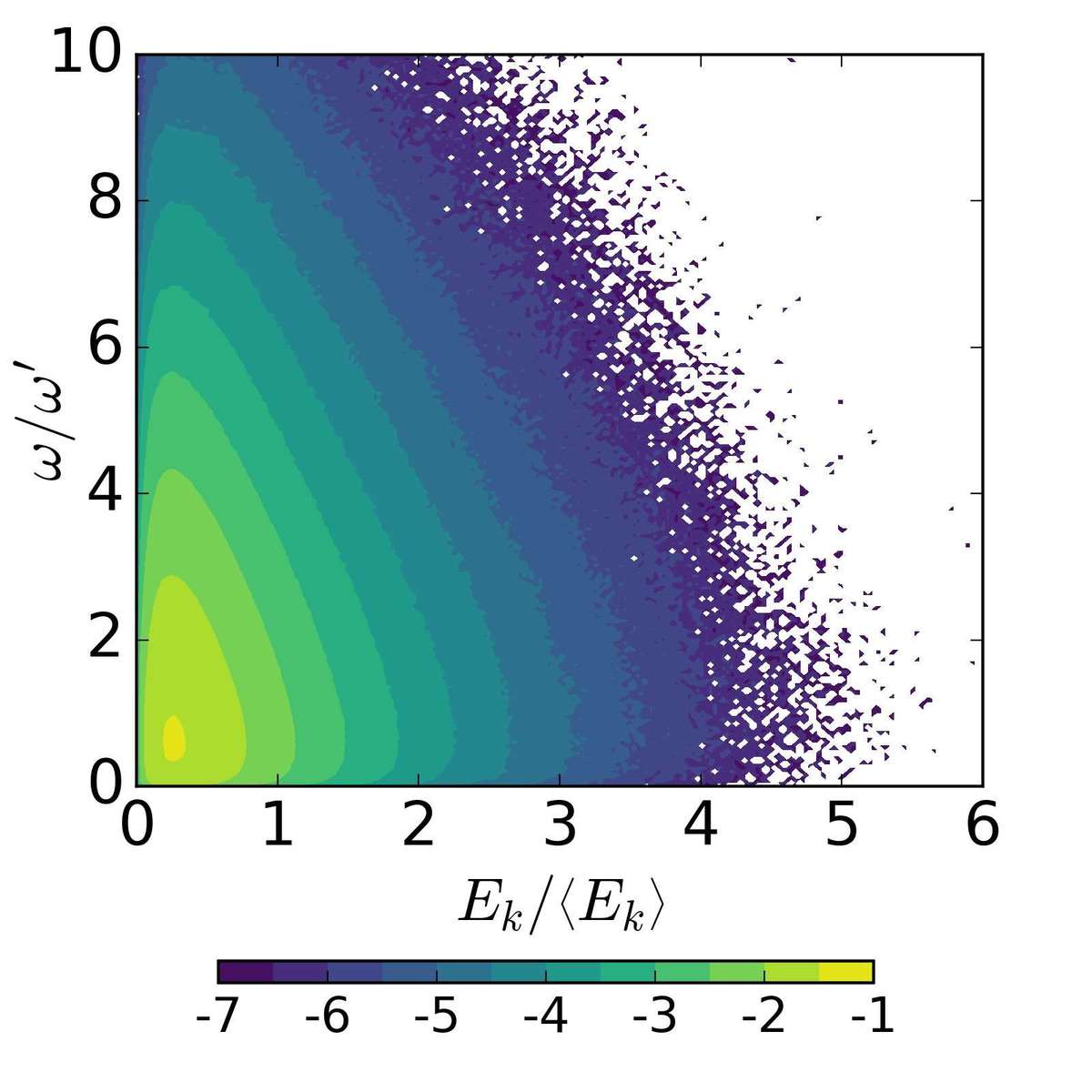}}

  	\caption{}
  \end{subfigure}\quad
\begin{subfigure}{0.49\linewidth}
  \centerline{\includegraphics[width=\linewidth]{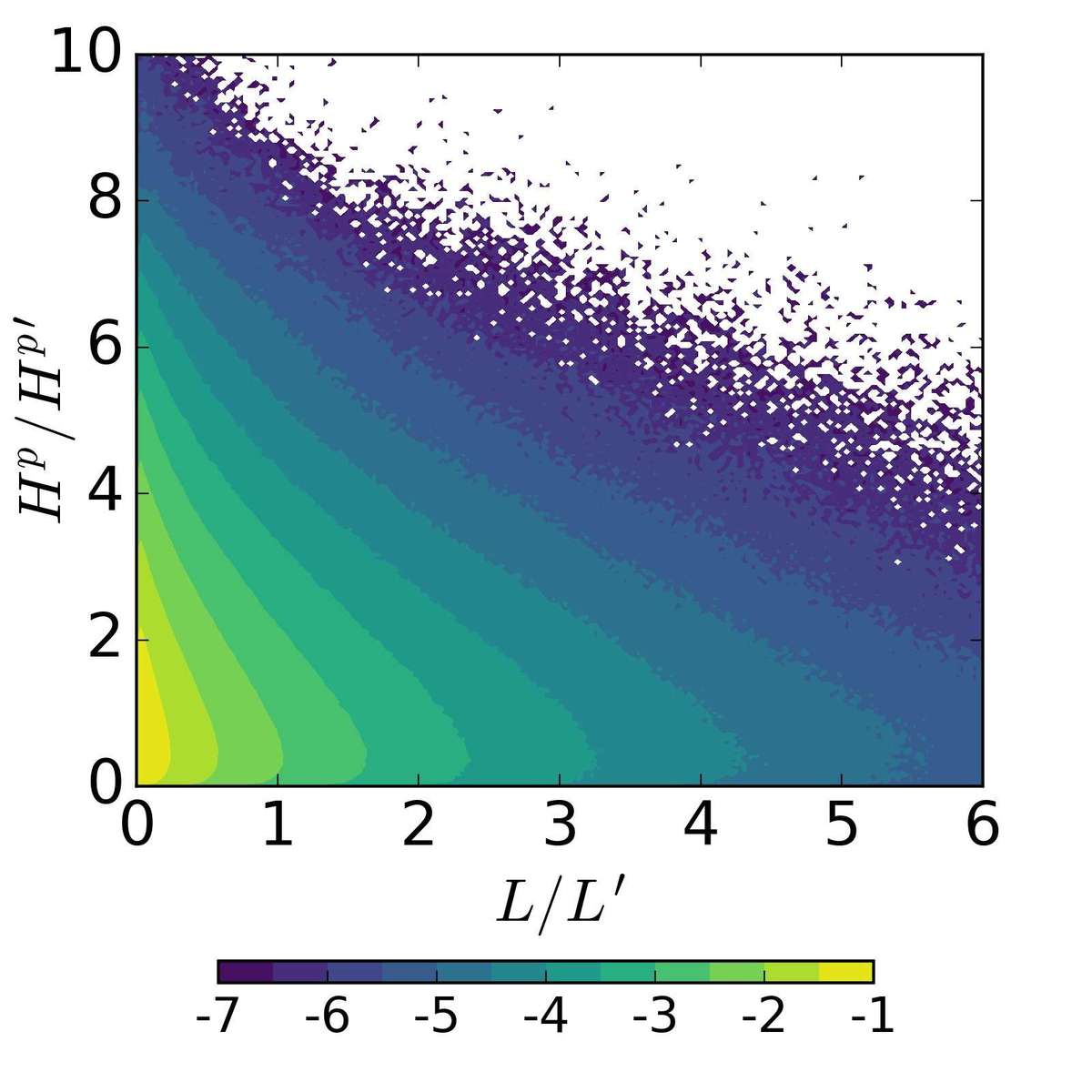}}

  	\caption{}
  	  \end{subfigure}
  \caption{Joint-PDFs of $L$, $E_k$, $H^p$ and $\omega$. The colors show logarithmically spaced values.}
  \label{fig:JointPDFCorrelations}
\end{figure}

The spatial distribution of the correlations, relative to each other and other turbulence quantities, like $E_k$ and $\omega$ (as shown in figure \ref{fig:CorrelationsSpatialDistribution}), can be quantified with the joint-PDFs of pairs of variables. \sid{Figure \ref{fig:JointPDFCorrelations} shows these distributions (with logarithmically spaced colour contours), that have been averaged over 20 field realizations, each separated by $50\tau_k$.}

Figure \ref{fig:JointPDFCorrelations}(a) shows the joint-PDF of $L$ and $E_k$. Since the two fields coincide strongly, they are highly correlated. Large values of $E_k$ also yield large values of $L$. This happens because, to recall, $L$ identifies flow regions that are comprised of velocity vectors that are (i) well-aligned and (ii) have a high magnitude. It so turns out, that regions of high $E_k$ are all well-aligned, and there are no regions of high $E_k$ with disordered velocity vectors. The slight asymmetry of the PDF towards $L$ shows that $L$ attains higher values, relative to $L^\prime$, than $E_k$ does relative to $\ang{E_k}$. This asymmetry hints that an increase in $E_k$ leads to stronger alignment of the velocity vectors, which yields higher $L$ values. 

Figure \ref{fig:JointPDFCorrelations}(b) shows the joint-PDF of $H^p$ and $\omega$. The two fields are again strongly correlated, while the probability of occurrence of large-valued $\omega$ is higher than large-valued $H^p$. Higher values of $\omega$ are invariably associated with high $H^p$ values, which shows that the flow around high $\omega$ regions always has a swirling motion. 

The relative spatial organization of $E_k$ and $\omega$ is shown in figure \ref{fig:JointPDFCorrelations}(c). \sid{The two quantities reflect intermediate sized, kinetic energy containing inertial structures, and small-scale, high vorticity structures, respectively. The joint-PDF shows that high values of the two quantities are \textit{mutually exclusive} in space}, i.e. the fields are anti-correlated. For instance, the probability of finding a high $E_k$ region, which also has a high $\omega$ value, is negligible. A similar anti-correlated distribution is found for $H^p$ and $L$, in figure \ref{fig:JointPDFCorrelations}(d). \sid{High values of $H^p$ coincide with regions of low $L$. This, together with the strong correlation between $L$ and $E_k$ in figure \ref{fig:JointPDFCorrelations}(a), shows that swirling-flow regions do not correspond with high kinetic energy structures, and vice-versa. This is consistent with the similarity between $\lvec$ and $\lsvec$, since swirling-flow structures with high kinetic energy would be associated with large values of $L^s$, which reaffirms that there are no ``large eddies'' (i.e large swirling-flow regions with high kinetic energy).}

To quantify the volume fraction and degree of spatial overlap between different fields, we construct the joint-CDFs, which are functions of the integration limits $\alpha_1$, $\alpha_2$, $\beta_1$ and $\beta_2$, on the fields $f_1$ and $f_2$ comprising a particular joint-PDF. This is defined as
\begin{equation}
\mathrm{CDF}\left(f_1, f_2 ; \alpha_1, \alpha_2, \beta_1,\beta_2 \right) = \int_{\beta_1}^{\beta_2} \left[ \int_{\alpha_1}^{\alpha_2}\mathrm{PDF}(f_1, f_2) df_1\right] df_2
\end{equation}
The CDF can directly be interpreted, \sid{in a statistical sense,} as the volume fraction of the region defined by the integration limits, as follows
\begin{equation}
\mathrm{CDF}\left(f_1, f_2 ; \alpha_1, \alpha_2, \beta_1,\beta_2 \right) = \frac{V\left\lbrace \left( \alpha_1 \leq f_1 \leq \alpha_2 \right)\cap \left( \beta_1 \leq f_2 \leq \beta_2 \right) \right\rbrace}{V_t}
\end{equation}
where $V$ is the intersection volume \sid{between regions} where the conditions $\alpha_1 \leq f_1 \leq \alpha_2$ and $\beta_1 \leq f_2 \leq \beta_2$ are both met, while $V_t$ is the total volume. The volume fraction of a single field, within prescribed threshold limits, can also be statistically quantified with the CDF, as follows
\begin{equation}
\mathrm{CDF}\left(f_1, f_2 ; \alpha_1, \alpha_2, 0,\infty \right) = \frac{V \left( \alpha_1 \leq f_1 \leq \alpha_2 \right)} {V_t}
\end{equation}
This is because the region $0 \leq f_2 \leq \infty$ corresponds to the total volume $V_t$, hence $V\left\lbrace \left( \alpha_1 \leq f_1 \leq \alpha_2 \right) \cap V_t \right\rbrace \equiv V\left( \alpha_1 \leq f_1 \leq \alpha_2 \right)$. The CDF can be used to evaluate the degree of spatial inclusivity between fields, $R(\widetilde{f}_1, \widetilde{f}_2)$, which can be defined as
\begin{equation}
R(\widetilde{f}_1, \widetilde{f}_2) = \frac{\mathrm{CDF} \left( f_1, f_2; \alpha_1, \alpha_2, \beta_1,\beta_2 \right)}{ \mathrm{CDF} \left( f_1, f_2; \alpha_1, \alpha_2, 0, \infty \right) }
\end{equation}
where $\widetilde{f}_1$ and $\widetilde{f}_2$ are conditionally sampled $f_1$ and $f_2$ fields, i.e. the region $\widetilde{f}_1 \equiv \alpha_1 \leq f_1 \leq \alpha_2$ and the region $\widetilde{f}_2 \equiv \beta_1 \leq f_2 \leq \beta_2$. The numerator on the right hand side gives the volume fraction of the intersection region $\widetilde{f}_1\cap \widetilde{f}_2$, while the denominator gives the volume fraction of $\widetilde{f}_1$. Hence, the fraction denotes the degree of inclusivity of the region $\widetilde{f}_1$ in the region $\widetilde{f}_2$. Conversely, $R(\widetilde{f}_2, \widetilde{f}_1)$ gives the inclusivity of $\widetilde{f}_2$ in $\widetilde{f}_1$.

\begin{figure}
	\centering
	\begin{subfigure}{0.45\linewidth}
  \centerline{\includegraphics[width=\linewidth]{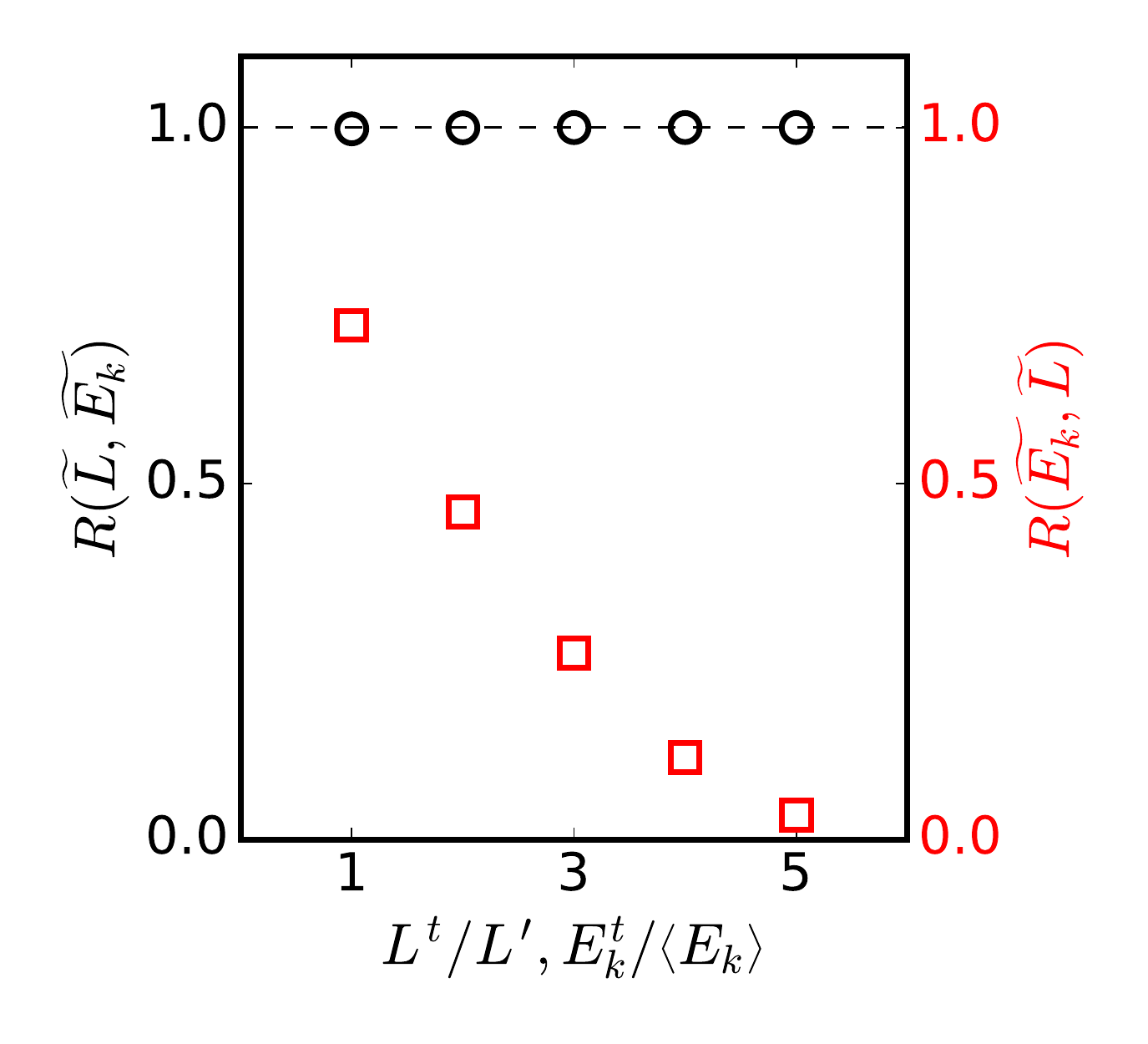}}

  	\caption{Inclusivity of $\widetilde{L}$ and $\widetilde{E_k}$.}
  \end{subfigure}\hspace{1em}
    	\begin{subfigure}{0.45\linewidth}
  \centerline{\includegraphics[width=\linewidth]{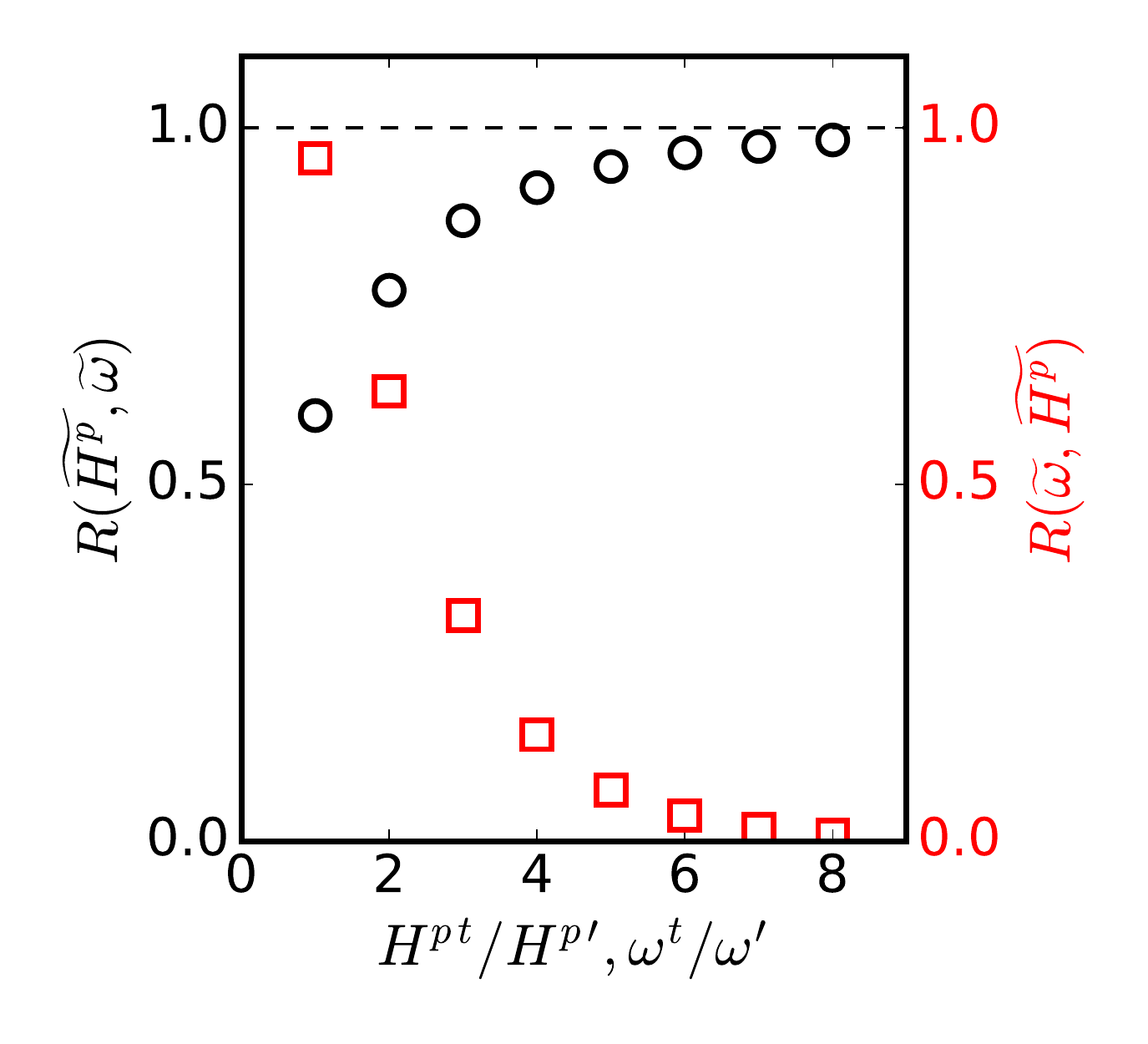}}

  	\caption{Inclusivity of $\widetilde{\omega}$ and $\widetilde{H^p}$.}
  	\end{subfigure}
    
  	\begin{subfigure}{0.4\linewidth}
  \centerline{\includegraphics[width=\linewidth]{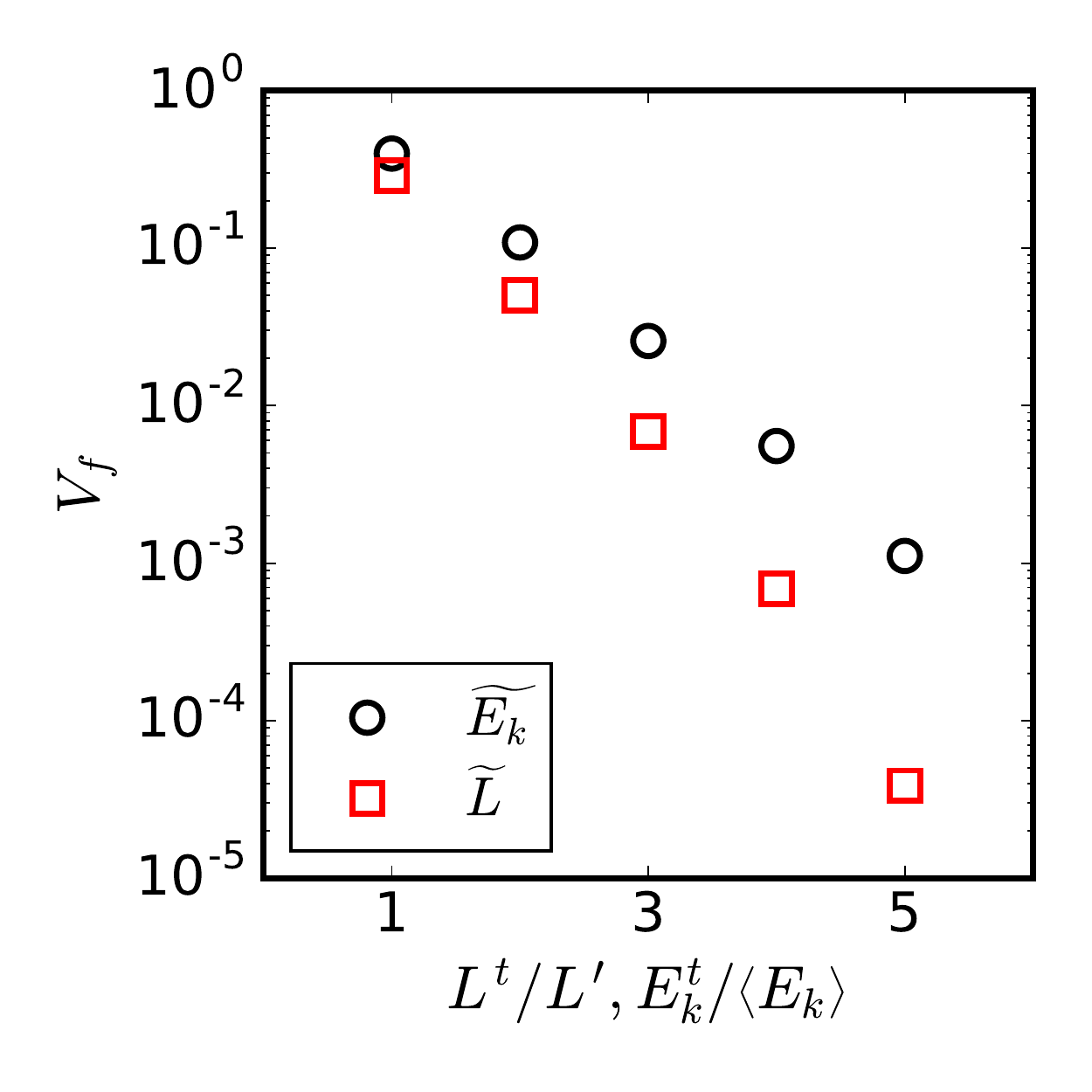}}

  	\caption{Volume fraction $V_f$ of $\widetilde{L}$ and $\widetilde{E_k}$.}
  	\end{subfigure}\hspace{1em}
  	  	\begin{subfigure}{0.4\linewidth}
  \centerline{\includegraphics[width=\linewidth]{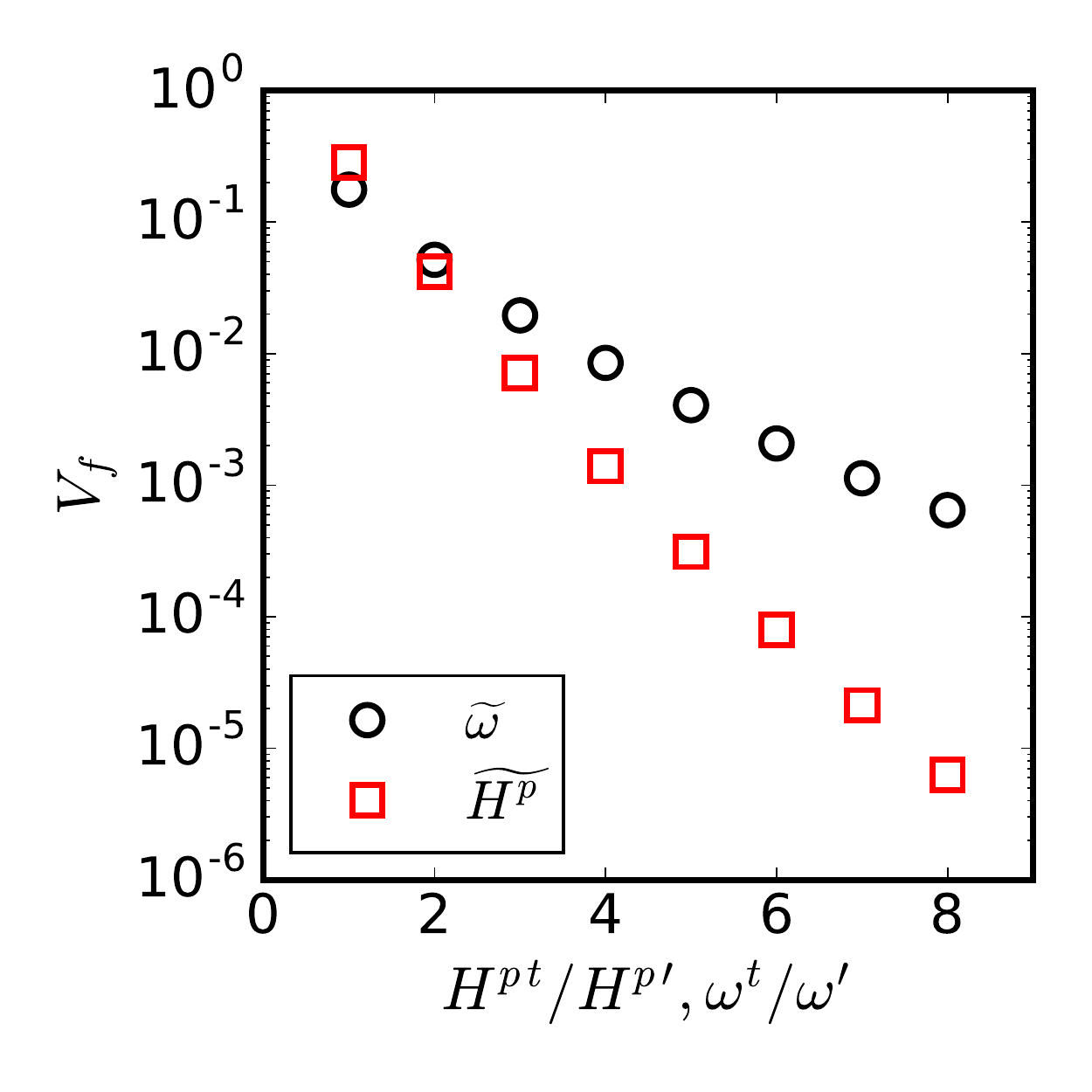}}

  	\caption{Volume fraction $V_f$ of $\widetilde{\omega}$ and $\widetilde{H^p}$.}
  	\end{subfigure}
  	
  \begin{subfigure}{0.4\linewidth}
  \centerline{\includegraphics[width=\linewidth]{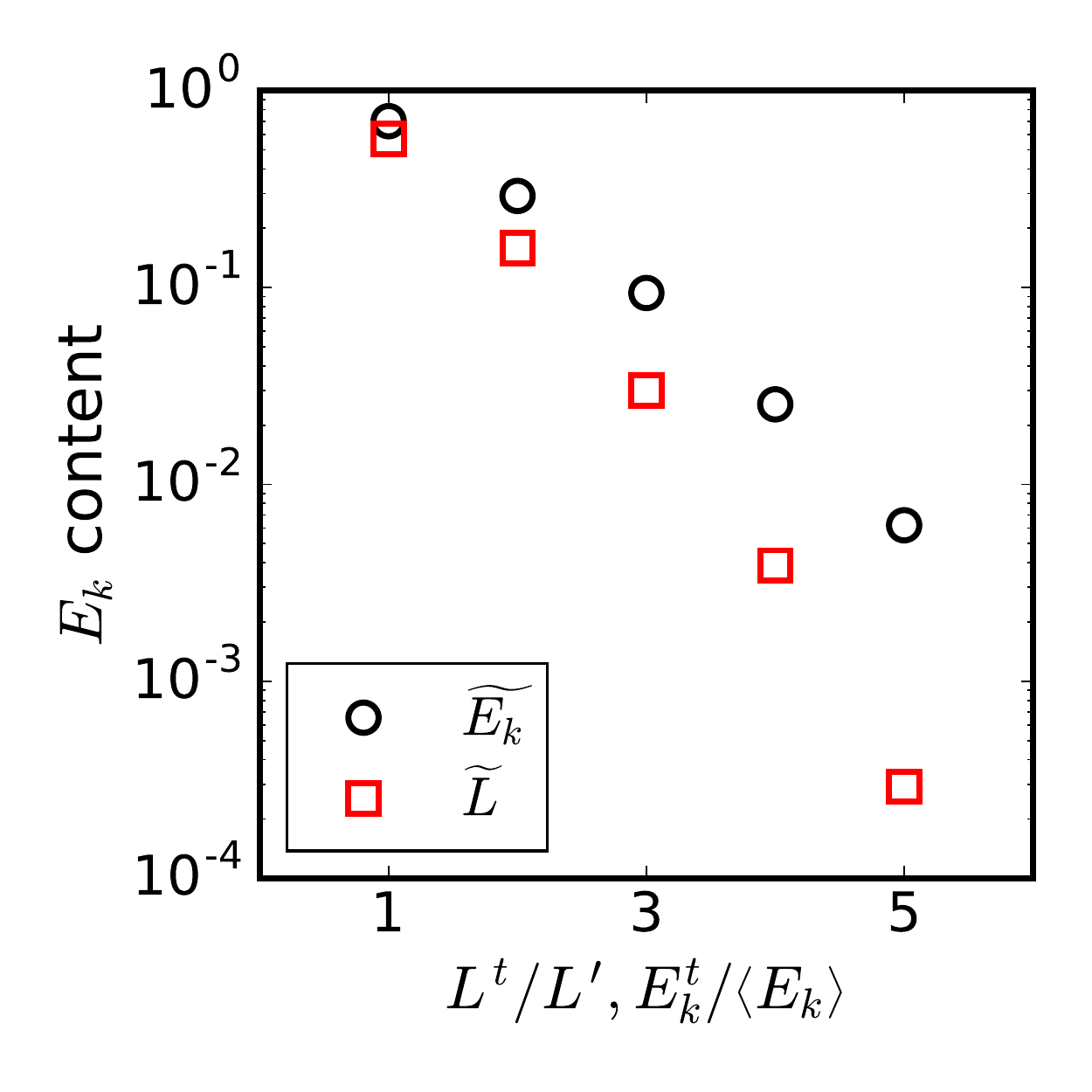}}

  	\caption{Kinetic energy fraction contained in $\widetilde{L}$ and $\widetilde{E_k}$ regions.}
  \end{subfigure}\quad
     \begin{subfigure}{0.4\linewidth}
  \centerline{\includegraphics[width=\linewidth]{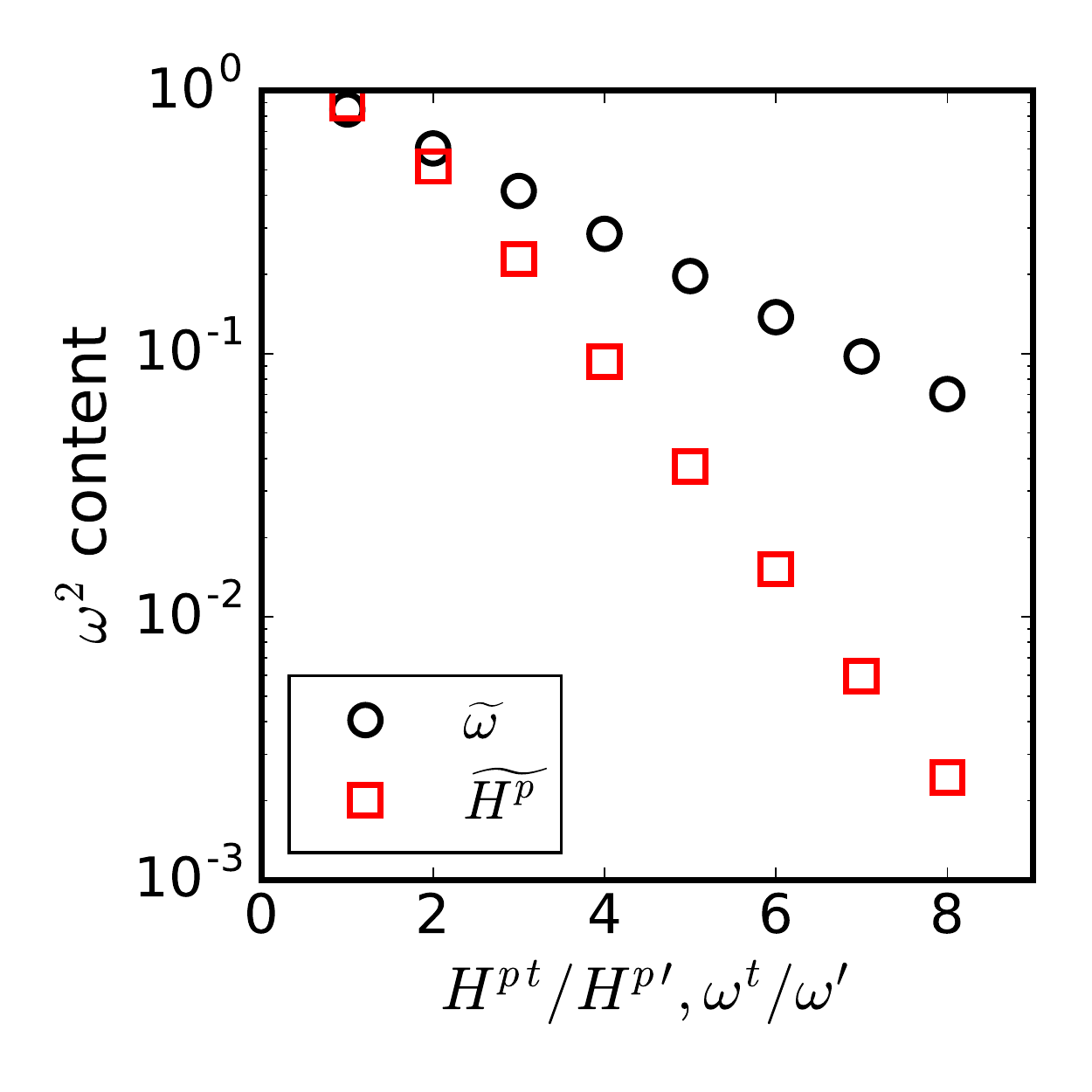}}

  	\caption{Enstrophy fraction contained in $\widetilde{\omega}$ and $\widetilde{H^p}$ regions.}
  \end{subfigure}
  \caption{Statistics of inclusivity between (a) $\widetilde{L}$ and $\widetilde{E_k}$, and (b) $\widetilde{H^p}$ and $\widetilde{\omega}$ regions, where each region, say $\widetilde{\psi}$, is defined as the thresholded $\psi$ field in the range $\widetilde{\psi}\equiv \psi_1 \leq \psi \leq \psi_2$. Panels (c) and (d) show the volume fractions of different thresholded regions of the fields, and panels (e) and (f) show the kinetic energy and enstrophy contents of the regions, respectively.}
  \label{fig:Intersection-Correlations}
\end{figure}

Figure \ref{fig:Intersection-Correlations}(a) shows the inclusivity of the fields $L$ and $E_k$, i.e. $R(\widetilde{L},\widetilde{E_k})$ and $R(\widetilde{E_k},\widetilde{L})$. The regions $\widetilde{L}\equiv L^t \leq L \leq \infty$ and $\widetilde{E_k}\equiv E_k^t \leq E_k \leq \infty$, where the thresholds $L^t$ and $E_k^t$ are values of $L^\prime$ and $\ang{E_k}$. The $L$ field is found to remain completely enclosed within the corresponding $E_k$ regions, since $R(\widetilde{L},\widetilde{E_k}) = 1$. \sid{This can be understood from the fact that highly organized velocity with a high amplitude yields high $L$ values, albeit, the kernels of the $L$ correlation that are associated with a large region of velocity organization are themselves (relatively) smaller. This becomes more pronounced at higher levels of the $L$ and $E_k$ fields, as higher $L$ values reflect both high $E_k$ and larger organization. Note that higher levels of $L$ give a good indication of the flow organization and velocity magnitude, $L$ and $E_k$ are not directly comparable, since $L$ is a non-local measure of structure while $E_k$ is a point criterion. $R(\widetilde{E_k},\widetilde{L})$ is found, conversely, to become successively smaller at higher threshold values, further reflecting that high $E_k$ regions occupy larger spatial regions than high $L$. This is also reflected in the volume fractions $V_f$ of $L$ and $E_k$, calculated as $\mathrm{CDF}\left(L, E_k;L^t, \infty, 0, \infty \right)$ and $\mathrm{CDF}\left(L, E_k; 0, \infty, E_k^t, \infty \right)$, respectively, as shown in figure \ref{fig:Intersection-Correlations}(c). At increasing threshold levels, the $L>L^t$ field occupies smaller volume fractions in comparison to $E_k>E_k^t$. Lastly, the kinetic energy content of the thresholded $L$ and $E_k$ regions is shown in figure \ref{fig:Intersection-Correlations}(e). Regions corresponding to $E_k\geq \ang{E_k}$, $E_k\geq 2\!\ang{E_k}$, $E_k\geq 3\!\ang{E_k}$ occupy $40\%$, $10\%$ and $2.5\%$ of the total volume (panel (c)), respectively, and contain $70\%$, $30\%$ and $9\%$ of the total kinetic energy. Similarly, regions corresponding to $L\geq L^\prime$, $L\geq 2L^\prime$ and $L\geq 3L^\prime$ occupy $30\%$, $5\%$ and $0.7\%$ of the total volume, while containing $55\%$, $15\%$ and $3\%$ of the total kinetic energy.}

Figure \ref{fig:Intersection-Correlations}(b) shows the inclusivity of the fields $H^p$ and $\omega$, i.e. $R(\widetilde{H^p},\widetilde{\omega})$ and $R(\widetilde{\omega},\widetilde{H^p})$, where the regions are defined as $\widetilde{H^p}\equiv {H^p}^t \leq H^p \leq \infty$ and $\widetilde{\omega}\equiv {\omega}^t \leq \omega \leq \infty$. The fraction of the $H^p$ field contained inside $\omega$ regions increases at higher threshold values. This shows, first, that the $H^p$ field occupies successively smaller spatial regions at higher thresholds. Secondly, high $H^p$ values are invariably found \textit{inside} high $\omega$ regions, reaffirming that \sid{strong vorticity is associated with surrounding regions of swirling motion}. This can also be seen from figure \ref{fig:JointPDFCorrelations}(b), where the lower bound on the value of $H^p$ increases with $\omega$. At low threshold values, $R(\widetilde{H^p},\widetilde{\omega})$ is low, which shows that the $H^p$ field at low values occupies more space in comparison to low $\omega$. This is reflected in $R(\widetilde{\omega},\widetilde{H^p})$, which has high values at low $\omega$. It is further confirmed in figure \ref{fig:Intersection-Correlations}(d), which shows that $H^p\geq{H^p}^\prime$ occupies $26\%$ of the volume, while $\omega\geq\omega^\prime$ occupies $17\%$ of the volume. Further, the volume fraction occupied by the $H^p$ field decays much faster than $\omega$, when thresholded at successively higher values. Lastly, panel (f) shows that regions of ${H^p}\geq{H^p}^\prime$, ${H^p}\geq 3{H^p}^\prime$ and $H^p\geq 5{H^p}^\prime$ contain $85\%$, $22\%$ and $3\%$ of the total enstrophy.
 
\begin{figure}
  \begin{subfigure}{0.45\linewidth}
  \centerline{\includegraphics[width=\linewidth]{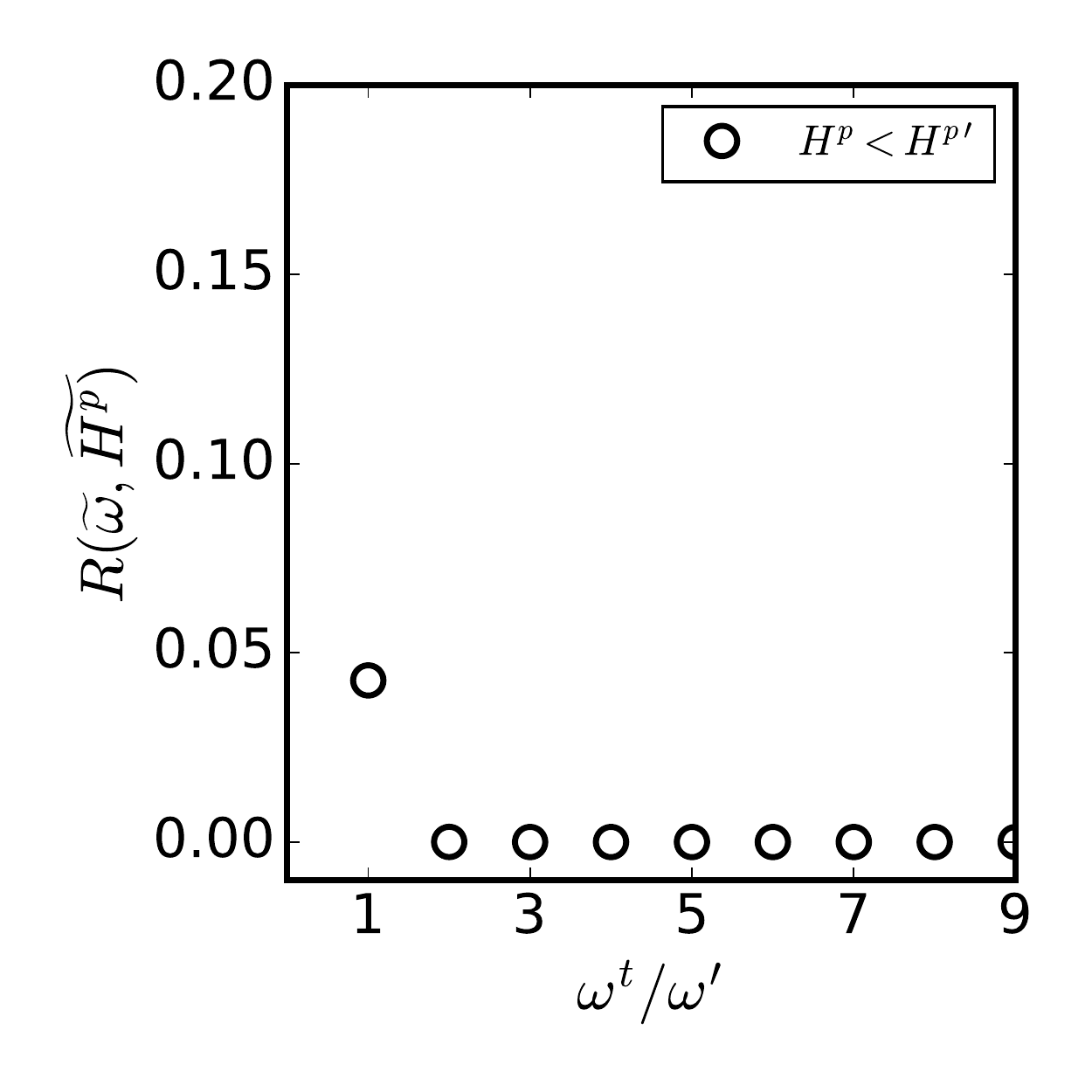}}

  	\caption{}
  \end{subfigure}\quad
\begin{subfigure}{0.45\linewidth}
  \centerline{\includegraphics[width=\linewidth]{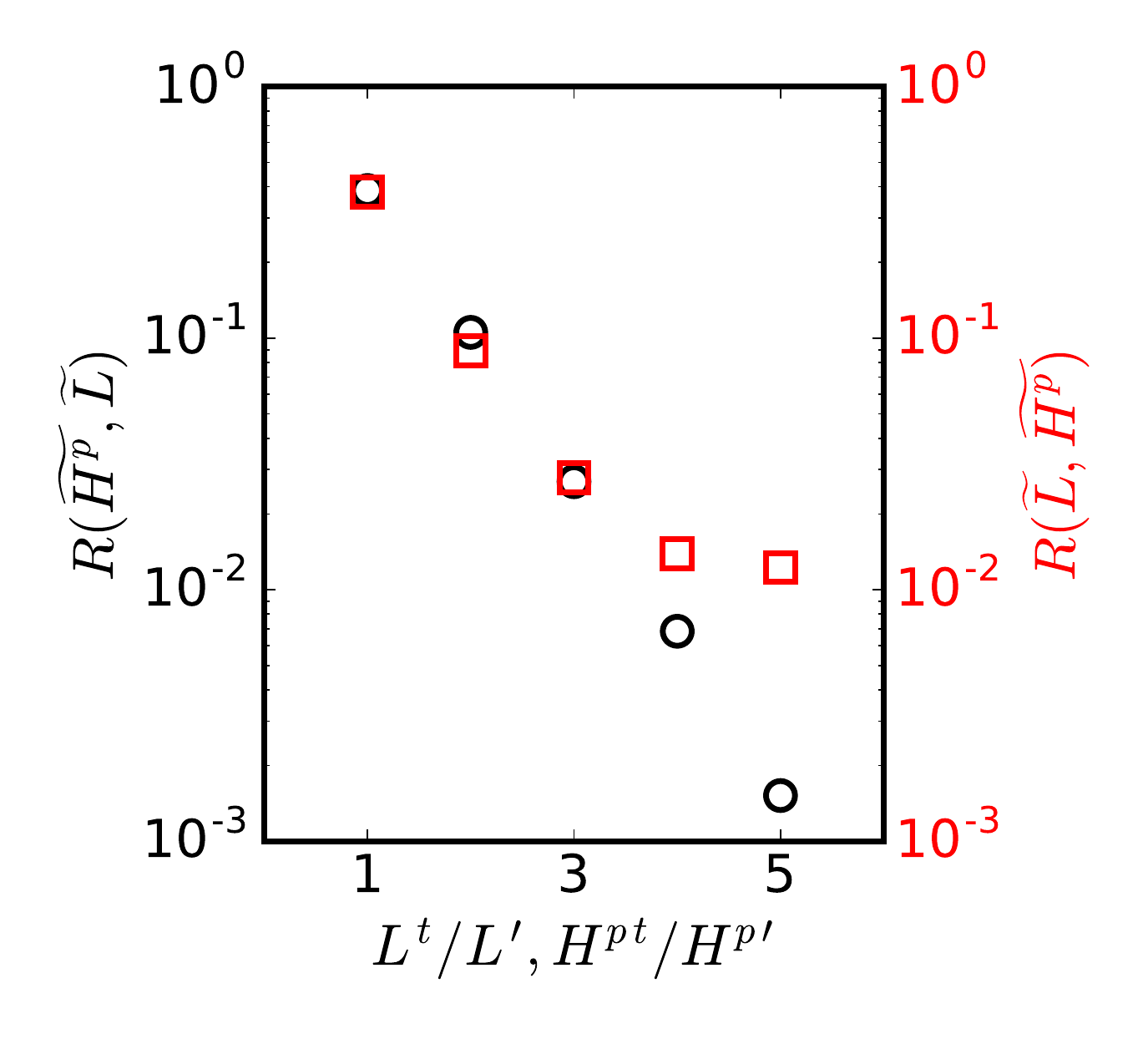}}

  	\caption{}
  \end{subfigure}\quad
  \caption{Inclusivity of (a) high $\omega$ regions in low $H^p$ regions and (b) high $H^p$ regions and high $L$ regions.}
  \label{fig:Intersection-Correlations2}
\end{figure}

Finally, figure \ref{fig:Intersection-Correlations2}(a) shows $R(\widetilde{\omega},\widetilde{H^p})$, the inclusivity of regions of high vorticity ($\widetilde{\omega}\equiv \omega^t \leq \omega \leq \infty$) with regions of low $H^p$ ($\widetilde{H^p}\equiv 0 \leq H^p \leq {H^p}^\prime$). For all instances of $\omega^t \geq \omega^\prime$, the intersection volume of $\widetilde{\omega}$ with $\widetilde{H^p}$ goes to zero. This confirms that there are no high $\omega$ regions in the flow field that are not associated with swirling motion in their vicinity. \sid{Figure \ref{fig:Intersection-Correlations2}(b) shows the inclusivity of the $L$ and $H^p$ fields, i.e. $R(\widetilde{H^p},\widetilde{L})$ and $R(\widetilde{L},\widetilde{H^p})$, for increasing threshold values. $R(\widetilde{H^p},\widetilde{L})$ shows that the correlation kernels become increasingly \textit{spatially exclusive} at higher threshold values, which can also be seen from the joint-PDFs in figure \ref{fig:JointPDFCorrelations}(d), where the $L$ and $H^p$ fields are anti-correlated. This reflects the spatial exclusivity of highly organized kinetic energy jets and vorticity induced swirls. Interestingly, $R(\widetilde{L},\widetilde{H^p})$ coincides with $R(\widetilde{H^p},\widetilde{L})$ for the first few threshold levels, showing that the volume fractions of $L$ and $H^p$ kernels are comparable at these levels. Beyond that, the $L$ field becomes (relatively) more inclusive in the $H^p$ field (around $L>3L^\prime$ and $H^p>3\hpp$), where the volume fraction of the kernels of $L$ is smaller than the volume fraction of the kernels of $H^p$. This is because the $H^p$ field can attain very high values (see figure \ref{fig:TurbulenceCorrelation-HCorrArea}c) in comparison to the $L$ field (see figure \ref{fig:TurbulenceCorrelation-LCorr}c), and the volume fraction decay of $L$ is steeper than that of $H^p$.}

\section{Flow structures in homogeneous isotropic turbulence}\label{sec:FlowStructure}
\subsection{Introduction}
\sid{So far, we found using the various correlations that turbulence fields comprise (at least) two distinct structures. One of them corresponds to regions of high kinetic energy, which are found to coincide with regions of high $L$, since they have a jet-like structure with parallel streamlines. The other corresponds to regions of high enstrophy with swirling velocity in the neighbourhood (which confirms the description of these structures according to  \citet{she1990intermittent,jimenez1993structure}). We further also quantified the statistics and spatial distribution of these coherent flow regions.}

\sid{In this section, we focus on individual flow structures by showing instances of the exact flow field in regions of high $L$ or high $H^p$ correlation. This shows, visually, how regions of \textit{instantaneous} coherence manifest in the flow field. Next, we investigate the Biot-Savart composition of these instantaneous structures, which allows us to address two crucial questions regarding the composition of turbulence velocity fields. First, we determine the extent to which structures yielding high $L$ or high $H^p$ are ``self-inducing'' in a Biot-Savart sense, as opposed to being ``externally-induced'' (i.e. being non-locally induced by the surrounding vorticity). Note that by ``self-induction'' we imply a flow field within a region $\mathcal{R}_\Omega$ being generated by the vorticity in $\mathcal{R}_\Omega$, while an ``externally-induced'' structure in $\mathcal{R}_\Omega$ is generated by the Biot-Savart contribution of the vorticity outside of $\mathcal{R}_\Omega$ (as illustrated in figure \ref{fig:schematicBiotSavart}). Secondly, we show both qualitatively and quantitatively, what is the relative contribution of different levels of vorticity in inducing these structures, and the total velocity field in general. This approach allows us to ``disentangle'' the velocity field into its various Biot-Savart components. A brief note on some practical concerns regarding the calculation of the Biot-Savart velocity field is presented in appendix \ref{app:BiotSavart}. Our analysis reveals the distribution of vorticity contributions to the generation of the velocity field, which gives insight regarding the organization of turbulence, with hints regarding the dynamics of the turbulence process.}

We first focus on individual flow structures, following which, we consider the Biot-Savart contributions from a statistical perspective. Finally, we summarize the picture of turbulence organization that emerges from this study.

\subsection{Individual flow structures}
\sid{We first look at regions of high magnitude $L$ by identifying isolated contours of the $L$ field. High $L$ regions were found to coincide with regions of high $E_k$, hence these are the ``energy containing'' structures. In figure \ref{fig:TurbulenceCorrelation-LCorr}(b), we find that the level of $L=2.5L^\prime$ marks regions of high correlation, which occupy $\approx 2\%$ of the volume while containing $\approx 10\%$ of the total kinetic energy. Figure  \ref{fig:TurbulenceCorrelation-LCorr}(a) also shows that the $L>2.5L^\prime$ field forms separate, individual regions, which can be considered distinctly. At lower correlation levels ($L<1.5L^\prime$), the $L$ regions become more connected, while at higher levels ($L>3.5L^\prime$), the $L$ regions become very small (occupying less than $0.2\%$ of the volume). We hence select $L=2.5L^\prime$ regions for the present analysis, while the results remain essentially similar for slightly different values of $L$.}

\begin{figure}
  \begin{subfigure}{0.9\linewidth}
  \centerline{\includegraphics[width=\linewidth]{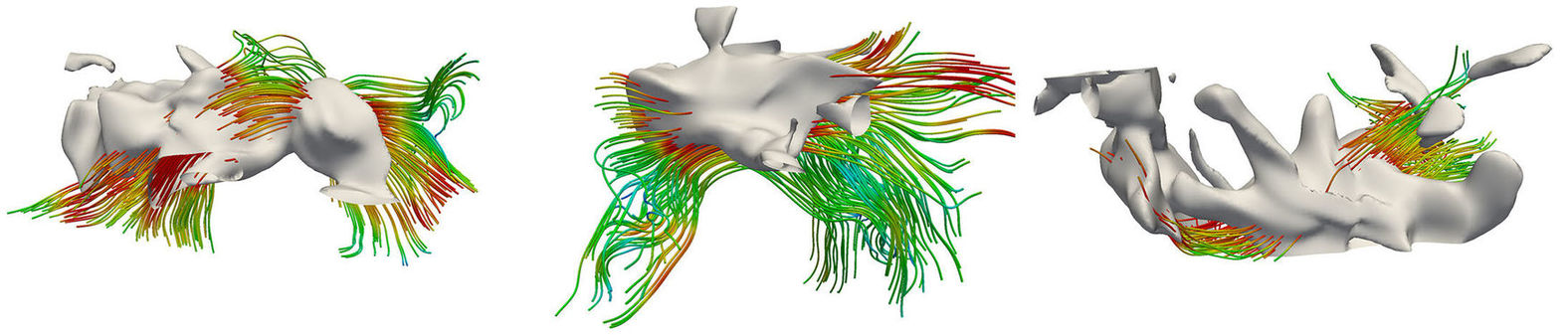}}

  	\caption{Contours of $L=2.5L^\prime$ along with local streamlines of the total velocity field.}
  \end{subfigure}
  
  \begin{subfigure}{0.9\linewidth}
  \centerline{\includegraphics[width=\linewidth]{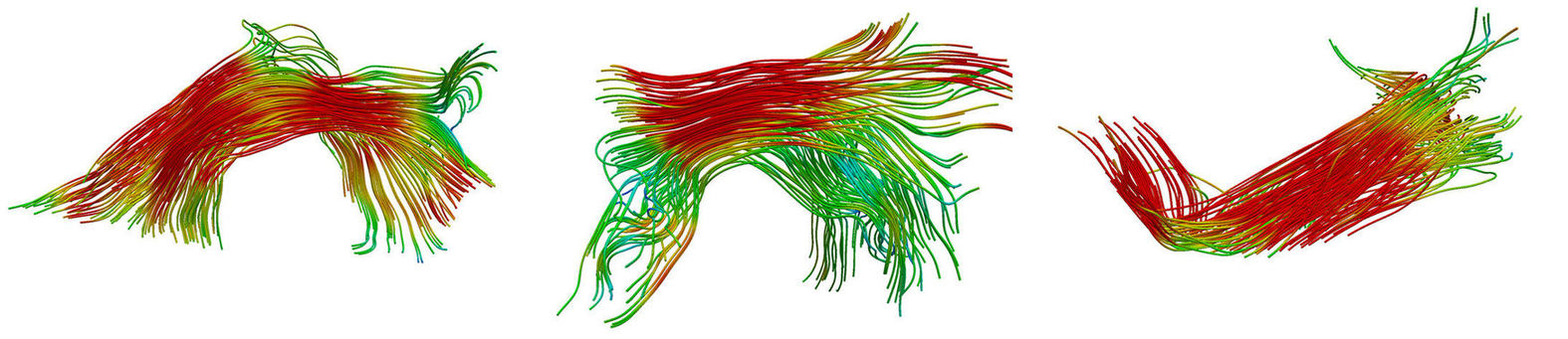}}

  	\caption{Streamlines of the total velocity.}
  \end{subfigure}
  
  \begin{subfigure}{0.9\linewidth}
  \centerline{\includegraphics[width=\linewidth]{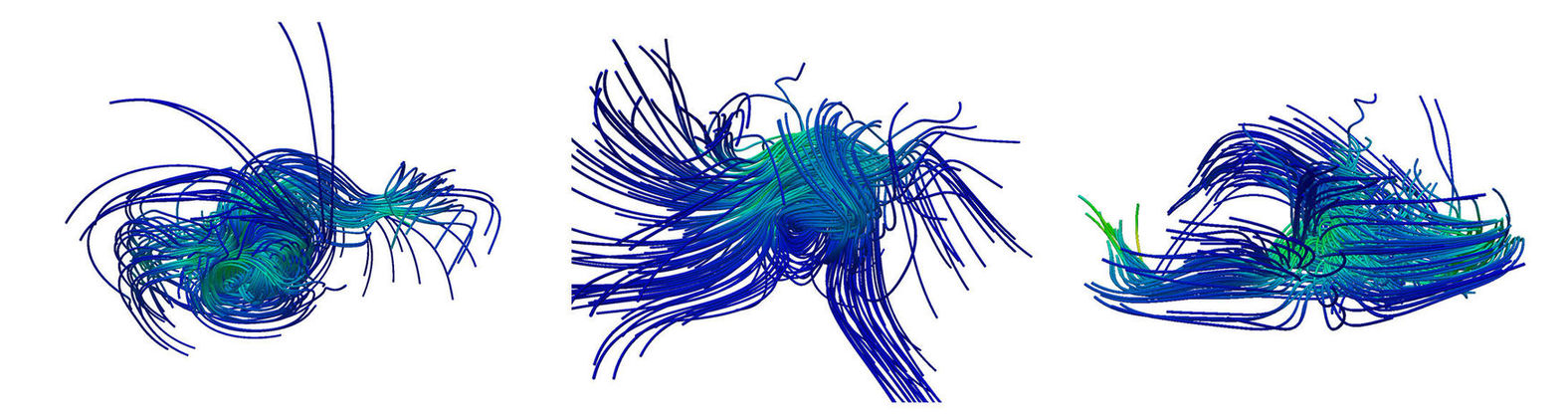}}

  	\caption{Streamlines of the self-induced velocity of $L\geq 2.5L^\prime$ regions.}
  \end{subfigure}
  
  \begin{subfigure}{0.9\linewidth}
  \centerline{\includegraphics[width=\linewidth]{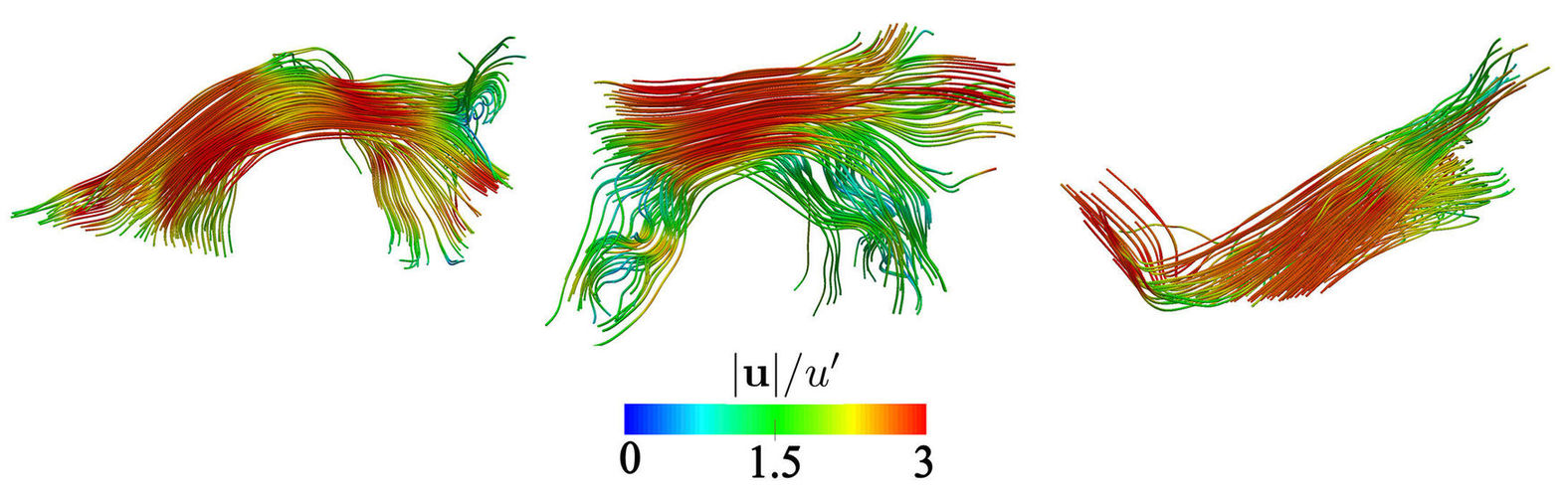}}

  	\caption{Streamlines of the externally-induced velocity.}
  \end{subfigure}  
  \caption{\sid{The flow structure in regions of high $L$ is found to be \textit{jet-like}. Panel (a) shows individual instances of $L=2.5L^\prime$ contours, together with streamlines of the local flow, which have also been shown separately in panel (b). Panel (c) shows the self-induced flow of the high $L$ regions, while panel (d) shows the externally-induced flow. Colorbar shows the velocity magnitude.}}
  \label{fig:StructureLCorr}
\end{figure}

Figure \ref{fig:StructureLCorr}(a) shows three individual, isolated contours of $L=2.5L^\prime$, along with the local flow streamlines \sid{(which have been initialized from points distributed within a small region around the core of the correlation kernels). Panel (b) shows the flow streamlines alone, where we find that these regions comprise well aligned, parallel streamlines, with indeed a \textit{jet-like} flow structure. Since we found that high values of $L$ identify regions of high $E_k$, we can conclude that high $E_k$ flow structures are jet-like. The streamlines diverge into more chaotic patterns away from the correlation kernels, which shows that the coherence of these structures is localized. This is an interesting finding, as in general we do not find large structures associated with high $E_k$ - which is the classical ``large eddy'' perspective. Instead, a distribution of relatively smaller, locally jet-like flow regions are found to populate the flow field in regions of high $E_k$.}

\sid{Next, we reconstruct the self-induced Biot-Savart velocity field, using all $L \geq 2.5L^\prime$ regions. Note that for calculating the self-induced velocity of a single structure, in principle, only the vorticity contained in that region should be used. This is possible, however, it requires additional clustering (or segmentation) of the correlation field to isolate regions, which adds a further complexity to the calculation. Since the identified $L$ correlation regions remain isolated in space, and moreover have low levels of vorticity (refer to table \ref{tab:LHpSelfExt}), their Biot-Savart contribution (which decays with the square of the distance), is negligible outside the correlation regions. Hence, in practice, the self-induced velocity field around any high $L$ region is essentially the same even upon reconstructing the velocity using all $L \geq 2.5L^\prime$ regions.}

Figure \ref{fig:StructureLCorr}(c) shows streamlines of this self-induced velocity field, in the same regions as shown in panel (a). First, we find that the streamlines of the self-induced velocity have a very different structure than the streamlines of the total velocity in these regions. Further, the self-induced velocity field is also very weak (with amplitude roughly 10 times smaller than the total velocity). These two aspects together imply that high $L$ regions have a negligible level of self-induction, and are hence externally-induced, as shown in Figure \ref{fig:StructureLCorr}(d) (which shows the externally induced velocity field, that is calculated by subtracting the self-induced velocity field shown in panel c from the total velocity field shown in panel b). The externally induced velocity streamlines coincide very well with the total velocity streamlines, which shows clearly that high $L$ regions are structures that are \textit{externally} induced by the non-local vorticity (in the Biot-Savart sense). Note that these results are essentially similar for slightly different levels of the $L$ threshold used to determine the correlation kernels, i.e. in the range $2.0 \leq L \leq 3.0$. \sid{While much higher $L$ regions become infrequent and small in volume, they can be expected to also lead to a similar result. Higher $L$ regions will further have an even smaller self-induced contribution due to their lower levels levels of vorticity, since the two are anti-correlated (see figure \ref{fig:JointPDFCorrelations}). At much lower levels of $L$, however, the self-induced contribution can be expected to increase, as the correlation kernels of low $L$ levels occupy larger volumes, and begin to contain intermediate levels of vorticity.}

Next, we show the flow structure around high $H^p$ regions, which identify regions of strong vorticity with swirling motion in the neighbourhood. \sid{We tested a few different levels of $H^p$ for our analysis, and found that $H^p=5\hpp$ both represents a relatively high level of the correlation field (see figure \ref{fig:TurbulenceCorrelation-HCorrArea}), while having a meaningful volume, comprising distinct regions that can be identified individually. The $H^p=5\hpp$ level occupies $\approx 0.018\%$ of the volume (in this particular field snapshot) while containing $\approx 4\%$ of the total enstrophy. Much higher values, like $H^p > 7\hpp$, do not occupy significant fractions of the volume ($<0.001\%$). At lower levels, the $H^p$ field becomes more diffused than the vorticity field (see figure \ref{fig:Intersection-Correlations}b), and hence does not adequately represent regions of strong swirling motion which occur around regions of high vorticity. The present results are also similar for slightly different choices of the $H^p$ regions, for instance in the range $4 < H^p/\hpp < 6$.}

\begin{figure}
  \begin{subfigure}{0.9\linewidth}
  \centerline{\includegraphics[width=\linewidth]{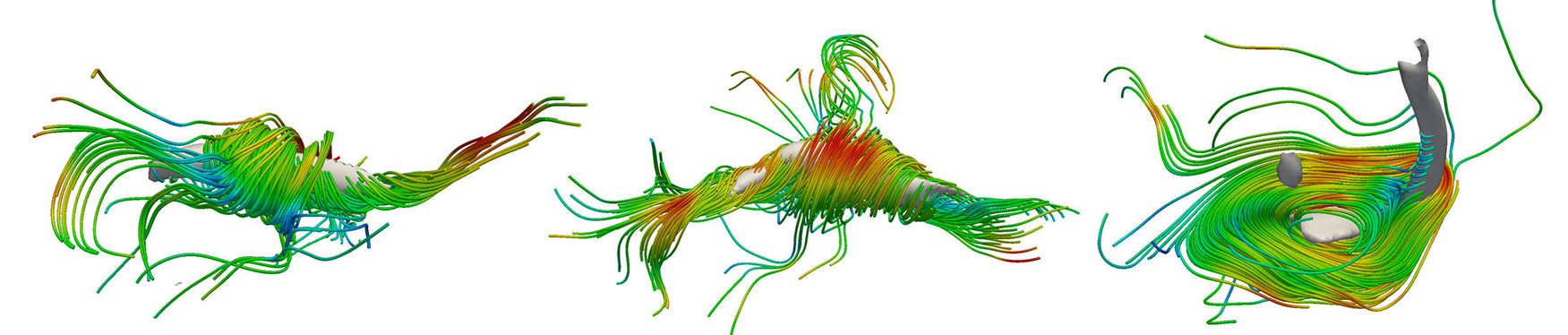}}

  	\caption{Contours of $H^p=5\hpp$ along with local streamlines of the total velocity field.}
  \end{subfigure}
  
  \begin{subfigure}{0.9\linewidth}
  \centerline{\includegraphics[width=\linewidth]{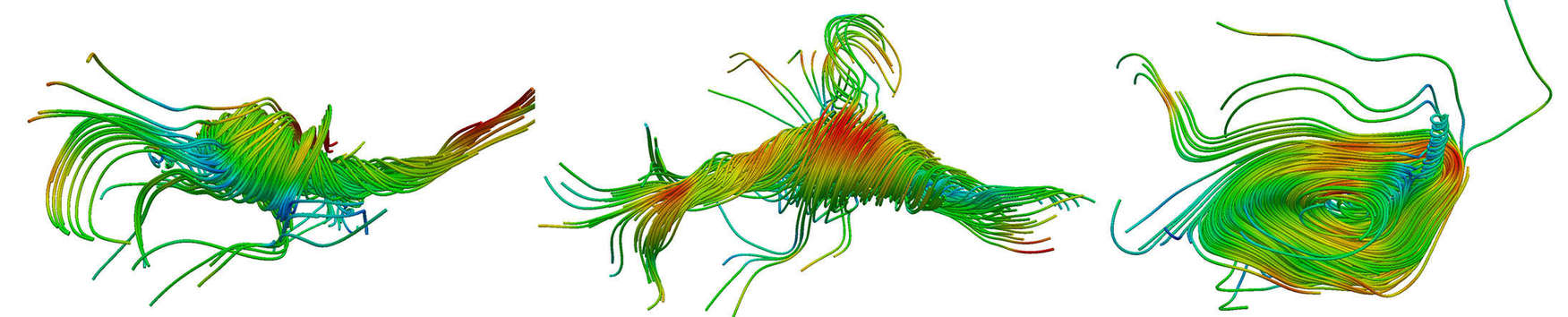}}

  	\caption{Streamlines of the total velocity.}
  \end{subfigure}
  
  \begin{subfigure}{0.9\linewidth}
  \centerline{\includegraphics[width=\linewidth]{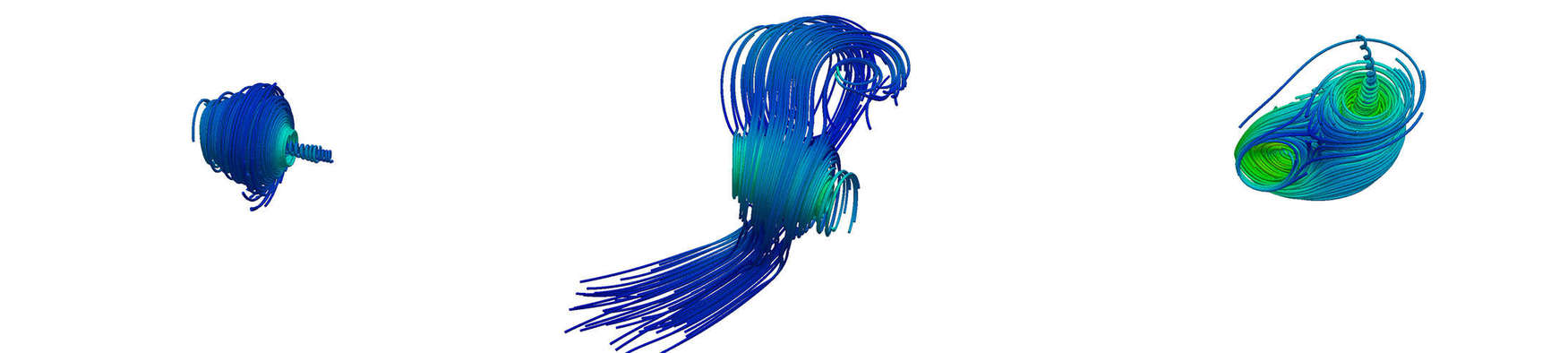}}

  	\caption{Streamlines of the self-induced velocity.}
  \end{subfigure}
  
  \begin{subfigure}{0.9\linewidth}
  \centerline{\includegraphics[width=\linewidth]{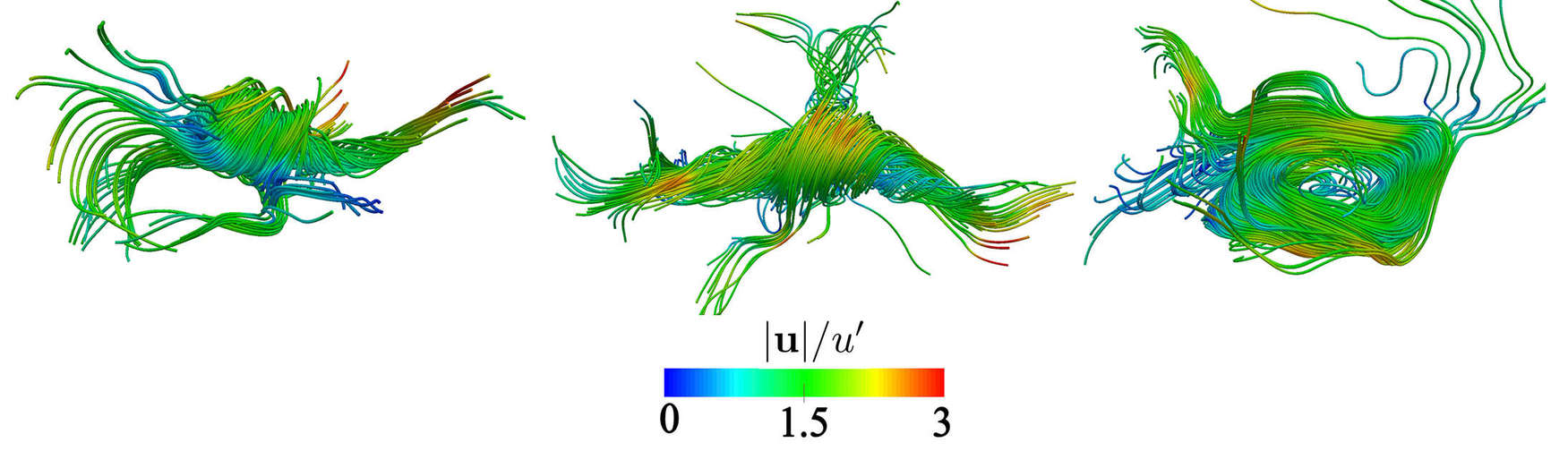}}

  	\caption{Streamlines of the externally-induced velocity.}
  \end{subfigure}  
  \caption{The flow around high $H^p$ regions, which also coincide with high $\omega$, is shown to be swirling. Panel (a) shows individual instances of high $H^p$ regions as identified by contours of $5\hpp$, together with local streamlines of the total flow, which alre also shown separately in panel (b). Panel (c) shows the self-induced flow, computed using the vorticity inside the $5\hpp$ contours, while panel (d) shows the externally-induced flow field. Colorbar shows the velocity magnitude.}
  \label{fig:StructureHp}
\end{figure}

Figure \ref{fig:StructureHp}(a) shows three instances of $H^p=5\hpp$ regions with the local flow streamlines (which have been initiated from a collection of points distributed near the core of each correlation kernel), while figure \ref{fig:StructureHp}(b) shows the streamlines alone. The first thing to note is that the velocity of these structures is mostly in the intermediate and low $E_k$ range (with very small regions of high $E_k$ emerging). Towards the core of these structures, the velocity field shows a strong swirling motion, while the flow decays into more disordered streamlines away from the core regions. \sid{Next, we calculate the self-induced velocity field of these structures. Here again, like in the analysis of high $L$ structures, we use all $H^p \geq 5\hpp$ regions to reconstruct the self-induced velocity field, instead of segmenting the field to isolate individual structures. Although high $H^p$ regions have higher vorticity, these regions are very small and isolated. Hence, the Biot-Savart influence of an individual high $H^p$ region does not extend far, and it is convenient to calculate the self-induced field using all $H^p \geq 5\hpp$ regions with a simple thresholding criterion.}

Figure \ref{fig:StructureHp}(c) shows the self induced velocity field, which forms purely swirling motion in the core regions, along with an instance of two vortices interacting in a figure-eight velocity pattern (right-most). This swirling velocity decays to a low amplitude away from the core region, which shows that the strong vorticity at the core influences the total velocity field only within a small region of influence (due to the rapid decay of the Biot-Savart contribution). The externally induced velocity streamlines in panel (d) resemble the total velocity streamlines outside the core of the $H^p$ regions, and is found to have an intermediate amplitude. This shows that some region \textit{around} the $5\hpp$ kernels also contains vorticity of a large enough amplitude to generate swirling motion in the neighbourhood. \sid{High $H^p$ structures are hence a superposition of self-inducing swirling flow, along with a background induced flow field which has a swirling component and a more disorganized structure away from the core regions.}

Performing this calculation for different thresholds of $H^p$ will change the relative contribution of the self-induced and externally-induced velocities to the total velocity. At lower $H^p$ values, the correlation kernel is larger, and it will contain more vorticity (in the high and intermediate ranges), and hence will have a larger self-induced Biot-Savart contribution.

To quantify the self-induced and externally-induced contributions to the total velocity field, we define a general measure $C$ using the Biot-Savart velocity field  $\mathbf{u}_{\mathrm{BS}}$ and the total velocity field $\mathbf{u}$ as  
\begin{equation}
C\left( \mathcal{R}_\Omega, \mathcal{R}_{\mathrm{BS}} \right) = \frac{ \ang{\mathbf{u}\cdot \mathbf{u}_{\mathrm{BS}}} }{ \ang{\mathbf{u}\cdot \mathbf{u}}}
\label{eq:corrBS}
\end{equation} 
Here $\mathcal{R}_\Omega$ denotes the region where $C$ is evaluated, which can be defined based upon a conditional sampling criterion, for instance $\mathcal{R}_\Omega \equiv L>L^t$, where $L^t$ is a threshold value for the correlation magnitude $L$. The regions of vorticity used to generate the Biot-Savart velocity field $\mathbf{u}_{\mathrm{BS}}$ is denoted by $\mathcal{R}_{\mathrm{BS}}$. These regions are also defined using a conditional sampling criterion. Lastly, $\ang{.}$ denotes averaging over the region $\mathcal{R}_\Omega$. We also use this measure to test the accuracy of the Biot-Savart reconstruction, by calculating $C(V,V)$, i.e. the correlation over the total volume $V$, of the total velocity field $\mathbf{u}$, with the Biot-Savart velocity field $\mathbf{u}_{\mathrm{BS}}$ generated using total the vorticity field. It is found that $C(V,V) \approx 0.99$, which means that our Biot-Savart reconstruction faithfully reproduces $99\%$ of the velocity field, while there is a numerical error of $\approx 1\%$.

In table \ref{tab:LHpSelfExt}, we quantify the self-induced and externally-induced Biot-Savart contributions to the velocity field within all $2.5L^\prime$ and $5\hpp$ kernels, using eq. \ref{eq:corrBS}. The contribution of the self-induced velocity of the high kinetic energy jets, i.e. $C(L\geq 2.5L^\prime,L\geq2.5L^\prime)$ is found to be $\approx 11\%$, while the externally-induced velocity dominates these structures with $\approx 88\%$ contribution. For the $5\hpp$ structures, the self-induced contribution i.e. $C(H^p\geq 5\hpp,H^p\geq 5\hpp)$ is $\approx 34\%$, while the externally-induced contribution is $\approx 66\%$. Note that we performed these calculations for regions contained \textit{inside} the correlation kernels, however, the kernels themselves are markers of larger regions of coherence \textit{around} them. Particularly, the $5\hpp$ regions occupy a small fraction of the volume ($\approx 0.018\%$), while they are associated with swirling motion \textit{outside} the kernel regions. This is why the velocity inside $5\hpp$ regions has a relatively low self-induced contribution, as it is significantly influenced by the strong vorticity which can be found in the immediate neighbourhood of the $5\hpp$ kernels \sid{(see figure \ref{fig:Intersection-Correlations}b, where the right $y-$axis shows that a large fraction of the high vorticity is contained outside $5\hpp$ regions).}

Lastly, we show in Appendix \ref{app:JHTD-Validation} that jet-like and swirling-flow structures correspond to high $L$ and high $H^p$ regions, respectively, also in the reference Johns Hopkins Turbulence Dataset.

\begin{table}
  \begin{center}
\def~{\hphantom{0}}
  \begin{tabular}{lccccc}
       Region & $V_f$ & Mean $\omega$ & Mean $E_k$ & BS-Self & BS-External\\[3pt]
       $L \geq 2.5L^\prime$ & $0.02$ & $0.78\omega^\prime$ & $3.67\left\langle E_k \right\rangle$ & 11.32\% & 88.65\% \\[3pt]
       $H^p \geq 5\hpp$ & $0.00018$ & $3.46\omega^\prime$ & $1.12\left\langle E_k \right\rangle$ & 34.15\% & 65.84\% \\[3pt]
  \end{tabular}
  \caption{\sid{Contributions of the self-induced and externally-induced flow to the total velocity within correlation kernels used to identify high kinetic energy jets and high enstrophy swirling regions, i.e. $2.5L^\prime$ and $5\hpp$, respectively.}}
  \label{tab:LHpSelfExt}
  \end{center}
\end{table}

\subsection{Relative contribution of vorticity levels in generating flow structures}
So far, we have shown that high kinetic energy jets are mostly externally induced coherent flow regions, while high enstrophy swirling regions are a superposition of self-induced flow and background induced flow. We now show, and quantify, the relative Biot-Savart contribution of different levels of the vorticity field in generating these structures, and the total velocity field in general. From the outset, one can expect that the weak vorticity range ($\omega_b \ll \omega^\prime$) will have a small or negligible Biot-Savart contribution in generating the velocity field, due to the low vorticity amplitude. Significant contributions can come from (i) the strong vorticity, i.e. $\omega_b \gg \omega^\prime$, which can intermittently assume very large amplitudes, albeit with a low volume fraction (\sid{see figure \ref{fig:TurbulenceData-PDF}c}) and (ii) the intermediate background vorticity, i.e. $\omega \sim \omega^\prime$, which has a lower magnitude, but permeates most of the volume and can hence have, when combined, a significant Biot-Savart contribution.

\begin{table}
  \begin{center}
\def~{\hphantom{0}}
  \begin{tabular}{lcccc}
       Bin & & $\omega_l \leq \omega_b/\omega^\prime < \omega_h$  & Classification\\[3pt]
       Bin1 & & $0      \leq \omega_b/\omega^\prime < 0.1$    & Weak \\
       Bin2 & & $0.1  \leq \omega_b/\omega^\prime < 0.25$   & Weak\\
       Bin3 & & $0.25   \leq \omega_b/\omega^\prime < 0.5$    & Weak\\
       Bin4 & & $0.5    \leq \omega_b/\omega^\prime < 1.0$    & Intermediate Weak\\
       Bin5 & & $1.0    \leq \omega_b/\omega^\prime < 2.0$    & Intermediate High\\
       Bin6 & & $2.0    \leq \omega_b/\omega^\prime < 4.0$    & High\\
       Bin7 & & $4.0    \leq \omega_b/\omega^\prime < \infty$ & Very High
  \end{tabular}
  \caption{Vorticity field divided into seven logarithmicallly spaced bins, along with an indicative classification based upon the vorticity amplitude.}
  \label{tab:bins}
  \end{center}
\end{table}

We divide the vorticity field in a range of bins, where each bin contains vorticity $\omega_b$ in the range $\omega_l \leq \omega_b/\omega^\prime < \omega_h$, with the lower and higher vorticity limits $\omega_l$ and $\omega_h$. \sid{To determine the precise range of the bins, we start with the original classification of $\omega \geq 4\omega^\prime$ as ``strong'' vorticity by \cite{she1990intermittent}. This is roughly the start of the deviation from Gaussian in the vorticity PDF, see \ref{fig:TurbulenceData-PDF}(c) (note that since we show the PDF of vorticity components, the deviation from Gaussian occurs around $\approx 4\omega^\prime/\sqrt{3}$, since $\omega_x^\prime \approx \omega_y^\prime \approx \omega_z^\prime \approx \omega^\prime/\sqrt{3}$). Hence we form a vorticity bin for the range $\omega \geq 4\omega^\prime$. Since almost the entire volume, i.e. $\approx 99\%$, has vorticity lower than $4\omega^\prime$, we need to has a number of separate vorticity bins for the range $\omega < 4\omega^\prime$. We do this by successively dividing the upper limit of vorticity for each bin by a factor of 2, in total creating 7 bins, which have been detailed in table \ref{tab:bins}. We also give a classification for these bins based upon the vorticity amplitude, for purpose of further discussion. The choice of these $7$ bins, albeit arbitrary, gives a good representation of the different vorticity levels with a sufficient separation of scale in the vorticity magnitude, while highlighting the small region around the peak of the vorticity PDF.}

\begin{figure}
  \centerline{\includegraphics[width=\linewidth]{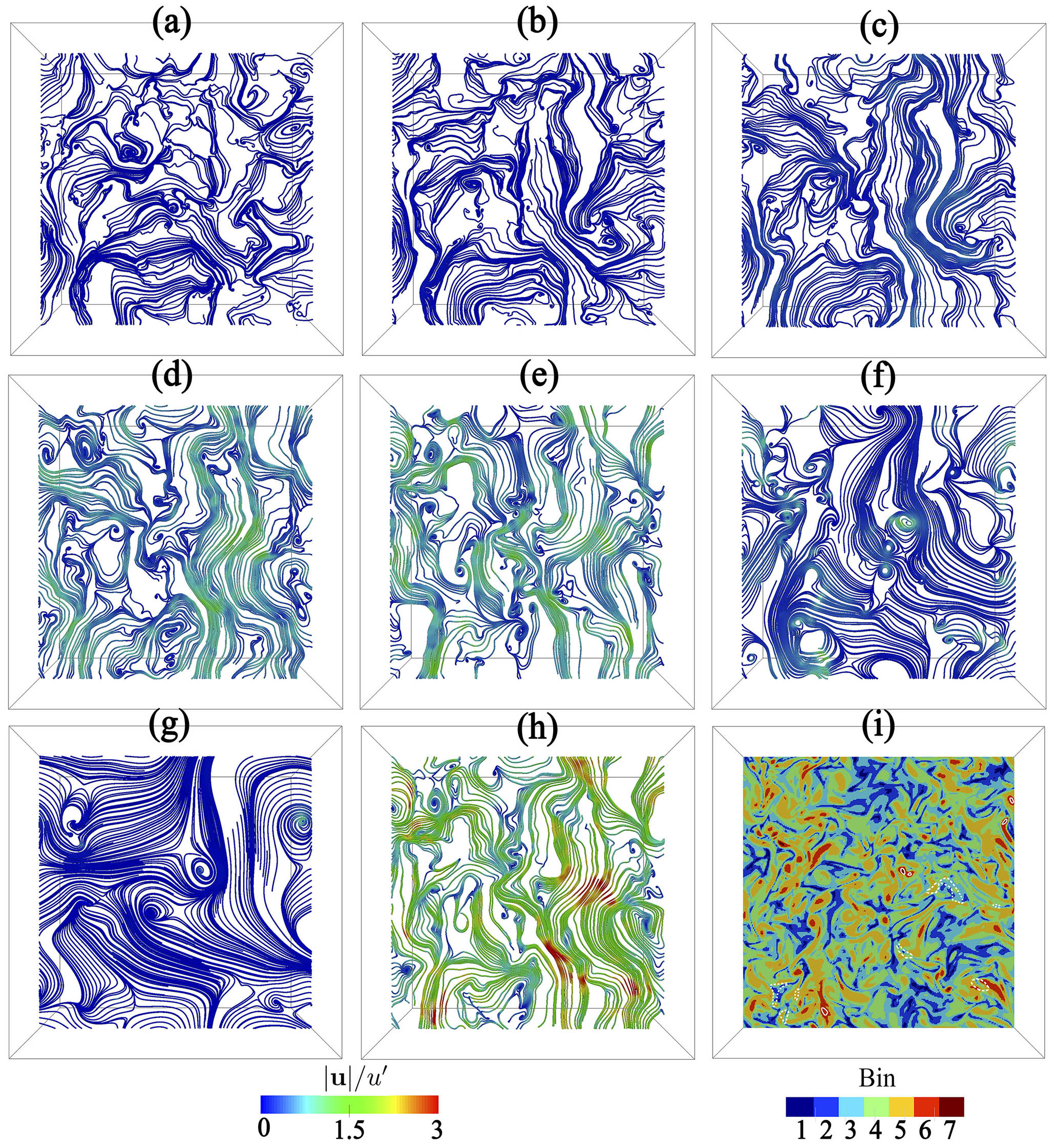}}

  \caption{Planar velocity streamlines (in the entire $xy-$plane, at an arbitrary $z$ location), have been shown for Bin1 through Bin7 going from panels (a) through (g), while panel (h) shows the streamlines of the total velocity field. \sid{Panel (i) shows the vorticity field coloured by the bin number, along with contours of $2.5L^\prime$ (dashed white lines) and $5\hpp$ (solid white lines). Colorbar shows the velocity magnitude.}}
  \label{fig:VortBins-Streamlines}
\end{figure}

We then compute the Biot-Savart velocity field, corresponding to each vorticity bin individually, while the sum of all these Biot-Savart fields together gives the total velocity field. Figure \ref{fig:VortBins-Streamlines} shows planar velocity streamlines (in the $xy-$plane, at an arbitrary $z$ location, \sid{which is representative of the planar streamline structure found throughout the data}), for Bin1-Bin7 going from panels (a)-(g), while panel (h) shows the streamlines of the total velocity field. \sid{Further, panel (i) shows the planar field of 7 vorticity bins, coloured by number, along with $2.5L^\prime$ regions shown as dashed contour lines and $5\hpp$ regions shown as solid contour lines (the latter, being very small, can be seen upon zooming into the figure). We first consider the total velocity streamlines in panel (h). The overall structure appears disorderly, with mostly $E_k$ in the intermediate range. A few small regions contain high $E_k$ jets, where the streamlines  become well aligned. Other regions of aligned streamlines, albeit with intermediate $E_k$, can also be seen. The total streamline structure does not seem to result from a single (or a few) ``large eddies'', as for instance could be seen in the Taylor-Green velocity structure (see figure \ref{fig:taylorGreen-Flow}b). A few small regions with swirling velocity, with intermediate and weak $E_k$, can also be seen. The vorticity bins in panel (i) also appear mostly disorderly. Most of the volume is occupied by Bin4 and Bin5, which have a very convoluted, fragmented structure, and there is no clear large-scale organization of the vorticity, which is perhaps the reason why a large-scale organization does not emerge in the velocity field (as happens in the case of the Taylor-Green velocity field, where a few large-scale velocity structures are associated with an underlying pattern in the vorticity field). Further, the contours of $2.5L^\prime$ show that these regions mostly coincide with intermediate and weak vorticity, showing that the jet-like flow emerges in regions of weak vorticity, surrounded by regions of intermediate (and occasionally, strong) vorticity. Contours of $5\hpp$ are found coinciding with strong vorticity bins, i.e. Bin6 and Bin7. It can be seen that the $5\hpp$ contours are surrounded by relatively larger regions of high vorticity, corresponding to Bin6 - which further explains why the self-induced flow due to $5\hpp$ regions (see figure \ref{fig:StructureHp}) is weak and does significantly contribute to the swirling-flow, which receives a large contribution from the surrounding strong vorticity.}

Panel (a) shows that Bin1 generates disordered velocity streamlines, with a negligible amplitude, as can be expected from the low vorticity magnitude in this bin. Panels (b) and (c), correspnding to Bin2 and Bin3, show that the streamlines due to these bins begin to have a weak organization, similar in pattern to the streamlines due to Bin4 and Bin5, which resemble some of the larger features of the total velocity field shown in panel (h). The magnitude of the streamlines due to Bin2 and Bin3, however, is still weak. Since Bin3 surrounds Bin4 and Bin5, it is not surprising that the velocity generated by them has a similar shape in many regions. Panel (d) and (e), i.e. Bin4 and Bin5, most resemble the total velocity field in panel (h), both in structure and velocity magnitude. Panel (f) shows Bin6, which has small, localized, regions of intermediate velocity amplitude, and larger regions of ordered streamlines with a low amplitude. Finally, panel (g) shows the streamlines for Bin7, corresponding to the very high vorticity range ($\omega/\omega^\prime \geq 4.0$). \sid{At this range, the vorticity regions become spatially isolated and occupy very small volumes. The associated velocity field shows large-scale structures, which look like a more disordered Taylor-Green flow pattern. However, due to the very low velocity magnitude, these structures do not emerge in the total velocity field in panel (h), as they are overwhelmed by the contributions from the other bins.} Bin7, hence, influences the velocity field \textit{only} in the immediate neighbourhood of the regions of high vorticity. The precise contribution of each bin to generating the velocity field is quantified in section \ref{sec:BiotSavartStatistics}.

\begin{figure}
  \centerline{\includegraphics[width=\linewidth]{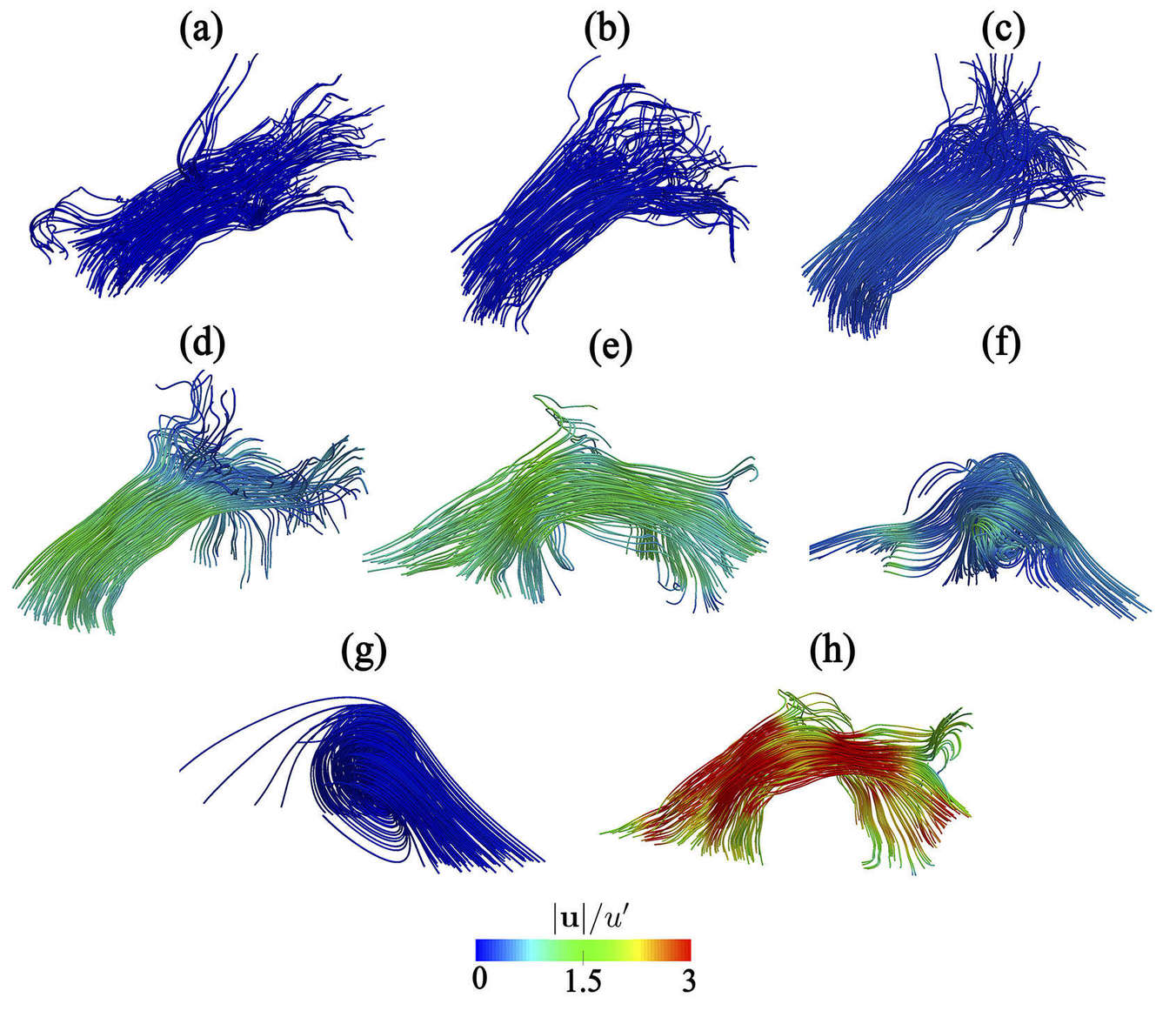}}

  \caption{A single $2.5L^\prime$ high kinetic energy region is highlighted, where panels (a) to (g) show the contribution from Bin1 to Bin7 to the generation of the total velocity field shown in panel (h).} 
  \label{fig:LCorrBins}
\end{figure}

We next look at instances of individual flow structures, along with the Biot-Savart contribution of each vorticity bin in generating this structure. Figure \ref{fig:LCorrBins} \sid{focuses on the first $2.5L^\prime$ structure shown in figure \ref{fig:StructureLCorr}(a)}. Panels (a) to (g) show the Biot-Savart contribution of Bin1 to Bin7 in generating the total velocity structure, which is shown in panel (h). We find that the most significant contribution comes from Bin4 and Bin5, i.e. panels (d) and (e), both of which represent bins with $\omega \sim \omega^\prime$. The streamlines resemble the structure of the high $E_k$ jet in panel (h), and also have an intermediate level of velocity amplitude. Bin6 also has a small contribution to the total velocity structure. Bin1 and Bin2 have a negligible contribution, while Bin3 has a mild contribution. Lastly, it is interesting to find that Bin7, which corresponds to the higher levels of vorticity, also has a negligible contribution to the generation of the kinetic energy jet. This result also reflects the spatial exclusivity between high kinetic energy and high enstrophy. \sid{From figures \ref{fig:JointPDFCorrelations}(a) and \ref{fig:JointPDFCorrelations}(c), we know that the joint distribution of $L$ and $\omega$ is anti-correlated. The very localized, high vorticity regions, do not extend their influence on the velocity field beyond their immediate neighbourhoods.} These results show that high kinetic energy jets are induced by non-local intermediate vorticity contributions, in the range of $\omega \sim \omega^\prime$. 

\begin{figure}
  \centerline{\includegraphics[width=\linewidth]{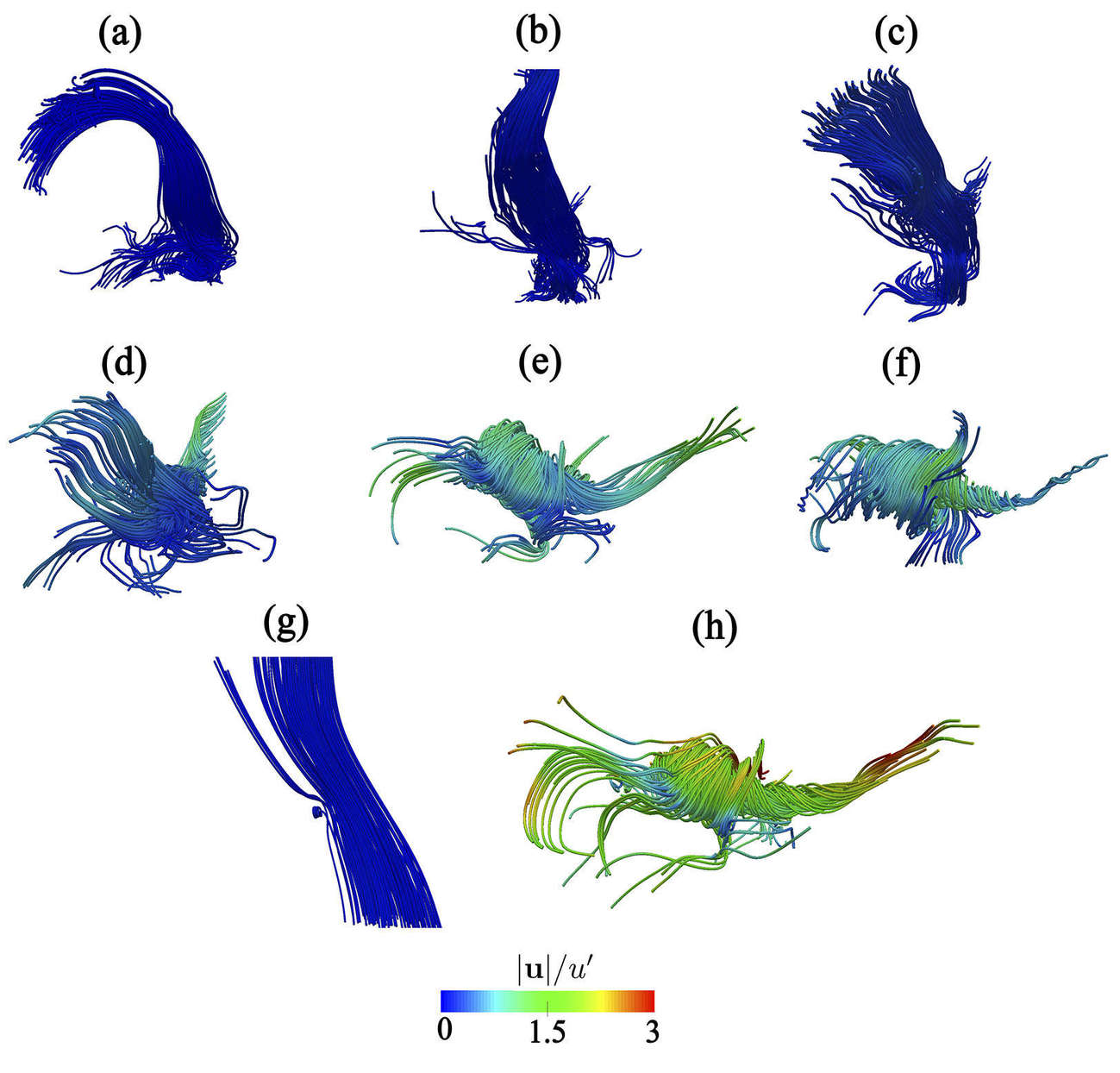}}

  \caption{A single $5\hpp$ swirling flow region is highlighted, where panels (a) to (g) show the contribution from Bin1 to Bin7 to the generation of the total velocity field shown in panel (h).}
  \label{fig:HpBins}
\end{figure}

Figure \ref{fig:HpBins} shows the first $5\hpp$ swirling flow structure from figure \ref{fig:StructureHp}(a). The Biot-Savart contribution from Bin1 to Bin7 is shown in panels (a) to (g). Bin5 and Bin6 in panels (e) and (f) are found to contribute most significantly to the swirling flow region. Bin1 to Bin3 in panels (a) to (c) have a negligible contribution, while Bin4 in panel (d) has a relatively mild velocity and can hence be expected to have only a small contribution to the generation of the total velocity. \sid{Bin7 in panel (g), corresponding to the highest range of vorticity, is also found to generate a very weak velocity field, and contribute negligibly to the swirling flow region in this example. Note that this range of vorticity,  $\omega>4\omega^\prime$, occurs intermittently. From the first structure shown in \ref{fig:StructureHp}(c), we know that this particular $5\hpp$ region self-induces a swirling velocity field, albeit with low magnitude. Since the Bin7 contribution is found to be negligible, we can conclude that this structure contains vorticity in the range of Bin6 and lower. In general, high $H^p$ structures will tend to have high vorticity (as evident from the joint-PDF in figure \ref{fig:JointPDFCorrelations}), while the higher amplitude regions will become exponentially smaller in size. Some of the high $H^p$ regions will indeed have contributions from Bin7, which becomes clear in section \ref{sec:BiotSavartStatistics} where we quantify the vorticity bin contributions statistically.}

\subsection{Statistics of the Biot-Savart contributions}\label{sec:BiotSavartStatistics}
\begin{figure}
\begin{subfigure}{0.5\linewidth}
  \centerline{\includegraphics[width=\linewidth]{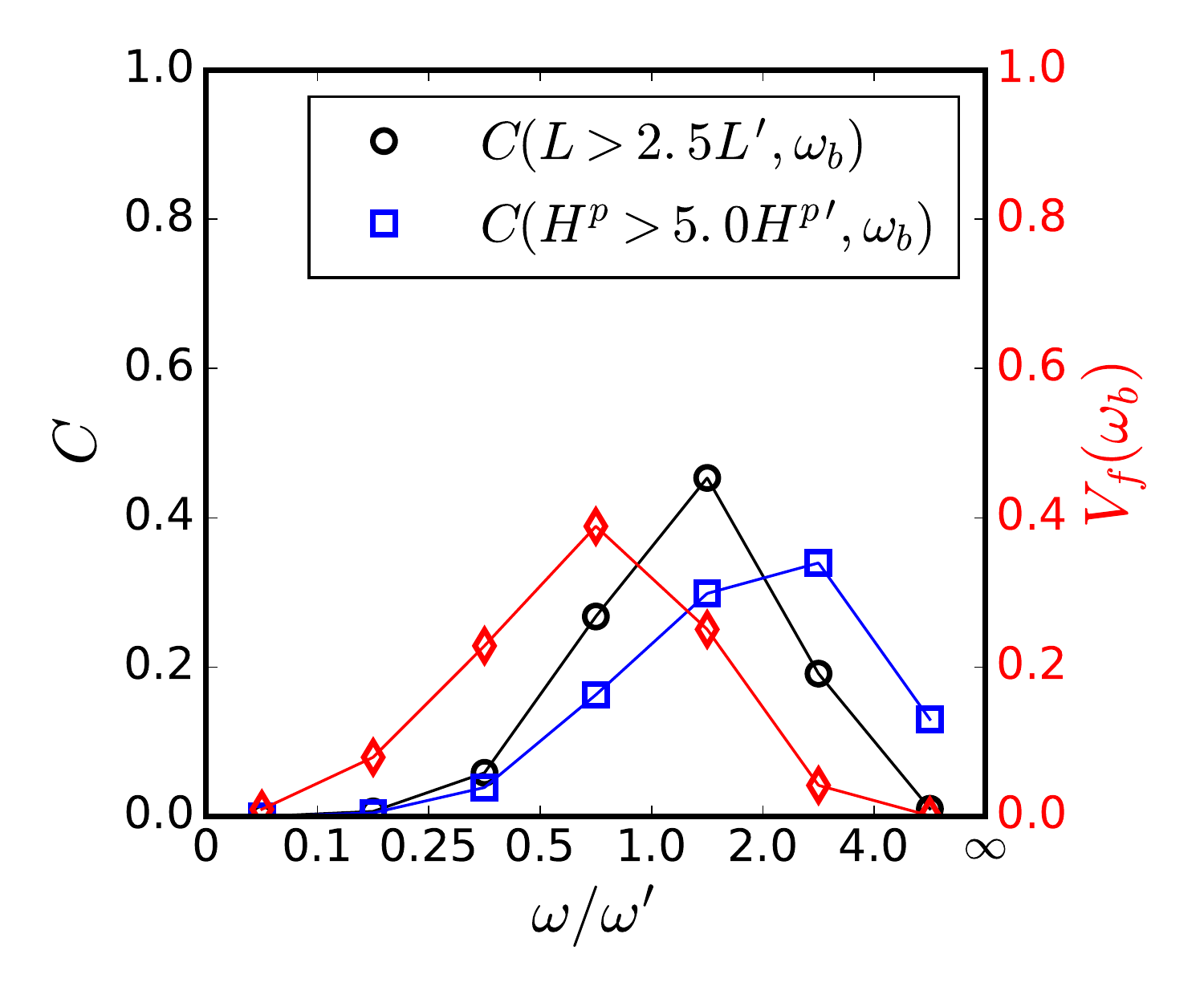}}

  	\caption{}
\end{subfigure}\quad
\begin{subfigure}{0.5\linewidth}
  \centerline{\includegraphics[width=\linewidth]{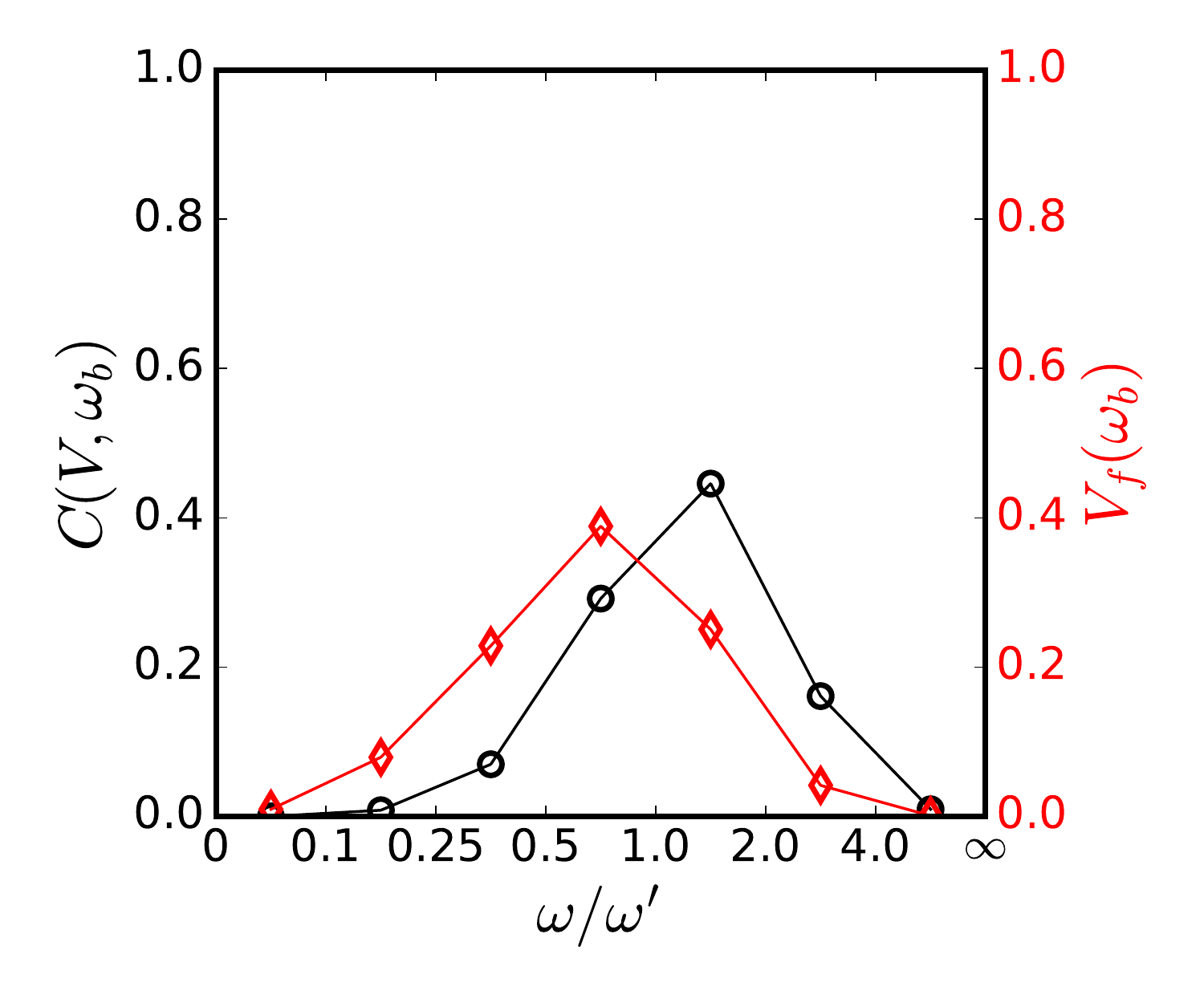}}

  	\caption{}
\end{subfigure}
\caption{Distribution of vorticity bin contributions to (a) the flow \textit{within} $L$ and $H^p$ regions, i.e. 
  	$C(L>2.5L^\prime,\omega_b)$ and $C(H^p>5{H^p}^\prime,\omega_b)$, respectively, and (b) the generation of the total flow field, in the entire volume $V$, i.e. $(C(V,\omega_b))$. The left $y-$ axis shows the correlation $C$, while the right $y-$axis shows the volume fraction of the vorticity bin $V_f(\omega_{\mathrm{b}})$.}
  \label{fig:BiotSavartFieldCorrelation-VortBudget}
\end{figure}

So far, the results show that both kinetic energy jets \sid{(identified by $L>2.5L^\prime$) and swirling flow regions (identified by $H^p>5\hpp$)} have a significant Biot-Savart contribution from the intermediate range vorticity. This finding is interesting, since it implies that the bulk of the volume, which contains vorticity of a relatively mild magnitude ($\omega\sim\omega^\prime$), is significant in determining the structure of the velocity field, possibly everywhere. We quantify this by calculating the correlations $C(L>2.5L^\prime, \omega_b)$ and $C(H^p > 5\hpp, \omega_b)$, along with $C(V,\omega_b)$, which have been shown in figure \ref{fig:BiotSavartFieldCorrelation-VortBudget}. Note that the $x-$axis in these figures shows the limits of consecutive vorticity bins, and the points representing each bin, which have been placed in between their corresponding limits, have been connected with lines to guide the eye. Further, the second $y-$axis (on the right) shows the volume fraction (in red) of each vorticity bin $V_f(\omega_b)$, to give a complete picture, which includes the Biot-Savart contribution of each vorticity bin along with the volume it occupies.

The black curve (circles) in figure \ref{fig:BiotSavartFieldCorrelation-VortBudget}(a) shows $C(L>2.5L^\prime, \omega_b)$. The highest contribution to the induction of high $L$ kinetic energy jets, of $\approx 45\%$, comes from Bin5 i.e. $1 \leq \omega_b/\omega^\prime < 2$, which occupies $\approx 25\%$ of the volume. Bin4 i.e. $0.5 \leq \omega_b/\omega^\prime < 1$ contributes $\approx 26\%$ while occupying $\approx 40\%$ of the volume. The only other significant contribution is from Bin6 i.e. $2 \leq \omega_b/\omega^\prime < 4$, of $\approx 20\%$, which occupies a much smaller volume ($<4\%$). Strong vorticity, i.e.  $\omega_b/\omega^\prime \geq 4$, has a negligible Biot-Savart contribution to the generation of the kinetic energy jets. This again reflects the spatial exclusivity of high kinetic energy and high vorticity regions, where the latter occupies isolated, small regions of the volume. It is interesting to find that strong vorticity, which forms the long tails of the vorticity PDF, is insignificant in the Biot-Savart sense. Further, as expected, weak vorticity i.e $\omega_b/\omega^\prime < 0.5$ also has a negligible contribution. 

The blue curve (squares) in \ref{fig:BiotSavartFieldCorrelation-VortBudget}(a) shows $C(H^p > 5\hpp,\omega_b)$, where the distribution of vorticity bin contributions is found shifted towards the higher vorticity ranges, as can be expected. \sid{The intersection between the distribution of bin-wise volume fraction and bin-wise Biot-Savart contribution, in comparison to $L>2.5L^\prime$ is also shifted towards higher vorticity levels.} In this case, Bin6 i.e. $2 \leq \omega_b/\omega^\prime < 4$ has the highest contribution ($\approx 34\%$) to generating flow inside $5\hpp$ regions, while occupying only $4\%$ of the volume. This is closely followed by the contribution from Bin5 of $\approx 30\%$. Strong vorticity in the range of $\omega_b/\omega^\prime \geq 4$ is also found to contribute $\approx 13\%$, i.e. a small amount as was anticipated in the discussion regarding figure \ref{fig:HpBins}. One difference between the distribution of vorticity bin contributions for $2.5L^\prime$ and $5\hpp$ is that the former (black curve in \ref{fig:BiotSavartFieldCorrelation-VortBudget}a) appears relatively narrower and higher peaked than the latter (blue curve in \ref{fig:BiotSavartFieldCorrelation-VortBudget}a), which has a broader distribution. \sid{This reflects the fact that the velocity field in high $H^p$ swirling flow regions has a background induced contribution, and a degree of self-induction due to the high vorticity levels.}

Trends in the distribution of vorticity bin contributions when considering the entire velocity field ($C(V,\omega_b)$), shown in figure \ref{fig:BiotSavartFieldCorrelation-VortBudget}(b), are found to be the same as those for $C(L>2.5L^\prime,\omega_b)$. It can be summarized that vorticity in the range of $0.5 \leq \omega_b/\omega^\prime < 1.0$ (Bin4) occupies the maximum volume, but due to a lower amplitude has a slightly smaller Biot-Savart contribution than vorticity in the range $1 \leq \omega_b/\omega^\prime < 2$ (Bin5), which has vorticity of a higher amplitude, but occupyies a lower fraction of the volume. Together, the vorticity in these two bins, which is essentially at the level of $\omega^\prime$ (and may be termed \textit{intermediate}), both permeates most of space, and generates most of the velocity field. The strong vorticity range i.e. $\omega/\omega^\prime \geq 4$ remains insignificant for generating the velocity field, except in the immediate vicinity of high vorticity regions.

\subsection{Summary}
\sid{In summary, regions with high $L$, which were found to coincide strongly with high $E_k$ (see figure \ref{fig:JointPDFCorrelations}a), form localized jet-like coherent flows, while containing very low levels of vorticity. Moreover, the coherence in these flow structures is externally induced, with most dominant contributions of the intermediate background vorticity. The high $H^p$ regions, that coincide with high enstrophy (see figure \ref{fig:JointPDFCorrelations}b) are interspersed in a more or less random manner through the flow field, and they do not add up together, in a Biot-Savart sense, to give rise to larger high $E_k$ structures. The flow in the vicinity of these regions has a swirling motion, and is due to a superposition self-induced swirling flow and an intermediate background vorticity induced flow. These results give a clear picture of the flow structures that together comprise turbulence fields, along with their vorticity composition.}

\section{Conclusions and outlook}
In this paper, we developed mathematical tools for identifying instantaneous, spatial, structures in vector fields associated with turbulent flows. By taking coherence to mean correlation, in this context, we began by generalizing the usual two-point correlation tensor, i.e. a statistical concept, to an instantaneous and deterministic correlation measure. This measure could, in principle, be reduced to a correlation manifold (surface) around each spatial point, while, for simplicity, we sample this manifold along an arbitrary orthogonal base, yielding a three-tuple correlation field. We then propose different correlation measures using the velocity and vorticity fields, aimed at identifying simple flow patterns.

\sid{For instance, we defined $\lvec$ to identify structures with well aligned (parallel) velocity streamlines, with a high kinetic energy, along with an analogous measure $\g$ for the vorticity field. Correlations $\lsvec$ and $\gs$ were defined to identify symmetries/anti-symmetries in the velocity and vorticity fields, respectively. Finally, in the vein of the the Biot-Savart law, correlation the $\hp$ (and also $\h$) was defined to identify regions of vorticity that induce swirling motion in their neighbourhood.}

We tested these correlation measures against canonical flows like Oseen vortices, Taylor-Green vortices and Burgers vortices, and then applied them to datasets from direct numerical simulations of incompressible, homogeneous isotropic turbulence. Further, \sid{reconstructing the velocity field using the Biot-Savart law in different coherent flow regions allowed us to disentangle the contributions from different levels and regions of the vorticity field in inducing flow structures, which revealed interesting aspects regarding the composition of the velocity field}. Our main findings are summarized below:

\begin{enumerate}
\item The velocity field has \sid{at least} two distinct coherent flow types, which correspond to the high turbulence kinetic energy ($E_k$) and the high enstrophy ($\omega^2$) structures.

\item \sid{High $E_k$ structures yield high values of the $\lvec$ correlation, as we find that the velocity in these regions is \textit{jet-like}, i.e. it comprises well-aligned streamlines with a high velocity magnitude. The similarity between $\lvec$ and $\lsvec$ shows the lack of larger symmetries in the velocity field. Interestingly, in the flow considered, we do not find clearly discernible ``large-eddies'', which is how high kinetic energy structures have been traditionally conceived. We emphasize that we analyze only homogeneous isotropic turbulence where the energy injection scale is not correlated with any timescale of the flow. In other kinds of flows, for instance channel flow, where the energy injection has a spatio-temporal structure, the situation might be different. Lastly, the joint distribution of $L$ and $E_k$, further, shows that increasing levels of $E_k$ lead to an increase in the local flow organization.}

\item \sid{High $\omega^2$ structures are found to yield high values for $\g$, $\gs$ and $\hp$ correlations. High $\g$ values show that these regions form \textit{vorticity jets}, which comprise well-aligned vorticity streamlines with a high amplitude. The coincidence of $\g$ and $\gs$ shows that there are no larger symmetries or patterns in the vorticity field. The predominantly high values of $\hp$ in these regions confirms that the flow surrounding high $\omega^2$ structures invariably has a \textit{swirling-motion}, as has been qualitatively described before \citep{she1990intermittent,she1991structure,jimenez1993structure}, which is further corroborated by the joint distribution of $H^p$ and $\omega$.}

\item \sid{The statistics of the spatial distribution of the correlations show that the jet-like and swirling-flow structures are spatially exclusive. This hints at the dynamical separation between high kinetic energy structures and the high enstrophy structures, which represent different parts of the Navier-Stokes dynamics. Since we focused on instantaneous structures, we did not investigate how they evolve in time.}

\item \sid{The Biot-Savart reconstruction of the velocity field shows that the high $L$, high $E_k$, jets are not self-inducing (with a low self-induced contribution of $\approx 11\%$), as their vorticity content is low. This shows that the coherence of these jets is induced by \textit{non-local} vorticity contributions. Reconstructing the total velocity field using logarithmically spaced vorticity bin contributions shows that the kinetic energy jets are mostly induced by $\omega \sim \omega^\prime$, which permeates the volume and has an intermediate magnitude. Strong vorticity, which occurs intermittently and is spatially distant from high $E_k$ regions, has a negligible contribution to the induction of kinetic energy jets.} 

\item \sid{High $\omega^2$ swirling flow structures are found to be the result of a superposition of background vorticity induced flow ($\approx 65\%$), and self-induced flow $\approx 35\%$. In a bin-wise sense, a significant Biot-Savart contribution ($\approx 34\%$) to the generation of these structures comes from the $2 \leq \omega/\omega^\prime < 4$ range. The two bins corresponding to intermediate, background vorticity, i.e. $0.5 \leq \omega/\omega^\prime < 2$, together contribute $\approx 50\%$.}

\item \sid{Considering the entire velocity field, we find a lack of clear large-scale flow organization. Flow streamlines appear, overall, disorganized, while being interspersed with localized regions of jet-like and swirling flow. The jet-like regions have intermediate to high levels of $E_k$, with intermediate to low levels of $\omega^2$. The opposite holds for the swirling flow regions, which have intermediate to high $\omega^2$, with intermediate to low $E_k$. Note that at intermediate values, the distinction begins to blur and structures can overlap.} 

\item \sid{The vorticity field, when viewed bin-wise, also shows a disorganized structure. Most of the volume is occupied by a convoluted and fragmented intermediate vorticity field, i.e. $0.5 < \omega/\omega^\prime < 2$, which is also found to most significantly induce the velocity field everywhere ($\approx 80\%$), in the Biot-Savart sense. Vorticity in this range corresponds to the narrow, (almost) Gaussian peak of the vorticity distribution, while ignoring the long intermittency tails which correspond to extreme vorticity. Strong (and extreme) vorticity ($\omega \geq 4\omega^\prime$) occupies very small regions that appear mostly in isolation. These regions generate a large-scale flow pattern, resembling a disordered Taylor-Green flow, which, due to its weak amplitude, does not contribute to the total velocity field. The influence of strong $\omega$ regions to the velocity is limited to their immediate neighbourhood. Usually, most research has focused on extreme vorticity, for instance due to its influence on mixing and particle dispersion. It turns out that the bulk of the flow, and, in particular, the kinetic energy containing regions, are completely impervious to extreme vorticity. The weak vorticity, i.e $\omega < 0.5 \omega^\prime$, as can be expected, does not contribute much to the generation of the velocity field.}
\end{enumerate}

\begin{figure}
  \centerline{\includegraphics[width=\linewidth]{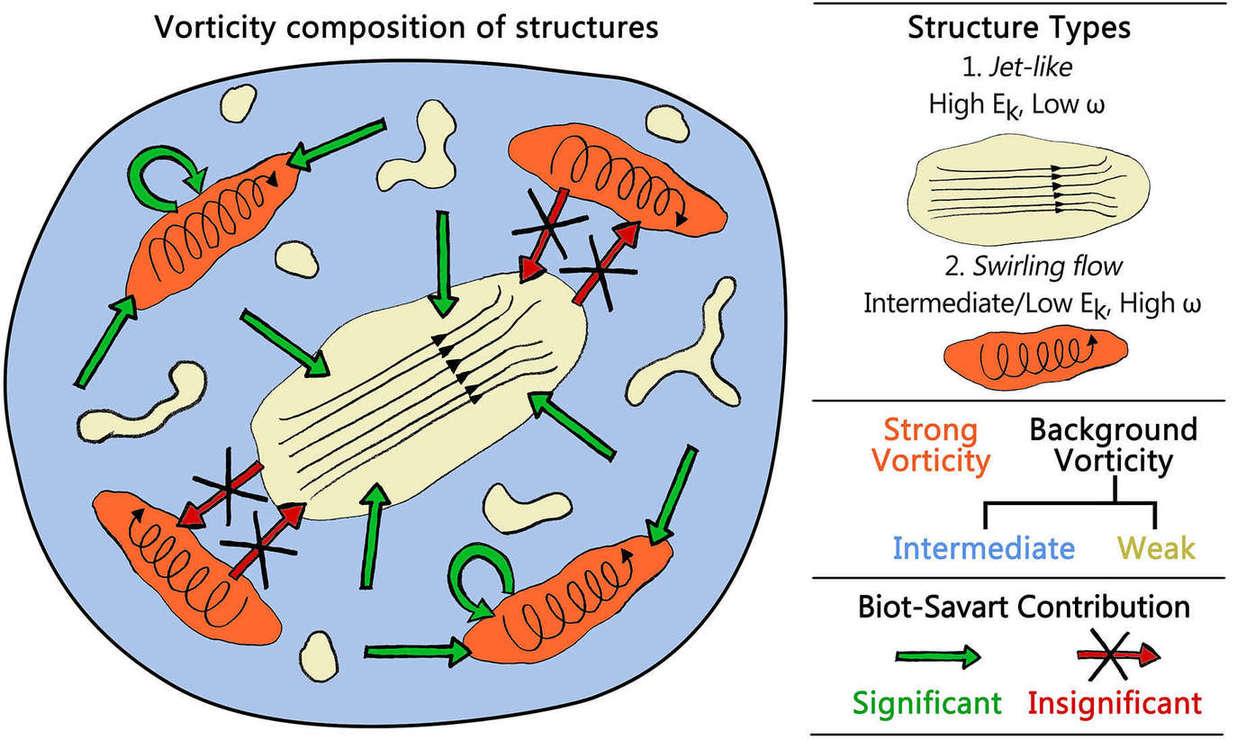}}

  \caption{Schematic of the organization of turbulence structures along with the Biot-Savart contributions that generate them. The vorticity ranges are colour coded with strong vorticity (orange), intermediate background vorticity (blue) and weak vorticity (off white). The kinetic energy structures are \textit{jet-like}, and do not have a significant self-generation due to their low vorticity. Their coherence is induced (in a Biot-Savart sense), almost entirely, by non-local, intermediate background vorticity contributions (shown in blue), while the contribution from strong vorticity to their generation is negligible. Strong vorticity regions have a local \textit{swirling motion} in the velocity field. These structures are a superposition of an intermediate background vorticity generated flow and a self-induced accentuation of the swirling flow.}
  \label{fig:schematicStructure}
\end{figure}

\sid{The structural view of turbulence fields, as evident from our results, is illustrated in figure \ref{fig:schematicStructure}. Particularly, the coherence of the kinetic energy containing jets being induced by non-local vorticity shows that these structures are not self-determining. The traditional ``cascade'' perspective of turbulence dynamics explains the phenomena as being dominated by high kinetic energy, large-scale structures, which in-turn determine the structure and generation of successively smaller scales via some ``eddy break-up'' mechanism, dating back to the idea of \citet{richardson1922weather}. First, we do not find kinetic energy containing ``large eddies'' in real-space velocity fields of homogeneous isotropic turbulence, and instead the flow is found to consist of a distribution of localized kinetic energy jets, which corroborates a \textit{non-hierarchical} flow organization. Secondly, most of the velocity field, which is traditionally believed to comprise a range of ``coherent scales'' (corresponding to the inertial range), is also generated, in a Biot-Savart sense, by the permeating, intermediate range background vorticity. Finally, small scale swirling flow regions are found to be a superposition of background induced flow and self-induced flow, hence being both externally and self-determined to varying degrees.}

\sid{Within the limits of our work, we show that an alternative paradigm for (homogeneous isotropic) turbulence can be considered. The overall organization of turbulence, along with its statistical features, likely \textit{emerges} from the combined contribution of the global vorticity field, while being dominated by the permeating intermediate background vorticity in the range $\omega \sim \omega^\prime$. This is contary to the usual view that large-scale kinetic energy structures dominate the dynamics. Further investigation of the lifecycle of the velocity and vorticity structures, along with identification of the typical force-field structures that drive the dynamics (i.e. structures associated with the pressure gradient and viscous stress fields), will help illumine or dispel notions regarding the existence of hierarchical coherent structures (corresponding to the inertial range), that has been expounded in different representational spaces, while having been elusive in the physical (real) space where the phenomenon of turbulence occurs.} 

\sid{Within the scope of this work, we did not investigate Reynolds number effects on the form and composition of the coherent flow regions. It can be expected that in homogeneous isotropic turbulence high $E_k$ regions will manifest as jets, while high $\omega$ regions will be surrounded by swirling flow, as we indeed verify also in the reference dataset from the Johns Hopkins Turbulence Dataset (see Appendix \ref{app:JHTD-Validation}), which has a higher $Re_\lambda$. The \textit{fraction} of Biot-Savart contribution from different vorticity levels to the generation of flow structures might have some degree of Reynolds number dependence. At higher $Re_\lambda$ the small scales of turbulence will become finer, with higher intermittency in the vorticity field. Although the volume fraction occupied by increasingly higher levels of $\omega$ will become successively smaller, and hence their Biot-Savart contribution to the velocity field should be minimal, except in their immediate neighbourhoods. Since we performed the bin-wise vorticity contribution analysis upon normalizing the vorticity by its root-mean-square value, we hypothesize that the trends in the Biot-Savart contribution are likely similar. This is because a higher $Re_\lambda$ flow will also have a higher $\omega^\prime$, hence the vorticity contribution \textit{relative} to $\omega^\prime$ may not change significantly, however, a full investigation of this shall be left for the future. Finally, a more detailed Biot-Savart analysis can also be performed, which considers both the magnitude of the vorticity in a region $\mathcal{R}_1$ and its distance from another region $\mathcal{R}_2$, when estimating the Biot-Savart contribution of $\mathcal{R}_1$ on $\mathcal{R}_2$. This will paint a fuller picture of the Biot-Savart composition of the velocity field and the spatial organization of coherent structures.}

The tools presented in this paper (or modified versions of them) can be readily applied to identifying structures in any scalar or vector fields (not just turbulence) like pressure, strain or eigenvector distributions, electromagnetic fields, to different dimensional data-sets, or be recast as space-time-correlations to study the spatio-temporal nature of coherence. We believe these tools, combined with the Biot-Savart construction, open a new door into studying the dynamics of turbulence from the perspective of its constituent structures, and may pave the way towards a new structural description of turbulence organization.

\section*{Acknowledgements}
SM would like to thank Dr. Zhao Wu (Johns Hopkins University) for providing access to the JHTD turbulence database. SM would also like to thank Dr. Jason Picardo (IIT-Bombay, India), for some useful discussions and suggestions.

\section*{Declaration of Interests}
The authors report no conflict of interest.

\appendix
\section{}\label{app:BiotSavart}
\sid{Here we describe a few practical aspects regarding the computation of the Biot-Savart integral over periodic domains. As described in section \ref{sec:BiotSavart}, for the case of incompressible flow in a periodic domain, the Helmholtz decomposition reduces to $\mathbf{u} = \mathbf{C}$ (eq. \ref{eq:Helmholtz}), with $\mathbf{C}$ given by the Biot-Savart formula as}

\begin{equation}
\mathbf{C}(\mathbf{x}) = \frac{1}{4\pi}\int_V \frac{\boldsymbol{\omega}\times \mathbf{r}}{|\mathbf{r}|^3}\mathrm{d} V^\prime
\end{equation}

\sid{For a triperiodic domain, it is important to note that all points in the volume $V$ are equivalent, as there is no distinct location like a ``real boundary'' or a ``center'' in the flow. This becomes important when calculating the Biot-Savart integral for points near to the ``simulation boundary''.}

\sid{We perform the Biot-Savart integral in a radially symmetric manner, for which we first tesselate the simulation domain $V$ (which has an edge length of $N$) along each direction. We then consider the volume $\widetilde{V} = (4/3)\pi r^3$, around the \textit{fictitious} center of the original cube, for the integral, i.e. the contribution of $\mathbf{C}$ to $\mathbf{u}$. We test different values of the radius $r$, to check how closely does $\mathbf{C}$ recover the original velocity field $\mathbf{u}$ (while, in principle, a large enough $r$ will reproduce the velocity field within $V$ to $100\%$ accuracy, discounting numerical errors). We find that an integral over $r = N/\sqrt{2}$ generates a Biot-Savart velocity field of $\approx 95\%$ accuracy in comparison to the original velocity field, while an integral over $r=N$ generates a Biot-Savart velocity of $\approx 99\%$ accuracy. Hence, for this paper we use a Biot-Savart reconstruction over $r=N$ (with $\widetilde{V} = (4/3)\pi N^3$).}

\sid{An alternative calculation was also tested, where \textit{each} point in the domain is treated as a \textit{fictitious center}, and a radially symmetric region of radius $r$ is created around it by tesselating the periodic cube. This region is then used to calculate the Biot-Savart velocity for the central point. This method was also found to yield an accuracy of approximately $99\%$ in reconstructing the velocity field, while being computationally more expensive.}

\section{}\label{app:JHTD-Validation}
In this section, we apply the correlations to a reference dataset of homogeneous isotropic turbulence from the Johns Hopkins Turbulence Databases (JHTD) \citep{perlman2007data,li2008public}, to show that, qualitatively, the results we presented using our in-house code are similar to a different turbulence dataset. We use the forced isotropic turbulence dataset, which is a pseudo-spectral simulation performed on $1024^3$ nodes, with a Taylor Reynolds number of $Re_\lambda \sim 433$ (note that our intention is not to study the effect of the different $Re_\lambda$ in the reference dataset). The correlations are calculated by integrating over $\Lambda \approx \lambda$, which was shown to be a reasonable choice in section \ref{sec:ChoiceOfLambda}.

\begin{figure}
  \begin{subfigure}{0.475\linewidth}
  \centerline{\includegraphics[width=\linewidth]{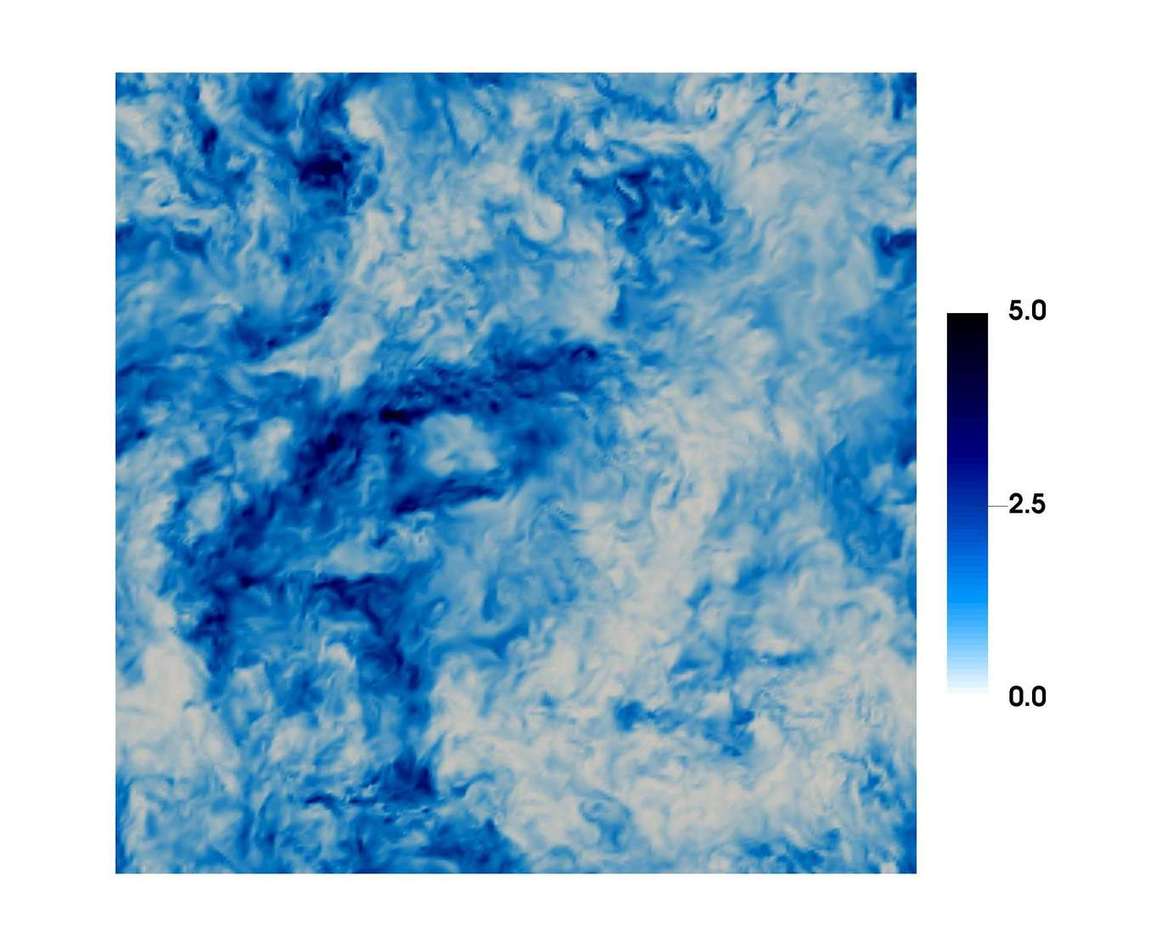}}

  	\caption{$E_k$ normalized with $\ang{E_k}$, shown at an arbitrary crossectional plane.}
	\end{subfigure}\quad
	\begin{subfigure}{0.475\linewidth}
  \centerline{\includegraphics[width=\linewidth]{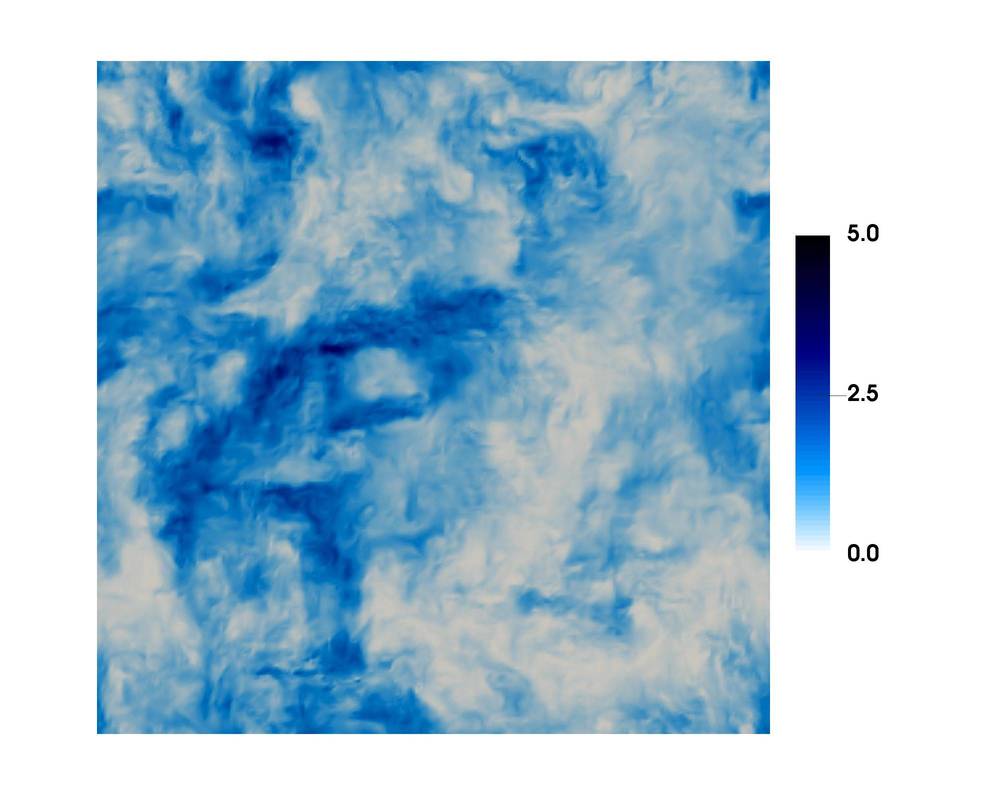}}

  	\caption{Amplitude of $\lvec$ normalized with $L^\prime$, shown at the same crossectional plane.}
  \end{subfigure}

  \begin{subfigure}{0.475\linewidth}
  \centerline{\includegraphics[width=\linewidth]{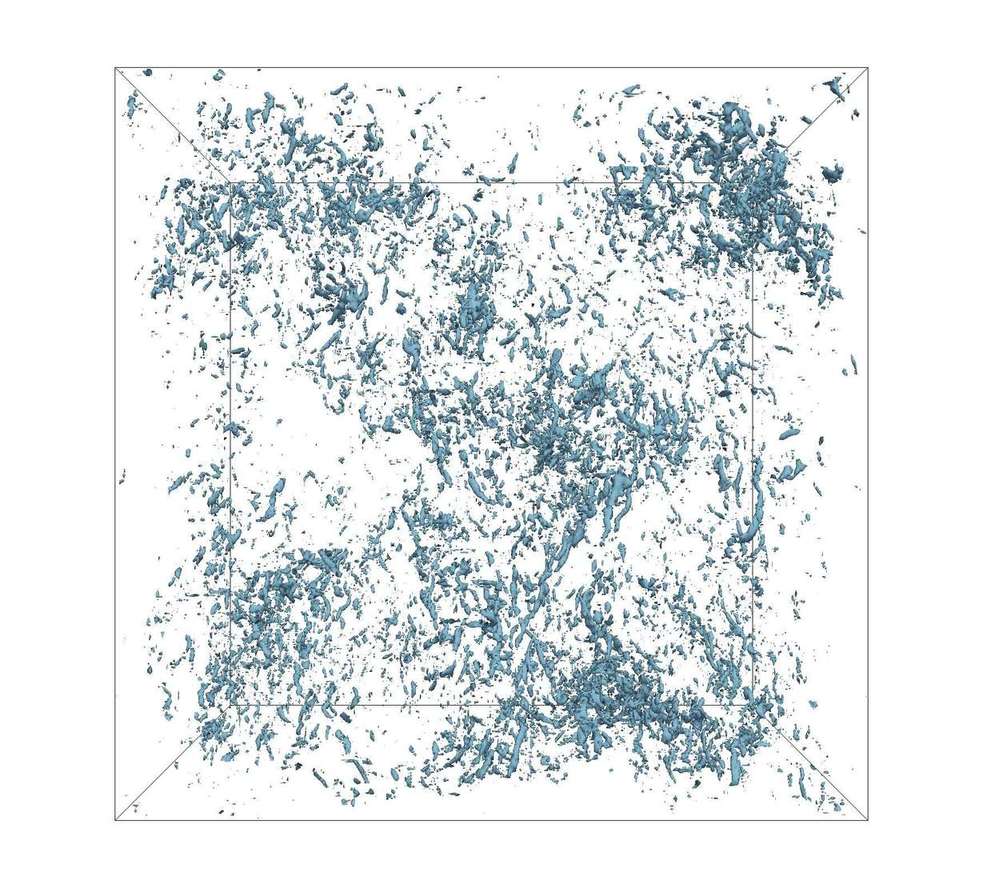}}

  	\caption{Contours of $\omega$ at the level $3\omega^\prime$.}
  \end{subfigure}\quad
\begin{subfigure}{0.475\linewidth}
  \centerline{\includegraphics[width=\linewidth]{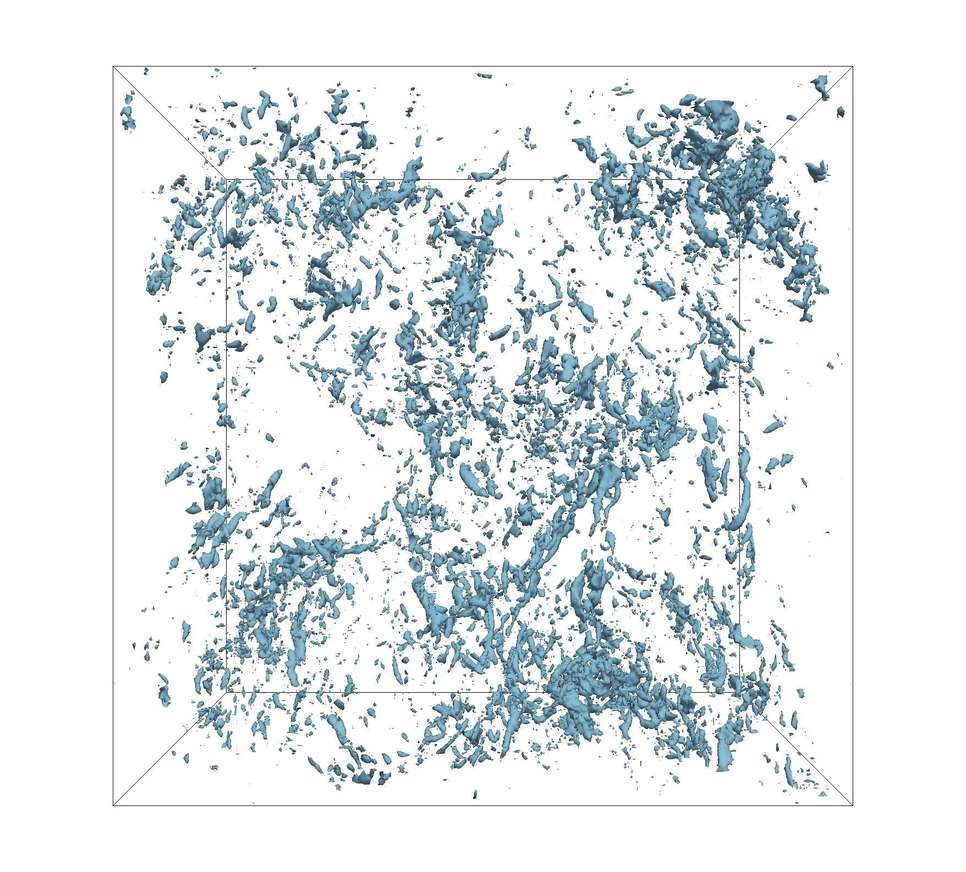}}

  	\caption{Contours of $H^p$ at the level $5\hpp$.}
 \end{subfigure}
  \caption{\sid{Flow field snapshots from the JHTD isotropic turbulence dataset are shown along with correlations $\lvec$ and $\hp$. High kinetic energy $E_k$ regions in (a) consistently yield high $L$ values in (b). Contours of high levels of vorticity in (c) visually coincide strongly with the kernels of the $H^p$ field in (d).}}
  \label{fig:JHTD-Fields}
\end{figure}

Figure \ref{fig:JHTD-Fields} shows the kinetic energy $E_k$ field (normalized by $\ang{E_k}$) in panel (a) and the $L$ field (normalized by $L^\prime$) in panel (b). \sid{Regions of high kinetic energy are also found to consistently yield high values of $L$, while the $L$ field appears smoother, similar to the results presented in figures \ref{fig:TurbulenceCorrelation-LCorr}.  Panel (c) shows contours of the vorticity field $\omega$, and panel (d) shows contours of the $H^p$ field. Visually, there is a strong coincidence between the $\omega$ contours and the kernels of the $H^p$ field, showing that high vorticity regions are associated with swirling velocity in their neighbourhood. Note that the JHTD simulation is performed at a much higher $Re_\lambda$, and has a domain size four times larger along each direction in comparison to our simulation, due to which the fine scale vorticity structures appear to be smaller in the JHTD reference dataset.}

\begin{figure}
  \begin{subfigure}{0.475\linewidth}
  \centerline{\includegraphics[width=\linewidth]{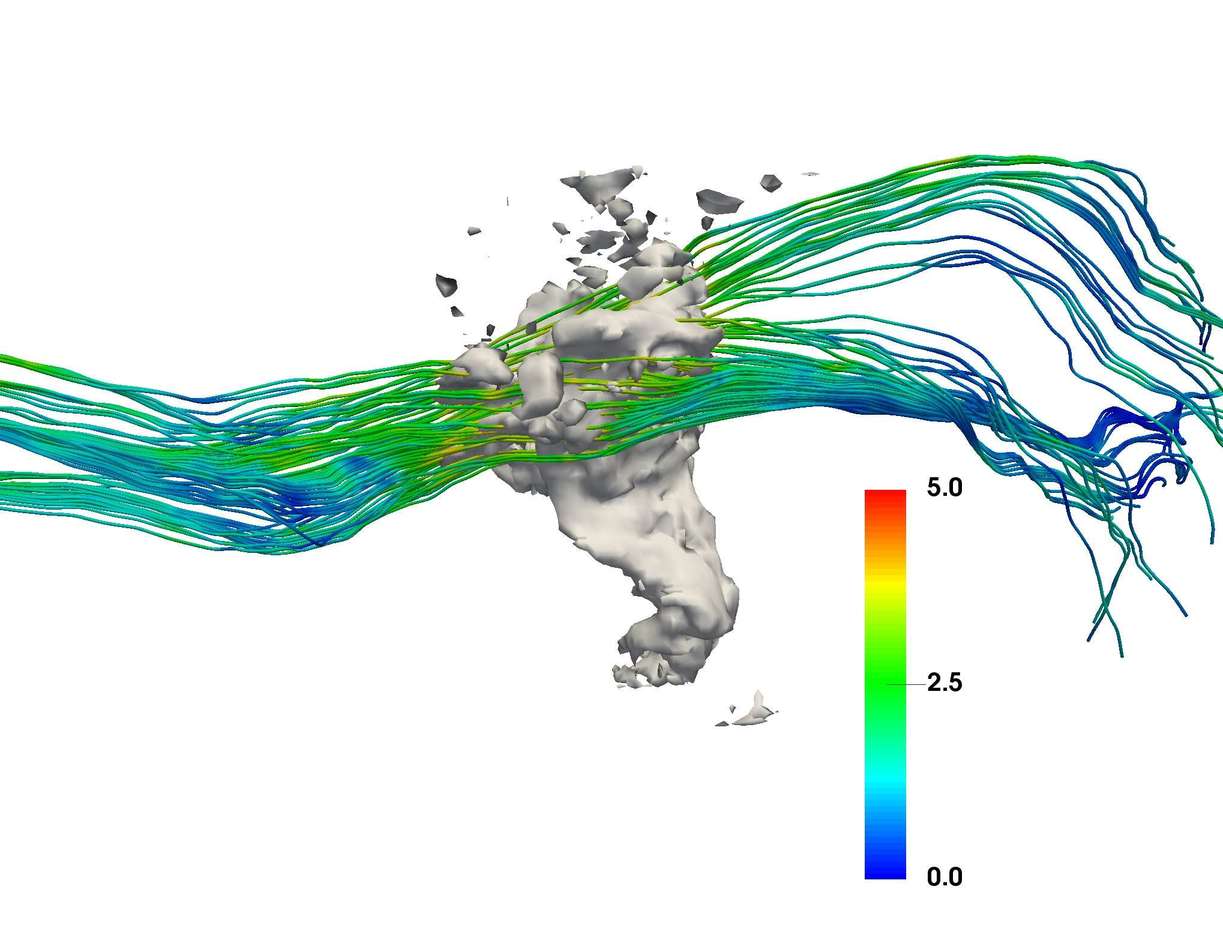}}

  	\caption{A single $L=2.5L^\prime$ region along with local flow streamlines.}
  \end{subfigure}\quad
\begin{subfigure}{0.475\linewidth}
  \centerline{\includegraphics[width=\linewidth]{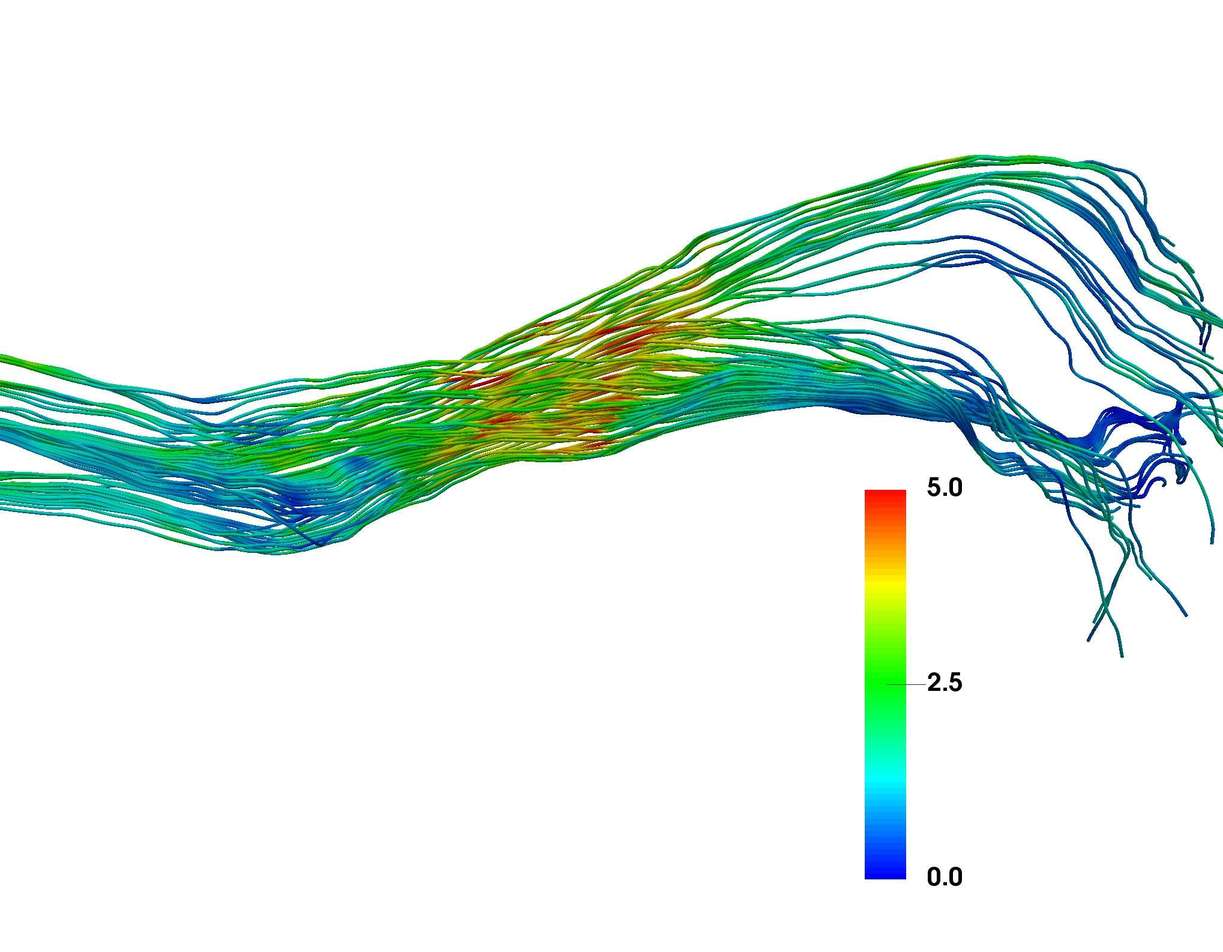}}

  	\caption{Local flow streamlines shown separately.}
  \end{subfigure}
    \caption{The structure of high $L$ regions in the JHTD dataset, which coincide with high $E_k$, are also found to be jet-like. The streamlines have been coloured by the normalized kinetic energy $E_k/\ang{E_k}$.}
  \label{fig:JHTD-Streamlines-Ek}
\end{figure}

\sid{We also verify, qualitatively, that the flow structures around high $L$ and high $H^p$ kernels in the JHTD dataset are similar to those described in section \ref{sec:FlowStructure}. Figure \ref{fig:JHTD-Streamlines-Ek}(a) shows a single contour of $L=2.5L^\prime$ along with the local streamlines (coloured by $E_k/\ang{E_k}$), while the streamlines have been separately shown in panel (b). The local flow structure is found to be jet-like, with a strong parallel alignment of the velocity streamlines near the core of the kernel, which is a region coinciding with the highest kinetic energy, and slightly more disorganized streamlines away from the correlation kernel.}

\begin{figure}
  \begin{subfigure}{0.475\linewidth}
  \centerline{\includegraphics[width=\linewidth]{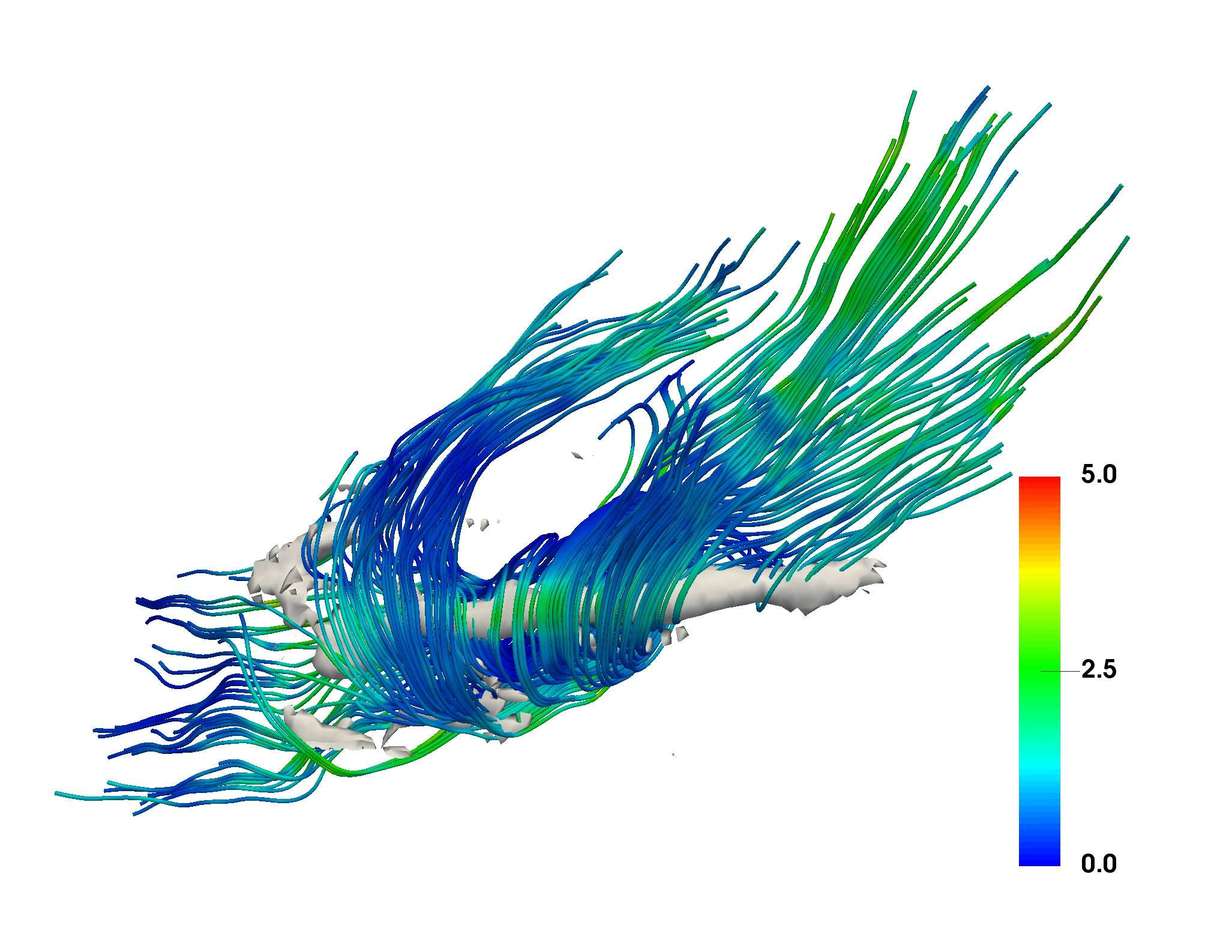}}

  	\caption{A single $H^p = 5\hpp$ region along with local flow streamlines.}
  \end{subfigure}\quad
\begin{subfigure}{0.475\linewidth}
  \centerline{\includegraphics[width=\linewidth]{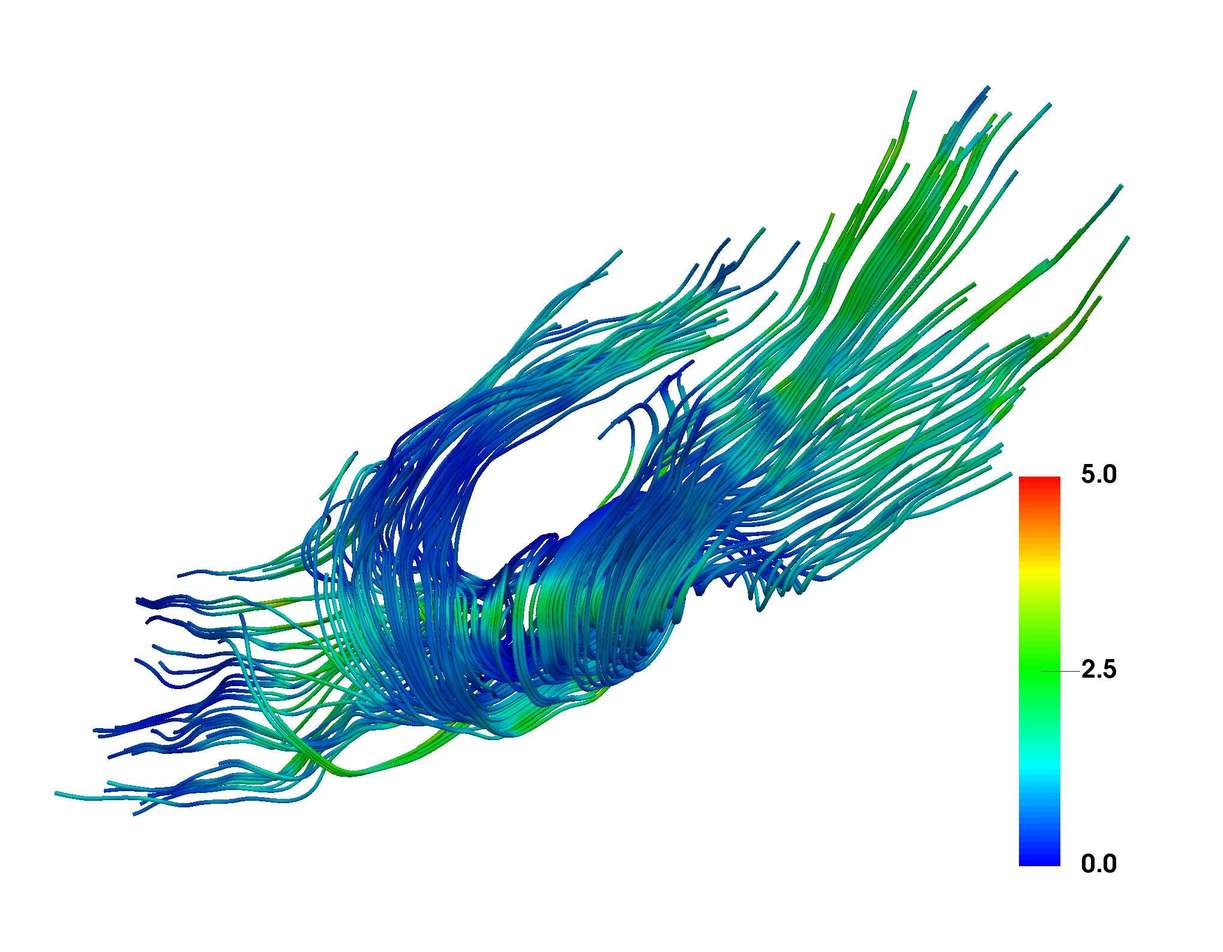}}

  	\caption{Local flow streamlines shown separately.}
  \end{subfigure}
    \caption{The flow around high $\hp$ regions is found to be swirling with intermediate to low $E_k$. The streamlines have been coloured by $E_k/\ang{E_k}$.}
  \label{fig:JHTD-Streamlines-Hp}
\end{figure}

Figure \ref{fig:JHTD-Streamlines-Hp} shows the flow structure in the neighbourhood of an isolated $5\hpp$ kernel in panel (a), along with the streamlines shown separately in panel (b). The velocity field has swirling flow, of intermediate to low $E_k$, which is also consistent with the results shown in section \ref{sec:FlowStructure}. 

\bibliographystyle{jfm}

\bibliography{coherence-arxiv-ed}

\begin{thebibliography}{37}
\expandafter\ifx\csname natexlab\endcsname\relax\def\natexlab#1{#1}\fi
\def\au#1{#1} \def\ed#1{#1} \def\yr#1{#1}\def\at#1{#1}\def\jt#1{\textit{#1}}
  \def\bt#1{#1}\def\bvol#1{\textbf{#1}} \def\vol#1{#1} \def\pg#1{#1}
  \def\publ#1{#1}\def\arxiv#1{#1}\def\org#1{#1}\def\st#1{\textit{#1}}

\bibitem[Alexakis \& Biferale(2018)]{alexakis2018cascades}
{\sc \au{Alexakis, A.} \& \au{Biferale, L.}} \yr{2018}  \at{Cascades and
  transitions in turbulent flows}.  \jt{Physics Reports} .

\bibitem[Alvelius(1999)]{alvelius1999random}
{\sc \au{Alvelius, K.}} \yr{1999}  \at{Random forcing of three-dimensional
  homogeneous turbulence}.  \jt{Physics of Fluids}  \bvol{11}~(7),
  \pg{1880--1889}.

\bibitem[Argoul {\em et~al.\/}(1989)Argoul, Arneodo, Grasseau, Gagne, Hopfinger
  \& Frisch]{argoul1989wavelet}
{\sc \au{Argoul, F.}, \au{Arneodo, A.}, \au{Grasseau, G.}, \au{Gagne, Y.},
  \au{Hopfinger, E.J.} \& \au{Frisch, U.}} \yr{1989}  \at{Wavelet analysis of
  turbulence reveals the multifractal nature of the richardson cascade}.
  \jt{Nature}  \bvol{338}~(6210),  \pg{51}.

\bibitem[Biferale {\em et~al.\/}(2011)Biferale, Perlekar, Sbragaglia,
  Srivastava \& Toschi]{biferale2011lattice}
{\sc \au{Biferale, L.}, \au{Perlekar, P.}, \au{Sbragaglia, M.}, \au{Srivastava,
  S.} \& \au{Toschi, F.}} \yr{2011} A lattice boltzmann method for turbulent
  emulsions.  \bt{In {\em Journal of Physics: Conference Series\/}}, ,
  \vol{vol. 318},  \pg{p. 052017}. IOP Publishing.

\bibitem[Chong {\em et~al.\/}(1990)Chong, Perry \& Cantwell]{chong1990general}
{\sc \au{Chong, M.S.}, \au{Perry, A.E.} \& \au{Cantwell, B.J.}} \yr{1990}
  \at{A general classification of three-dimensional flow fields}.  \jt{Physics
  of Fluids A: Fluid Dynamics}  \bvol{2}~(5),  \pg{765--777}.

\bibitem[Deguchi \& Hall(2014)]{deguchi2014canonical}
{\sc \au{Deguchi, K.} \& \au{Hall, P.}} \yr{2014}  \at{Canonical exact coherent
  structures embedded in high reynolds number flows}.  \jt{Philosophical
  Transactions of the Royal Society of London A: Mathematical, Physical and
  Engineering Sciences}  \bvol{372}~(2020),  \pg{20130352}.

\bibitem[Dubief \& Delcayre(2000)]{dubief2000coherent}
{\sc \au{Dubief, Y.} \& \au{Delcayre, F.}} \yr{2000}  \at{On coherent-vortex
  identification in turbulence}.  \jt{Journal of turbulence}  \bvol{1}~(1),
  \pg{011--011}.

\bibitem[Dubrulle(2019)]{dubrulle2019beyond}
{\sc \au{Dubrulle, B.}} \yr{2019}  \at{Beyond kolmogorov cascades}.
  \jt{Journal of Fluid Mechanics}  \bvol{867}.

\bibitem[Farge \& Pellegrino(2001)]{farge2001coherent}
{\sc \au{Farge, M.} \& \au{Pellegrino, G.and~Schneider, K.}} \yr{2001}
  \at{Coherent vortex extraction in 3d turbulent flows using orthogonal
  wavelets}.  \jt{Physical Review Letters}  \bvol{87}~(5),  \pg{054501}.

\bibitem[Frisch(1995)]{frisch1995turbulence}
{\sc \au{Frisch, U.}} \yr{1995} {\em Turbulence: the legacy of AN
  Kolmogorov\/}.  \publ{Cambridge university press}.

\bibitem[Haller(2005)]{haller2005objective}
{\sc \au{Haller, G.}} \yr{2005}  \at{An objective definition of a vortex}.
  \jt{Journal of fluid mechanics}  \bvol{525},  \pg{1--26}.

\bibitem[Haller(2015)]{haller2015lagrangian}
{\sc \au{Haller, G.}} \yr{2015}  \at{Lagrangian coherent structures}.
  \jt{Annual Review of Fluid Mechanics}  \bvol{47},  \pg{137--162}.

\bibitem[Hussain(1986)]{hussain1986coherent}
{\sc \au{Hussain, A.K.M.F.}} \yr{1986}  \at{Coherent structures and
  turbulence}.  \jt{Journal of Fluid Mechanics}  \bvol{173},  \pg{303--356}.

\bibitem[Jeong \& Hussain(1995)]{jeong1995identification}
{\sc \au{Jeong, J.} \& \au{Hussain, A.K.F.M.}} \yr{1995}  \at{On the
  identification of a vortex}.  \jt{Journal of fluid mechanics}  \bvol{285},
  \pg{69--94}.

\bibitem[Jim{\'e}nez {\em et~al.\/}(1993)Jim{\'e}nez, Wray, Saffman \&
  Rogallo]{jimenez1993structure}
{\sc \au{Jim{\'e}nez, J.}, \au{Wray, A.A.}, \au{Saffman, P.G.} \& \au{Rogallo,
  R.S.}} \yr{1993}  \at{The structure of intense vorticity in isotropic
  turbulence}.  \jt{Journal of Fluid Mechanics}  \bvol{255},  \pg{65--90}.

\bibitem[Li {\em et~al.\/}(2008)Li, Perlman, Wan, Yang, Meneveau, Burns, Chen,
  Szalay \& Eyink]{li2008public}
{\sc \au{Li, Y.}, \au{Perlman, E.}, \au{Wan, M.}, \au{Yang, Y.}, \au{Meneveau,
  C.}, \au{Burns, R.}, \au{Chen, S.}, \au{Szalay, A.} \& \au{Eyink, G.}}
  \yr{2008}  \at{A public turbulence database cluster and applications to study
  lagrangian evolution of velocity increments in turbulence}.  \jt{Journal of
  Turbulence} ~(9),  \pg{N31}.

\bibitem[Lozano-Dur{\'a}n \& Jim{\'e}nez(2014)]{lozano2014time}
{\sc \au{Lozano-Dur{\'a}n, A.} \& \au{Jim{\'e}nez, J.}} \yr{2014}
  \at{Time-resolved evolution of coherent structures in turbulent channels:
  characterization of eddies and cascades}.  \jt{Journal of Fluid Mechanics}
  \bvol{759},  \pg{432--471}.

\bibitem[Moin \& Mahesh(1998)]{moin1998direct}
{\sc \au{Moin, P.} \& \au{Mahesh, K.}} \yr{1998}  \at{Direct numerical
  simulation: a tool in turbulence research}.  \jt{Annual review of fluid
  mechanics}  \bvol{30}~(1),  \pg{539--578}.

\bibitem[Moisy \& Jim{\'e}nez(2004)]{moisy2004geometry}
{\sc \au{Moisy, F.} \& \au{Jim{\'e}nez, J.}} \yr{2004}  \at{Geometry and
  clustering of intense structures in isotropic turbulence}.  \jt{Journal of
  fluid mechanics}  \bvol{513},  \pg{111--133}.

\bibitem[Mukherjee {\em et~al.\/}(2019)Mukherjee, Safdari, Shardt, Kenjeres \&
  Van~den Akker]{mukherjee2019droplet}
{\sc \au{Mukherjee, S.}, \au{Safdari, A.}, \au{Shardt, O.}, \au{Kenjeres, S.}
  \& \au{Van~den Akker, H.E.A.}} \yr{2019}  \at{Droplet-turbulence interactions
  and quasi-equilibrium dynamics in turbulent emulsions}.  \jt{Journal of Fluid
  Mechanics}  \bvol{878},  \pg{221--276}.

\bibitem[Peacock \& Dabiri(2010)]{peacock2010introduction}
{\sc \au{Peacock, T.} \& \au{Dabiri, J.}} \yr{2010} Introduction to focus
  issue: Lagrangian coherent structures.

\bibitem[Peacock \& Haller(2013)]{peacock2013lagrangian}
{\sc \au{Peacock, T.} \& \au{Haller, G.}} \yr{2013}  \at{Lagrangian coherent
  structures: The hidden skeleton of fluid flows}.  \jt{Physics today}
  \bvol{66}~(2),  \pg{41--47}.

\bibitem[Pearson {\em et~al.\/}(2004)Pearson, Yousef, Haugen, Brandenburg \&
  Krogstad]{pearson2004delayed}
{\sc \au{Pearson, B.R.}, \au{Yousef, T.A.}, \au{Haugen, N.E.L},
  \au{Brandenburg, A.} \& \au{Krogstad, P.}} \yr{2004}  \at{Delayed correlation
  between turbulent energy injection and dissipation}.  \jt{Physical Review E}
  \bvol{70}~(5),  \pg{056301}.

\bibitem[Perlekar {\em et~al.\/}(2012)Perlekar, Biferale, Sbragaglia,
  Srivastava \& Toschi]{perlekar2012droplet}
{\sc \au{Perlekar, P.}, \au{Biferale, L.}, \au{Sbragaglia, M.}, \au{Srivastava,
  S.} \& \au{Toschi, F.}} \yr{2012}  \at{Droplet size distribution in
  homogeneous isotropic turbulence}.  \jt{Physics of Fluids}  \bvol{24}~(6),
  \pg{065101}.

\bibitem[Perlman {\em et~al.\/}(2007)Perlman, Burns, Li \&
  Meneveau]{perlman2007data}
{\sc \au{Perlman, E.}, \au{Burns, R.}, \au{Li, Y.} \& \au{Meneveau, C.}}
  \yr{2007} Data exploration of turbulence simulations using a database
  cluster.  \bt{In {\em Proceedings of the 2007 ACM/IEEE conference on
  Supercomputing\/}},  \pg{p.~23}. ACM.

\bibitem[Phillips(1933)]{phillips1933vector}
{\sc \au{Phillips, Henry~Bayard}} \yr{1933} {\em Vector analysis\/}.

\bibitem[Richardson(1922)]{richardson1922weather}
{\sc \au{Richardson, L.F.}} \yr{1922}  \at{Weather prediction by numerical
  process}.  \jt{Cambridge University Press} .

\bibitem[She {\em et~al.\/}(1990)She, Jackson \& Orszag]{she1990intermittent}
{\sc \au{She, Z.S.}, \au{Jackson, E.} \& \au{Orszag, S.A.}} \yr{1990}
  \at{Intermittent vortex structures in homogeneous isotropic turbulence}.
  \jt{Nature}  \bvol{344}~(6263),  \pg{226}.

\bibitem[She {\em et~al.\/}(1991)She, Jackson \& Orszag]{she1991structure}
{\sc \au{She, Z.S.}, \au{Jackson, E.} \& \au{Orszag, S.A.}} \yr{1991}
  \at{Structure and dynamics of homogeneous turbulence: models and
  simulations}.  \jt{Proceedings of the Royal Society of London. Series A:
  Mathematical and Physical Sciences}  \bvol{434}~(1890),  \pg{101--124}.

\bibitem[Sirovich(1987)]{sirovich1987turbulence}
{\sc \au{Sirovich, L.}} \yr{1987}  \at{Turbulence and the dynamics of coherent
  structures. i. coherent structures}.  \jt{Quarterly of applied mathematics}
  \bvol{45}~(3),  \pg{561--571}.

\bibitem[Ten~Cate {\em et~al.\/}(2004)Ten~Cate, Derksen, Portela \& Van
  Den~Akker]{ten2004fully}
{\sc \au{Ten~Cate, A.}, \au{Derksen, J.J.}, \au{Portela, L.M.} \& \au{Van
  Den~Akker, H.E.A.}} \yr{2004}  \at{Fully resolved simulations of colliding
  monodisperse spheres in forced isotropic turbulence}.  \jt{Journal of Fluid
  Mechanics}  \bvol{519},  \pg{233--271}.

\bibitem[Ten~Cate {\em et~al.\/}(2006)Ten~Cate, Van~Vliet, Derksen \& Van~den
  Akker]{ten2006application}
{\sc \au{Ten~Cate, A.}, \au{Van~Vliet, E.}, \au{Derksen, J.J.} \& \au{Van~den
  Akker, H.E.A.}} \yr{2006}  \at{Application of spectral forcing in
  lattice-boltzmann simulations of homogeneous turbulence}.  \jt{Computers \&
  fluids}  \bvol{35}~(10),  \pg{1239--1251}.

\bibitem[Tsinober(2014)]{tsinober2014essence}
{\sc \au{Tsinober, A.}} \yr{2014} {\em The essence of turbulence as a physical
  phenomenon\/}, ,  \vol{vol.~10}.  \publ{Springer}.

\bibitem[Waleffe(1992)]{waleffe1992nature}
{\sc \au{Waleffe, F.}} \yr{1992}  \at{The nature of triad interactions in
  homogeneous turbulence}.  \jt{Physics of Fluids A: Fluid Dynamics}
  \bvol{4}~(2),  \pg{350--363}.

\bibitem[Waleffe(1997)]{waleffe1997self}
{\sc \au{Waleffe, F.}} \yr{1997}  \at{On a self-sustaining process in shear
  flows}.  \jt{Physics of Fluids}  \bvol{9}~(4),  \pg{883--900}.

\bibitem[Waleffe(2001)]{waleffe2001exact}
{\sc \au{Waleffe, F.}} \yr{2001}  \at{Exact coherent structures in channel
  flow}.  \jt{Journal of Fluid Mechanics}  \bvol{435},  \pg{93--102}.

\bibitem[Wu {\em et~al.\/}(2007)Wu, Ma \& Zhou]{wu2007vorticity}
{\sc \au{Wu, J.Z.}, \au{Ma, H.Y.} \& \au{Zhou, M.D.}} \yr{2007} {\em Vorticity
  and vortex dynamics\/}.  \publ{Springer Science \& Business Media}.

\end{thebibliography}

\end{document}